\documentclass{elsart}

\newcommand{\vv}[1]{\mbox{\boldmath{$#1$}}}

\newcommand{\vt}[1]{\mbox{\tiny\boldmath{$#1$}}}
 \usepackage{graphics}
 \usepackage{graphicx}
 \usepackage{epsfig}
\usepackage{subfigure}
\usepackage{amssymb}
\begin{document}
\ifx\href\undefined\else\hypersetup{linktocpage=true}\fi 
\begin{frontmatter}
%
%
  \title{Random Network Models and Quantum Phase Transitions in Two
    Dimensions\thanksref{label5}}
\thanks[label5]{dedicated to Klaus von Klitzing on the occasion of the
    25th anniversary of the Quantum Hall Effect}


\author[label1]{B. Kramer}
\footnote{permanent address: 
International University Bremen, Campusring 1, 28759
  Bremen, Germany}
\author[label2]{T. Ohtsuki}
\author[label1]{S. Kettemann}
\address[label1]{I. Institut f\"ur Physik, Universit\"at Hamburg,
  Jungiusstra\ss{}e 9, 20355 Hamburg, Germany} 
\address[label2]{Department of
  Physics, Sophia University, Kioi-cho 7-1, Chiyoda-ku, Tokyo, Japan}

\begin{abstract}
  An overview of the random network model invented by Chalker and Coddington,
  and its generalizations, is provided.  After a short introduction into the
  physics of the Integer Quantum Hall Effect, which historically has been the
  motivation for introducing the network model, the percolation model for
  electrons in spatial dimension 2 in a strong perpendicular magnetic field
  and a spatially correlated random potential is described. Based on this, the
  network model is established, using the concepts of percolating probability
  amplitude and tunneling. Its localization properties and its behavior at the
  critical point are discussed including a short survey on the statistics of
  energy levels and wave function amplitudes. Magneto-transport is reviewed
  with emphasis on some new results on conductance distributions.
  Generalizations are performed by establishing equivalent Hamiltonians. In
  particular, the significance of mappings to the Dirac model and the two
  dimensional Ising model are discussed. A description of renormalization
  group treatments is given. The classification of two dimensional random
  systems according to their symmetries is outlined. This provides access to
  the complete set of quantum phase transitions like the thermal Hall
  transition and the spin quantum Hall transition in two dimension. The
  supersymmetric effective field theory for the critical properties of network
  models is formulated. The network model is extended to higher dimensions
  including remarks on the chiral metal phase at the surface of a multi-layer
  quantum Hall system.
\end{abstract}
\end{frontmatter}

\newpage
\tableofcontents

\newpage
\section{Introduction}
\label{sec:introduction}

At the end of the seventies of the past century, many thought that solid state
physics had matured to such a degree that practically everything important had
been discovered. Superconductivity seemed to be well understood with the
transition temperatures ceasing to increase further. Semiconductor physics had
developed almost into an engineering discipline. No significant further
progress of the field of solid state physics was predicted for the foreseeable
future. It was widely believed in the community that the development more or
less had come to an end. In this situation, a completely new and by no one
foreseen phenomenon was discovered in the magneto-transport properties of a
commercial electronic device, the Silicon MOSFET. The {\em Integer Quantum
  Hall Effect} was found.

This experimental discovery, together with several others that came roughly at
the same time, opened a completely new area of solid state research and during
the forthcoming years initiated a truly novel view on the field of condensed
matter. This concerns the quantum mechanical properties of disordered and
interacting electronic solid state systems on mesoscopic scales of which the
Quantum Hall Effect is only one example, though a very prominent one. Until
today, researchers in the field of the Quantum Hall Effect continue to produce
new surprises, perhaps not on daily, but certainly on monthly time scales. In
many cases, these concern only at first glance the smaller area of the
quantum Hall phenomenon. Often, as in the case of the fractional Quantum Hall
Effect, the discovered phenomena later turn out to be of much wider importance
than foreseen at the time of their discovery.

In this review article\footnote{In contrast to the short overview that has
  been published earlier [B. Kramer, S. Kettemann, T. Ohtsuki, Physica E {\bf
    20}, 172 (2003); cond-mat/0309115], the present article is supposed to
  provide a much more self-contained --- as much as it is possible ---
  overview of the derivation of the network model and its quantum mechanical
  properties, amended by some recent results. Much emphasis is on
  generalizations of the model and many connections to other models and
  descriptions of quantum phase transitions, especially in two dimensions.} we
want to describe one example of such a development, namely the discovery and
further development of a theoretical model which originally was designed to
describe a special aspect of the integer Quantum Hall Effect, namely the
localization-delocalization quantum phase transition in a Landau band in the
limit of long-range randomness. Later, the model --- although at first glance
very specialized and restricted --- has been shown to be able to account for
many phenomena in a much wider class of systems, namely the disorder-induced
quantum phase transitions in seemingly {\em all} of the different presently
considered disordered systems. The model, invented by Chalker and Coddington
in 1988, describes a {\em random network} of currents.

In the following sections of this introductory chapter, we first provide
shortly some insight into, and understanding of the integer Quantum Hall
Effect as it has been originally detected. Our considerations will be based on
the so-called localization model. This is used for describing some fundamental
physical aspects of the Integer Quantum Hall Effect, the interplay between a
high --- so-called {\em quantizing} --- magnetic field, and the disorder in
the system due to impurities. This interplay can lead to a
localization-delocalization transition in a two dimensional electronic system.
To the best of our present knowledge, this appears to be a paradigm of a
genuine {\em quantum phase transition}. Together with the gauge argument first
proposed by R. B. Laughlin, this can explain not only the very existence of
plateaus in the Hall conductivity as a function of the electron density, and
the simultaneous vanishing of the magneto-conductivity, but also the
quantization in integer units of $e^{2}/h$. The Chalker-Coddington network
model is now widely accepted as one possibility for describing the fundamental
physics behind the quantum Hall phase transition. While the latter is
nevertheless still waiting for a complete and quantitative theoretical
description with predictive power, especially including the precision aspect,
the network model of Chalker and Coddington seems to have acquired more
fundamental importance also in other fields of solid state physics like
superconductivity and magnetism, as will be seen below.

\subsection{The Discovery of the Quantum Hall Effect}
\label{sec:qhe}

\begin{figure}[htbp]
\begin{center}
{\includegraphics[width=0.8\linewidth]{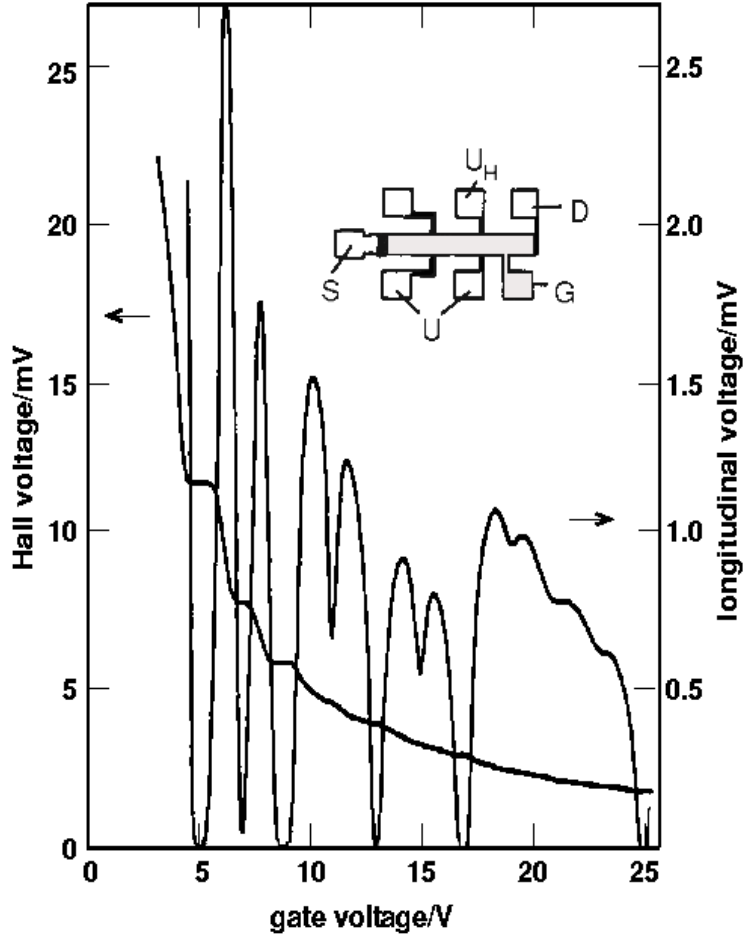}}
 \caption{The Integer Quantum Hall Effect \protect\cite{kdp80}. 
    The Hall voltage $U_{\rm H}$ (left axis) shows plateaus in regions of the
    gate voltage $U_{\rm g}$ where the source drain voltage $U$ (right axis)
    is vanishingly small. The experiment was done on a Si-MOSFET of length
    400\,$\mu$m and width 50\,$\mu$m at $B=18$\,T and at a temperature
    $T=1.5$\,K.   
\label{fig:1}}
\end{center}
\end{figure}

The Quantum Hall Effect has been discovered by Klaus von Klitzing in 1980 when
working as a guest researcher at the {\em High Magnetic Field Laboratory} of
the Max-Planck-Gesellschaft in Grenoble \cite{kdp80} (Fig.~\ref{fig:1}). He
was investigating the electronic transport in a Silicon MOSFET subject to a
high magnetic field of about 18\,T flux density at temperatures near 1\,K. To
his great surprise, he found that the Hall resistance $R_{\rm H}$ --- the
ratio between the Hall voltage across the two dimensional electron
inversion layer in the transistor, and the source-drain current, ($I=1\mu$\,A)
--- shows extremely well-defined plateaus when changing the gate voltage
$U_{\rm g}$. In the gate voltage regions of these Hall plateaus, he found the
longitudinal magneto-resistance to be vanishingly small. Most strikingly, he
was able to identify the values of the Hall resistances of the plateaus as
integer fractions of $h/e^{2}$
\begin{equation}
  \label{eq:hallplateaus}
  R_{\rm H}=\frac{1}{j}\,\frac{h}{e^{2}} \qquad (j=1,2,3,\ldots).
\end{equation}
Here, $h$ is the Planck constant and $e$ the elementary charge. 

Klaus von Klitzing found that the relative uncertainty of the plateau values
was much better than $10^{-5}$ in the very first experiments. Thus, as a
surprise, despite of the presence of strong disorder and electron interactions
in the MOSFET, the effect appeared to be very promising for measuring the {\em
  Sommerfeld constant} $\alpha=(\mu_{0}c_{0}/2)(e^{2}/h)$
($c_{0}=299\,792\,458$\,ms$^{-1}$ speed of light in vacuum, $\mu_{0}=4\pi\cdot
10^{-7}$\,NA$^{-2}$ permeability of vacuum) independently from optics. Also,
von Klitzing immediately realized the importance of his finding for
metrological applications, i.e. the realization, dissemination and maintenance
of the ''Ohm'', the unit of the electrical resistance.  Eventually, this led
to the re-definition of the ''Ohm'' in terms of the {\em von Klitzing
  constant} $R_{\rm K}=25\,812.8085\,\Omega$ in 1990. Meanwhile, the
experimental reproducibility of the quantized values of the Hall resistance
has been improved to values better than 10$^{-9}$. This is of special
importance in view of the metrological applications \cite{bk92}.
Some excellent reviews about the Integer Quantum Hall Effect can be found in
the literature \cite{a87,pg90,jvfh94,kvk96}.

The discovery of the von Klitzing effect stimulated intense experimental and
theoretical research in many international laboratories. A particularly
important discovery was made only two years later by the group of Daniel C.
Tsui, Horst Stormer and Arthur C. Gossard at Bell Laboratories \cite{tsg82}.
In this experiment, a GaAs/AlGaAs-heterostructure which contained the dopant
impurities only far from the electron inversion layer was used. As a
consequence, these samples had a very high electron mobility in the inversion
layer at the interface of a few hundred thousand cm$^{2}$/Vs. The researchers
detected quantization of the Hall conductivity at fractional multiples of
$e^{2}/h$ at temperatures of a few 0.1\,K.

By improving the sample fabrication technology, more and more such additional
fractional features were uncovered. Hitherto, electron mobilities of more than
10,000,000\,cm$^{2}/\mathrm{V}\mathrm{s}$ have been achieved and several
dozens of the fractionally quantized features in the Hall resistance (and
their counterparts in the magneto-resistance) are known now \cite{stg99}.
They have been associated with hierarchies of novel correlated electron states
induced by the electron-electron interaction \cite{l83,cp95,sp97}. The
existence of these states goes far beyond the traditional phenomenological
Fermi liquid picture for electrons in metallic systems. Thus, the {\em
  Fractional Quantum Hall Effect} has opened a new field in the physics of
{\em correlated} electrons, but is not the subject of this review article.

\subsection{The Localization Model}
\label{sec:models}

According to the theory of the classical Hall effect the Hall resistivity of
independent particles with the charge $e$ is a monotonic function of the
magnetic flux density $B$ and the number density $n_{e}$ of the charges. This
can be easily seen by using the Drude friction model for diffusion of charges
in external crossed electric and magnetic fields \cite{detal98}. It yields for
the components of the conductivity tensor
\begin{equation}
  \label{eq:drudesigma}
  \sigma_{xx}=\frac{\sigma_{0}}{1+(\omega_{B}\tau)^{2}}\,, 
\qquad \sigma_{yx}=\omega_{B}\tau\sigma_{xx}\,.
\end{equation}
Here, $\sigma_{0}=e^{2}n_{e}\tau/m^{*}$ is the Drude conductivity. The mean
free time $\tau$ of the particles due to scattering is assumed to contain all
microscopic processes. The effective mass $m^{*}$ contains the effect of the
lattice of atoms and interactions. The quantity $\omega_{B}=eB/m^{*}$ is the
cyclotron frequency. The resistance components are obtained by inverting the
conductivity tensor
\begin{equation}
  \label{eq:druderesistance}
  \rho_{xx}=\frac{\sigma_{xx}}{\sigma_{xx}^{2}+\sigma_{yx}^{2}}\,,
\qquad
  \rho_{xy}=\frac{\sigma_{yx}}{\sigma_{xx}^{2}+\sigma_{yx}^{2}}\,.
\end{equation}
For strong magnetic field ($B\to\infty$) $\sigma_{xx}\propto
B^{-2}\tau^{-1}\to 0$.  Correspondingly, since $\sigma_{yx}\neq 0$, the
magneto-resistivity $\rho_{xx}\propto \sigma_{xx}\to 0$. The Hall resistivity
is
\begin{equation}
  \label{eq:druderesistivity}
  \rho_{xy}:=\rho_{\rm H}=\frac{1}{\sigma_{yx}}=\frac{B}{en_{e}}\,.
\end{equation}
As the von Klitzing experiment contradicts this result, it is clear that the
quantum nature of the two dimensional electron system in the MOSFET subject to
the high magnetic field must be the reason for the quantization of the Hall
resistivity at sufficiently low temperatures.

For the Landau model for a single spinless electron in two dimensions in a
perpendicular homogeneous magnetic field $\vv{B}=\nabla\times \vv{A}$, the
vector potential given in the Landau gauge $\vv{A}=(0,Bx,0)$, the
Schr\"odinger equation is
\begin{equation}
  \label{eq:landaumodel}
H_{0}\psi_{nk}(x,y):=
\frac{1}{2m^{*}}\left(\frac{\hbar}{i}\nabla + e\vv{A}\right)^{2}
\psi_{nk}(x,y)=E_{nk}\psi_{nk}(x,y)\,.
\end{equation}
It is solved by the Landau states
\begin{equation}
  \label{eq:landaustates}
\langle x,y|nk\rangle:=
\psi_{nk}(x,y)=\frac{1}{L^{1/2}}\,e^{iky}\,\phi_{n}(x-X_{k})   
\end{equation}
where $X_{k}=-k\ell_{B}^{2}$ and $\ell_{B}:= \sqrt{\hbar/eB}$ the magnetic
length. The eigenvalues are the Landau levels
\begin{equation}
  \label{eq:landaulevels}
  E_{nk}=\hbar\omega_{B}\left(n+\frac{1}{2}\right)\,.
\end{equation}
They are associated with the normalized eigenfunctions
\begin{equation}
  \label{eq:hermite}
\phi_{n}(x)=\frac{1}{\sqrt{\pi^{1/2}\ell_{B}n!}}\,\exp{\left(\frac{-x^{2}}
{2\ell_{B}^{2}}\right)}\,
H_{n}\left(\frac{x}{\ell_{B}}\right)  
\end{equation}
with the Hermite polynomials $H_{n}$ ($n=0,1,2,3,\ldots$). The Landau states
are degenerate with respect to the wave number $k$ with a degree per unit area
of 
\begin{equation}
  \label{eq:degeneracy}
  n_{B}=\frac{eB}{h}=\frac{B}{\Phi_{0}}\,,
\end{equation}
the density of flux quanta, $\Phi_{0}=h/e$, in the system. This degeneracy can
be easily understood by considering the maximum possible number of Landau wave
functions (Eq.~(\ref{eq:landaustates})) that can be contained in a system of
the size $L^{2}$ \cite{detal98},
\begin{equation}
  \label{eq:maxnumber}
  X_{k}^{\rm max}=\frac{2\pi \ell_{B}^{2}}{L}n_{B}=L\,.
\end{equation}

An important quantity is the {\em filling factor} which is defined by the
ratio of the electron number density and the density of flux quanta,
\begin{equation}
  \label{eq:filling}
  \nu_{B}=\frac{n_{e}}{n_{B}}\,.
\end{equation}
Integer filling factors $\nu_{B}=j$ correspond then to $j$ completely filled
Landau levels. At the corresponding electron densities, the Hall resistance is
an integer fraction of $h/e^{2}$ (cf. Eq.~(\ref{eq:druderesistivity})).

However, this does {\em not} explain the existence of the wide plateaus in
$\rho_{xy}(n_{e})$ since upon increasing $n_{e}$ the Fermi level jumps between
the Landau levels. In order to generate the observed wide plateaus, it is
necessary to keep the Fermi level continuously varying {\em between} the
Landau levels when changing $n_{e}$, but {\em without} changing the
resistivity values of the plateaus.

\begin{figure}
\begin{center}
{\includegraphics[width=0.8\linewidth]{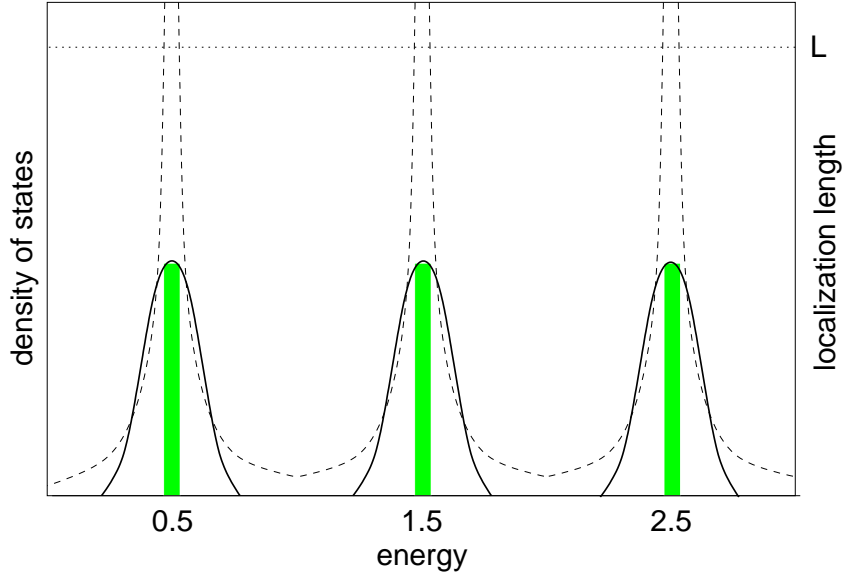}}
\end{center}
\caption{Schematic picture of the density of states 
  (full line, arb. units) and the localization lengths (dashed line, arb.
  units) as a function of the energy (in units of the cyclotron energy) in the
  localization model. When introducing a static random potential into the
  Landau model, Landau levels are broadened into bands consisting of
  ''incompressible'' spectral regions that correspond to localized states in
  the band tails, and ''compressible'' regions of effectively delocalized
  states (shaded) near the band centers. In the latter regions, the
  localization length exceeds the system size $L$. At finite temperature, the
  system size is replaced by the temperature dependent phase coherence length
  $L_{\phi}$.}
  \label{fig:2}
\end{figure}

A mechanism that can account for this pinning without affecting transport is
localization of the wave functions due to disorder induced by the presence of
the impurities in the system \cite{anderson58,aa81,ta81}. By introducing
randomness into the Landau model of Eq.~(\ref{eq:landaumodel}),
\begin{equation}
  \label{eq:randomness}
  H=H_{0}+V(\vv{r})
\end{equation}
the degeneracy of the Landau levels is removed.  Here, $V(\vv{r})$ is a
randomly varying potential which is usually defined via its statistical
properties. The Landau levels are broadened into Landau bands by the disorder,
with localized eigenstates occurring in the band tails. These localized states
can pin the Fermi level at the corresponding eigenenergies. A qualitative
picture of the density of states of such a model Hamiltonian is shown in
Fig.~\ref{fig:2}.

Localized states correspond to random wave functions with envelopes that are
exponentially decaying at large distances from some localization center
\cite{lr85,km93}. This implies exponentially decaying correlation functions
which indicates that the localized states cannot contribute to dc-transport at
zero temperature. This means that the magneto-conductivity, and
correspondingly the magneto-resistivity, vanish for filling factors
corresponding to the energy regions of the localized states, while the Hall
conductivity, and correspondingly the Hall resistivity, stay at a constant
value.

On the other hand, non-localized --- extended --- states near the band
centers, which do not decay exponentially and are spread randomly across the
entire system, can account for electron transport. For filling factors close
to half integer numbers, which correspond to the energy regions near the
centers of the Landau bands, one expects peaks in the magneto-resistivity with
widths that reflect the widths of the energy regions of the extended states.
Simultaneously, the Hall resistivity is expected to change from one plateau
value to the next. Often, these spectral regions are denoted as {\em
  compressible} since the Fermi energy hardly changes with increasing electron
number while the localized states are associated with {\em incompressible
  spectral regions} where the Fermi level strongly changes when varying the
electron number \cite{m96}.

As long as the Fermi level stays within the region of localized states, the
zero temperature conductivity components are not changed. Assuming that the
extended states provide the correctly quantized values of the Hall
conductivity, and simultaneously a vanishing magneto-conductivity, one obtains
the experimentally observed behavior of the conductivity. However, {\em that}
the Hall conductance remains correctly quantized {\em exactly} at integer
multiples of $e^{2}/h$ in the regions of localization, needs closer
consideration. Laughlin's gauge argument, which we will shortly discuss below,
is supposed to provide an explanation.
  
All of the analytical and numerical results obtained until recently are
consistent with the picture that at the absolute zero of temperature the
localization length diverges only at specific energies close to the centers of
the Landau bands with a universal exponent $\nu$ which is independent of the
specific form of the randomness and the band index \cite{hk90,h95},
\begin{equation}
  \label{eq:000}
  \xi(E-E_{n})=\frac{\xi_{n}}{|E-E_{n}|^{\nu}}
\end{equation}
where the constant $\xi_{n}$ depends on microscopic details of the randomness
and on the Landau band index $n$. It will be one of the main tasks of the
following sections to explain this critical behavior in some detail. 

Table \ref{critexp} contains a representative selection of the exponents
determined numerically. If the exponent were universal the quantum Hall
effect could be considered as a paradigm of a genuine universal quantum phase
transition.
\begin{table}[htbp]
\begin{center}
\vspace{5mm}
\begin{tabular}{p{1.7cm}p{3.9cm}p{4.6cm}p{1.7cm}}\hline\hline
  $\nu$  & model & method&reference\\\hline
  $\infty$        &short-range impurities&self-consistent perturbation&
\protect\cite{o82}\\
  $\approx 2$     &Peierls tight binding&transfer matrix scaling&
\protect\cite{skm84}\\
  $\approx 2.0$   &short-range impurities&recursive Green function&
\protect\cite{aa85,aa85a}\\
  $2.35(3)$  &random Landau matrix&recursive Green function&
\protect\cite{hk90,h95}\\
  $2.3(1)$   &random Landau matrix&recursive Green function&
\protect\cite{m90}\\
  $2.4(2)$     &random Landau matrix&recursive Green function&
\protect\cite{h92}\\
  $2.4(1)$     &finite range impurities&Chern number scaling&
\protect\cite{hb92}\\
  $\approx 2.3$   &spin orbit scattering&Thouless number scaling&
\protect\cite{hamg95}\\
  $\approx 2$     &double layer system&Thouless number scaling&
\protect\cite{sm96}\\
  $\approx 2$     &random matrix model&scaling of level statistics&
\protect\cite{ook96}\\
  $2.5(5)$    &Chalker-Coddington&transfer matrix scaling&
\protect\cite{cc88}\\
  $2.4(2)$  &random saddle points&transfer matrix scaling&
\protect\cite{lwk93}\\
  $2.5(5)$    &Chalker-Coddington type&real space renormalization&
\protect\cite{gr97}\\
  $2.39(1)$   &Chalker-Coddington type&real space renormalization&
\protect\cite{crsr01,crr03,cr03}\\
$2.5(4)$     &super spin chain&density matrix renormalization&
obtained from \protect\cite{kondev}\\
$2.33(3)$&counter-propagating chiral Fermions&Monte Carlo&
\protect\cite{Ziqiang}\\
\hline
\end{tabular}
\end{center}
\vspace{5mm}
\caption{Critical exponents $\nu$ of the quantum Hall phase transition 
in the lowest Landau band obtained by various theoretical methods 
(numbers in parentheses denote the uncertainty in the last digit 
of the exponent).}
\label{critexp}
\vspace{12mm}
\end{table}

An important issue is how to detect this critical exponent in an experiment.
As in an experiment the size of the two dimensional electron system is finite,
it is intuitively clear that the singularities of the localization length near
the band centers cannot be resolved. Furthermore, at non-zero temperature,
inelastic processes will lead to phase breaking scatterings between different
eigenstates \cite{lr85,km93}. The latter can be described by a mean,
temperature dependent phase breaking time $\tau_{\phi}(T)$ which can be
assumed to increase with decreasing temperature according to
\begin{equation}
  \label{eq:coherencetime}
  \tau_{\phi}(T)\propto \frac{1}{T^{p}}
\end{equation}
with an exponent $p$ of order 1. During the time interval $\tau_{\phi}$, the
electron can be considered as moving diffusively under the influence of the
impurity scattering. This suggests to define a phase coherence length
\begin{equation}
  \label{eq:coherencelength}
  L_{\phi}(T)=\sqrt{D\tau_{\phi}(T)}
\end{equation}
with a disorder-induced diffusion constant $D$ that is related to the
dc-conduc\-ti\-vi\-ty via the Einstein relation $\sigma=e^{2} D(E_{\rm
  F})\rho(E_{\rm F})$ where $\rho(E_{\rm F})$ is the density of states at the
Fermi level.

In order to connect the singular behavior of the localization length with the
results of a transport experiment it is important to note that a localized
state appears extended if its localization length does exceed the system size
$L$ or, at non-zero temperature, the above phase coherence length $L_{\phi}$
(Fig.~\ref{fig:2}). This implies that the widths of the peaks in the
magneto-conductivity at the centers of the Landau bands are given by the
condition $\xi(\Delta E)={\rm min}\{L,L_{\phi}(T)\}$. The resulting
characteristic temperature behaviors \cite{hk90,h95} have been found to
be consistent with experimental data
\cite{wetal88,khk91,eetal93,hetal91,ketal00,hetal02}.

\begin{figure}[htpb]
\begin{center}
{\includegraphics[width=0.6\linewidth]{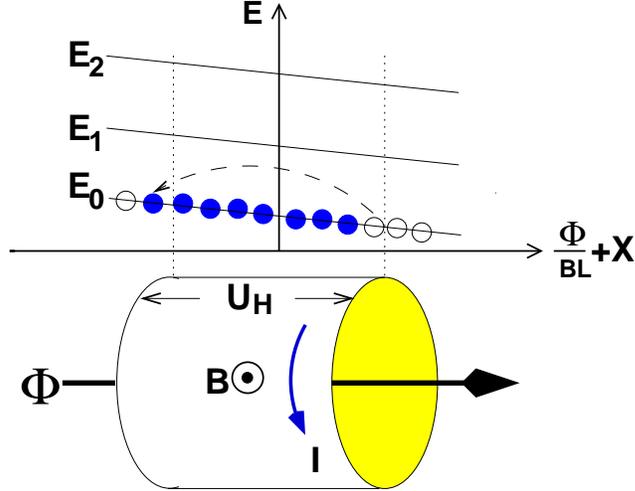}}
\end{center}
\caption{Illustrating Laughlin's gauge argument: introducing a magnetic flux
  $\Phi$ along the axis of the cylinder corresponds to a shift of the wave
  number in azimuthal direction by $e\Phi/\hbar L$. A change of the gauge flux
  by $\Delta \Phi=\Phi_{0}$ corresponds to a shift of exactly $2\pi/L$. This
  leads to a corresponding shift in $X_{k}$ that is equivalent to transfering
  one electron per Landau band from one edge of the cylinder to the other that
  corresponds to a change in energy of $eU_{\rm H}$.}
  \label{gaugefig:3}
\end{figure}

One still needs an argument as to why the Hall conductance is exactly
quantized at integer multiples of $e^{2}/h$ in spite of the presence of the
impurity potential. Such an argument has been pioneered by Laughlin
\cite{l81}. The idea is to relate the current in an ideally metallic cylinder,
subject to a homogeneous magnetic field perpendicular to the cylinder's
surface, to the change of the total electronic energy $E(\Phi)$ when
adiabatically changing a magnetic flux piercing the cylinder along its axis
(Fig.~\ref{gaugefig:3}),
\begin{equation}
  \label{eq:gaugeargument}
  I=\frac{\Delta E}{\Delta \Phi}\,.
\end{equation}
Introducing a gauge flux $\Phi$ along the axis of the cylinder corresponds to
a shift of the wave number in azimuthal direction by $e\Phi/\hbar L$. This can
also be considered as a change in the boundary condition.

A change of the flux by $\Delta \Phi=\Phi_{0}$ corresponds to a shift of
exactly $2\pi/L$ and leads to a corresponding shift in $X_{k}$ (cf.
Eq.~(\ref{eq:landaustates})) that is equivalent to transfering one electron
per Landau band from one edge of the cylinder to the other.\footnote{Strictly
  speaking, for a two dimensional system with periodic boundary conditions in
  one direction and fixed boundary conditions in the perpendicular direction,
  one finds a non-degenerate energy spectrum that differs from the Landau
  spectrum by the presence of edge states \protect\cite{detal98,h82}. However,
  for the present qualitative argument, this is not important.}  When
occupying $j$ Landau bands, a transfer of $j$ electrons is associated with one
flux quantum. Due to the presence of a Hall voltage $U_{\rm H}$ across the
cylinder, the Landau levels are no longer degenerate, but on a straight line
with a slope given by the Hall electric field $U_{\rm H}/L$
(Fig.~\ref{fig:3}). The energy change caused by the flux change is then
$\Delta E=jeU_{\rm H}$, and the Hall current
\begin{equation}
  \label{eq:hallcurrent}
 I=j\,\frac{e^{2}}{h}U_{\rm H}\,, 
\end{equation}
consistent with integer quantization of the Hall conductance.

As presented, the argument seems to hold only for the ideal Landau Hamiltonian
applied to a system of finite width. However, it can also be used in the
presence of disorder \cite{skm84}. To understand this, one has to have in mind
that on the one hand, by definition, localized states are insensitive to
changes in the boundary conditions or, equivalently, flux changes. On the
other hand, the flux sensitivity of the extended states is enhanced, in order
to compensate for the localized states
\cite{a87,pg90,jvfh94,p81,b83,c83,jp84}. This compensation is assured to be
exact such that the final result for the current is the same as in
Eq.~(\ref{eq:hallcurrent}).

Despite there exist several experimental features apparently related to
electron interaction
\cite{lg83,getal85,eetal85,hzc86,metal86,as92,petal96,sak97,cbf99}, it seems
that the integer quantization of the Hall conductance can be understood within
the localization model without taking into account the correlations between
the electrons \cite{jvfh94}.

On the other hand, for explaining the fractionally quantized Hall effect,
interaction and correlation effects are generally accepted to constitute the
necessary ingredients for generating the {\em excitation gaps} between the
many-electron ground states associated with rational filling factors.
Nevertheless, in order to generate the fractional {\em plateaus} in
the Hall conductivity, disorder is believed to be essential in pinning the
chemical potential, in close analogy to the Integer Quantum Hall Effect.

\subsection{Plan of the Article}

In summary, the Quantum Hall Effect --- consistent with all of the presently
available experiments --- seems to be an exact and very fundamental
phenomenon. One needs for its physical understanding, and in particular for a
theory with predictive power, not only interactions and correlations in a two
dimensional many-body system but in particular, and somewhat
counter-intuitively --- in view of the precision aspect --- {\rm the presence
  of a certain amount of disorder}. It is one of the main purposes of this
article to review our present understanding, why, in spite of and due to the
disorder, the Quantum Hall Effect is the most exact phenomenon known in low
dimensional electron systems. We will restrict ourselves to the {\em Integer}
Quantum Hall Effect, ignoring interactions completely.

Especially, we will discuss the localization concept by introducing the
so-called random network model originally invented by Chalker and Coddington
\cite{cc88} in order to describe the localization of quantum states in a
random potential with long-range spatial correlations in the presence of a
strong, ''quantizing'' magnetic field. Later, we will discuss several
generalizations of the network model which underline the general importance of
the phenomenon.

We will proceed in the following chapters by pursuing a strategy which we hope
will enable non-specialists in the field to understand the main ideas in a
self-contained way. We will always start from elementary accessible facts that
form the background of more complicated relations before we eventually
describe to some extent the issues that are more at the frontier of present
research. As the field has been, and still is, extremely rapidly growing, we
cannot guarantee that all of the available results have been included,
especially the most recent ones. We have attempted our best to include at
least references to all of the important results. We apologize in advance
to all colleagues, whose research efforts might not be sufficiently
highlighted according to their importance or even not be included at all.

The plan of the paper is as follows. In the next chapter, the so-called
percolation model for an electron in a spatially slowly varying random
potential landscape is introduced in some detail. It can be viewed as
the backbone of the random network model. If it is supplemented by quantum
effects, it can be considered as a precursor of the random network model of
Chalker and Coddington for describing localization in the presence of a
quantizing magnetic field. 

The latter will be introduced in chapter \ref{sec:randomnetwork}. In chapter
\ref{sec:chalkercoddington} the localization properties of the model will be
discussed in some detail. Strangely enough, technical difficulties seem to
have prevented a high-precision determination of the critical exponent up to
now for the original Chalker-Coddington random network model. Chapter
\ref{sec:transport} contains some results for the transport properties,
especially near the quantum critical point.

In the chapter \ref{sec:hierarchy}, we investigate the renormalization group
approach for describing the critical properties in some detail. Especially, we
will emphasize here that one can use this as a starting point for the
definition of a truncated Chalker-Coddington model which can be treated
numerically exactly and allows for determining the critical exponent with high
precision.

Chapters \ref{sec:hamiltonians}, \ref{sec:ising}, \ref{sec:symmetries} and
\ref{sec:susy} contain extensions to a number of equivalent Hamiltonians, to
systems with other symmetries, and to an equivalent field theoretical
formulation of the random network model of Chalker and Coddington,
respectively. In these chapters, we have attempted to provide some flavor of
the power and importance of the model in fields other than the quantum Hall
phase transition.  While we will review much numerical and experimental
evidences for the universality of the quantum Hall transition, any attempts to
classify it, like other two dimensional phase transitions, based on the
conformal invariance at the critical point have failed so far.

In chapter \ref{sec:susy} we review a very recent suggestion, namely that the
quantum Hall critical point may belong to a new class of critical points being
described by a supersymmetric conformal field theory
\cite{zirnconform,tsvelikconform}.  This is based on the fact that an
anisotropic version of the Chalker-Coddington model as well as the random
Landau model at the critical point can both be mapped on the Hamiltonian of a
chain of antiferromagnetic superspin chains. We review its derivation, and
the progress which has been achieved towards the characterization of the
quantum Hall transition that way, although an analytical calculation of its
critical exponents is still missing.

Generalizations to several layers and higher dimensions are briefly discussed
in the chapter \ref{sec:higherdimensions} before we conclude by summarizing
the status of the field and comment on possible future developments.
 
\section{The Percolation Model}
\label{sec:percolationmodel}

We start by describing the physics of the so-called percolation model for the
Quantum Hall Effect. In this model, which is valid in the limit of a {\em very
  high magnetic field}, the quantum mechanical wave functions are assumed to
percolate along the equipotential lines of a slowly varying random
potential landscape, in analogy to a classical percolating fluid in a random
system \cite{sa94}. Using this picture of a percolating wave function, one can
understand why localized wave functions may exist, and, in particular, why
there may be isolated critical points in the energy spectrum of the
Hamiltonian where the localization length diverges. This high-field limit is
often denoted as the classical percolation limit, although what is
percolating is probability amplitude and not a classical fluid. In the
quantum mechanical version of the model, tunneling of probability amplitude
between the equipotential lines is taken into account whenever they get close
to each other in space. In this version, the necessary competing ingredients
--- quantum tunneling and interference --- are included such that a generic
universal quantum mechanical localization-delocalization transition can be
described.

The percolation model provides the physical background of the
Chalker-Cod\-ding\-ton network model --- a generic model which is assumed to
describe the {\em universal} quantum mechanical properties of non-interacting
electrons in two dimensions in the presence of a random potential subject to a
strong perpendicular magnetic field.

\subsection{Wave Functions in a Spatially Correlated Random Potential 
in a Strong Magnetic Field}
\label{sec:correlatedrandomness}

The origin of the random potential in the plane of the inversion layer
$V(x,y)$ of the electrons in a GaAs/AlGaAs heterostructure or a MOSFET
transistor is the impurities in the semiconductor material, especially as a
consequence of the doping. These impurities are distributed randomly at some
distance from the inversion layer. Thus, only the Coulomb tails of their
potentials do influence the dynamics of the charges. This implies that the
distribution of the potential energy $V(x,y)$ is long-range correlated in
space. For convenience, we assume in the following that the spatial average of
the random potential vanishes. This can always be achieved by a suitable
choice of the zero of energy. For the correlator we assume
\begin{equation}
  \label{eq:correlationfunction}
  \overline{V(x',y')V(x,y)}:= W^{2}C(x'-x,y'-y)\,,
\end{equation}
with the correlation function $C(x,y)$ having an exponential or Gaussian decay
length $\ell_{\rm c}$ much longer than the magnetic length $\ell_{B}$. It is
also assumed that the probability distribution of the potential, $P[V(x,y)]$ is
symmetric and homogeneous, i.e. independent of the origin of coordinate
system.

\begin{figure}
\begin{center}
\subfigure[]{\includegraphics[width=0.50\linewidth]{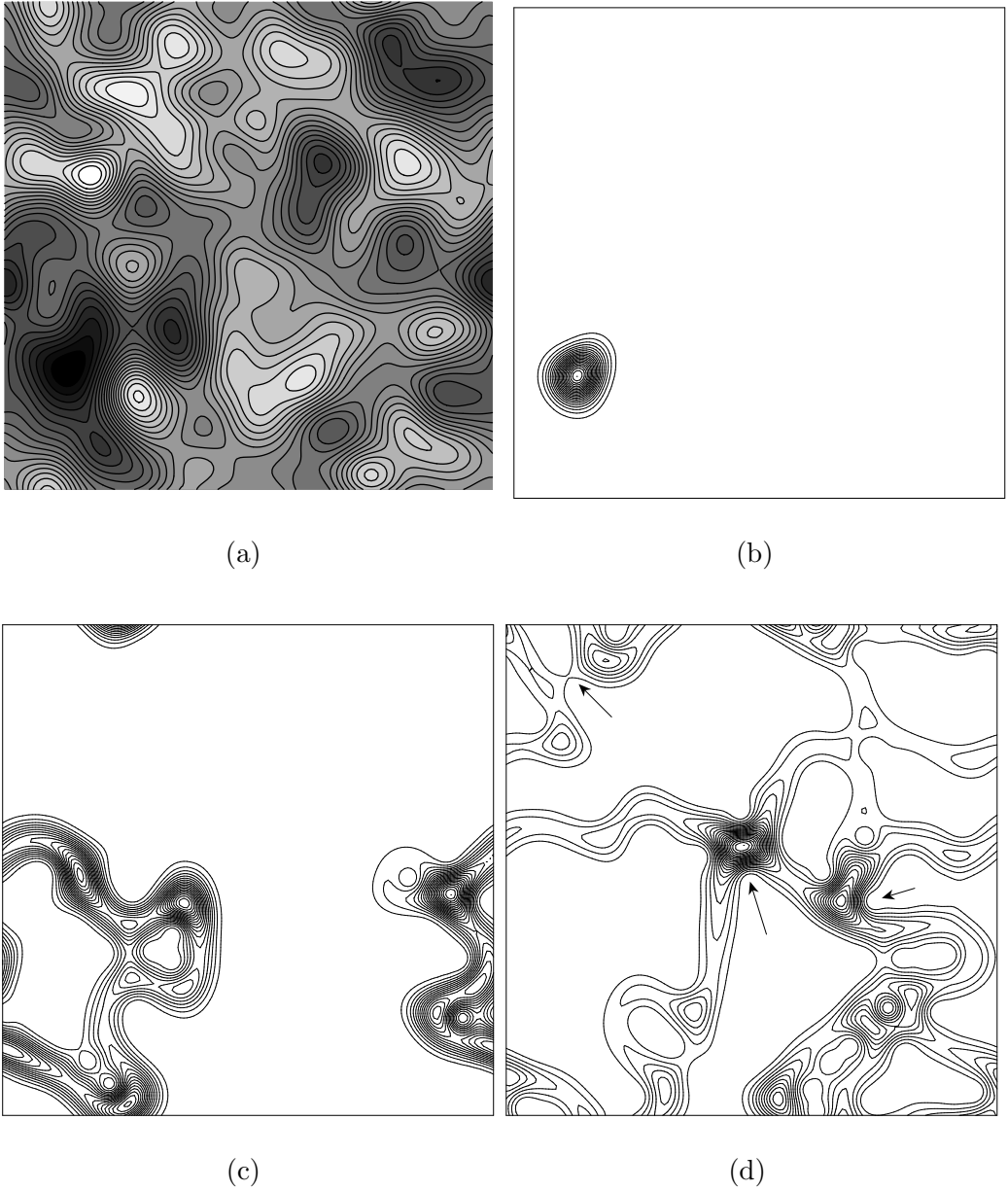}}
\subfigure[]{\includegraphics[width=0.485\linewidth]{wavecont1.ps}} 
\subfigure[]{\includegraphics[width=0.485\linewidth]{wavecont2.ps}} 
\subfigure[]{\includegraphics[width=0.485\linewidth]{wavecont3.ps}} 
\end{center}
  \caption{Long-range correlated random potential
    obtained as superposition of randomly placed Gaussian potentials with a
    width $\ell_{\rm c}=2\ell_{B}$ and some examples of eigenfunctions. (a)
    gray scale plot of the potential landscape with equipotential lines
    indicated; white: high potential, black: low potential; (b)-(d) the moduli
    of a selection of characteristic wave functions corresponding to (b) a low
    energy in the tail of a Landau Band, localized in a deep potential valley;
    (c) an intermediate energy in the Landau band, still localized and
    following mainly an equipotential line; (d) an energy near the center of
    the Landau band, extending essentially along equipotential lines,
    occasionally inter-connected via tunneling near the saddle points of the
    potential (indicated by arrows).}
  \label{fig:3}
\end{figure}

Figure \ref{fig:3} shows an example of such a long-range correlated random
potential, together with several examples of the corresponding wave
functions. They were obtained by solving numerically the Schr\"odinger
equation
\begin{equation}
  \label{eq:disorderedhamiltonian}
  \left[H_{0}+V(x,y)\right]\psi_{\nu}(x,y)=E_{\nu}\psi_{\nu}(x,y)
\end{equation}
in the basis of the Landau states. An important characteristic feature to be
kept in mind is that the wave function amplitudes are essentially non-zero
only along equipotential lines of the potential. In addition, when two
equipotential lines get close to each other near a saddle point of the
potential, the amplitude becomes high {\em across} the saddle point. This can
be taken as an indication of quantum tunneling as indicated in Fig.
\ref{fig:3} by arrows.

The qualitative features of the wave functions in Fig.~\ref{fig:3} may be
taken as the motivation for introducing the concept of a percolating
probability amplitude which we will now describe. The resulting model is
important because it allows for qualitative understanding of localization in
two dimensional electron systems in the presence of a strong magnetic field.
It allows also to understand why, within each Landau band, there must be an
energy where the localization length diverges. This energy corresponds to the
percolation critical point at which the wave function can percolate throughout
the entire system. The basic physical ingredients of the percolation model may
then be used to establish as an idealized version the random quantum network
which allows for systematically dealing with certain universal localization
and transport phenomena, especially in the quantum Hall critical regime.

In order to introduce the essential idea, we consider the Schr\"odinger
equation (\ref{eq:disorderedhamiltonian}) in the limit $\ell_{\rm c}\gg
\ell_{B}$ that can always be achieved for very large magnetic field,
$B\to\infty$ \cite{t76}. In this limit, we will eventually also fulfill the
condition $\hbar \omega_{B}\gg W$ such that the disorder-induced mixing
of the Landau bands can be neglected. 

It is of advantage to introduce new coordinates \cite{jvfh94,t76,kmh65}
\begin{equation}
  \label{eq:centercoordinate}
  X=-k\ell_{B}^{2},\qquad Y=-i\ell_{B}^{2}\frac{\rm d}{{\rm d}X}
:= \frac{\ell_{B}^{2}}{\hbar}P\,.
\end{equation}
They are called center-of-motion coordinates. This is reasonable since $X$ is
the center of the Gaussian wave packet in the $x$-direction of the Landau
states (Eq.~(\ref{eq:landaustates})).  Furthermore, one observes that $(i{\rm
  d}/{\rm d}k)^{p}$ plays the role of $\langle y^{p}\rangle$ by calculating
the matrix elements of $y^{p}$ in the Landau states ($p$ integer). At finite
magnetic field, the electrons can be viewed as describing cyclotron orbits
with radius $\ell_{B}$ around the center of motion. For $\ell_{B}\ll \ell_{\rm
  c}$ the diameter of the cyclotron orbit becomes vanishingly small such that
$(x,y)$ can be replaced by center of motion $(X,Y)$.\footnote{Note that this
  limit has to be taken only after the thermodynamic limit $L\to\infty $ has
  been performed which is necessary to make $X$ a continuous variable.}  The
center-of-motion coordinates fulfill commutation relations similar to position
and momentum operators,
\begin{equation}
  \label{eq:commutators}
 [X,Y]=i\ell_{B}^{2}\,. 
\end{equation}

We start by writing the solutions of the Schr\"odinger equation for the system
of the size $L$ in the representation of the Landau states
\begin{equation}
  \label{eq:expansion1}
  \Psi(x,y)=\sum_{nk}C_{n}(-\ell_{B}^{2}k)\,\psi_{nk}(x,y)
:= \sum_{nX}\langle xy|nX\rangle\,C_{n}(X)\,. 
\end{equation}
This gives
\begin{equation}
  \label{eq:landauschrodinger}
  \sum_{X'}\left(E_{n}\delta_{X,X'}+\langle nX|V|nX'\rangle \right)
C_{n}(X')=EC_{n}(X)
\end{equation}
with inter-Landau band couplings neglected as a consequence of the high field
limit, index $k$ on the energy $E_{n}$ of the Landau level omitted due to
degeneracy, and the Kronecker delta $\delta_{X,X'}$. The potential matrix
elements are
\begin{equation}
  \label{eq:potentialmatrix}
\langle nX|V|nX'\rangle=
\frac{1}{L}\int{\rm d}x{\rm d}y \,\phi_{n}(x-X)\,V(x,y)\,\phi_{n}(x-X')
e^{iy(X-X')/\ell_{B}^{2}}\,.   
\end{equation}
Since the potential is slowly varying on the scale of $\ell_{B}$ it may be
expanded into a power series near some arbitrary $y_{0}(x)$ in
Eq.~(\ref{eq:potentialmatrix}). With this, one easily verifies that
\begin{equation}
  \label{eq:powerseries}
  \int{\rm
  d}y\,V(x,y)e^{iy(X-X')/\ell_{B}^{2}}=V\left(x,-i\ell_{B}^{2}\frac{\rm
  d}{{\rm d}X'}\right)
\frac{1}{L}\int {\rm d}y e^{-iy(X-X')/\ell_{B}^{2}}\,.
\end{equation}
Finally, the integration with respect to $x$ may be approximated by assuming
$x\approx X$ in the potential since at high magnetic field
$|\phi_{n}(x-X)|^{2}$ can be assumed to be very well localized near $X$ within
a distance $\ell_{B}$.  Now Eq.~(\ref{eq:landauschrodinger}) becomes
\begin{equation}
  \label{eq:randomschrodinger}
\left.V\left(X,Y(X')\right)C(X')\right|_{X=X'}\approx E C(X)\,.
\end{equation}
Here, $E$ is the energy corresponding to the equipotential line, and the index
$n$ has been suppressed. Furthermore, $E_{n}$ has been assumed to be the zero
of the energy. This is justified since the distance between the Landau bands
increases $\propto B$ and the disorder broadening of the bands is only
$\propto \sqrt{B}$ such that at high field the couplings between the bands due
to disorder can be neglected.

The random Schr\"odinger equation (\ref{eq:randomschrodinger}) can
approximately be solved by using the semi-classical WKB
(Wentzel-Kramers-Brillouin) Ansatz
\begin{equation}
  \label{eq:ansatz}
  C(E,X)\propto \exp{\left[-i\int^{X}{\rm d}X'k(E,X')\right]}\,,
\end{equation}
neglecting the derivatives of the local wave number $k(E,X)$ with respect to
$X$. This is possible for sufficiently smooth randomness such that second
order, and higher order derivatives in the expansion may be omitted. Under
these assumptions, the dispersion $k(E,X)$ fulfills the equation
\begin{equation}
  \label{eq:dispersion}
E=V\left[X,\ell_{B}^{2}k(E,X)\right]\,.  
\end{equation}
This establishes a most remarkable result, namely that the eigenfunctions can
be considered as the superpositions of the Landau states that are associated
with the equipotential line associated with the energy $E$.

\subsection{The Form of the Semiclassical Wave Functions}
\label{subsec:formofwavefunction}

The general form of these wave functions which propagate on the equipotential
lines may be conjectured by considering the elementary example of a two
dimensional harmonic potential $V(x,y)=V_{0}(x^{2}+y^{2})$. This represents
approximately the situation far from the Landau band center near a local
minimum or maximum of the potential. Of course, this model can be solved
exactly. However, it is instructive to look at it using the above percolation
viewpoint.

The equipotential lines are the circles
$X^{2}+\ell_{B}^{4}k^{2}(E,X)=E/V_{0}:= \epsilon$.  By applying
Eq.~(\ref{eq:randomschrodinger}) and calculating $C(X)$ from
Eq.~(\ref{eq:ansatz}) one obtains
\begin{equation}
  \label{eq:harmoniccoefficient}
  C(X)\propto \exp{\left[-\frac{i\epsilon_{m}}{2\ell_{B}^{2}}
\arccos{\left(\frac{X}{\sqrt{\epsilon_{m}}}\right)}
\mp\frac{iX}{2\ell_{B}^{2}} \sqrt{\epsilon_{m}-X^{2}}\right]}\,.
\end{equation}
The energy is given by the condition that $C(X)$ must be a periodic function
of $X$. This implies for the closed integral over the equipotential circle
\begin{equation}
  \label{eq:bohr-sommerfeld}
\oint {\rm d}X'\,k(E,X')=\frac{\pi \epsilon_{m}}
{\ell_{B}^{2}}:=
2\pi m \qquad (m\,\, {\rm integer}).   
\end{equation}
This is consistent with the result one obtains for the lowest Landau level in
the symmetric gauge with the definition $V_{0}:= e^{2}B_{0}^{2}/8m^{*}$, and
$m$ being the angular momentum quantum number \cite{ll79}.

By inserting the result Eq.~(\ref{eq:harmoniccoefficient}) into the expression
for the wave function Eq.~(\ref{eq:expansion1}) one gets for $\ell_{B}^{2}\ll
\hbar\omega_{B}/V_{0}$, i.e. neglecting the inter-Landau level coupling, an
expression for the wave function which is a reasonable approximation for
$X/\sqrt{\epsilon_{m}}<1$,
\begin{eqnarray}
  \label{eq:percolationwave}
  \psi(x,y)&\propto& \int_{-\sqrt{\epsilon_{m}}}^{\sqrt{\epsilon_{m}}}
{\rm d}X\,\exp{\left[-\frac{iX}{2\ell_{B}^{2}}
\left(2y\pm\sqrt{\epsilon_{m}-X^{2}}\right)\right]}\,
\exp{\left[-\frac{(x-X)^{2}}{2\ell_{B}^{2}}\right]}\nonumber\\
&&\qquad\qquad\qquad\qquad\qquad\qquad\qquad\qquad
\,\times\exp{\left[-\frac{i\epsilon_{m}}{2\ell_{B}^{2}}\varphi_{m}(X)\right]}
\end{eqnarray}
with $\varphi_{m}(X)=\arccos(X/\sqrt{\epsilon_{m}})$. Approximate evaluation
of the integral gives
\begin{equation}
  \label{eq:percolationstate}
  \psi(x,y)\propto \exp{\left[-\frac{ixy}{2\ell_{B}^{2}}\right]}\,
\exp{\left[-\frac{1}{8\ell_{B}^{2}}
\left(y\pm\sqrt{\epsilon_{m}-x^{2}}\right)^{2}\right]}\,
e^{-ik_{m}u}\,.
\end{equation}
The wave number corresponds to $k_{m}:= m/\sqrt{\epsilon_{m}}$ and
$u=\varphi \sqrt{\epsilon_{m}}$ is the azimuthal coordinate on the circle with
the radius $\sqrt{\epsilon_{m}}$. The phase factor mediates a gauge
transformation to the symmetric gauge. The Gaussian factor localizes the wave
function along the circle within an interval of the width $\approx
2\sqrt{2}\ell_{B}$.

\begin{figure}[htbp]
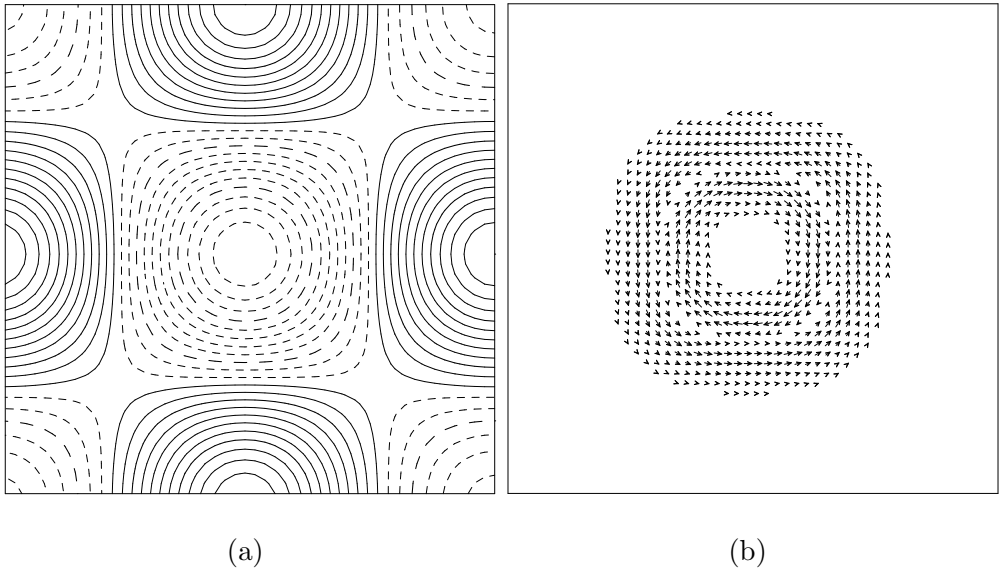

\begin{center}
\subfigure[]{\includegraphics[width=0.47\linewidth]{localpot.ps}} 
\subfigure[]{\includegraphics[width=0.47\linewidth]{localcurrent.ps}} 
\end{center}
  \caption{
    (a) Scheme of a potential profile near the local potential minimum, and
    (b) the current density of a state localized along an equipotential line.
    The current density circles within a region of diameter $2\ell_{B}$ around
    the equipotential line. Only the net current flows along the equipotential
    line.}
  \label{fig:4}
\end{figure}

This suggests that in general the wave functions corresponding to the
equipotential lines can be written in the form \cite{t83,pj82}
\begin{equation}
  \label{eq:general form}
  \psi(x,y)\propto f(v)\,e^{i\kappa(u,v)u}
\end{equation}
with a local wave number $\kappa(u,v)$ that depends on coordinates $u$ and $v$
which parameterize the distance along and perpendicular to the equipotential
line $V(x,y)=E$, respectively. The function $f(v)$ is non-vanishing only
within a region of the approximate width $\ell_{B}$ along this line. Locally,
the wave function has the form of a wave propagating along the equipotential
line. This is associated with an equilibrium current density 
\begin{equation}
  \label{eq:equilibriumcurrent}
\vv{j}\propto\frac{1}{m^{*}}
\vv{e}_{u}|f(v)|^{2}\hbar \kappa(u,v)  
\end{equation}
that yields a net current along
the equipotential line
\begin{equation}
  \label{eq:current1}
\vv{j}\propto \frac{\hbar \kappa(u,v)}{m^{*}}\,\vv{e}_{u} \,.  
\end{equation}
Perpendicular to the direction of $u$, the current density obviously must
vanish (Fig.~\ref{fig:4}). Classically, the net current is produced by the
cyclotron motion of the electron which drifts along the equipotential line in
the strong magnetic field and the local electric field of the potential that
is directed perpendicular to the equipotential line (Fig.~\ref{fig:4}).
This visualizes that only superpositions of those Landau states can contribute
to the eigenstates which are located near the equipotential line at the
corresponding eigenenergy. 

The quantum mechanical problem of solving the Schr\"odinger equation of a
particle in a high magnetic field and a random potential has thus been
replaced by the task of finding the equipotential lines in the random
potential landscape. This corresponds to a classical percolation problem
\cite{s79}. Recently, another connection between quantum Hall plateau
transitions and percolation, based on the classical limit of quantum kinetic
equations, has been discussed \cite{gz2001,m2002}.

\subsection{Localization in the Percolation Limit}
\label{sec:localization}

The localization properties in this limit of a very strong magnetic field may
now be easily discussed \cite{t83,kl82}. Consider the smoothly varying
landscape of the random potential (Fig.~\ref{fig:3}). When the energy is very
low (or very high), the corresponding equipotential lines are closed
trajectories captured within the minima (or maxima) of the potential
(Fig.~\ref{fig:3}b). When the energy increases, these closed trajectories
begin to meander around several of the minima (or maxima) (Fig.~\ref{fig:3}c).

The wave functions that correspond to these equipotential lines are
necessarily localized superpositions of Landau states. They are localized
exponentially as has been shown in \cite{exponentiallandaustates}. Only at a
certain critical energy, say $E_{\rm c}$, close to the center of a Landau
band, an equipotential line can percolate through the whole system. Only at
this energy the wave function can propagate throughout the entire system in
the limit of infinite system size (Fig.~\ref{fig:3}d). The energy $E_{\rm c}$
corresponds to the percolation threshold of the classical percolation problem.
For symmetric distribution of the potential, $P(V)=P(-V)$, the critical energy
is at the center of the Landau band, $E_{\rm c}=0$.

Thus, the percolation picture allows immediately for a very important
conclusion: since the percolating equipotential lines are closed for all
energies except the critical energy of the percolation threshold, all of the
wave functions must be localized except for the one associated with the
critical energy.

Let us define the localization length $\xi_{\rm p}(E)$ as the correlation
length of a percolating equipotential line. Then, percolation theory says that
$\xi_{\rm p}(|E-E_{\rm c}|)$ must increase according to a power law as
$|E-E_{\rm c}|$ decreases,
\begin{equation}
  \label{eq:percolationexponent}
  \xi_{\rm p}(E-E_{\rm c})\propto \frac{1}{|E-E_{\rm c}|^{\nu_{\rm p}}}\,,
\end{equation}
with a universal exponent $\nu_{\rm p}=4/3$ that has been calculated exactly
\cite{sd87}.

\subsection{Tunneling Correction to the Percolating Wave Functions}
\label{sec:quantumcorrection}

When two equipotential lines get close to each other near a saddle point of
the potential, tunneling processes will occur. In order to determine the
localization length, these have to be taken into account \cite{ms88}. For a
symmetric distribution of the potential, the tunneling, however, will not
change the position of the critical energy.

A general definition of the localization length in terms of the Green function
is \cite{km93}
\begin{equation}
  \label{eq:localizationlength1}
  \gamma(E):= \frac{1}{\xi(E)}=-\lim_
{|\vt{r}-\vt{r}'|\to\infty}
\frac{\langle\ln |G(\vv{r},\vv{r}';E)|\rangle}{|\vv{r}-\vv{r}'|}
\end{equation}
with $\langle\ldots\rangle$ denoting the ensemble average with the above
probability distribution of the potential, $P$. It is well known
\cite{aalr79,hikami,wegner,w89,elk}, that in weakly disordered systems,
defined by $k_{\rm F} l \gg 1$, where $l$ is the disorder induced mean free
path and $k_{\rm F}$ the Fermi wave number, the disorder averaged electron
wave function amplitude $\langle\psi ({\bf x},t)\rangle$ decays on length
scales of the order of $l$, since the random phase shifts associated with the
scattering at the impurities are averaged out. This destroys the information
on multiple scattering, and thus on localization. In order to describe
localization, it is therefore necessary to average over disorder functions
containing higher moments of the propagator $G$, like the expression used in
Eq.~(\ref{eq:localizationlength1}). Furthermore, it is well-established that
this quantity is self-averaging, namely its ensemble average coincides with
its most probable value \cite{km93} in the thermodynamic limit. Therefore, one
can calculate instead of the ensemble average the spatial average for a given
realization of the random potential. We will now use this definition of the
localization length for estimating the correction to the critical exponent,
$\nu_{\rm p}$, of the percolation model due to tunneling at the saddle points
of the potential.

Using the above implicit dispersion relation, Eq.~(\ref{eq:dispersion}), we
first determine the Green function in the mixed representation
$(X,k(E_{0},X))$ where $E_{0}$ denotes the energy of an equipotential line,
\begin{equation}
  \label{eq:gf}
  G(X,k(E_{0},X);E)=\frac{1}{E-E_{0}(k,X)-i\eta}\,,
\end{equation}
with $\eta$ a positive infinitesimal real number.
By Fourier transforming with respect to $k$ one can determine the behavior of
the Green function along the $y$-axis which is sufficient for estimating the
asymptotic behavior at $|y|\to \infty$, 
\begin{equation}
  \label{eq:fouriertransformedgf}
G(X,y;E)=\frac{1}{2\pi}\int {\rm d}k\,e^{-iky}\,G(X,k;E)\,,   
\end{equation}
one obtains
\begin{equation}
  \label{eq:gfresidua}
  G(X,y;E)\propto\sum_{j}e^{-ik_{j}(E,X)y}
=\sum_{j}e^{-i{\rm Re}\,k_{j}y}e^{-{\rm Im}\,k_{j}(E,X)|y|}
\end{equation}
where $k_{j}(E,X)$ are the complex roots of Eq.~(\ref{eq:dispersion}) for
given $X$ and $E=E_{0}$. The asymptotic exponential decay of the Green
function for $|y|\to\infty$ is given by the spatial average of the smallest
${\rm Im}k_{j}(E,X)>0$,
\begin{equation}
  \label{eq:decayrate}
  \gamma(E)=\langle{{\rm min}_{j}\,{\rm Im}\,k_{j}(E,X)}\rangle_{P}\,.
\end{equation}

\begin{figure}[htbp]
\begin{center}
 {\includegraphics[width=0.47\linewidth,height=0.2\textheight]
{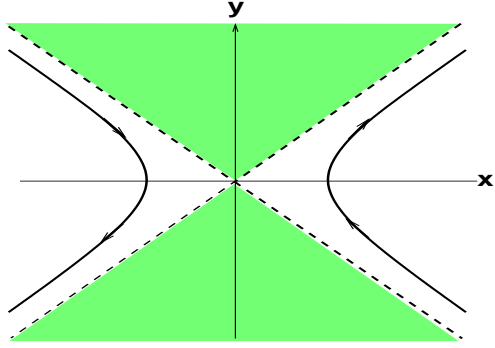}} 
\end{center}
  \caption{Equipotential trajectories of a
    saddle point potential for an energy below the saddle point potential.
    Dashed lines: equipotential lines exactly at the energy of the saddle
    point; shaded: region with potential energy higher than the saddle
    point energy $E_{\rm c}$ (after \protect\cite{fh87}).}
  \label{fig:saddletrajectory}
\end{figure}

As described above (compare Fig.\ref{fig:3}), we need to consider saddle point
regions of the random potential near $E_{\rm c}$, since predominantly it will
be here where the wave functions are connected between different equipotential
lines via tunneling, and for the critical behavior, energies close to $E_{\rm
  c}$ are important. Near a saddle point $(X_{0},k_{0})$, the potential can be
expanded (Fig.~\ref{fig:saddletrajectory}),
\begin{equation}
  \label{eq:saddlepointpotential}
V(X,\ell_{B}^{2}k)=E_{\rm c}-a^{2}(X-X_{0})^{2}+
b^{2}\ell_{B}^{2}(k-k_{0})^{2}\,.  
\end{equation}
>From $E=V(X,\ell_{B}^{2}k)$ one obtains the solution
\begin{equation}
  \label{eq:wavenumberatsaddlepoint}
  k-k_{0}=\pm\frac{1}{b\ell_{B}}\sqrt{E-E_{\rm c}+a^{2}(X-X_{0})^{2}}\,.
\end{equation}
Due to symmetry, it is sufficient to consider, say, $E<E_{\rm c}$. Then, ${\rm
  Im}k\neq 0$ only for $E_{\rm c}-E>a^{2}(X-X_{0})^{2}$
\begin{equation}
  \label{eq:imaginaryk}
  {\rm Im}k(E,X)=
\frac{1}{b\ell_{B}}\sqrt{E_{\rm c}-E-a^{2}(X-X_{0})^{2}}\,.
\end{equation}
Averaging this with respect to $X$ yields
\begin{equation}
  \label{eq:saddledecay}
\gamma(E)\propto\int_{X_{-}}^{X_{+}}{\rm d}X\,
\sqrt{E_{\rm c}-E-a^{2}(X-X_{0})^{2}}\propto E_{\rm c}-E  
\end{equation}
with $X_{\pm}=X_{0}\pm\sqrt{E_{\rm c}-E}/a$.

The asymptotic exponential decay for energies near $E_{\rm c}$ can be
estimated by using the following heuristic argument. The equipotential lines
representing the wave functions form percolation clusters with an energy
dependent correlation length $\xi_{\rm P}(E)\sim |E-E_c|^{-4/3}$.
Delocalization of the wave functions associated with these clusters can occur
via tunneling through the saddle points that connect different equipotential
clusters. The number of saddle points will be proportional to the size of the
clusters at energy $E$, namely $\propto \xi_{\rm P}(E)$. Thus, the probability
of finding the saddle point with the smallest $\gamma$ at energy $E$
connecting two adjacent clusters will be $\ell_{B}/\xi_{\rm P}(E)$. The saddle
point with the {\em smallest} $\gamma$ will dominate the asymptotic behavior
of the wave functions. All of the contributions of the other saddle points
will be exponentially suppressed with the diameter of the clusters.
Therefore, we finally can write for the average inverse localization length
the power law
\begin{equation}
  \label{eq:localizationlength2}
  \gamma(E)\propto \frac{1}{\xi_{\rm P}(E)}|E-E_{\rm c}|\propto
|E-E_{\rm c}|^{4/3+1}\,.
\end{equation}
According to this qualitative argument, the critical exponent of the
localization length in a quantum Hall system including the effect of quantum
tunneling should be $\nu=7/3$, without interactions.

It has taken considerable theoretical \cite{hk90,h95,aa85,h86,h90,h94} and
experimental \cite{wetal88,khk91,eetal93,hetal91} efforts to determine
quantitatively this exponent for two dimensional disordered Landau system
without interaction. The most accurate numerical values obtained so far are
$\nu=2.35\pm 0.03$ by using a random Landau matrix model \cite{h95},
$\nu=2.33\pm 0.03$ from a Monte Carlo calculation for chiral fermions
\cite{Ziqiang}, and a renormalization group approach based on a random network
model $\nu=2.39\pm 0.01$ \cite{crsr01,crr03,cr03} (cf. Tab.~\ref{critexp}).
Numerical investigations were done for a variety of completely different
models, including white noise \cite{hk90,aa85} as well as long-range
correlated \cite{cc88,h90} randomness and including also higher Landau bands
\cite{h95,h94}. All of the results are consistent with these values within
error bars. This strongly suggests that the critical exponent of the quantum
phase transition associated with the Quantum Hall Effect is indeed universal,
and does not depend on the microscopic details of the model used\footnote{For
  very long range correlated randomness there recently seems to be evidence
  that $\nu $ changes \cite{smk03}.}.  Although the near coincidence of this
exponent with the value $7/3$, as obtained from the simple heuristic argument
outlined above, seems compelling we caution that an analytical derivation of
the critical exponents at the quantum Hall transition is still lacking. Recent
progress towards achieving this goal is reviewed in chapter \ref{sec:susy}.

\section{The Random Network Model}
\label{sec:randomnetwork}

The results obtained in the above discussed high-magnetic field limit may be
used not only for estimating the critical behavior but also as a starting
point for constructing a model that contains all of the necessary physical
ingredients --- backscattering and tunneling --- for describing the quantum
critical points near the centers of the Landau levels and the corresponding
transport quantities. If the critical behavior could be shown to be universal
such a model should be of great importance. This would be especially true, if
one could use the model as a starting point for more rigorous theoretical
formulations. 

As argued above, percolation of probability amplitude along equipotential
lines as well as tunneling of amplitude between equipotential lines near the
saddle points of the potential have to be taken into account on an equal
footing when determining critical properties. This suggests to construct a
model consisting of a regular lattice of saddle points that are connected via
links along which probability amplitude can propagate.

\subsection{The Scattering Wave Functions Associated with a Saddle Point}
\label{subsec:wavefunctions}

Before doing so, it is instructive to consider saddle point tunneling more
formally from the scattering point of view, and establish an exact expression
for the transmission probability. As a side remark, we note that this is yet
another illuminating example of a quantum wave function that can be treated
exactly in the high magnetic field limit \cite{fh87}.

The most simple saddle point is defined by the quadratic potential considered
in the previous chapter with equal coefficients $a^{2}=b^{2}\equiv U$
\cite{fh87}
\begin{equation}
  \label{eq:saddlepoint}
  V_{\rm sp}(x,y)=E_{\rm c}+U(y^{2}-x^{2})\,.
\end{equation}
The classical trajectories in such a potential for energies below $E_{\rm c}$
are shown in Fig.~\ref{fig:saddletrajectory}. For the corresponding quantum
mechanical wave functions the considerations of the previous sections apply
when neglecting tunneling between the trajectories. This can be justified for
energies well below (or well above) $E_{\rm c}$. However, when $E\approx
E_{\rm c}$ this is no longer the case.

The total Hamiltonian including the saddle point
\begin{equation}
  \label{eq:saddlepointhamiltonian}
  H=H_{0}+V_{\rm sp}(x,y)=
\frac{1}{2m^{*}}\left(\frac{\hbar}{i}\nabla + e\vv{A}\right)^{2}
+V_{\rm sp}(x,y)
\end{equation}
is quadratic in the variables. It can be diagonalized in analogy with in the
case of the two dimensional harmonic oscillator.

However, here we follow the very illuminating method used in
\cite{fh87}. For the vector potential, we assume the symmetric gauge,
$\vv{A}=(B/2)(-y,x)$. We perform first a unitary transformation that
brings the Hamiltonian into a separable form.

By introducing the new variables
\begin{eqnarray}
 a_x&=&\frac{x}{2\ell_B}+\ell_B\frac{\partial}{\partial x}\,,\\
 a_y&=&\frac{y}{2\ell_B}+\ell_B\frac{\partial}{\partial y}\, ,
\end{eqnarray}
that fulfill the commutation relations
\begin{eqnarray}
  \label{eq:commutator1}
  [a_x,a_x^\dagger]&=&[a_y,a_y^\dagger]=1\,,\\
  \left[a_x,a_y\right]&=&[a_x^{\dagger},a_y^\dagger]=0\,,
\end{eqnarray}
the Hamiltonian becomes
\begin{eqnarray}
 H&=&\frac{\hbar\omega_{B}}{2}(a_x^\dagger a_x+a_y^\dagger a_y+1)+
\frac{\hbar\omega_{B}}{2i}(a_x^\dagger a_y-a_y^\dagger a_x)+ \nonumber\\
&&\qquad\qquad\qquad\qquad
+\gamma[(a_y+a_y^\dagger)^2-(a_x+a_x^\dagger)^2]+E_{\rm c}\,,
\end{eqnarray}
where $\gamma=U \ell_B^2$. The unitary transformation
\begin{equation}
       \left(\begin{array}{c}a_x\\ a_y\end{array}\right)=
\left(
\begin{array}{cc}
i\cos\phi & \sin\phi\\
-\sin\phi & -i \cos\phi
\end{array}
\right)
      \left(\begin{array}{c}b_1\\ b_2\end{array}\right)\,,
\end{equation}
with $\tan(2\phi)=-\hbar\omega_{B}/4 \gamma$, transforms the Hamiltonian into
a sum of two independent terms, $H_{1}(b_1,b_1^\dagger)$ and
$H_{2}(b_2,b_2^\dagger)$,
\begin{eqnarray}
 H_{1}&=&
  (\begin{array}{cc}b_1^\dagger & b_1\end{array})
\left(
\begin{array}{cc}
\frac{\hbar\omega_{B}}{4}-\hbar\Omega&
\gamma\\
\gamma &
\frac{\hbar\omega_{B}}{4}-\hbar\Omega\end{array}
\right)
      \left(\begin{array}{c}b_1\\b_1^\dagger\end{array}\right) \\
&&\nonumber\\
 H_{2}&=&
   (\begin{array}{cc}b_2^\dagger & b_2\end{array})
\left(
\begin{array}{cc}
\frac{\hbar\omega_{B}}{4}+\hbar\Omega&
-\gamma\\
-\gamma &
\frac{\hbar\omega_{B}}{4}+\hbar\Omega\end{array}
\right)
      \left(\begin{array}{c}b_2\\b_2^\dagger\end{array}\right) ,
\end{eqnarray}
where $\hbar\Omega=\sqrt{\gamma^2+\left({\hbar\omega_{B}}/{4}\right)^2}$, and
\begin{eqnarray}
  \label{eq:commutator2}
 [b_1,b_1^\dagger]&=&[b_2,b_2^\dagger]=1\,,\\
\left[b_1,b_2\right]&=&[b_1,b_2^\dagger]=0\,. 
\end{eqnarray}

These can be diagonalized by a Bogoliubov transformation,
\begin{equation}
 \left(
\begin{array}{cc}
b_i\\b_i^\dagger
\end{array}
\right)
= \left(
\begin{array}{cc}
\cosh\theta_i&\sinh\theta_i \\
\sinh\theta_i&\cosh\theta_i
\end{array}
\right)
 \left(
\begin{array}{cc}
c_i\\c_i^\dagger
\end{array}
\right),
\end{equation}
with
$ \tanh(2\theta_1)=\gamma^{-1}(-{\hbar\omega_{B}}/{4}+\hbar\Omega)$,
and
$\tanh(2\theta_2)=\gamma(\hbar\omega_{B}/{4}+\hbar\Omega)^{-1}$,
which gives for the Hamiltonian
\begin{equation}
 H=E_1\left(c_1^2+{c_1^\dagger}^2\right)+
E_2\left(c_2^\dagger c_2+\frac{1}{2}\right)+E_{\rm c},
\end{equation}
with 
\begin{eqnarray}
  \label{eq:commutator3}
 [c_1,c_1^\dagger]&=&[c_2,c_2^\dagger]=1\,, \\
\left[c_1,c_2\right]&=&[c_1,c_2^\dagger]=0\,, 
\end{eqnarray}
and the energy eigenvalues
\begin{eqnarray}
E_1&=&\sqrt{\gamma^2-\left(\frac{\hbar\omega_{B}}{4}-
\hbar\Omega\right)^2}\, , \\
\nonumber\\
E_2&=&2\sqrt{\left(\frac{\hbar\omega_{B}}{4}+
\hbar\Omega\right)^2-\gamma^2}\, .
\end{eqnarray}

Finally, we introduce variables
\begin{eqnarray}
  \begin{array}{cc}
X=\frac{1}{\sqrt{2}i}(c_1^\dagger-c_1),&\quad 
s=\frac{1}{\sqrt{2}}(c_2^\dagger+c_2),\\
P=\frac{1}{\sqrt{2}}(c_1^\dagger+c_1),&\quad
p=\frac{1}{\sqrt{2}i}(c_2-c_2^\dagger), 
  \end{array}
\end{eqnarray}
and obtain
\begin{equation}
 H=H_{1}+H_{2}=E_1\left(P^2-X^2\right)
 +\frac{1}{2}E_2\left(p^2+s^2\right)+E_{\rm c} \,,
\end{equation}
with the commutation relations 
\begin{eqnarray}
  \label{eq:commutator4}
 [X,P]&=&[s,p]=i\,,\\
&&\nonumber\\
\left[s,X\right]&=&[s,P]=[p,X]=[p,P]=0,. 
\end{eqnarray}
Thus, apart from the factor $\ell_{B}^{2}/\hbar$ in
Eq.~(\ref{eq:centercoordinate}), $H_{1}$ corresponds to the center-of-motion
part of the total Hamiltonian, and $H_{2}$ is a one dimensional harmonic
oscillator which implies that the wave function is harmonically confined in
the direction of $s$.

The eigenfunctions of the Hamiltonian can now be factorized
\begin{equation}
 \Psi(X,s)=\phi(X)\psi_n(s),
\end{equation}
where $\psi_n$ corresponds to the $n$--th harmonic oscillator level. 

In order to arrive at a scattering solution of the Schr\"odinger equation, we
prepare a initial wave packet at $\langle X^2 \rangle \approx \langle P^2
\rangle\gg 1$.  Since
\begin{equation}
 \frac{x}{\ell_B}=\sqrt{2}(\alpha_1 X-\beta_2 s) \, ,\,\quad
 \frac{y}{\ell_B}=\sqrt{2}(\beta_1 P+\alpha_2 p)
\end{equation}
with
\begin{equation}
 \alpha_i=\cos\phi\,e^{-\theta_i}\, ,\quad
 \beta_i=-\sin\phi\,e^{\theta_i} \,\quad (i=1,2),
\end{equation}
we have asymptotically
\begin{equation}
 \frac{\langle x\rangle}{\alpha_1}\approx 
\frac{\langle y\rangle}{\beta_1} \, .
\end{equation}
Since $\hbar\omega_{B}\gg\gamma$, 
$\beta_1/\alpha_1\approx -1$, and the wave packet is
centered near an asymptote of the equipotential line of $V_\mathrm{sp}$ in the
upper left quadrant of the $(x,y)$-plane (Fig.~\ref{fig:saddletrajectory}).

In order to find the transmission probability, we need now to construct a
scattering wave function. The eigenfunction $\phi$ satisfies
\begin{equation}
H_{1}\phi(X)=E_1(P^2-X^2)\phi(X)
=\left[E-\left(n+\frac{1}{2}\right) E_2-E_{\rm c}\right]\phi(X) ,
\end{equation}
or equivalently
\begin{equation}
 \left(
\frac{{\rm d}^2}{{\rm d} X^2}+X^2+\epsilon \right)\phi(X)=0
\end{equation}
with the energy parameter
\begin{equation}
  \label{eq:energyparameter}
\epsilon=\frac{E-\left(n+{1}/{2}\right)E_2-E_{\rm c}}{E_1}\,.  
\end{equation}
In the limit of high magnetic field, $E_2=\hbar\omega_{B}$ and $E_1=\gamma$.
Therefore, $\epsilon$ measures the energy deviation from the saddle point
energy $E_{\rm c}$ in a Landau band normalized by a typical potential
strength.

This Schr\"odinger equation is discussed in detail in \cite{mf53}. The
eigenfunctions even and odd in $X$ are
\begin{eqnarray}
 \Phi_{+}&=&\quad e^{-i X^2/2}
F\left(\frac{1+i\epsilon}{4}\left|\frac{1}{2}\right|iX^2\right)\, ,\nonumber\\
&&\\
 \Phi_{-}&=&Xe^{-i X^2/2}
F\left(\frac{3+i\epsilon}{4}\left|\frac{3}{2}\right|i X^2\right)\,,\nonumber
\end{eqnarray}
where $F$ is a confluent hypergeometric function. For large $|X|$ they can be
written in the form
\begin{eqnarray}
  \label{eq:evenodd}
  \Phi_{+}(X)&=&
\frac{\Gamma(1/2)}{\Gamma(Z_{+})}\,
e^{-i(\pi/2) Z_{+}^{*}}\,f(Z_{+}^{*};X)e^{iX^{2}/2} \quad\,+ {\rm c.c.}
\nonumber\\
&&\\
\Phi_{-}(X)&=&
\frac{\Gamma(3/2)}{\Gamma(Z_{-})}\,
e^{-i(\pi/2) Z_{-}^{*}}\,f(Z_{-}^{*};X)Xe^{iX^{2}/2}+ c.c.\nonumber
\end{eqnarray}
with the definitions
 $f(z;X)=|X|^{-2z}$, $Z_{-}=(3+i\epsilon)/4$, $Z_{+}=(1+i\epsilon)/4$.

The currents in these states are related to the derivatives
\begin{equation}
  \label{eq:fertigcurrents}
\frac{\partial}{i\partial X}e^{\pm iX^{2}/2} =\pm Xe^{\pm iX^{2}/2}\,.  
\end{equation}
Therefore, each first term in Eqs. (\ref{eq:evenodd}) corresponds to a
currents in the positive $X$-direction. The conjugate complex term represents
a current in the negative $X$-direction.

These solutions are linearly superposed, $\phi_{\rm t}=A\Phi_{+}+B\Phi_{-}$,
to form the scattering wave functions which must fulfill boundary conditions
such that for $X>0$ only a transmitted wave function exists with the current
flowing away from the origin. This leads to the relation
\begin{equation}
   \label{eq:condition1}
  A\,\frac{\Gamma(1/2)}{\Gamma(Z_{+}^{*})}\,e^{i\pi/8}
+B\,\frac{\Gamma(3/2)}{\Gamma(Z_{-}^{*})}\,e^{i3\pi/8}=0\,.
\end{equation}
One finds after some algebra for the transmitted wave for large values of
$X<0$
\begin{equation}
  \label{eq:scatteringstates1}
  \phi_{\rm t}(X)=f(Z_{+};X)\,
e^{-\pi\epsilon/8}\,e^{iX^{2}/2}\,
\left[A\,\frac{\Gamma(1/2)}{\Gamma(Z_{+})}e^{-i\pi/8}
+B\,\frac{\Gamma(3/2)}{\Gamma(Z_{-})}e^{-i3\pi/8}\right]\,,
\end{equation}
and for the incoming wave
\begin{equation}
\label{eq:scatteringstates}
\phi_{\rm i}(X)=f(Z_{+};X)\,
e^{-\pi\epsilon/8}\,
e^{-iX^{2}/2}
\left[A\,\frac{\Gamma(1/2)}{\Gamma(Z_{+}^{*})}e^{i\pi/8}
-B\,\frac{\Gamma(3/2)}{\Gamma(Z_{-}^{*})}e^{i3\pi/8}\right]\,.
\end{equation}
This guarantees that for $X>0$ no incoming wave exists.  The transmission
probability is defined by
\begin{equation}
  \label{eq:transmissionprobability}
  T(\epsilon)=\lim_{X\to\infty}
\frac{|\phi_{\rm t}(X)|^{2}}{|\phi_{\rm i}(-X)|^{2}}\,.
\end{equation}
One gets from the above equations by using the relations for the
$\Gamma$-functions
\begin{equation}
  \label{eq:gammarelation}
   \Gamma(z)^*=\Gamma(z^*)\, ,\quad
\Gamma\left(\frac{1}{4}+iy\right)
\Gamma\left(\frac{3}{4}-iy\right)
=\frac{\sqrt{2}\pi}{\cosh\pi y+i\sinh\pi y},
\end{equation}
 the final result
\begin{equation}
  \label{eq:result}
  T(\epsilon)=\frac{1}{1+\exp{(-\pi\epsilon})}\,.
\end{equation}
For $\epsilon\to -\infty$, well below the saddle point energy, the
transmission probability vanishes. The incoming wave is completely reflected.
For $\epsilon\to +\infty$, the transmission probability approaches unity.
Exactly at the saddle point energy, $\epsilon=0$, one has $T(0)=1/2$. For
energies near the saddle point, one can expand 
\begin{equation}
  \label{eq:expansion}
T(\epsilon)\approx \frac{1}{2}+\frac{\pi}{4}\epsilon + \cdots\,.  
\end{equation}

\subsection{Parameterizing the Scattering at a Saddle Point}
\label{subsec:saddletransmission}

As we have seen, the percolating wave functions of the electron carry an
equilibrium current density. Near the points where percolating wave functions
approach each other closely, quantum tunneling takes place.

A natural model for the description of the percolating quantum states can then
be obtained by considering a network of current loops, occasionally
inter-connected via tunneling near the saddle points of the potential
(Fig.~\ref{fig:3}). The simplest choice is a regular network. Disorder can be
introduced either by randomizing the relative phases of the current loops
and/or the heights of the saddle points. The distribution of the wave function
amplitude is then closely related to that of the corresponding quantum
currents and the asymptotic behavior is essentially determined by the coherent
interplay of the quantum transmission and scattering near the saddle points.

\begin{figure}[htbp]
\begin{center}
\subfigure[]{\includegraphics[width=0.47\linewidth]{potschema.ps}} 
\subfigure[]{\includegraphics[width=0.47\linewidth]{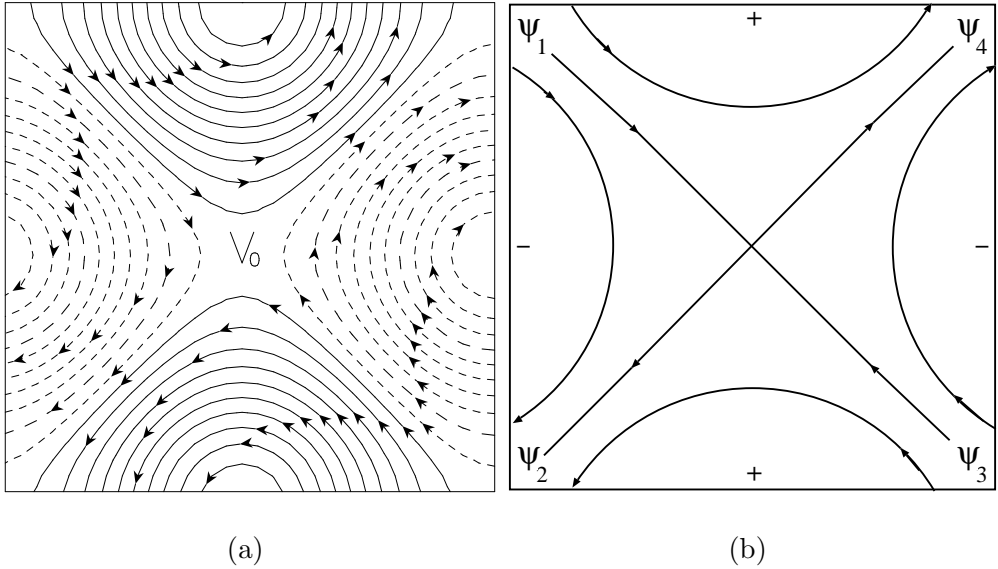}} 
\end{center}
  \caption{(a) Potential landscape near a saddle 
    point with adjacent maxima (full equipotential lines) and minima
    (dashed equipotential lines). (b) The corresponding idealized potential
    landscape with incoming ($\psi_{1}$ and $\psi_{3}$) and outgoing
    ($\psi_{2}$ and $\psi_{4}$) current amplitudes flowing along equipotential
    lines. Note that the saddle point energy is denoted here by $V_{0}$.}
  \label{fig:5a}
\end{figure}

Let us consider the region close to a specific saddle point, say $E_{\rm c}$.
The potential landscape can be modeled by two neighboring adjacent potential
maxima and minima (Fig.~\ref{fig:5a}). We assume that the directions of the
currents along the equipotential lines in the potential minima and maxima are
clockwise and counter clockwise, respectively. Let us consider energies close
$E_{\rm c}$. There are four equipotential lines that enter the region of the
saddle point. The total current entering and leaving the saddle point region
must be conserved (Fig.~\ref{fig:5a}).

The effect of the saddle point potential can be considered as a quantum
transmission problem as described above \cite{cc88}. In more simplified terms,
we assume that the amplitudes which correspond to the currents on the boundary
of the saddle point region are $\psi_{1}\ldots \psi_{4}$ with the amplitudes
of incoming and outgoing currents $\psi_{1}$, $\psi_{3}$ and $\psi_{2}$,
$\psi_{4}$, respectively. The formal relationship between them can be
described by a scattering matrix ${\bf S}$,
\begin{equation}
\left(
\begin{array}{c}
\psi_{2}\\
\psi_{4}
\end{array}
\right)
= {\bf S} 
\left(
\begin{array}{c}
\psi_{1}\\
\psi_{3}
\end{array}
\right).
\label{eq:scattering_node}
\end{equation}
In general, the ${\bf S}$ matrix can be written as \cite{mpk88,beenakker97}
\begin{equation}
{\bf S}=
\left(
\begin{array}{cc}
{\rm e}^{-i\varphi_2}&0 \\
0& {\rm e}^{i\varphi_4}
\end{array}
\right)
\left(
\begin{array}{cc}
-r&t \\
t&r
\end{array}
\right)
\left(
\begin{array}{cc}
{\rm e}^{i\varphi_1}& 0\\
0&{\rm e}^{-i\varphi_3}
\end{array}
\right),
\label{eq:decomposition}
\end{equation}
where $0\le r\le 1,0\le  t \le 1$, $r^2+t^2=1$, and
$r^2$  and $t^2$ are the probabilities of the current $\psi_{1}$ to be
reflected to $\psi_{2}$ and to be transmitted to $\psi_{4}$, respectively.
Note that the current conservation relation implies
\begin{equation}
|\psi_{1}|^2+|\psi_{3}|^2=|\psi_{2}|^2+|\psi_{4}|^2
\label{eq:current_conservation}
\end{equation}
which is always valid due to the unitarity of the scattering matrix ${\bf S}$.

For iterative numerical calculations, it is useful to introduce the transfer
matrix ${\bf T}$ that is equivalent to ${\bf S}$. It relates the amplitudes on
the left, $(\psi_{1},\psi_{2})^\mathrm{T}$, to those on the right hand side of
the saddle point, $(\psi_{4},\psi_{3})^\mathrm{T}$ ($\mathrm{T}$ denotes the
transposed).
\begin{equation}
\left(
\begin{array}{c}
\psi_{4}\\
\psi_{3}
\end{array}
\right)
= {\bf T} 
\left(
\begin{array}{c}
\psi_{1}\\
\psi_{2}
\end{array}
\right)\,.
\label{eq:transfer_node}
\end{equation}
Using Eqs. (\ref{eq:scattering_node}) and
(\ref{eq:decomposition}) one finds the result
\begin{equation}
{\bf T}=
\left(
\begin{array}{cc}
{\rm e}^{i\varphi_4}&0 \\
0 & {\rm e}^{i\varphi_3}
\end{array}
\right)
\left(
\begin{array}{cc}
1/t&r/t \\
r/t&1/t
\end{array}
\right)
\left(
\begin{array}{cc}
{\rm e}^{i\varphi_1}& 0\\
0&{\rm e}^{i\varphi_2}
\end{array}
\right)\,.
\label{eq:decomposition_t}
\end{equation}
For completeness, we only mention here, that due to the current conservation,
the $2\times 2$-matrix ${\bf T}$ must fulfill the symmetry condition (cf.
chapter \ref{sec:symmetries})
\begin{equation}
\label{eq:condition}
{\bf J}={\bf T}^{\dagger}{\bf J}{\bf T} \qquad {\rm with} \qquad
{\bf J}=\left(
\begin{array}{cc}
1 & 0\\
0 & -1
\end{array}
\right)\,.
\end{equation}

It is useful to relate the present description to the microscopic saddle point
model introduced above. By using Eq.~(\ref{eq:condition}), and imposing the
correct asymptotic behavior from the scattering geometry in Fig.~\ref{fig:5a},
one can write the reflection and transmission amplitudes $r$ and $t$ in
terms of a single parameter
\begin{equation}
r:=\frac{1}{\cosh\Theta}
\label{eq:theta_r}
\end{equation}
and by the current conservation relation $t=\sqrt{1-r^2}$,
\begin{equation}
  \label{eq:theta_t}
t=\tanh\Theta\,. 
\end{equation}
Equation~(\ref{eq:decomposition_t}) can be written in terms of the variable
$\Theta$,
\begin{eqnarray}
{\bf T}&=&
\left(
\begin{array}{cc}
{\rm e}^{i\varphi_4}&0 \\
0 & {\rm e}^{i\varphi_3}
\end{array}
\right)
\left(
\begin{array}{cc}
\mathrm{cotanh}\Theta&\mathrm{cosech}\Theta \\
\mathrm{cosech}\Theta&\mathrm{cotanh}\Theta
\end{array}
\right)
\left(
\begin{array}{cc}
{\rm e}^{i\varphi_1}& 0\\
0&{\rm e}^{i\varphi_2}
\end{array}
\right) \nonumber\\
&&\nonumber \\&&\nonumber \\
&=&
\left(
\begin{array}{cc}
{\rm e}^{i\varphi_4}&0 \\
0 & {\rm e}^{i\varphi_3}
\end{array}
\right)
\,\left(
\begin{array}{cc}
\cosh\Theta'&\sinh\Theta' \\
\sinh\Theta'&\cosh\Theta'
\end{array}
\right)
\,\left(
\begin{array}{cc}
{\rm e}^{i\varphi_1}& 0\\
0&{\rm e}^{i\varphi_2}
\end{array}
\right)\,.
\end{eqnarray}
$\Theta'$ and $\Theta$ being related via the equation
\begin{equation}
\label{eq:duality}
\sinh\Theta \sinh\Theta'=1 \,.
\end{equation}

By comparing with Eq.~(\ref{eq:result}) one finds that the dimensionless
parameter $\Theta$ must depend monotonically on the energy of the
equipotential line. It characterizes completely the transmission properties of
the saddle point. For $E\ll E_{\rm c}$ and $E\gg E_{\rm c}$, $\Theta\ll 1
(r\approx 1)$ and $\Theta \gg 1 (t\approx 1)$, respectively. This is because
for $E\ll E_{\rm c}$, $\psi_{1}\approx \psi_{2}$
for almost all $\psi_{3}$ and $\psi_{4}$,
and for $E\gg E_{\rm c}$, $\psi_{1}\approx \psi_{4}$. 
One can determine the energy
dependence of $\Theta$ by comparing with the above microscopic result near
$\epsilon=0$ where $T(0)=1/2:= |t|^{2}=\tanh^{2}\Theta_{\rm c}$ and
expanding for $\Theta\approx\Theta_{\rm c}$ since the model is only suitable
for the region near the saddle point energy, $E_{\rm c}$. One finds (cf.
Eq.~(\ref{eq:expansion}))
\begin{equation}
  \label{eq:theta_c}
  \Theta(\epsilon)\approx \Theta_{\rm c}+\frac{\pi\epsilon}{2\sqrt{2}}
 +\cdots
\end{equation}
with the saddle point value $\Theta_{\rm c}=\ln(1+\sqrt{2})$.

The characteristic feature of this model for the transmission through a saddle
point is that incident and transmitted channels are locally separated. The
interplay between the saddle point potential and the high magnetic field
introduces a spatial separation of the incoming and outgoing channels. This
makes the model particularly suited for describing the critical localization
features in the quantum Hall region.

\subsection{Establishing the Random Network of Saddle Points}
\label{subsec:ccrandomnet}

Starting from the smooth random landscape of the potential, the network model
could now be defined from a random system of circular, localized wave
functions of the type Eq.~(\ref{eq:general form}), equivalent to circular
equilibrium currents. These would correspond to states associated with
randomly distributed sites. For technical convenience, however, it is
preferable to assume the circular wave functions to have random phases and
being associated with the sites of a regular lattice. In addition, it is
reasonable to assume that nearest-neighbored wave functions are
connected by tunneling contacts described by the above transfer matrices
${\bf T}$ (cf. Eq.~(\ref{eq:decomposition_t})). This enables the particles
to hop from one circular state to another. A delocalization mechanism for the
total wave function is introduced in this way.

The model constructed is the analogue at high magnetic field of a two
dimensional Anderson model for a disordered system \cite{km93,mk81,mk83}. The
site states of the latter are replaced by the circular currents of the former,
and the hopping amplitudes by the unitary transfer matrices. As in the
Anderson model, the network contains the necessary competing ingredients ---
localization within the circular current states and tunneling between them ---
for describing a localization-delocalization phase transition. In the case of
the Anderson model, there is no phase transition in two dimensions. As we have
already seen above, the present network model must show a singularity in the
localization properties which represents a quantum phase transition point.
This will be discussed in more detail in the next chapter.

\begin{figure}[hbtp]
\includegraphics[width=0.9\linewidth]{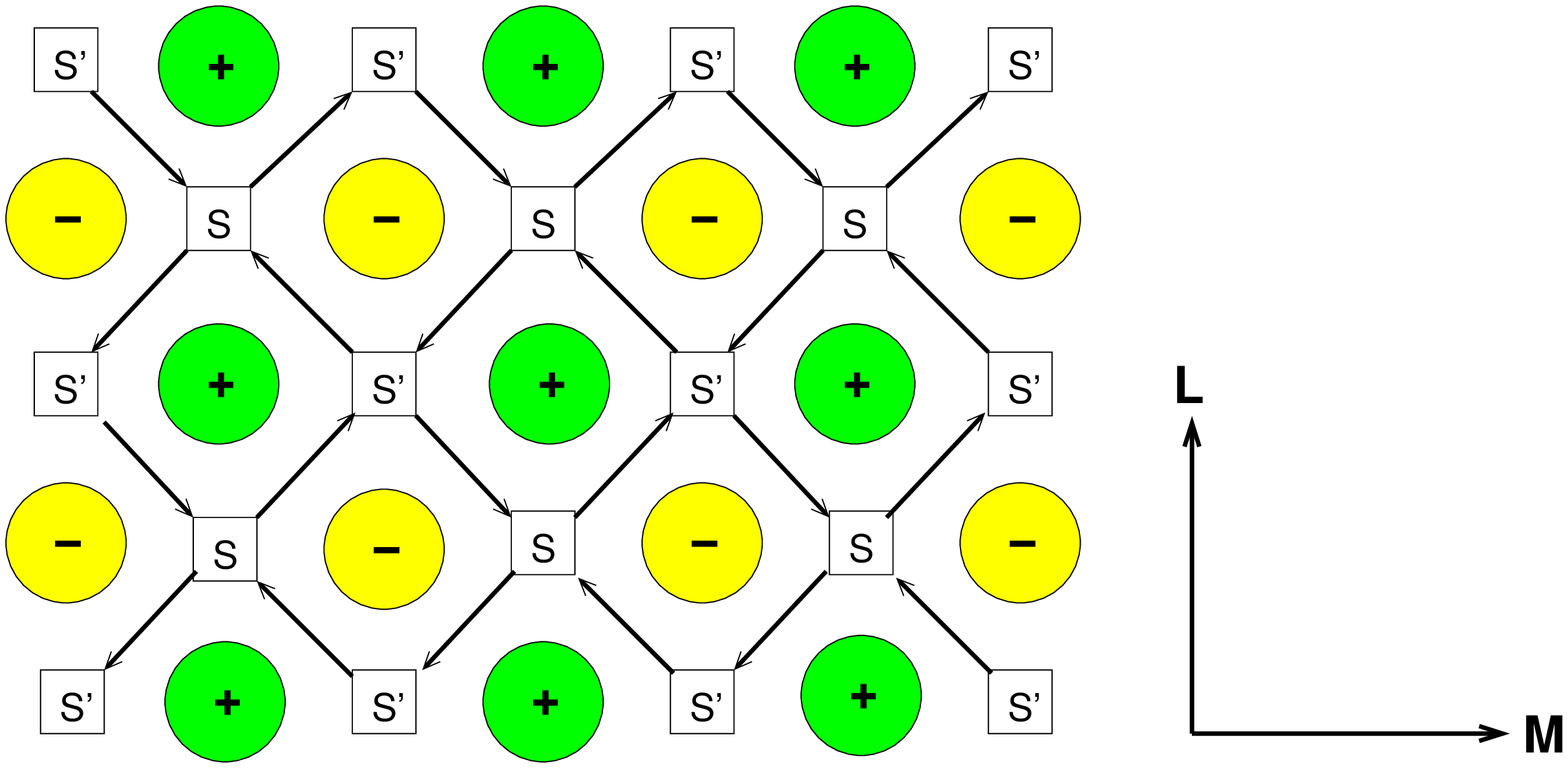}

\vspace{-6cm}
\hspace{9.2cm}
\includegraphics[width=0.3\linewidth]{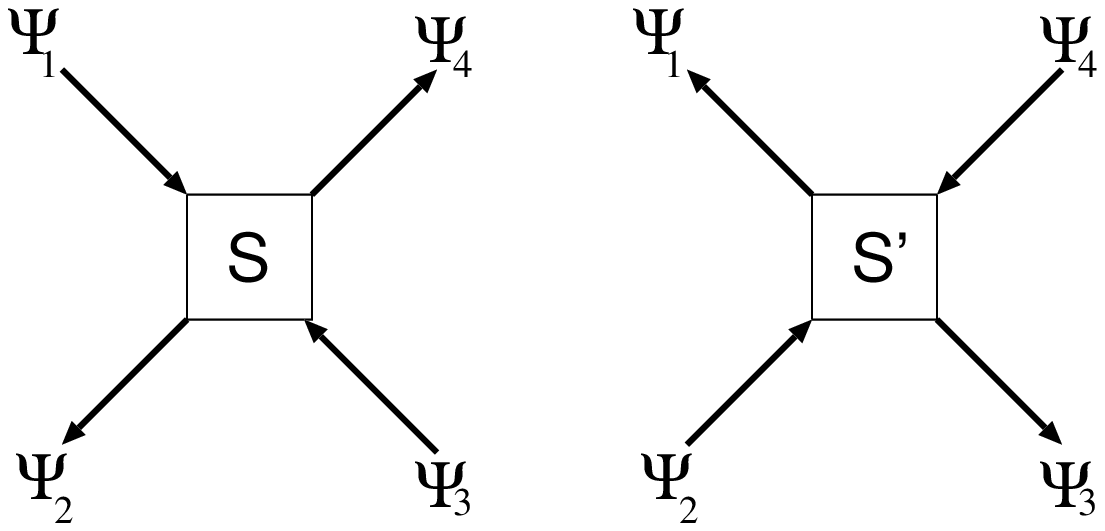}
\vspace{4.5cm}
\caption{Left: The network model in the quantum Hall regime:
  $+$ and $-$ indicate the hills and valleys of the potential, respectively.
  The scattering at the saddle points is characterized by {\bf S} matrices,
  {\bf S} and {\bf S}'. The arrows indicate the directions of the guiding center motion.
  Right: The scattering described by {\bf S}' differs from that of {\bf S}
  by $\pi/2$ rotation in the incoming and outgoing channels; $L$ width of the
  system, $M$ length of the system in the direction of the repeated
  application of the transfer matrix.}
\label{fig:network}
\end{figure}

But let us first complete the model by providing a more formal description,
especially suitable for numerical work. We consider a two dimensional
rectangular geometry with $2M\times 2L$ current loops on a square lattice
(Fig.~\ref{fig:network}).  The structure of the current flow near the nodes
described by ${\bf S}'$ is rotated by $\pi/2$ as compared with that of the
nodes described by ${\bf S}$ (Fig.~\ref{fig:network}). We have
\begin{equation}
\left(
\begin{array}{c}
\psi_{1}\\
\psi_{3}
\end{array}
\right)
= {\bf S}'
\left(
\begin{array}{c}
\psi_{2}\\
\psi_{4}
\end{array}
\right).
\label{eq:sprime_node}
\end{equation}
with the $\pi/2$ rotated scattering matrix
\begin{equation}
{\bf S}'=
\left(
\begin{array}{cc}
{\rm e}^{-i\varphi_1'}&0 \\
0& {\rm e}^{i\varphi_3'}
\end{array}
\right)
\left(
\begin{array}{cc}
-t&r \\
r&t
\end{array}
\right)
\left(
\begin{array}{cc}
{\rm e}^{i\varphi_2'}& 0\\
0&{\rm e}^{-i\varphi_4'}
\end{array}
\right),
\label{eq:decomposition2}
\end{equation}
that results in a corresponding transfer matrix ${\bf T}'$
relating
$(\psi_{4}, \psi_{3})^t$ to $(\psi_{1}, \psi_{2})^t$, 
\begin{equation}
{\bf T}'=
\left(
\begin{array}{cc}
{\rm e}^{i\varphi_4'}&0 \\
0 & {\rm e}^{i\varphi_3'}
\end{array}
\right)
\left(
\begin{array}{cc}
1/r&t/r \\
t/r&1/r
\end{array}
\right)
\left(
\begin{array}{cc}
{\rm e}^{i\varphi_1'}& 0\\
0&{\rm e}^{i\varphi_2'}
\end{array}
\right)\,,
\label{eq:decomposition_tprime}
\end{equation}
or in terms of $\Theta$,
\begin{equation}
{\bf T}'=
\left(
\begin{array}{cc}
{\rm e}^{i\varphi_4'}&0 \\
0 & {\rm e}^{i\varphi_3'}
\end{array}
\right)
\left(
\begin{array}{cc}
\cosh\Theta&\sinh\Theta \\
\sinh\Theta&\cosh\Theta
\end{array}
\right)
\left(
\begin{array}{cc}
{\rm e}^{i\varphi_1'}& 0\\
0&{\rm e}^{i\varphi_2'}
\end{array}
\right)\,.
\end{equation}

The transfer matrix ${\bf T}_{L}^{M}$ that relates the amplitudes on the left of the
sample $(\psi_{0,1},\psi_{0,2},\cdots,\psi_{{0,2L}})^t$ to those on the right
$(\psi_{M,1},\psi_{M,2},\cdots,\psi_{{M,2L}})^t$ ($L$ integer) is then defined
as follows. We divide the system along the direction of $M$ into slices, each
containing $L$ of subsequent saddle point scatterers and the $2L$
corresponding channels (Fig.~\ref{fig:network}). The latter are assumed to
have completely random phases. The amplitudes at the end of the system can
then be written in terms of a transfer matrix\footnote{We use here the
  somewhat unconventional notation that $M$ and $L$ denote the length and
  width of the system, respectively. The reason is that when discussing the
  numerical scaling in the next chapter, we prefer to use $L$ for the length
  scaling instead of $M$.}
\begin{equation}
  \label{eq:transfermatrix}
  \left(
    \begin{array}{c}
\psi_{M,1}\\
\psi_{M,2}\\
\cdot\\
\cdot\\
\cdot\\
\psi_{M,2L}
    \end{array}
\right)={\bf T}_{L}^M\, \left(
    \begin{array}{c}
\psi_{0,1}\\
\psi_{0,2}\\
\cdot\\
\cdot\\
\cdot\\
\psi_{0,2L}
    \end{array}
\right)
\end{equation}
with the product of the transfer matrices
\begin{equation}
{\bf T}_{L}^M ={\bf T}^{(M)} {\bf T}^{(M-1)} \cdots 
{\bf T}^{(2)}{\bf T}^{(1)}
\label{eq:product_transfer}
\end{equation}
where
\begin{equation}
{\bf T}^{(k)}=
{\bf V}_1^{(k)} {\bf V}_2 {\bf V}_3^{(k)} {\bf V}_4 \,.
\end{equation}
Here the matrix ${\bf V}_4$ is given by
\begin{equation}
({\bf V}_4)_{i,j}=\left\{\begin{array}{lll}
\cosh\Theta' & \quad i=j=n,& \quad n=1,\cdots,2L\\
\sinh\Theta' &
\left\{\begin{array}{l}
i=j+1=2n,\\
i=j-1=2n-1,
\end{array}\right. 
                         & \quad n=1,2,\cdots,L \\
0 &{\rm otherwise} &
           \end{array}
\right.
\end{equation}
and ${\bf V}_2$ by
\begin{equation}
({\bf V}_2)_{i,j}=\left\{\begin{array}{lll}
\cosh\Theta & \quad i=j=n, &\quad n=2,3,\cdots,2L-1\\
\sinh\Theta & 
\left\{\begin{array}{l}
i=j+1=2n+1, \\
i=j-1=2n,
\end{array}
\right.
            &\quad n=1,2,\cdots,L-1                         
           \end{array}
\right.
\end{equation}
Assuming periodic boundary conditions in the transverse direction
gives 
\begin{equation}
\begin{array}{ccc}
({\bf V}_2)_{1,1}&=({\bf V}_2)_{2L,2L}&=\cosh\Theta \\
({\bf V}_2)_{1,2L}&=({\bf V}_2)_{2L,1}&=\sinh\Theta
\end{array}
\label{eq:pbc}
\end{equation}
while for fixed boundary conditions one has to use
\begin{equation}
\begin{array}{ccc}
({\bf V}_2)_{1,1} &=({\bf V}_2)_{2L,2L}&=1 \\
({\bf V}_2)_{1,2L}&=({\bf V}_2)_{2L,1}&=0 .
\end{array}
\label{eq:fbc}
\end{equation}
All other matrix elements of ${\bf V}_2$ are 0.

When propagating from one node to the other, the probability amplitudes gain
phase factors. This is contained in the matrix elements of ${\bf V}_1^{(k)}$
and ${\bf V}_3^{(k)}$,
\begin{equation}
\label{eq:108}
({\bf V}_l^{(k)})_{i,j}=
\delta_{ij}\,{\rm e}^{i\phi_{i}^{(k)}}\qquad (l=1,3)\,.
\end{equation}
Since the distance between the nodes is random, we assume that
$\phi_{i}^{(k)}$ are independent and uniformly distributed between $[0,2\pi)$.
So far, this is the only source of randomness in the model. Note that the
phases $\varphi$ and $\varphi'$ in Eqs.~(\ref{eq:decomposition_t}) and
(\ref{eq:decomposition_tprime}) can be included in the $\phi$'s in
Eq.(\ref{eq:108}).

In this form, the transfer matrix method \cite{km93,mk81,mk83} has been
applied to the model for estimating the critical behavior of the localization
length in the quantum Hall regime \cite{cc88} to be described in more detail
in the following chapter.

Attempting some completeness, we mention at this point that some versions of
the network, which essentially reproduce features of the underlying classical
percolation problem, and with random saddle point potential values $E_{\rm c}$
have been used to obtain detailed information about the longitudinal
conductivity and several classical percolation properties
\cite{lwk93,j91,web96,m99}. A classical version of the regular network
consisting of coupled metallic wires has been used to calculate the Hall
conductance. Quantization in integer (positive and negative) units of
$e^{2}/h$ has been predicted consistent with earlier findings for the pure
quantum Hall case without any disorder \cite{k93,tkn82}. The model has been
also used to study the localization problem in two dimensions in the presence
of a random magnetic field \cite{lck94,kfl95}. As our emphasis will be on the
universal features of the network model in the forthcoming chapters, we will
not go into the details of these works.

\section{The Localization-Delocalization Transition in the Network Model}
\label{sec:chalkercoddington}

In this chapter we will explore the localization properties of the original
Chalker-Coddington network introduced in the previous Chapter
\ref{sec:randomnetwork}. As this has been a key issue for establishing the
model, this will be done in some detail. Necessarily, for obtaining
quantitative information about the critical behavior, numerical methods will
be used. We attempt to provide a complete overview of the statistical
properties of the energy spectrum and the wave functions near the critical
point, as far as it is presently available.

The model has been the subject of numerous numerical studies which are all
similar in spirit but different in the details
\cite{lwk93,lc94,wlw94,kha95,cd95}. We provide here results obtained recently
by the numerical scaling method \cite{mk81,mk83} which have been undertaken in
order to improve the treatment of corrections to scaling. For obtaining
reliable and precise results for the critical exponent, the latter has been
shown to be decisively important \cite{h94,so99}. To our great disappointment,
however, as we will show below, up to now it has not been achieved to remove
corrections to scaling to such a degree that the precision of the exponent of
the Chalker-Coddington network model can compete with the results obtained for
other models in particular the random Landau model, see Tab. 1!

We will also discuss the critical properties of the wave functions as well as
the eigenenergy statistics. Due to its simplicity, the network model enables
us to investigate them in detail, and because of universality, they are
supposed to be generally valid for two dimensional electron systems in high
perpendicular magnetic fields.

To study the localization-deloca\-li\-za\-tion transition, we investigate the
scaling properties of the localization length in a quasi-one dimensional long
strip for which the localization length is calculated by the transfer matrix
method and apply the finite size scaling method developed earlier for
estimating the asymptotic value in the thermodynamic limit. The value of the
critical exponent $\nu$ in the network model \cite{cc88,lwk93,eb98,jw98} is
consistent with, though considerably less accurate than, earlier results
obtained for several very different models such as the random Landau model
with a white noise potential \cite{aa85,aa85a} and the random matrix model
including spatial correlations of the randomness \cite{hk90,h95}. This
supports strongly that the model of Chalker and Coddington belongs to the same
universality class as the models considered previously, in spite of the
intriguingly large corrections to scaling which defy the high accuracy in the
estimate of the critical exponents.

\subsection{Numerical Scaling at the Localization-Delocalization Transition}

To determine the critical behavior quantitatively, we use the numerical finite
size scaling approach. We define a quantity $F_{L}$ which is a function of a
set of parameters ${x_i}$ characterizing the system, such as the Fermi energy,
parameters characterizing the disorder, as the variance and the correlation
length, the magnetic field, and the system size $L$ \cite{cardy96},
\begin{equation}
F_{L}=f(\{x_i\},L) .
\end{equation}
We assume the existence of a scaling law such that 
$F_{L}$ can be expressed as
\begin{equation}
F_{L}=F(\chi L^{1/\nu},\phi_1 L^{y_1},\phi_2 L^{y_2},\cdots)\, ,
\label{eq:scaling}
\end{equation}
with $\chi$ being the relevant scaling variable and $\phi_i$ denoting the
irrelevant ones. The latter nevertheless can cause corrections to scaling as
long as the system size is finite. Therefore, in a numerical calculation they
must not be ignored. These variables characterize distances from the critical
point, $\nu$ is the critical exponent and $y_i<0$ are the exponents of the
irrelevant scales. Eventually, in the limit of infinite system size, only the
relevant scaling variable survives, and Eq.~(\ref{eq:scaling}) becomes
\begin{equation}
f_{L}=F_1\left(\frac{L}{\xi}\right) ,
\label{eq:fss}
\end{equation}
with $\xi\sim \chi^{-\nu}$. The quantity $\chi$ as a function of some control
parameter, say $x$, can be expanded near the critical point $x_{\rm c}$
\begin{equation}
\chi=\chi_1 (x-x_{\rm c})+\chi_2 (x-x_{\rm c})^2+\cdots .
\label{eq:chi_expansion}
\end{equation}

Which quantity should be used as the scaling variable $F_{L}$? It should be a
quantity that shows a singularity in the limit of infinite system size as one
crosses the localization-delocalization critical point. Furthermore, it should
be a quantity to be determined numerically easily, precisely and effectively.
There are several such quantities, such as for instance the level spacing
distribution \cite{zk97} or the conductance (see below). Here, we consider the
renormalized localization length in the {\em finite} system (MacKinnon-Kramer
variable), $\Lambda$, introduced previously \cite{mk81} which we will now
define.

Consider a very long strip of width $L$. This is a quasi-one dimensional
system and all the states are expected to be localized. The localization
length $\lambda(L; x_{1},x_{2},\ldots)$ is a function of the width of the
strip, $L$. Now let us define
\begin{equation}
\Lambda(L; x_{1},x_{2},\ldots) = \frac{\lambda(L; x_{1},x_{2},\ldots)}{L} \,.
\end{equation}
If the system parameters are such that $\lambda(L\to\infty;
x_{1},x_{2},\ldots)$ remains finite, the system is in the localized regime and
$\Lambda(L\to\infty; x_{1},x_{2},\ldots)\to 0$. The localization length is
then given by $\xi=\lambda(L\to\infty; x_{1},x_{2},\ldots)$. On the other
hand, if $\lambda(L\to\infty; x_{1},x_{2},\ldots)$ increases faster than
$\propto L$, the system is in the delocalized state, and $\lambda(L\to\infty;
x_{1},x_{2},\ldots)$, the correlation length in the metallic regime,
corresponds to the inverse of the dc-conductivity \cite{mk81,mk83}. The
critical point is defined by the condition $\Lambda(L\to\infty; x_{1\rm
  c},x_{2\rm c},\ldots)={\rm const}=\Lambda_{\rm c}$ (critical
MacKinnon-Kramer variable).

The most efficient way to calculate $\lambda(L)$ is the transfer matrix
method \cite{km93}. We define the product of the transfer matrices
Eq.~(\ref{eq:product_transfer}) and consider the limit
\begin{equation}
{\Gamma}=\lim_{M\rightarrow \infty}({\bf T}_L^M {\bf T}_L^{M\dagger})^{1/2M}.
\label{eq:lyapunov}
\end{equation}
The theorem by Oseledec \cite{oseledec68,r82,k85,cl90} guarantees that
$\Gamma$ has always positive eigenvalues, which are denoted as
$\exp(\pm\gamma_i)$ where $\gamma_i (> 0)$ can be interpreted as the
exponential change of the wave function for a single slice. The smallest value
of $\gamma_i$ is the inverse of the quasi-one dimensional localization length
\cite{km93},
\begin{equation}
\frac{1}{\lambda(L)}=\min \{\gamma_i\}\,.
\label{eq:lyapunov2}
\end{equation}
This definition can be shown to be equivalent to the one in terms of the
exponential decay of the Green function which could also be used to study
$\lambda(L)$ \cite{mk93}. If one is interested only in the localization length
the transfer matrix method has been shown to be superior.

\subsection{Numerical Results near the Quantum Critical Point}

The two dimensional electron systems in high magnetic fields are characterized
by many control parameters such as Fermi energy, magnetic fields as well as
parameters describing the properties of randomness.  In the Chalker-Coddington
model, however, all the informations are contained in a single parameter, the
transmission at the saddle point.  In the actual simulation, it is convenient
to use as control parameter
\begin{equation}
x=-\ln\sinh\Theta\,,
\label{eq:x_theta}
\end{equation}

With this choice of $x$, we obtain from Eq. (\ref{eq:theta_r})
\begin{equation}
\label{eq:tofx}
t=\frac{1}{\sqrt{{\rm e}^{2x}+1}}\, \quad\quad
r=\frac{1}{\sqrt{{\rm e}^{-2x}+1}}\,.
\end{equation}
Due to the particle-hole symmetry ($r\leftrightarrow t$) the localization
length is an even function of $x$.  Comparing Eqs.~(\ref{eq:result}) and
(\ref{eq:tofx}) we note that $x$ can be interpreted as the energy measured
from the center of the Landau band.

In the actual simulation, we need to simulate not too small systems so that
only a single relevant scaling variable and at most one irrelevant variable
are sufficient for fitting the data. In this case, the scaling form
Eq.~(\ref{eq:scaling}) reads
\begin{equation}
\Lambda(L)=F(\chi L^{1/\nu},\phi L^{y})\,,
\label{eq:lambda_scaling}
\end{equation}
where $\chi$ is related to $x$ via Eq.~(\ref{eq:chi_expansion}).  We consider
in the calculation the region $x\ll 1$ so that $\chi\sim x$ and $\phi=
\mathrm{const}$.

As noted above, due to particle-hole symmetry $\Lambda$ is an even analytic
function of $x$ as long as $L$ is finite. By assuming $xL^{1/\nu}$ to be
sufficiently small in order to truncate the expansion after the second order,
Eq.~(\ref{eq:lambda_scaling}) can be expressed as
\begin{equation}
\Lambda(L)=\Lambda_{\rm c}+a_1(xL^{1/\nu})^2+a_2(xL^{1/\nu})^4
+b_0 L^y+c_0 L^{2y}+\cdots ,
\label{eq:scalingexpansion}
\end{equation}
Previously, this non-linear fitting scheme has been working perfectly well for
the three dimensional Anderson transition \cite{so99,so97,sok00,smo01} as well
as in the case of two dimensional systems with strong spin-orbit coupling
\cite{aso02,aso04}.
\begin{figure}[hbtp]
\vspace{0.5cm}
\begin{center}
\includegraphics[width=0.6\textwidth,angle=270]{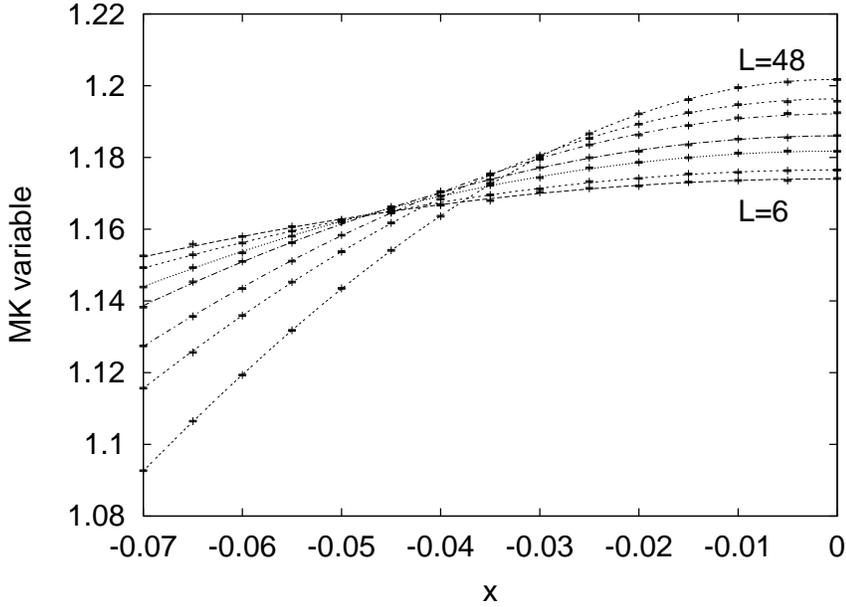}
\caption{MacKinnon-Kramer variable
  $\Lambda(L)=\lambda(L)/L$ as a function of $x$ for $L=6,8,12,16,24,32$ and
  48.  The relative uncertainty of the data is 0.02$\%$. The curves are results
  of the fitting according to Eq. (\ref{eq:scalingexpansion}) with
  $\Lambda_{\mathrm{c}}=1.23, \nu=2.48$ and $y=-0.53$. The fitting, however,
  is unstable, so the curves should be regarded as guides to the eye.  }
\label{fig:lambda_rawdata}
\end{center}
\vspace{5mm}
\end{figure}

High precision data are required for a precise determination of the exponent.
As seen in Fig.~\ref{fig:lambda_rawdata}, for such data the corrections to
scaling are not negligible \cite{eb98}.  Unfortunately, a stable fit that
takes into account these corrections has not yet been found
\cite{evers03,hs04}. As a result, there has not been any significant
improvement of the precision of the estimate of $\nu$ over the original
estimate of $\nu=2.5\pm 0.5$ by Chalker and Coddington \cite{cc88}.

Fluctuations of the transmission properties of the saddle points cause another
type of randomness in addition to the phase randomness assigned to the wave
function when traveling from one node to another. These fluctuations are
equivalent to randomness in the mass of the Dirac Hamiltonian to be described
later in the chapter \ref{sec:hamiltonians}.  It has been numerically
demonstrated that this effect gives larger --- but irrelevant --- corrections
to scaling \cite{eb98}.

\subsection{The Critical Properties of the Wave Functions}
\label{subsec:wf_qcp}

Exactly at the critical point $t=r=1/\sqrt{2}$ $(x=-\ln\sinh\Theta=0)$, the
wave function is delocalized. It is now well-established that due to the
divergence of the length scale, the wave function at the critical point shows
multi-fractal behavior \cite{aoki83,se84,lfsg94}. This self-similar structure
is in principle reflected in an anomalous temperature behavior of the
diffusion constant, the energy level statistics, the conductance distribution,
and many other properties.

\begin{figure}[btp]
\begin{center}\leavevmode
\includegraphics[width=5cm]{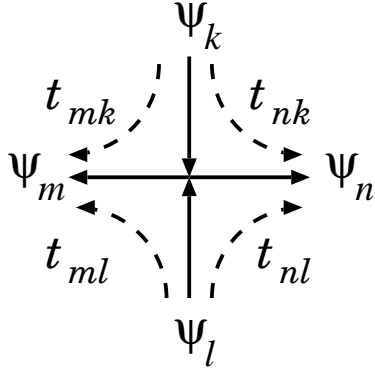}
\caption{The stationary wave function relation characterized by the
scattering matrix amplitudes. Incoming waves and outgoing waves are
$\psi_{l}$,$\psi_{k}$ and $\psi_{m}$,$\psi_{n}$, respectively.}
\label{fig:wf_smatrix}
\end{center}
\vspace{5mm}
\end{figure}

Since the Chalker-Coddington model is characterized by the scattering matrices
at saddle points rather than by the Hamiltonian, the wave function must be
determined from the scattering matrices. Let the amplitude on the link $l$
be $\psi_l$. As explained above, the stationary states of the model must
satisfy \cite{km95,eks89}
\begin{equation}
\psi_m=t_{mk}\psi_k+t_{ml}\psi_l,
\end{equation}
where $t_{mk}$ and $t_{ml}$ relate the amplitudes $\psi_k$
and $\psi_l$ to $\psi_m$ (Fig.~\ref{fig:wf_smatrix}).
This can be rewritten as
\begin{equation}
{\bf U}(E)\Psi=\Psi\,,
\label{eq:unitary_develop}
\end{equation}
with ${\bf U}(E){\bf e}_l=t_{ml}(E){\bf e}_{m}+t_{nl}(E){\bf e}_n$, ${\bf
  e}_i$ being the unit vector with the $i$th component unity,
$|t_{mk}|^2=T=1-|t_{ml}|^2 := 1/2$ at the critical point, and $\Psi$ the
  vector with the components $\psi_{m}$. Again, we assume
for simplicity that randomness enters only via the phases of the amplitudes.

Equation~(\ref{eq:unitary_develop}) has non-zero scattering solutions only for
certain discrete energies $E_{\nu}$, which determine the eigenstates and
eigenenergies of the system \cite{fertig88}. Solving
Eq.~(\ref{eq:unitary_develop}), however, is in general very complicated for a
disordered system. We will derive an alternative method by establishing
equivalent Hamiltonians later in the chapter \ref{sec:hamiltonians}.

For the moment, we proceed by calculating instead the return probability
\cite{hk99} from the equation of motion determined by
Eq.~(\ref{eq:unitary_develop}). We start by introducing the time $t_0$ during
which the incident wave packet passes the scatterer and measure time $t$ in
units of $t_0$.  We can then define the time evolution operator
\begin{equation}
{\bf U}(t)={\bf U}^t(E)
\end{equation}
where $t$ is an integer.

The probability $P(t)$ of return within the period of time $t$ 
is determined by the probability density of the
wave packet at time $t$ near a given scatterer, say located at the origin,
which was previously located there at $t=0$. This is defined by
\begin{equation}
P(t)=|\langle0|{\bf U}(t)| 0\rangle|^2\,.
\end{equation}
The return probability is related to the fractal dimension $D(2)$ defined by
the second moment of the probability density, the inverse participation number
$p_{2}(L)$
\begin{equation}
  \label{eq:d2}
p_{2}(L):=\langle|\Psi(\vv{r})|^{4}\rangle\propto \frac{1}{L^{2+D(2)}}\,.
\end{equation}
with $\langle \ldots \rangle $ denoting a configurational average.  If we
integrate $p_2(L)$ over the two dimensional system and then take the inverse,
we have $L^{D(2)}$ and we can interpret this quantity as the portion of space
where the wave function amplitude is significant.

Let the wave packet spread to the radius $r(t)$ after time $t$, the number of
sites occupied becomes $\sim r(t)^{D(2)}$. On the other hand, the
conductivity at the quantum Hall transition is finite, and from the Einstein
relation
\begin{equation}
\sigma=e^2\rho D\,,
\label{eq:einstein}
\end{equation}
($\rho$ the density of states per volume and $D$ the diffusion constant),
we see that $D$ is finite, hence $r(t)\sim t^{1/2}$.
One then finds
\cite{kpg92,hs94,bhs96,ko95,ko96}
\begin{equation}
P(t)\propto t^{-D(2)/2}\,.
\end{equation}
In general dimension the diffusion radius $r$ is proportional
to $t^{1/d}$\cite{ok97}, hence
\begin{equation}
P(t)\propto t^{-D(2)/d}\,.
\end{equation}
For  the spatial dimension $d=2$, we can estimate $D(2)=1.52\pm 0.06$
from numerical simulations \cite{hk99}.
This is in agreement with the results in the continuum model
\cite{aoki83,pj91} as well as in the tight binding model \cite{hs94}. 

A similar idea has been developed to calculate the local density of states
\cite{hk97} from which one can calculate the multi-fractal exponents $D(q)$
that are defined by the higher moments of the density,
\begin{equation}
  \label{eq:dq}
  p_{q}(L):= \langle |\Psi(\vv{r})|^{2q}\rangle 
\propto \frac{1}{L^{2+(q-1)D(q)}}
\propto\frac{1}{L^{2+\tau(q)}}\,,
\end{equation}
The moments of the density characterize the degree of localization of the wave
function. For plane waves, $p_{q}(L)\propto L^{-2q}\to 0$ in the thermodynamic
limit. For localized states $p_{q}(L)\propto L_{0}^{-2q}\to \mathrm{const}
\neq 0$.  Generally, the fractal dimensionalities $D(q)$ depend on $q$ and
$\tau(q)$ is non-linear in $q$. This is the celebrated multi-fractal behavior
\cite{ny03}.  It means that the complex structure of wave function at the
critical point can not be described by single dimension $D(2)$ but infinite
number of generalized dimensions are required to characterize it.

A convenient quantity to summarize the behavior of the multi-fractal exponents
is to introduce a Legendre transformation on $\tau(q)$ by defining
\cite{hjkp86,l90,dl91}
\begin{equation}
  \label{eq:legendre}
  \alpha=\frac{{\rm d}\tau(q)}{{\rm d}q}
\end{equation}
and the $f(\alpha)$-spectrum
\begin{equation}
  \label{eq:falpha}
  f(\alpha)=\alpha q -\tau(q)\,.
\end{equation}
The physical meaning of the quantity $f(\alpha)$ can be understood by relating
it to the probability distribution of the random amplitudes of $\Psi$,
$P_{\Psi}(|\Psi|^{2})$. Starting point is the identity between the latter and
the distribution, $P_{\alpha}(\alpha)$, of the random variable
\begin{equation}
  \label{eq:randomalpha}
\alpha:= -\frac{\ln{|\Psi|^{2}}}{\ln{L}},  
\end{equation}
\begin{equation}
  \label{eq:probablityofalpha}
  P_{\alpha}(\alpha){\rm d}\alpha = P_{\Psi}(|\Psi|^{2}){\rm d}|\Psi|^{2}\,.
\end{equation}
The inverse participation numbers are obtained in terms of these distributions
\begin{equation}
  \label{eq:averageparticipation}
  p_{q}(L)=\int{\rm d}|\Psi|^{2}\,|\Psi|^{2q}P_{\Psi}(|\Psi|^{2})
=\int{\rm d}\alpha \,e^{-\alpha q\ln{L}}P_{\alpha}(\alpha)\,.
\end{equation}
In order to obtain the correct dependence of $p_{q}(L)$ on the system size the
distribution of $\alpha$ must be of the form
\begin{equation}
  \label{eq:p(alpha)}
  P_{\alpha}(\alpha)\propto e^{[f(\alpha)-2]\ln L}\,.
\end{equation}
For very large $L$, the average in Eq.~(\ref{eq:averageparticipation}) is
dominated by the maximum of $f(\alpha)$ at $\alpha_0$. Thus,
$\alpha_0$ is the exponent of the system size dependence of the {\em
  typical} value of the probability,
\begin{equation}
  \label{eq:typical}
  (|\Psi|^{2})^{\rm typ}(L):= e^{\langle\ln(|\Psi|^{2})\rangle}
\propto \frac{1}{L^{\alpha_0}}
\end{equation}

\begin{figure}[htbp]
\vspace{8mm}
\hspace{7mm}
\includegraphics[width=11cm]{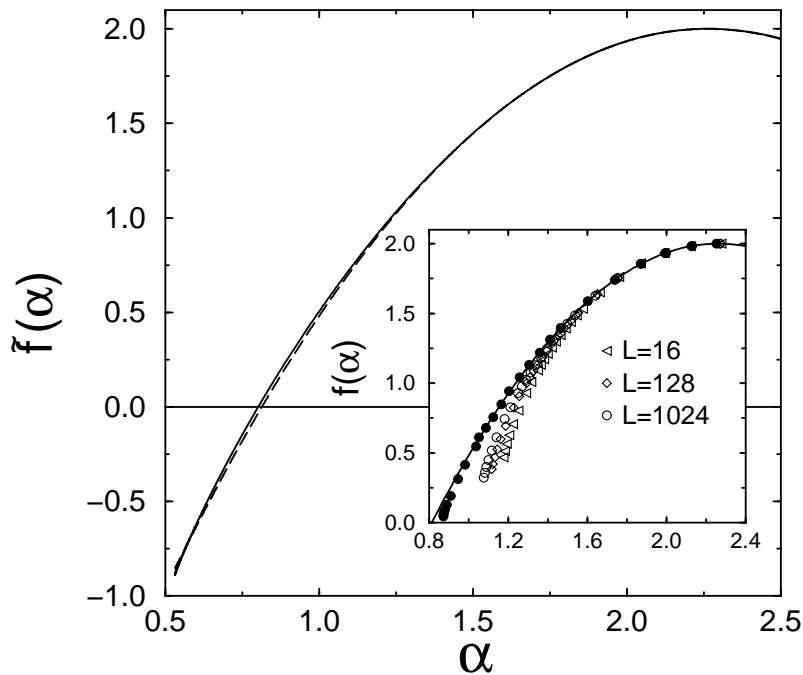}
    \caption{The $f(\alpha)$-spectrum at the quantum Hall transition 
      determined numerically (full line) and the result of an analytical
      conjecture (dashed line) with $\alpha_0=2.262$. Inset: Spectra
      of typical eigenfunctions for different system sizes $L$ and
      extrapolated with $L\to \infty$ [from Ref. \protect\cite{emm01}].}
    \label{fig:falphaspectrum}
\end{figure}

Figure~\ref{fig:falphaspectrum} shows some numerical results for $f(\alpha)$
together with the result obtained from the Dirac model with random vector
potential (\cite{lfsg94} see chapter \ref{sec:lsfg94} below) which is exactly
parabolic
\begin{equation}
  \label{eq:diracfalpha}
  f(\alpha)=2-\frac{(\alpha-\alpha_0)^{2}}{4(\alpha_0-2)}
\end{equation}
with $ \alpha_0=2+\Delta_{A}/\pi$ where $\Delta_{A}$ is the variance
of the random vector potential which is in that case equivalent to the random
phases associated with the links of the network.

\begin{figure}[htbp]
  \begin{center}
\includegraphics[width=12cm]{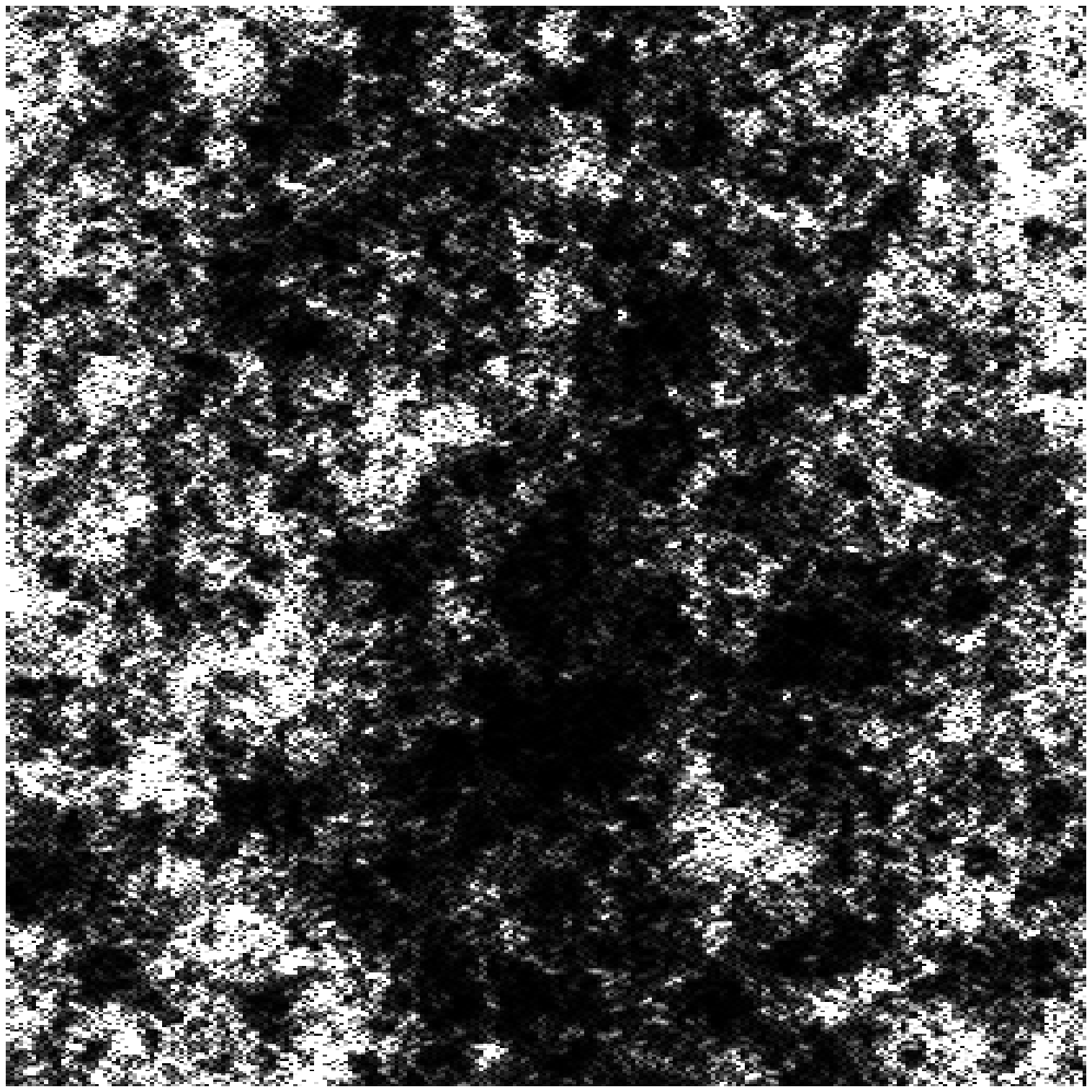}
   \caption[]{Squared amplitude of a critical wave 
     function in a Chalker-Coddington network of $256 \times 256$ saddle
     points. Darker areas denote lower square amplitude. No characteristic
     lengths scale can be identified [Figure taken from \protect\cite{km99}].
     }
\vspace{12mm}
    \label{fig:multifractalwave}
 \end{center}
\end{figure}

Again, this is in agreement with the diagonalization analysis \cite{pj91}.
Together with the analytical results from the Dirac model, which will be
described below, this provides strong evidence for the overall conjecture that
the Integer Quantum Hall Effect is a quantum critical phenomenon. A graphical
representation of a multi-fractal wave function at the critical point is shown
in Fig.~\ref{fig:multifractalwave}.

The fractal dimensionality of the wave function can be related to the
Mac\-Kin\-non-Kramer parameter $\Lambda_{\rm c}$ by assuming conformal
invariance \cite{cc88,cardy84,janssen94,janssen98,janssen01}. The quantity
$\alpha_0$, is related to $\Lambda_c$ by (see Sect. \ref{sec:pointcontact} for
more details)
\begin{equation}
\alpha_0=2+\frac{1}{\pi\Lambda_{\rm c}}\, ,
\label{eq:janssen}
\end{equation}
which is in excellent agreement with the numerical results
\cite{eb98,emm01,janssen01,dfj96}.

\subsection{Energy Level Statistics near the Critical Point}
\label{subsec:els_qcp}

Random matrix theory is a very powerful tool to characterize the properties of
complex systems \cite{mehta,haake02}. It can also be applied to the
localization-delocalization transition. In the metallic regime, the overlap of
the wave functions is strong, which results in strong level repulsion. In the
limit of large system size, the energy level spacing $s$ (in units of the mean
level spacing) is given approximately by the distribution function
\begin{equation}
P(s)\propto s^\beta e^{-A(\beta)s^2}\,.
\end{equation}
This is often denoted as the Wigner surmise. The value of the parameter
$\beta$ ($=1,2,4$) is determined by the symmetry of the system. The value of
the constant $A$ depends on $\beta$. When the system has both time reversal
and spin rotation symmetry, $\beta=1$. This is called the {\em orthogonal}
symmetry class. If the system has only time reversal symmetry but spin
rotational symmetry is broken by spin-orbit interaction, $\beta=4$. This
characterizes the {\em symplectic} class where level repulsion is strongest.
If $\beta=2$, time reversal symmetry is broken, irrespectively of whether or
not spin rotational symmetry is present. This is the {\em unitary} symmetry
class. Symmetries classes are indeed very important ingredients for
characterizing the universal properties of a quantum phase transitions. We will
discuss this in more detail later in the chapter \ref{sec:symmetries}.

On the other hand, when the states are localized, the correlations between
eigenenergies vanish in the limit of large system size. Then, the level
spacing distribution $P(s)$ is Poissonian, $P(s)=\exp(-s)$. When the system
size is finite, the correlations between energy values are still present and
the spacing distribution deviates from the Poissonian. The deviation from the
limit of infinite system size can be described by using a single parameter
scaling assumption Eq.~(\ref{eq:fss}) \cite{ook96,sssls93,azks88,oo93,zk97}
\begin{equation}
P(s)=f\left(s,\frac{L}{\xi}\right),
\label{eq:elsscaling}
\end{equation}
where $\xi$ is the localization length. For the ordinary metal-insulator
transition, the above equation has two branches corresponding to metallic and
insulating phases. In the Chalker-Coddington model, we do not have a true
metallic phase. Therefore, we expect to find only a single branch.

Exactly at the critical point, $\xi$ diverges and $P(s)$ becomes size
independent. It is neither a Poissonian nor it corresponds to the Wigner
surmise, but has both characteristics. It grows as $s^\beta$ for $s\ll 1$ as
in the case of the Wigner surmise and it decays according to $\exp(-As)$ for
$s \gg 1$ like the Poisson distribution \cite{zk97,evangelou94,koso96}.

We now want to obtain the energy level statistics from
Eq.~(\ref{eq:unitary_develop}). To achieve this, we need a method to extract
information about energy eigenvalues from the transfer matrix approach. This
requires the construction of an equivalent Hamiltonian which we will do below
in great detail. For the present purposes it is sufficient to consider the
following.
 
First we note that the eigenvalue equation of the unitary operator ${\bf
  U}(E)$ is
\begin{equation}
{\bf U}(E)\Psi_\nu=\e^{i\omega_\nu}\Psi_\nu \,.
\end{equation}
The eigenenergies $E_\nu$'s that correspond to the stationary states are
obtained from the condition $\exp[i\omega_\nu(E_\nu)]=1$.
\begin{figure}[btp]
\begin{center}\leavevmode
\includegraphics[width=10cm,height=7cm]{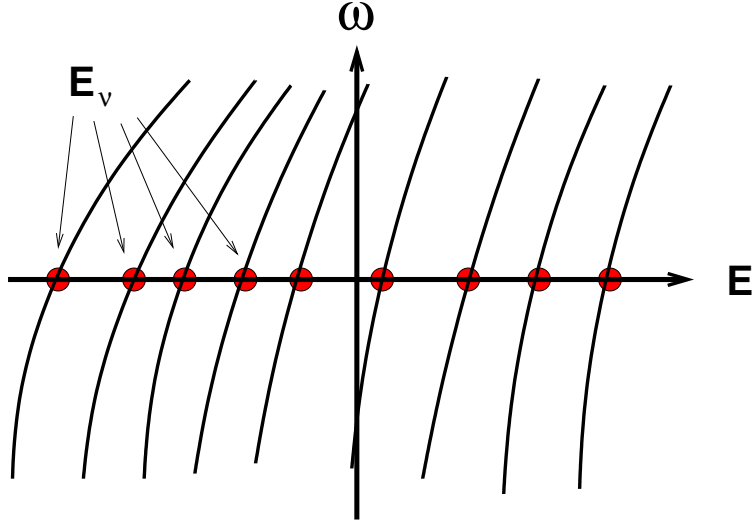}
\caption{Scheme of the quasi-energies
  $\omega_\nu$ as a function of $E$. The condition $\omega_\nu=0$ gives the
  eigenenergies, $E_\nu$.}
\label{fig:wf_smatrix1}
\end{center}
\end{figure}

The statistics of the energy levels obtained from the condition $\omega(E)=0$
is the same as that of levels obtained by $\omega=\Omega$, since the latter
correspond to the solution for $U'(E)=\e^{-i\Omega}U(E)$ which belongs to the
same universality class. This together with the level repulsion leads to the
conjecture that $\omega_\nu (0)$ obey the same statistics as $E_\nu$
\cite{km97}. The quantities $\omega_\nu (0)$ are often called quasi-energies.
This assumption greatly simplifies the numerical calculation. The results, for
example the form of $P(s)$ at the critical point, agree in fact very well with
those obtained for the continuum model \cite{oo95}.  Due to the simplicity of
the network model and the possibility of using the quasi-energy concept, the
level statistics can be investigated in detail.

Another important quantity to characterize a random sequence of energy levels
is the so-called number variance defined by \cite{mehta}
\begin{equation}
\Sigma_2 (N)=\langle (n-N)^{2} \rangle = \langle n^{2} \rangle- N^{2}
\end{equation}
where $n$ is the number of levels in a randomly chosen interval $N\Delta$
($\Delta$ average level spacing), $\langle\cdots\rangle$ denotes the
configurational average, and $N=\langle n\rangle$. In the insulating region,
$\Sigma_2(N)=N$. In the metallic region, it increases only logarithmically
$\Sigma_2(N)\sim \ln N$ due to the level repulsion that makes the spectrum
rigid. At the metal-insulator transition,
\begin{equation}
\lim_{N\rightarrow \infty}\frac{\Sigma_2 (N)}{N}=\chi ,
\end{equation}
with $0<\chi<1$ \cite{azks88}.

The value $\chi=1$ implies that the system is an insulator with no level
repulsion, and $\chi=0$ is equivalent to metallic behavior with maximal level
repulsion. The fact that $0<\chi$ means that the level repulsion is weakened
as compared to the metallic limit. This is due to the fact that the wave
function has a very sparse multi-fractal structure. Thus, $\chi$ reflects the
multi-fractal behavior of the wave functions. In fact, the quantity $\chi$ and
the fractal dimension $D(2)$ are related via \cite{ckl96}
\begin{equation}
\chi=\frac{d-D(2)}{2d} .
\label{eq:ckl_compressibility}
\end{equation}
Inserting $D(2)=1.52\pm 0.06$ and $d=2$ yields $\chi=0.120\pm 0.005$.  Klesse
and Metzler have estimated $\chi$ from the quasi-energies, $\chi=0.124\pm
0.006$, in agreement with Eq.~(\ref{eq:ckl_compressibility}) \cite{km97}.
This agreement, however, is not exact since Eq. (\ref{eq:ckl_compressibility})
holds only approximately as has been discussed in Refs.\cite{em00,em00a}.

\section{Linear Electrical Transport at Zero-Temperature}
\label{sec:transport}

The linear electrical conductance of a quantum system at zero temperature is
related to quantum mechanical transmission via the Landauer formula
\cite{landauer57,landauer75}. Thus, the network model is perfectly designed to
provide quantitative information about the linear conductance. If the model
describes the physics near the critical point, one can expect also that it is
especially suitable for the critical conductance. The study of the linear
conductance tells us the importance of the conductance distribution instead of
the averaged conductance, which converges to a form independent of the system
size.  The qualitative behavior of the conductance distribution is expected to
be valid even in different situations such as 4-terminal conductance
measurement.

\begin{figure}[htpb]
\begin{center}\leavevmode
\includegraphics[width=0.6\linewidth]{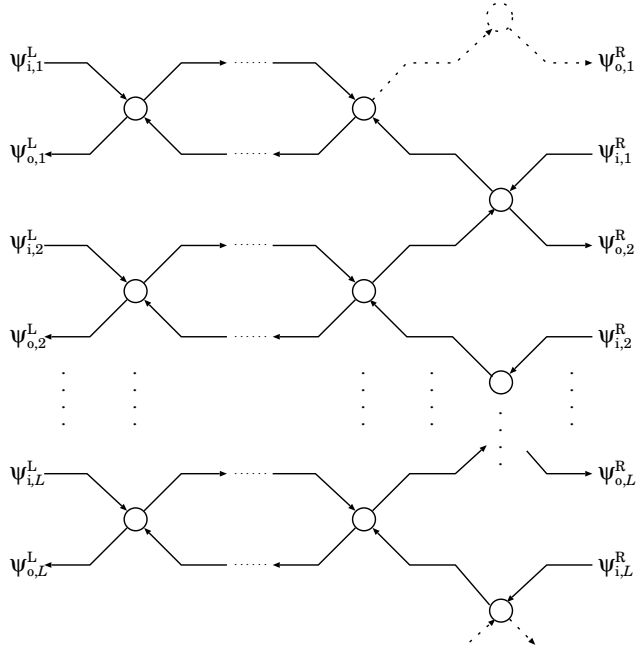}
\caption{
  The network model as modified for calculating the two terminal conductance.
  The amplitudes $\psi_{{\rm i},i}^{\rm L}$ and $\psi_{{\rm o},i}^{\rm L}$
  $(i=1,\cdots,L)$ represent the incoming and the outgoing currents on the
  left hand side, respectively, $\psi_{{\rm i},i}^{\rm R}$ and $\psi_{{\rm
      o},i}^{\rm R}$ $(i=1,\cdots,L)$ represent the amplitudes on the right
  hand side of the sample.}
\label{fig:cond_geometry}
\end{center}
\end{figure}

In this chapter we want to discuss the quantum conductance and its
distribution function at the critical point at absolute zero of the
temperature. For the conductance, we use a slightly modified random
network model (Fig.~\ref{fig:cond_geometry} and \cite{cf97}).

\subsection{The Transfer Matrix and the Conductance}
\label{sec:4.1}

The general wave function on the left hand side of the system is a
superposition of the incident and reflected amplitudes, $\{\psi_{{\rm
    i},i}^{\rm L}\}$ and $\{\psi_{{\rm o},i}^{\rm L}\}$, respectively. These
are related to those on the right hand side, $\{\psi_{{\rm i},i} ^{\rm R}\}$
and $\{\psi_{{\rm o},i}^{\rm R}\}$, via the transfer matrix of the system of
the lengths $M$ and the width $L$, ${\bf T}_L^M$,
\begin{equation}
\label{cond01}
\left(
\begin{array}{c}
\psi_{{\rm o},1}^{\rm R}\\
\psi_{{\rm i},1}^{\rm R}\\
\vdots\\
\psi_{{\rm o},L}^{\rm R}\\
\psi_{{\rm i},L}^{\rm R}
\end{array}
\right)
={\bf T}_L^M
\left(
\begin{array}{c}
\psi_{{\rm i},1}^{\rm L}\\
\psi_{{\rm o},1}^{\rm L}\\
\vdots\\
\psi_{{\rm i},L}^{\rm L}\\
\psi_{{\rm o},L}^{\rm L}
\end{array}
\right).
\end{equation}
By introducing an operator ${\bf U}$ which changes the order of the components,
\begin{equation}
\label{cond02}
\left(
\begin{array}{c}
\psi_{{\rm i},1}^{\rm L}\\
\psi_{{\rm o},1}^{\rm L}\\
\vdots\\
\vdots\\
\psi_{{\rm i},L}^{\rm L}\\
\psi_{{\rm o},L}^{\rm L}
\end{array}
\right)
={\bf U}
\left(
\begin{array}{c}
\psi_{{\rm i},1}^{\rm L}\\
\vdots\\
\psi_{{\rm i},L}^{\rm L}\\
\psi_{{\rm o},1}^{\rm L}\\
\vdots\\
\psi_{{\rm o},L}^{\rm L}
\end{array}
\right)\,, \qquad\quad
\left(
\begin{array}{c}
\psi_{{\rm o},1}^{\rm R}\\
\psi_{{\rm i},1}^{\rm R}\\
\vdots\\
\vdots\\
\psi_{{\rm o},L}^{\rm R}\\
\psi_{{\rm i},L}^{\rm R}
\end{array}
\right)
={\bf U}
\left(
\begin{array}{c}
\psi_{{\rm o},1}^{\rm R}\\
\vdots\\
\psi_{{\rm o},L}^{\rm R}\\
\psi_{{\rm i},1}^{\rm R}\\
\vdots\\
\psi_{{\rm i},L}^{\rm R}
\end{array}
\right),
\end{equation}
one can write
\begin{equation}
\label{cond03}
\left(
\begin{array}{c}
\vv{\psi}_{\rm o}^{\rm R}\\
\vv{\psi}_{\rm i}^{\rm R}
\end{array}
\right)=
{\bf U}^\dagger {\bf T} {\bf U}
\left(
\begin{array}{c}
\vv{\psi}_{\rm i}^{\rm L}\\
\vv{\psi}_{\rm o}^{\rm L}
\end{array}
\right)
\end{equation}
where $\vv{\psi}_{\rm o}^i (i={\rm R,L})$ denotes the vector of amplitudes
$(\psi_{{\rm o},1}^i,\psi_{{\rm o},2}^i,\cdots,\psi_{{\rm o},L}^i)^\mathrm{T}$
and $\vv{\psi}_{\rm i}^i (i={\rm R,L})$ is $(\psi_{{\rm i},1}^i,\psi_{{\rm
    i},2}^i,\cdots,\psi_{{\rm i},L}^i)^\mathrm{T}$ ($^\mathrm{T}$ denotes the
transposed vector).

The linear zero-temperature conductance $G$ is given by the quantum
transmission probability through the system \cite{landauer57,landauer75,fl81}.
\begin{equation}
\label{eq:landauer}
G=\frac{e^2}{h}{\rm Tr}\, {\bf t}{\bf t}^\dagger\,. 
\end{equation}
Thus, a relation between the transfer matrix and the transmission matrix
$t_{ij}$ ($i,j=1\ldots L$) is required. In the following paragraph this
relation will be established by writing the wave function on the left and
right hand sides of the scatterer in terms of $L\times L$ transmission and
reflection matrices ${\bf t}$ and ${\bf r}$ \cite{p84,pa86}.

When a flux of probability amplitude is injected into the $i$th scattering
channels from the left and from the right, represented by $\psi_{{\rm
    i},i}^{\rm L}$ and $\psi_{{\rm i},i}^{\rm R}$, respectively, the total
scattering wave function in the $i$th channels is a superposition of the
incident wave and the reflected and transmitted waves from all of the
channels. This is described by the scattering matrix ${\bf S}$
\begin{equation}
\label{eq:scattering}
\left(
\begin{array}{c}
\vv{\psi}_{\rm o}^{\rm L}\\
\vv{\psi}_{\rm o}^{\rm R}
\end{array}
\right)=
{\bf S}
\left(
\begin{array}{c}
\vv{\psi}_{\rm i}^{\rm L}\\
\vv{\psi}_{\rm i}^{\rm R}
\end{array}
\right)
\end{equation}
with
\begin{equation}
  \label{eq:scatteringmatrix}
  {\bf S}=
\left(
\begin{array}{ccc}
{\bf r}&\,\,&{\bf t}'\\
{\bf t}&\,\,&{\bf r'}
\end{array}
\right)\,.
\end{equation}
By solving Eq.~(\ref{eq:scattering}) for $\vv{\psi}_{\rm o}^{\rm R}$ and
$\vv{\psi}_{\rm i}^{\rm R}$ and inserting into Eq.~(\ref{cond03}) one obtains
straightforwardly
\begin{equation}
\label{eq:transfercomponents}
\widetilde{{\bf T}}=
{\bf U}^\dagger {\bf T} {\bf U}
=
\left(
\begin{array}{ccc}
{\bf t}-{\bf r}'{\bf t}'^{-1}{\bf r} &\,\,& {\bf r}'{\bf t}'^{-1} \\
-{\bf t}'^{-1}{\bf r} &\,\,& {\bf t}'^{-1}
\end{array}
\right):=
\left(
\begin{array}{ccc}
\widetilde{{\bf T}}_{11}&\,&\widetilde{{\bf T}}_{12}\\
\widetilde{{\bf T}}_{21}&\,&\widetilde{{\bf T}}_{22}
\end{array}
\right).
\end{equation}
In order to determine the $(L\times L)$-matrix ${\bf t}'$ that one can use
equivalently for the conductance instead of ${\bf t}$, it is sufficient to
generate $\widetilde{{\bf T}}_{12}$ and $\widetilde{{\bf T}}_{22}$ by
multiplying $\widetilde{{\bf T}}$ to the $2L\times L$ matrix which consists of
the zero-matrix ${\bf 0}_{L}$ in the upper half and the unit matrix ${\bf
  1}_{L}$ in the lower part,
\begin{equation}
\left(
\begin{array}{c}
\widetilde{{\bf T}}_{12}\\
\widetilde{{\bf T}}_{22}
\end{array}
\right)
=\widetilde{{\bf T}}\left(
\begin{array}{c}
{\bf 0}_L\\
{\bf 1}_L
\end{array}
\right).
\end{equation}

Multiplying iteratively ${{\bf T}}^{(k)}$ ($k=0,1,2,3,\ldots M$) is
numerically unstable. Thus, after each $m$ steps one must perform a
$QR$ decomposition,
\begin{equation}
{\bf T}^{(i\times m+m)} {\bf T}^{(i\times m+m-1)}
\cdots {\bf T}^{(i\times m+1)} {\bf V}^i=
{\bf V}^{i+1} {\bf \omega}^{i}
\end{equation}
where ${\bf V}^i$ are $(2L\times L)$ orthonormalized matrices and
${\bf \omega}^{i}$  are $(L\times L)$ upper triangular matrices.
This eventually yields
\begin{equation}
\left(
\begin{array}{c}
\widetilde{{\bf T}}_{12}\\
\widetilde{{\bf T}}_{22}
\end{array}
\right)
=\left(
\begin{array}{c}
{\bf A}\\
{\bf B}
\end{array}
\right) {\bf \omega}^k\cdots{\bf \omega}^2 {\bf \omega}^1,
\end{equation}
with matrices ${\bf \omega}^{i}$ generated during the iteration and $(L\times
L)$-matrices ${\bf A}$ and ${\bf B}$ the upper and lower blocks of $V^{M/m}$.

Using this result, one finds for the transmission and reflection matrices from
Eq.~(\ref{eq:transfercomponents}). Details and generalization can be found in
\cite{so05}.
\begin{equation}
{\bf t}'=({\bf \omega}^1)^{-1} ({\bf \omega}^2)^{-1} 
\cdots ({\bf \omega}^k)^{-1}{\bf B}^{-1} \quad , \quad
{\bf r}'={\bf A}{\bf B}^{-1}\,.
\end{equation}

\subsection{The Critical Conductance and its Statistics}
\label{sec:4.2}

The conductance $G$ is a strongly fluctuating quantity with a broad
distribution function. Therefore, not the conductance but its distribution
function $P(G)$ has to be considered.

In the metallic regime, the conductance is described by a normal distribution
with its variance independent of the details of the system as well as of the
system size. This is known as the phenomenon of reproducible universal
conductance fluctuations (UCF \cite{ls85,lsf87}). In the insulating regime,
the distribution function has a log-normal shape, reflecting the exponential
localization of wave functions.  At the critical point of the
localization-delocalization transition not only the variance but the whole
distribution function becomes size independent. This reflects again the scale
invariance of the quantum critical point
\cite{so97,shapiro90,shapiro90b,mk93,markos94,markos99}. Remarkably, however,
the critical distribution depends on the boundary conditions \cite{sok00}.

\begin{figure}[htbp]
\vspace{5mm}
\begin{center}
\subfigure[]{\includegraphics[width=0.47\linewidth]{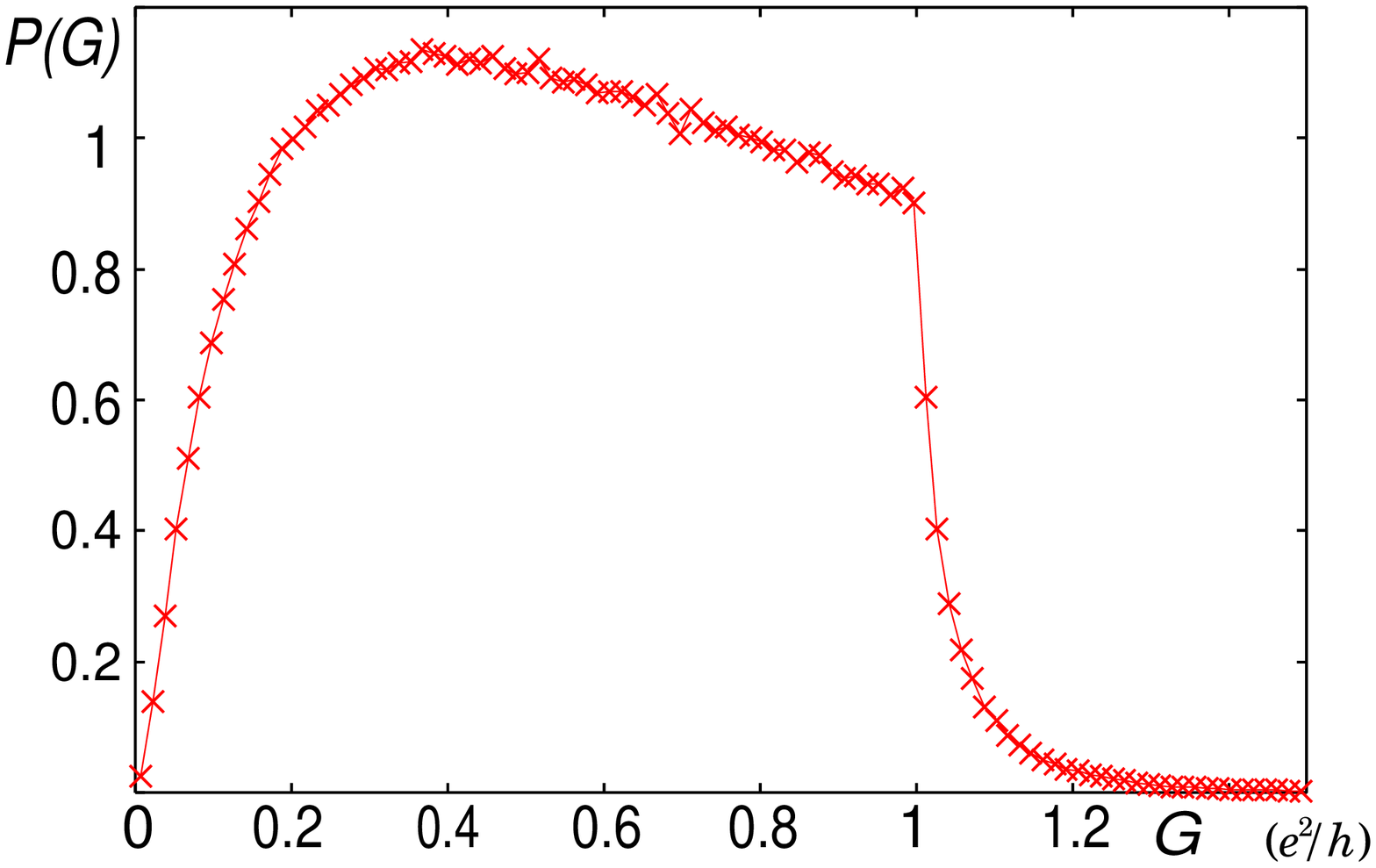}} 
\subfigure[]{\includegraphics[width=0.47\linewidth]{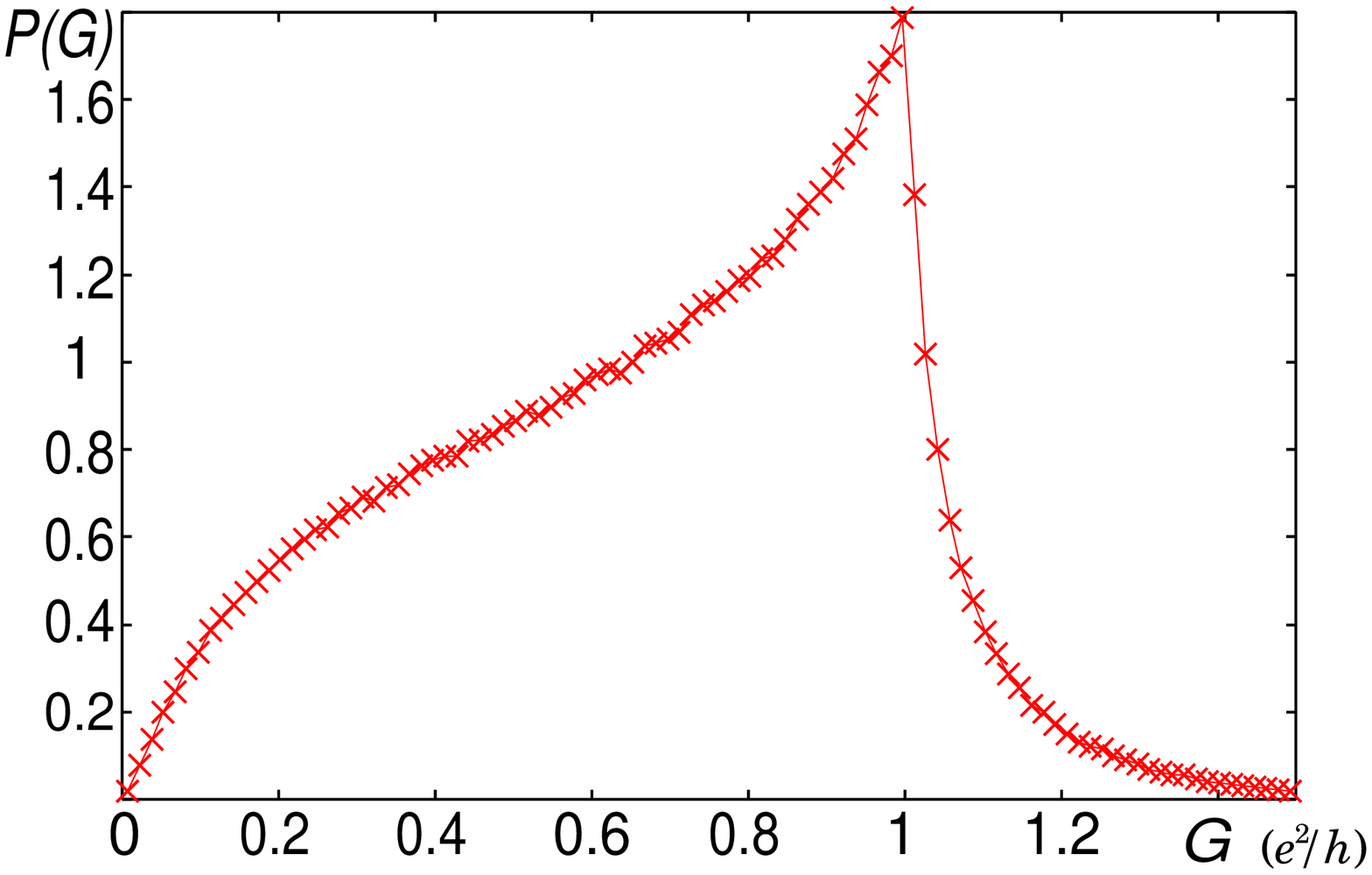}} 
\end{center}
\caption{Conductance distribution at the critical point.
  The number of incoming links is 128. The geometrical shape of
  the system is a square. Left: for periodic boundary conditions; right: for
  fixed boundary conditions. See also \cite{cf97}}  
\vspace{5mm}
  \label{fig:cond_dis}
\end{figure}

In the case of the quantum Hall transition, all the states except those at the
band center are localized, so that the conductance distribution is log-normal.
As expected, at the band center the conductance distribution function becomes
size independent. However, it has the peculiar form shown in
Fig.~\ref{fig:cond_dis}.
It neither is consistent with the log-normal distribution expected in the
localized regime nor reproduces the non-universal behavior of the moments
predicted for the metallic region \cite{wjl96}. It also depends strongly on
the boundary conditions. This is consistent with earlier results obtained for
the level statistics at the Anderson transition \cite{bszk96,bmp98}.

>From the distribution, one can calculate a configurationally averaged
conductance $\langle G\rangle$ and its moments \cite{osk04}. We consider here
as an example only the two terminal conductance for periodic boundary
conditions. We expect, in addition to corrections to scaling, effects of the
contacts \cite{bhmm97} and of the boundary conditions \cite{sok00}. Both of
them can be expected to give rise to contributions proportional to $L^{-1}$.
Therefore, we attempted to fit the average conductance to the scaling Ansatz
\begin{equation}
\langle G\rangle=G_c+a L^y +\frac{b}{L}\,.
\label{conductancefit}
\end{equation}
This fit yields the bulk ($L\rightarrow\infty$) conductance
\begin{equation}
G_c=(0.570\pm 0.02)\,\frac{e^{2}}{h}\quad, \quad y=-0.56\pm 0.05
\end{equation}
\begin{figure}[htbp]
\begin{center}
\includegraphics[width=0.75\linewidth]{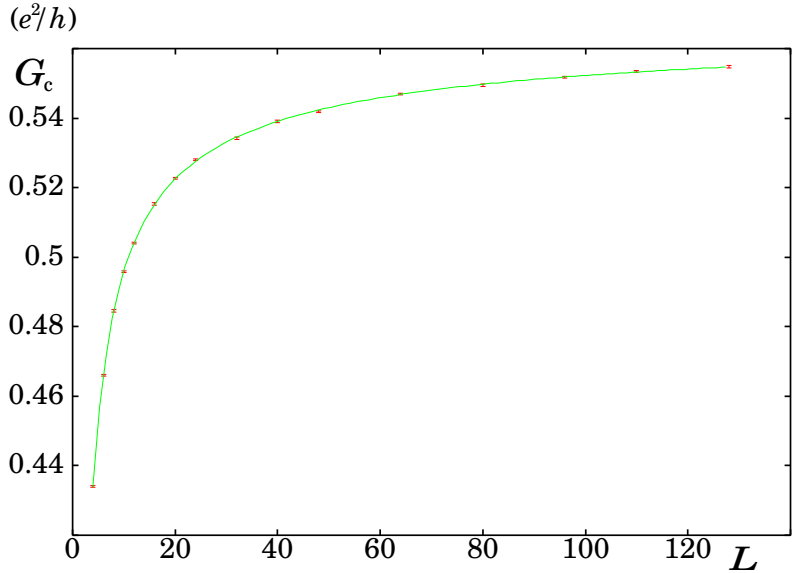}
\caption{
  The system size dependence of the ''bulk conductance'' at the critical
  point. Solid line: fit to the scaling law
  Eq.~(\protect\ref{conductancefit}), $G_{\mathrm{c}}=0.5702-0.2150
  L^{-0.5587}-\frac{0.1486}{L}$.}
\end{center}
\label{fig:gcvsl}
\end{figure}

The form of the distribution of the critical conductance is similar to that
obtained from simulating the conductance of the tight binding model in
magnetic fields \cite{wls98}. However, there are quantitative discrepancies.
For example, the two terminal conductance in units of $e^2/h$ is 0.506 for the
tight binding model \cite{wjl96}, but 0.57 for the Chalker-Coddington model.
This might be due to the inter-band coupling present in the tight binding
model.

In Fig.~\ref{fig:cond_dis}, we observe a continuous increase of $P(G)$ from 0
near $G=0$, and a kink near $G=e^2/h$. The kink is expected in one dimension
\cite{mwgg03,mwg03}, in the two dimensional symplectic ensemble
\cite{markos94,rms01} and in three dimensional disordered systems
\cite{markos99}. It is related to the fact that for one transport channel, one
can achieve at best unit transmission. Fluctuations of the conductance to
values larger than $e^2/h$ are related to the contributions from the
exponentially smaller contributions of the other transport channels.

\begin{figure}[h]
\begin{center}\leavevmode
\includegraphics[width=0.6\linewidth]{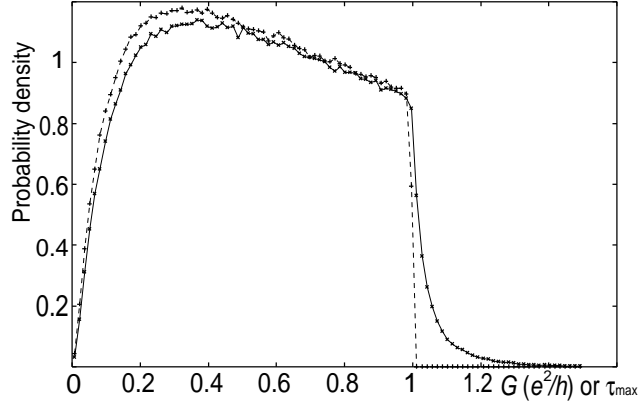}
\caption{
  The distribution of the critical conductance ($\times$) compared with the
  distribution of the largest transmission eigenvalue $\tau_\mathrm{max}$ (+)
  at the critical point. (After \cite{osk04})}
\label{fig_poftau}
\end{center}
\end{figure}

For better understanding the kinks in the distribution functions, the
transmission eigenvalues $\{\tau_i\}$ obtained by diagonalizing ${\bf tt}^\dagger$
have been analyzed in more detail \cite{osk04}. It has been found that for the
quantum Hall transition the critical conductance distribution is well
approximated by taking into account only the largest transmission eigenvalue
as demonstrated in Fig.~\ref{fig_poftau}.
This result suggests that the knowledge of the distribution of
$\tau_\mathrm{max}$ would enable us to approximately predict the transport
coefficients at the quantum Hall transition.

\begin{figure}[h]
\begin{center}\leavevmode
\includegraphics[width=0.6\linewidth]{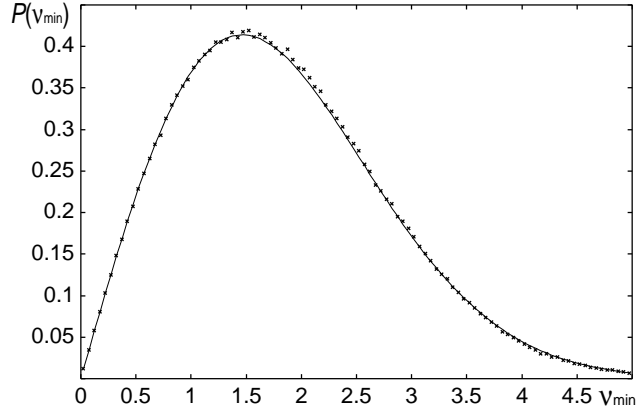}
\caption{
  The distribution of $\nu_\mathrm{min}$ compared with the Poissonian
  distribution $P(\nu_\mathrm{min})=2a\nu_\mathrm{min} \exp(-a
  \nu_\mathrm{min}^2)$ with $a$=0.233 [After \cite{osk04}].}
\label{fig_pofnumin}
\end{center}
\end{figure}

In order to further investigate $P(\tau_\mathrm{max})$, we transform to a new
variable $\nu_\mathrm{min}$ according to
\begin{equation}
\tau_\mathrm{max}=\frac{2}{\cosh\nu_\mathrm{min}+1}\,.
\label{eq:taumax}
\end{equation}
It has been suggested in \cite{sn93} that $\nu_\mathrm{min}^2$ is Poissonian
distributed,
\begin{equation}
P(\nu_\mathrm{min}^2)=\beta N\Omega e^{-\beta N\Omega\nu_\mathrm{min}^2}
\end{equation}
or equivalently,
\begin{equation}
P(\nu_\mathrm{min})=2\beta N\Omega\nu_\mathrm{min}
e^{-\beta N\Omega\nu_\mathrm{min}^2}.
\label{eq:wigner}
\end{equation}
In Fig. \ref{fig_pofnumin}, we show $P_\mathrm{c}(\nu_\mathrm{min})$ and
fitted the data to Eq.~(\ref{eq:wigner}) by assuming $\beta N \Omega=0.233$.

>From the very good agreement between $P(\nu_\mathrm{min})$ and the
Slevin-Nagao form \cite{sn93} one might conclude that the distribution of
$\{\nu_i\}$ is given by
\begin{equation}
P(\nu_1,\nu_2,...,\nu_N)
=C\prod_i {\rm e}^{-\beta \Omega\sum_i \nu_i^2}
\prod_{i<j} |\nu_i^2-\nu_j^2|^2 ,
\end{equation}
which defines the Laguerre ensemble \cite{sn93,sn94}. However, the
distribution of second and higher values, i.e. $\nu_i$ $(i=2,3,..)$ as
calculated from the Laguerre ensemble \cite{dn01} deviate from the result of
the present numerical simulation. This is not unexpected since the Laguerre
ensemble has been proposed to be a good approximation for the statistics of the
metallic diffusive regime. At the quantum Hall transition, the so-called
eigenvalue ''interaction'' term may be modified,
\begin{equation}
P(\nu_1,\nu_2,...,\nu_N)
=C\prod_i {\rm e}^{-\beta \Omega\sum_i \nu_i^2}
\prod_{i<j} f(|\nu_i^2-\nu_j^2|) .
\end{equation}
This form leads to the same distribution for $\nu_\mathrm{min}$ but different
distributions for higher $\nu_i (i=2,3,...)$ \cite{osk04}.

Before concluding this section, we mention that the energy correlation
function of the two terminal conductance has been analyzed in detail by
Jovanovic and Wang \cite{jw98}.

\subsection{Conformal Invariance at the Critical Point.}
\label{sec:pointcontact}

We have presented above strong numerical evidence for the scale invariance of
the wave functions, energy level statistics and the conductance distribution
at the critical point. Under a scale transformation, the length scale is
rescaled globally by some factor $b$ such that the coordinates of a given
point transform according to
\begin{equation}
  \label{eq:scaletrafo}
  \vv{r'}=b^{-1}\vv{r}\,.
\end{equation}
Scale invariance at a critical point means that correlation functions of
scaling variables $\phi_{j}(\vv{r})$ ($j=1,2,3\ldots$) are invariant under
such a transformation which generally can include also a rotation and a
translation
\begin{equation}
  \label{eq:scaleinvariance}
\langle \phi_1(\vv{r}_1)\phi_2(\vv{r}_2)\ldots\rangle
=
 \prod_j  b^{-h_j}
\langle \phi_1(\vv{r}_1')\phi_2(\vv{r}_2')\ldots\rangle\,.
\end{equation}
The exponents  $h_{j}$ are called scaling dimensions.

In analogy to classical phase transitions, one can expect not only that global
scale invariance holds but also that the more general concept of conformal
invariance \cite{cardy84,cardy96} applies to the quantum Hall
transition \cite{chalker88,cc88}. Generally, a conformal transformation
corresponds to {\em local} translations, rotations and dilatations of the
coordinates which preserve angles. 

In order to explain this in more detail, we consider the above general
correlation function in two dimensions,
$\langle\phi_1(z_1,z_1^*)\phi_2(z_2,z_2^*)\cdots\rangle$, with the scaling
variables $\phi_j$ ($j=1,2,3\ldots$) and the spatial variables represented as
$z_{j}=x_{j}+iy_{j}$ and its complex conjugate $z_{j}^*$. Now let us
consider an arbitrary analytic mapping 
\begin{equation}
  \label{eq:mapping}
 z'=w(z)\,. 
\end{equation}
Locally, close to some point $z_{0}$, which is equivalent to the infinitesimal
map
\begin{equation}
  \label{eq:infinitesimalmap}
  z'-z'_{0}=w'(z_{0})(z-z_{0})\,,
\end{equation}
with $w'(z_{0})={\rm d}w(z_{0})/{\rm d}z_{0}$. Comparing with
Eq.~(\ref{eq:scaletrafo}) one notes that the global scale factor $b^{-1}$ of
the scaling transformation corresponds to $w'(z_{0})$.

In analogy to Eq.~(\ref{eq:scaleinvariance}), conformal invariance of the
correlation functions can be defined by
\begin{eqnarray}
\langle \phi_1(z_1,z_1^*)\phi_2(z_2,z_2^*)\ldots\rangle
&=&\\\nonumber
&& \hspace{-1cm}\prod_i  w'(z_i)^{h_i}(w'(z_i)^{*})^{h_i'} 
\langle \phi_1(z_1',{z_1'}^{*})\phi_2(z_2',{z_2'}^{*})\ldots\rangle\,.
\end{eqnarray}
The exponents $h_i$ and $h_i'$ are real-valued conformal scaling dimensions
\cite{cardy96}. 

In order to show how powerful the concept of conformal invariance is for
extracting properties of the correlation functions we consider the
two-point correlation function of only one scaling variable. Conformal
invariance requires
\begin{equation}
\langle \phi(z_1,z_1^*)\phi(z_2,z_2^*)\rangle=
|w'(z_1)|^{\eta}|w'(z_2)^{*}|^{\eta} 
\langle \phi(z_1',{z_1'}^{*})\phi(z_2',{z_2'}^{*})\rangle\,.
\label{eq:twopointconformal}
\end{equation}
It is a straightforward exercise to show that this implies a power law decay
of the correlation function \cite{cardy96},
\begin{equation}
  \label{eq:powerlawdecay}
\langle \phi(z_1,z_1^*)\phi(z_2,z_2^*)\rangle\sim
|z_1-z_2|^{-2\eta}\,.  
\end{equation}

As a second example, we consider the mapping between the infinite two
dimensional plane and the surface of cylinder with circumference $L$,
\begin{equation}
 z'=\frac{L}{2\pi}\ln z\,.
\end{equation}
This is motivated by our above numerical studies in which we have used systems
of finite width $L$ and length $M\to \infty$. It is easy to get convinced that
variation of $x,y$ in the complete two dimensional plane implies $-\infty < x'
< \infty$ and $0 < y' <L$. This corresponds to an infinitely long strip of
width $L$ with periodic boundary conditions in the direction of $L$ which is
equivalent to a cylinder. The correlation function on the cylinder reads
\begin{eqnarray}
 \langle \phi(z_1',{z_1'}^{*})\phi(z_2',{z_2'}^{*})\rangle
&\sim&\\\nonumber
&&\hspace{-1cm}\frac{(2\pi/L)^{2\eta}}{\{2\cosh [2\pi(x_1'-x_2')/L]-2\cos
[2\pi(y_1'-y_2')/L]\}^\eta} \,.
\label{eq:cylinder}
\end{eqnarray}
Therefore, the quasi-one dimensional exponential decay length $\xi_{\rm cyl}$
and $\eta$ are related via
\begin{equation}
\label{eq:xicyl}
 \xi_{\rm cyl}=\frac{L}{2\pi\eta}\,.
\end{equation}

If we assume $\phi(z)$ to be the typical density of states defined via 
the square amplitude of the wave function,
\begin{equation}
\rho^\mathrm{typ}=\frac{(|\Psi|^2)^\mathrm{typ}}{\Delta E}\,
\end{equation}
where $\Delta E\sim L^{-2}$ is the mean level spacing, we obtain from
Eq.~(\ref{eq:typical}) 
\begin{equation}
  \label{eq:alpha-eta}
 \eta=\alpha_0-2\,. 
\end{equation}

The transfer matrix estimates the exponential decay length by averaging
$\log|\psi(0)\psi(z)|$. Therefore, it is the typical amplitude of the wave
function which is related to the decay length. Combining Eqs.~(\ref{eq:xicyl})
and (\ref{eq:typical})
\begin{equation}
 \xi_{\rm cyl}=\frac{L}{2\pi(\alpha_0 -2)}\,,
\end{equation}
which is the same as Eq.~(\ref{eq:janssen}) since $\Lambda_{\rm c}=2\xi_{\rm
  cyl}/L$. This example illustrates how powerful the requirement of the
conformal invariance is. Therefore, whether or not this symmetry holds for a
phase transition is very important and can provide useful information about
the properties near the critical point.

\begin{figure}[hbtp]
\begin{center}
{\includegraphics[width=0.7\linewidth]{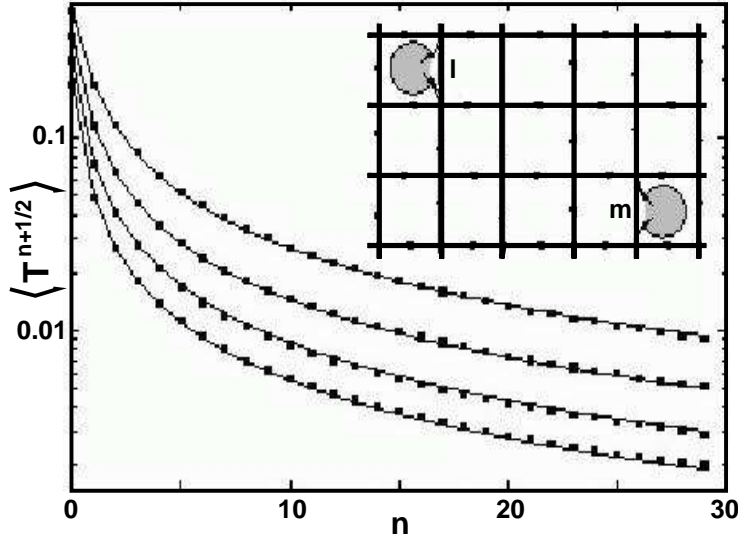}}
\caption{Average half-integer moments 
  $\langle T^{n+1/2}\rangle$ of the two-point contact conductance of the
  Chalker-Coddington model at the critical point as a function of $n$. Data
  sets (top to bottom) correspond to distances 5,10,15,20 (in units of the
  lattice constant) in a cylinder geometry with total length $M=100$ and
  circumference $L=10$. Solid lines are analytical results obtained by
  conformal invariance. For best fit an exponent $X_{\rm t}=0.54\pm 0.01$ was
  found. Inset: schematic view for the model. Two links, $l$ and $m$, are cut
  and connected to reservoirs $l$ and $m$ (after \cite{kz01}).}
 \label{fig:pointcontact}
\end{center}
\vspace{6mm}
\end{figure}

An important transport coefficient which can be used to clarify whether or not
the conformal invariance applies is the point-contact conductance
\cite{jmz99,kz01}. Consider a network with two links $l$ and $m$ cut
from the interior, and connected to two reservoirs
(Fig.~\ref{fig:pointcontact}).  Then we can define a $2\times 2$ ${\bf
  S}$ matrix
\begin{equation}
{\bf S}=\left(\begin{array}{cc}
S_{ll} & S_{lm}\\
S_{ml} & S_{mm}\\
\end{array}\right)
\end{equation}
and the point-contact conductance is given by (in units of $e^2/h$)
$T=|S_{lm}|^2$.
The ${\bf S}$ matrix is calculated by
\begin{equation}
S_{ij}=\langle i|  (1-{\bf U} {\bf P}_l {\bf P}_m)^{-1} {\bf U}
|j \rangle \,\,\,\, (i,j=l,m)
\end{equation}
where ${\bf U}$ is the time evolution operator introduced by
Eq.~(\ref{eq:unitary_develop}) and ${\bf P}_l=1-|l \rangle \langle l|$ and
${\bf P}_m=1-| m \rangle \langle m|$ the projection operators which describe
the nature of reservoirs (Fig. \ref{fig:pointcontact}).

The point-contact conductance $T$ is related to the wave function intensity at
$l$ and $m$, $|\Psi_l|^2$ and $|\Psi_m|^2$ via \cite{kz01}
\begin{equation}
 2\pi \rho(E)\langle
|\Psi_m|^2 f(|\Psi_m|^2/|\Psi_l|^2)
\rangle
=\langle F(T)\rangle\,,
\end{equation}
where
\begin{equation}
 F(T):=
\int_0^{2\pi}\frac{\mathrm{d}\phi}{2\pi}
f(T^{-1}|1-\mathrm{e}^{i\phi}
\sqrt{1-T}|^2)\,.
\end{equation}
Here $\rho(E)$ is the density of states and $\langle \cdots\rangle$
means the average over randomness.
Assuming $f(x)=-\ln x$, we have
\begin{equation}
 \langle \ln T\rangle=2\pi \rho (E)
\langle
|\Psi_m|^2 \ln(|\Psi_l|^2/|\Psi_m|^2)
\rangle\,.
\end{equation}
Thus one can evaluate the typical conductance $T^\mathrm{typ}=\exp(\langle \ln
T\rangle)$ very efficiently using only the information of the wave functions
in an isolated system.

At the critical point, in the infinite two dimensional plane $T^\mathrm{typ}$
is expected to decay as
\begin{equation}
 T^\mathrm{typ}\sim r^{-X_\mathrm{t}}\,,
\label{eq:pointconductancedecay}
\end{equation}
where $r$ is the distance between the links $l$ and $m$ and $X_\mathrm{t}$ a
critical exponent related to $\tau(q)$ (cf. Eq. \ref{eq:dq}) via
$\tau'(0)-\tau'(1)$ \cite{emm03}.

>From Eqs.~(\ref{eq:cylinder}) and (\ref{eq:pointconductancedecay}), the decay
of the typical conductance on a long cylinder surface becomes \cite{kz01}
\begin{equation}
 T_\mathrm{cyl}^\mathrm{typ}(x)
= \left|\frac{L}{2\pi}\sinh\left(
\frac{\pi x}{L}
\right)\right|^{-X_\mathrm{t}}\,,
\label{eq:confinvconductance}
\end{equation}
where $x$ is the distance between the links $l$ and $m$ along the cylinder
axis.

\begin{figure}[hbtp]

\begin{center}
{\includegraphics[width=0.7\linewidth]{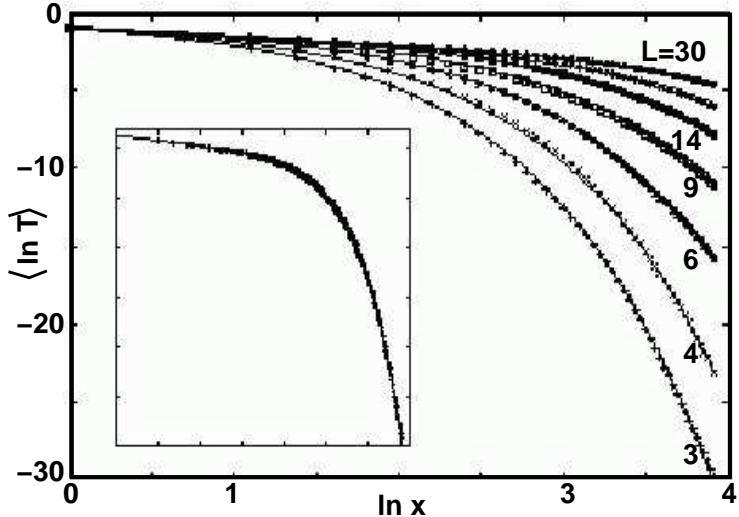}}
\caption{Logarithm of the typical two-point conductance 
  $\langle \ln T \rangle$ of the Chalker-Coddington network model at the
  critical point as a function of the logarithm of the distance $\ln |x|$ for
  system widths $L=3,4,6,9,14,20,30$ (bottom to top). Solid lines: analytical
  result obtained using conformal invariance,
  Eq.~(\ref{eq:confinvconductance}).  Inset: rescaled curves $\langle \ln
  T\rangle + X_{\rm t}\ln L$ versus $\ln|x/L| $ with $X_{\rm t}=0.57$ (after
  \cite{kz01}).}
 \label{fig:typconductance}
\end{center}
\vspace{6mm}
\end{figure}

Numerical results for the moments of $T(x)$, $\langle T(x)^{n+1/2}\rangle$
(Fig.~\ref{fig:pointcontact}) and the typical conductance
$T_\mathrm{cyl}^\mathrm{typ}(x)$ (Fig.~\ref{fig:typconductance}) coincide
with the behavior expected from the conformal invariance \cite{jmz99,kz01}.
This strongly supports the conjecture that the wave function at the quantum
Hall transition is conformally invariant \cite{chalker88}.

\section{The Renormalization Group Approach}
\label{sec:hierarchy}

In this chapter, we discuss the application of the real-space renormalization
group approach that is widely used in classical percolation and spin systems
\cite{cardy96,sa94} to the random network model. A closed set of
renormalization group equations is derived which describes the universal
distribution of the conductance. The numerical solution is used to estimate
the critical exponent of the localization length and the behavior of the
moments of the distribution function \cite{gr97,sr95,ajs97,jmmw98}. Amazingly
enough, by adopting a slightly different point of view, eventually one can
interpret the approach as providing an additional, independent precision
determination of the critical exponent. The agreement of this critical
exponent with the earlier results obtained for completely different models
confirms that the quantum Hall phase transition is indeed a universal critical
phenomenon.

The general idea is that near the quantum critical point all microscopic
details of the system eventually must become irrelevant since the correlation
length diverges with a universal exponent which, however, still can depend on
the fundamental global symmetries. In order to detect such universal behavior
one must perform a well-controlled thermodynamic limit. An example how to do
this in a controlled way has been given above when applying the transfer
matrix method and using the numerical scaling method. This procedure is
basically exact and suffers only from technical numerical errors, but it does
in general not provide analytical insight.

In classical phase transition theory, the renormalization group transformation
is a well-established procedure to approach the critical point. Generally,
this method consists of a sequence of unitary and subsequent scale
transformations. The former are used to diagonalize relatively small systems.
The latter scale the system back to the original length scale. From the
behavior of the coupling parameters under this repeatedly applied
transformation conclusions can be drawn on the critical behavior. In
principle, this method would also be exact, apart from the fact that in each
renormalization step couplings to states that are energetically far away from
the critical point are neglected. While this seems not to be crucial for
showing the very existence of a critical point, one needs quite substantial
numerical efforts in order to correctly obtain the critical behavior quantitatively.

There are many alternative possibilities for constructing the renormalization
group transformation which are all similar in spirit but may be different in
the details. We will discuss here two of them. The random network model of
Chalker and Coddington is an example in which the renormalization group
transformation can most advantageously be used.

\subsection{An Illustrative Example: the Tile Lattice}

\begin{figure}[hbtp]
\vspace{6mm}
\begin{center}
{\includegraphics[width=0.6\linewidth]{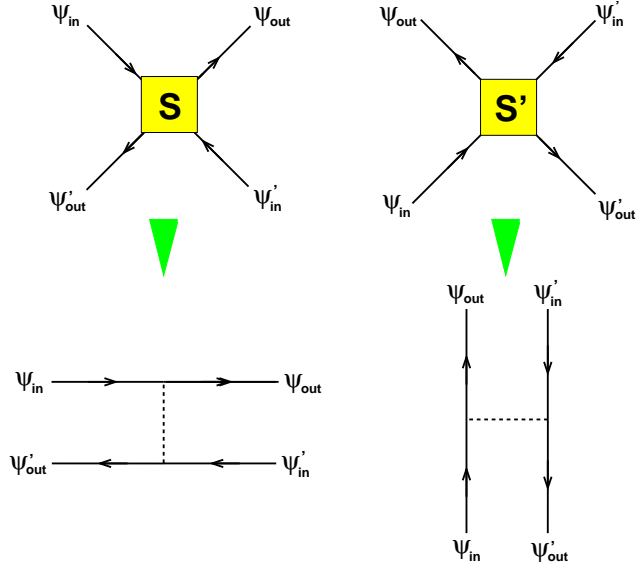}}
\caption{
  Representations of saddle point scattering. The scattering matrices ${\bf
    S}$ and ${\bf S}'$ relate outgoing with incoming flux amplitudes (top).
  The same information can be expressed by transfer matrices that relate top
  and bottom amplitudes (left) and left and right amplitudes (right).}
 \label{fig:saddlediagr}
\end{center}
\vspace{6mm}
\end{figure}

\begin{figure}[hbtp]
\vspace{6mm}
\begin{center}
\subfigure[]{\includegraphics[width=0.4\linewidth]{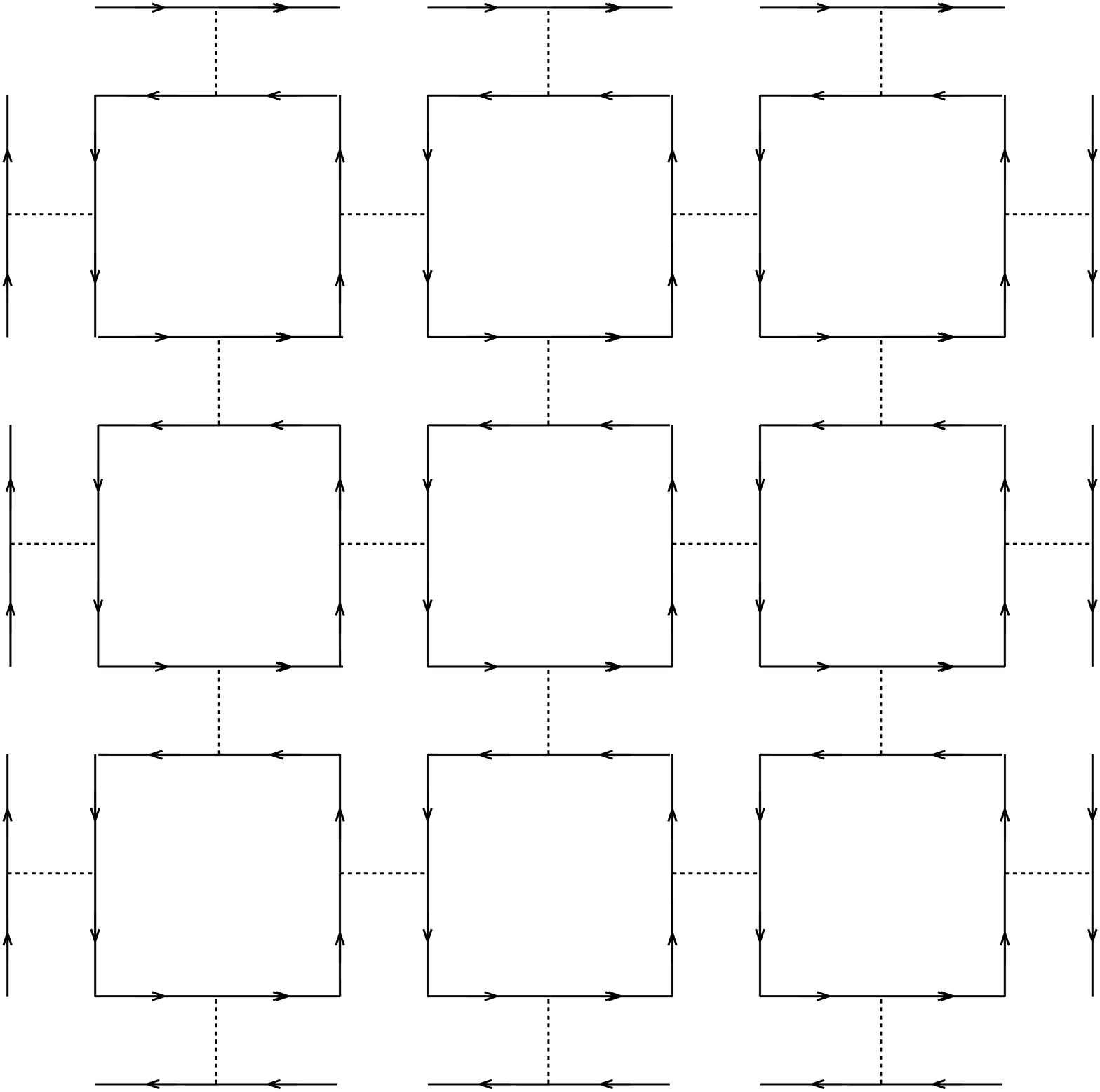}}
\hspace{1cm}
\subfigure[]{\includegraphics[width=0.4\linewidth]{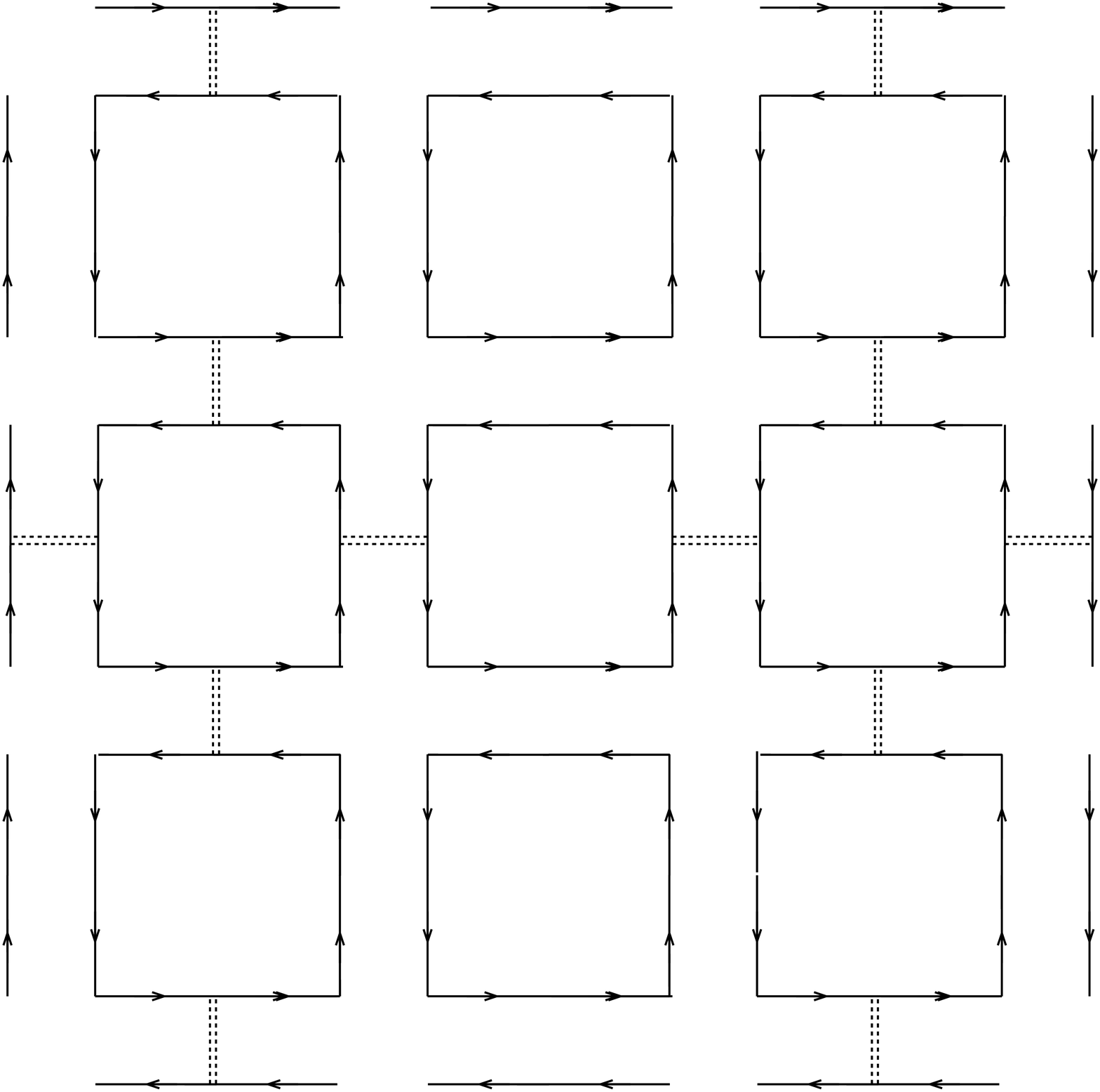}}
\subfigure[]{\includegraphics[width=0.36\linewidth]{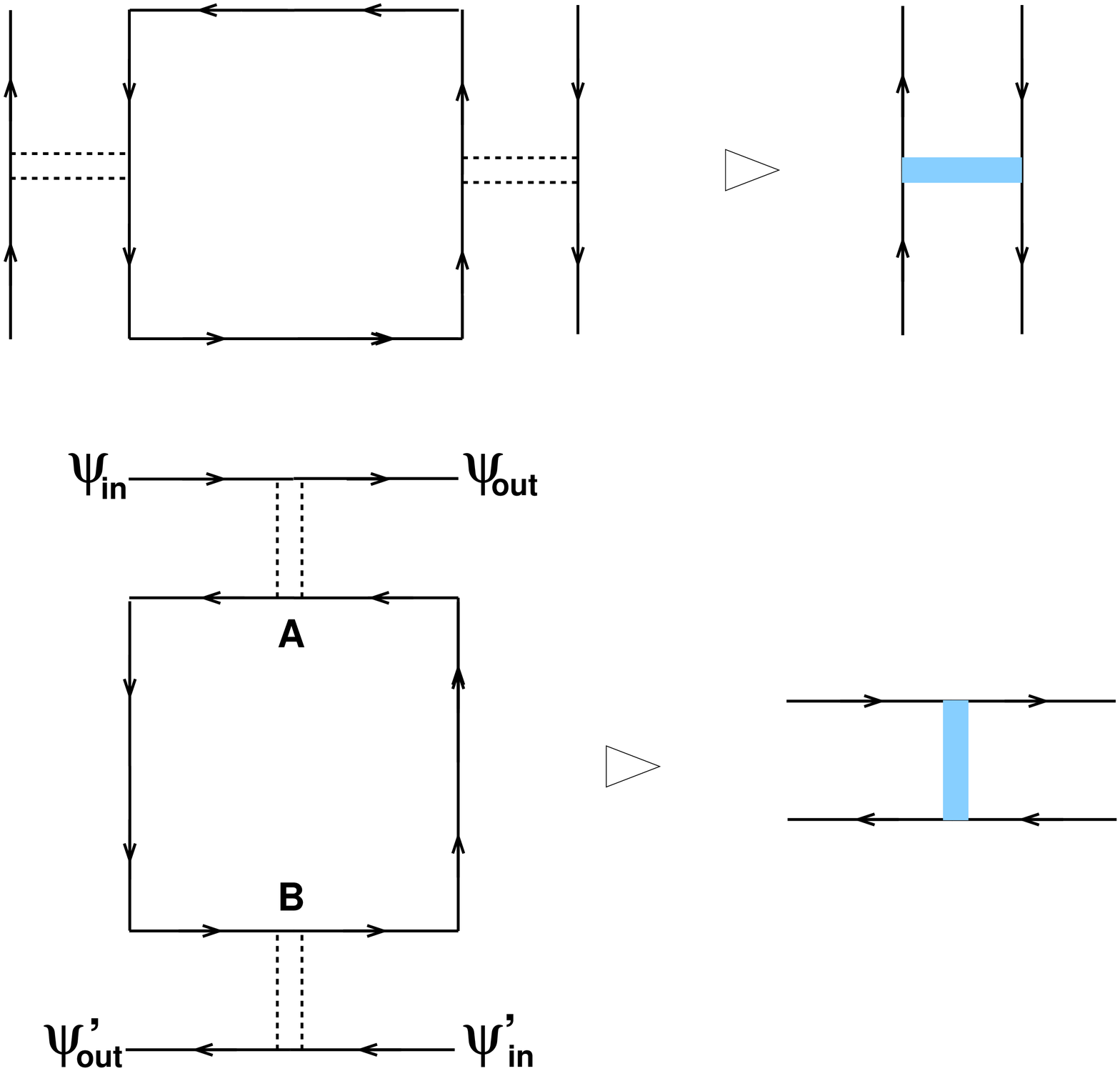}}
\hspace{1cm}
\subfigure[]{\includegraphics[width=0.4\linewidth]{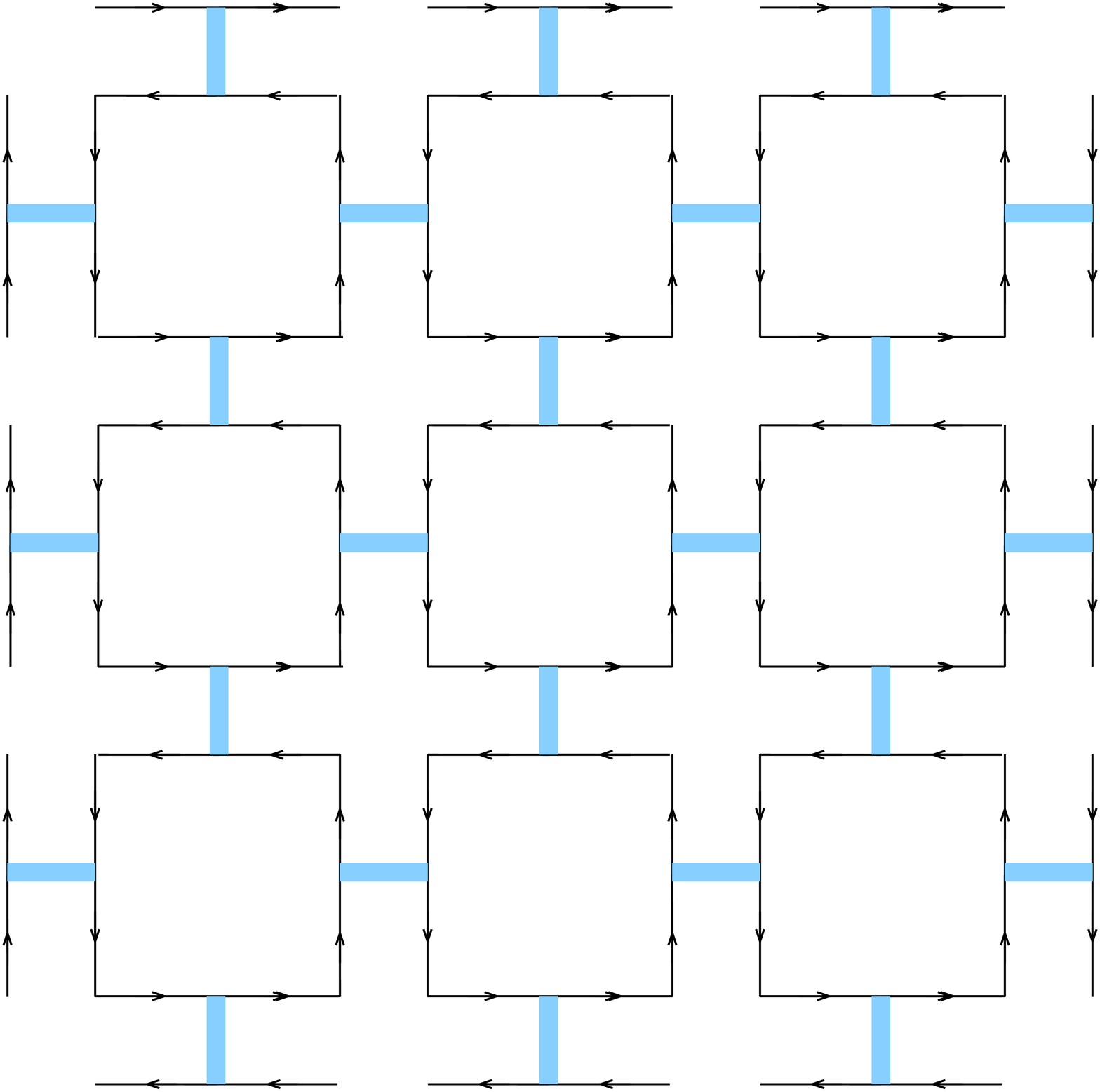}}
\caption{
  Illustration of the renormalization procedure on a tile lattice. (a) Lattice
  of original saddle points (b) Second-generation saddle point lattice where
  every other saddle point in the rows and columns are removed and the
  remaining saddle points are doubled. Isolated rings of amplitudes can be
  removed. (c) By renormalizing the saddle point as indicated one arrives (d)
  at a lattice that, apart from a scale transformation is structurally
  identical to the original one [after \protect\cite{ajs97}].}
\label{fig:tilenetwork}
\end{center}
\vspace{6mm}
\end{figure}

A particularly instructive example which resembles the procedure invented
originally by Migdal and Kadanoff is shown in Fig.~\ref{fig:tilenetwork}
\cite{ajs97}. For a better understanding of this procedure, we first introduce
different graphical representations of the scattering matrices of the saddle
points which are more convenient for the present purpose
(Fig.~\ref{fig:saddlediagr}).
The original scattering matrices ${\bf S}$ and ${\bf S}'$ relate outgoing and
incoming amplitudes, $\psi_{\rm out}$, $\psi_{\rm out}^{'}$ and $\psi_{\rm
  in}$, $\psi_{\rm in}^{'}$, respectively. This is represented as in the top
part of Fig.~\ref{fig:saddlediagr}. Rewriting the equations such that the
amplitudes on the right hand side of the saddle point are obtained as
functions of the amplitudes on the left, $\psi_{\rm in}^{'}$, $\psi_{\rm
  out}^{'}$ and $\psi_{\rm in}$, $\psi_{\rm out}$, respectively, one obtains
the usual transfer matrix representation which we have used so far
(Fig.~\ref{fig:saddlediagr}, right).  The saddle point vertex is now
represented by a horizontal dashed line.  For ${\bf S}$, the corresponding
representation is shown in the left part of Fig.~\ref{fig:saddlediagr}.
With this, the original random network of Fig.~\ref{fig:network} can be
graphically represented as shown in Fig.~\ref{fig:tilenetwork}\,a. For
simplicity, the system is assumed to consist of independent saddle points with
randomly varying saddle point energies. The phases associated with the links
between them are assumed to be completely random and independent.

Starting from this a renormalization group transformation may be constructed
by removing every other saddle point line in a each row and column, and
replacing the remaining single saddle point lines by two. Thereby, one
reaches the situation shown in Fig.~\ref{fig:tilenetwork}\,b. One observes that
every other closed loop of amplitudes now is completely disconnected from the
rest of the system and can be removed. On the other hand, each of the
remaining saddle points is replaced by two saddle points in series. By
combining these to new scattering centers as indicated in
Fig.~\ref{fig:tilenetwork}\,c one arrives at exactly the same lattice structure
as in Fig.~\ref{fig:tilenetwork}\,a but with renormalized saddle point
scattering centers (Fig.~\ref{fig:tilenetwork}\,d).

\begin{figure}[hbtp]
\vspace{6mm}
\begin{center}
{\includegraphics[width=0.6\linewidth]{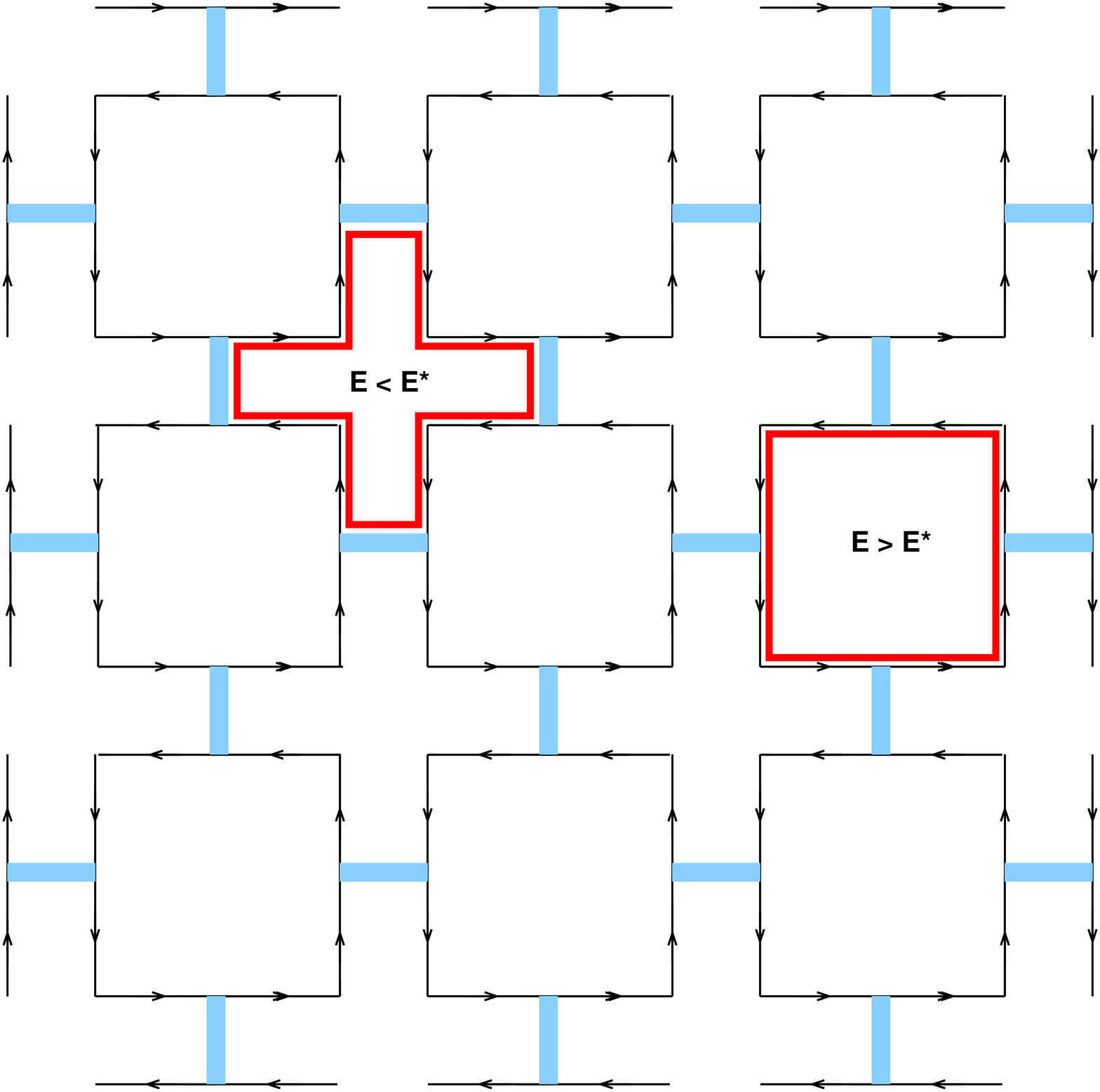}}
\caption{
  Illustration of localization in the renormalized network for energies below
  the critical point, $E<E^{*}$, where the transmission renormalizes to zero,
  and above the critical point, $E>E^{*}$, where the transmission renormalizes
  to one. Extended states can exist only at $E=E^{*}$ where the
  transmission has a non-trivial fixed point $T^{*}\neq 0$. }
\label{fig:tileloc}
\end{center}
\vspace{6mm}
\end{figure}

By iteratively applying this procedure, more and more of the network is
incorporated into the scattering properties of the scattering centers until
eventually one of the latter incorporates the whole system. If the energy is
lower than the saddle point energy $E^{*}$, the transmission probability will
eventually renormalize exponentially to zero. The system becomes completely
localized since the closed loops labeled with $E<E^{*}$
(Fig.~\ref{fig:tileloc}) will be completely disconnected. If on the other hand
the energy is higher than the saddle point energy, the transmission
probability will renormalize to one. The states are again localized since now
the loops labeled with $E>E^{*}$ in Fig.~\ref{fig:tileloc} become
disconnected. Thus, the transmission probability has two stable fixed points
$T=0$ and $T=1$ which correspond to the localized phases. The transmission
probability becomes independent of the ^^ ^^ size'' of the scatterer exactly at
the critical point $E^{*}$. Here, we have an {\em unstable} non-trivial fixed
point $T^{*}\neq 0$. Even an infinitesimally small deviation of the energy
from the critical point localizes the system completely.

For the tile lattice, the calculation of the scattering properties of a new
scattering center by combining two original ones is an easy but nevertheless
instructive exercise. The transmission matrix of the new scattering center
(Fig.~\ref{fig:tilenetwork}\,c) can be determined in terms of the transmission
matrices of the original scattering centers by performing simple matrix
multiplications. Consider, for example, the diagram at the bottom of
Fig.~\ref{fig:tilenetwork}\,c. First, one calculates the transmission of the
two pairs of saddle points in series at the top (A) and the bottom (B) of the
graph. Second, the two resulting effective scatterers are combined ^^ ^^ in
parallel'' in order to obtain the transfer of amplitudes between channels
($\psi_{\rm in}$, $\psi_{\rm out}$) and ($\psi_{\rm in}^{'}$, $\psi_{\rm
  out}^{'}$). Eventually, one finds for the {\em typical} total transmission
probability $T:= \exp{\langle\ln T\rangle}$ (with $T=t^{2}$ and $\langle\ldots
\rangle$ the ensemble average) the renormalization relation \cite{ajs97}
\begin{equation}
  \label{eq:2tiletransmission}
  T'= 2T^{2}-T^{4}:= f(T;2)
\end{equation}
where $T'$ is the typical transmission probability of the renormalized
scattering center.  

This scaling relation has three fixed points defined by
$T^{*}=2T^{*2}-T^{*4}$, two stable ones at $T^{*}=0$ and $T^{*}=1$ and an
unstable one at $T^{*}=(\sqrt 5-1)/2=0.618$. The stable fixed points
correspond to the localized states at energies away from the saddle point
energy. The unstable fixed point corresponds to the quantum critical point.
The localization length exponent is found by linearizing around the unstable
fixed point
\begin{equation}
  \label{eq:tileexponent}
  \nu = \frac{\ln 2}{\ln \lambda}
\end{equation}
with $\lambda = [\partial f/\partial T]_{T^{*}}=6-2\sqrt 5$. One gets
$\nu\approx 1.635$.

In the general case, where one saddle point vertex is replaced by $b$
original vertices, the result is
\begin{equation}
  \label{eq:btiletransmission}
  T'=1-(1-T^{b})^{b}:= f(T;b)\,.
\end{equation}
Assuming $b=1+\epsilon$ one obtains the infinitesimal Migdal-Kadanoff
transformation which has a fixed point at $T^{*}=1/2$ and an eigenvalue
$\lambda =1+2(1-\ln 2)\epsilon$. This gives an exponent 
\begin{equation}
  \label{eq:185}
\nu=\frac{\ln b}{\ln \lambda}=\frac{1}{2(1-\ln 2)}\approx 1.629  
\end{equation}
and the beta function
\begin{equation}
  \label{eq:betafunction}
  \beta(T):=\left.\frac{{\rm d}}{{\rm d}\epsilon}\right|_{\epsilon =0}
f(T;1+\epsilon)=T\ln T-(1-T)\ln(1-T)\,.
\end{equation}

Although the result for the critical behavior is far from being satisfactory,
it is nevertheless remarkable that the very existence of a non-trivial quantum
critical point is correctly predicted by the model even in the crudest
approximation for the renormalization group transformation. It is therefore
worthwhile to investigate whether one can obtain more reliable quantitative
results by refining the approximations.

\subsection{State-of-the-art results for the hierarchical lattice}

\begin{figure}[hbtp]
\vspace{6mm}
\begin{center}
\subfigure[]{\includegraphics[width=0.6\linewidth]{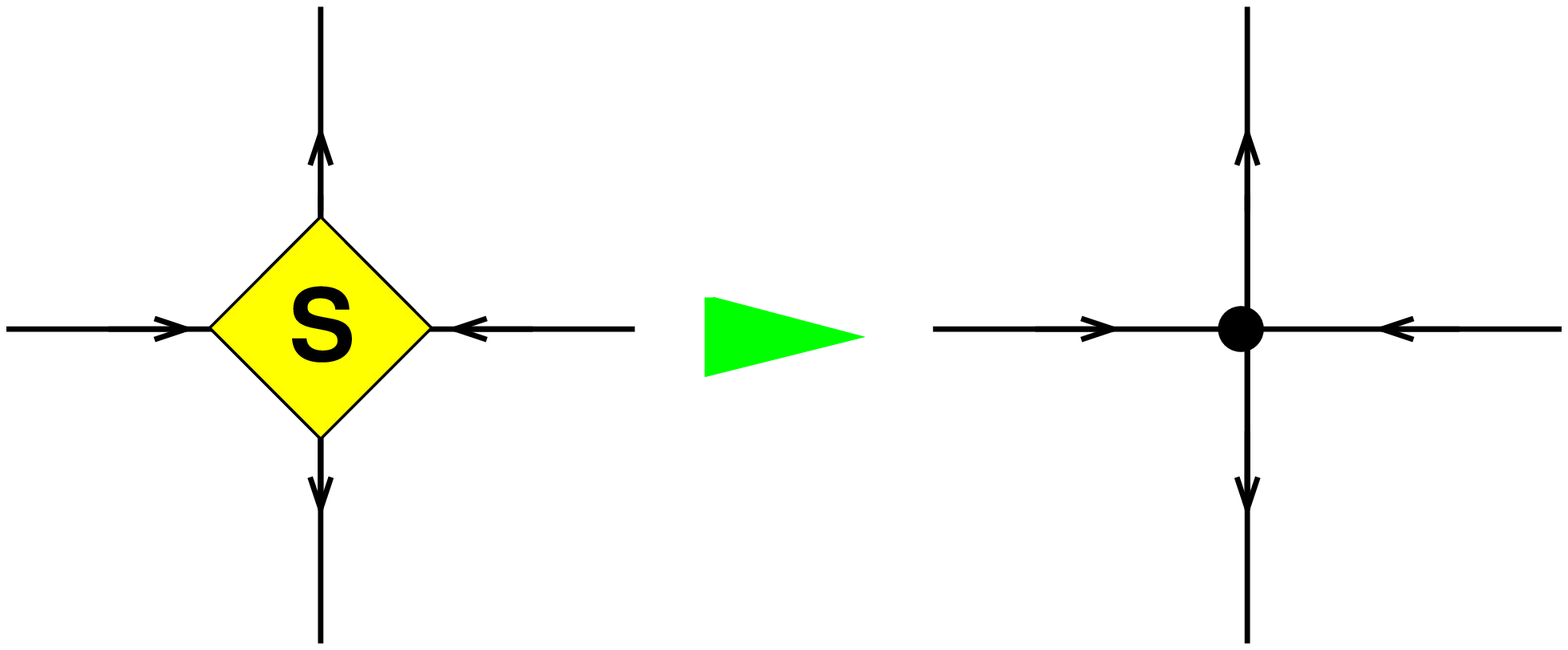}}
\subfigure[]{\includegraphics[width=0.45\linewidth]{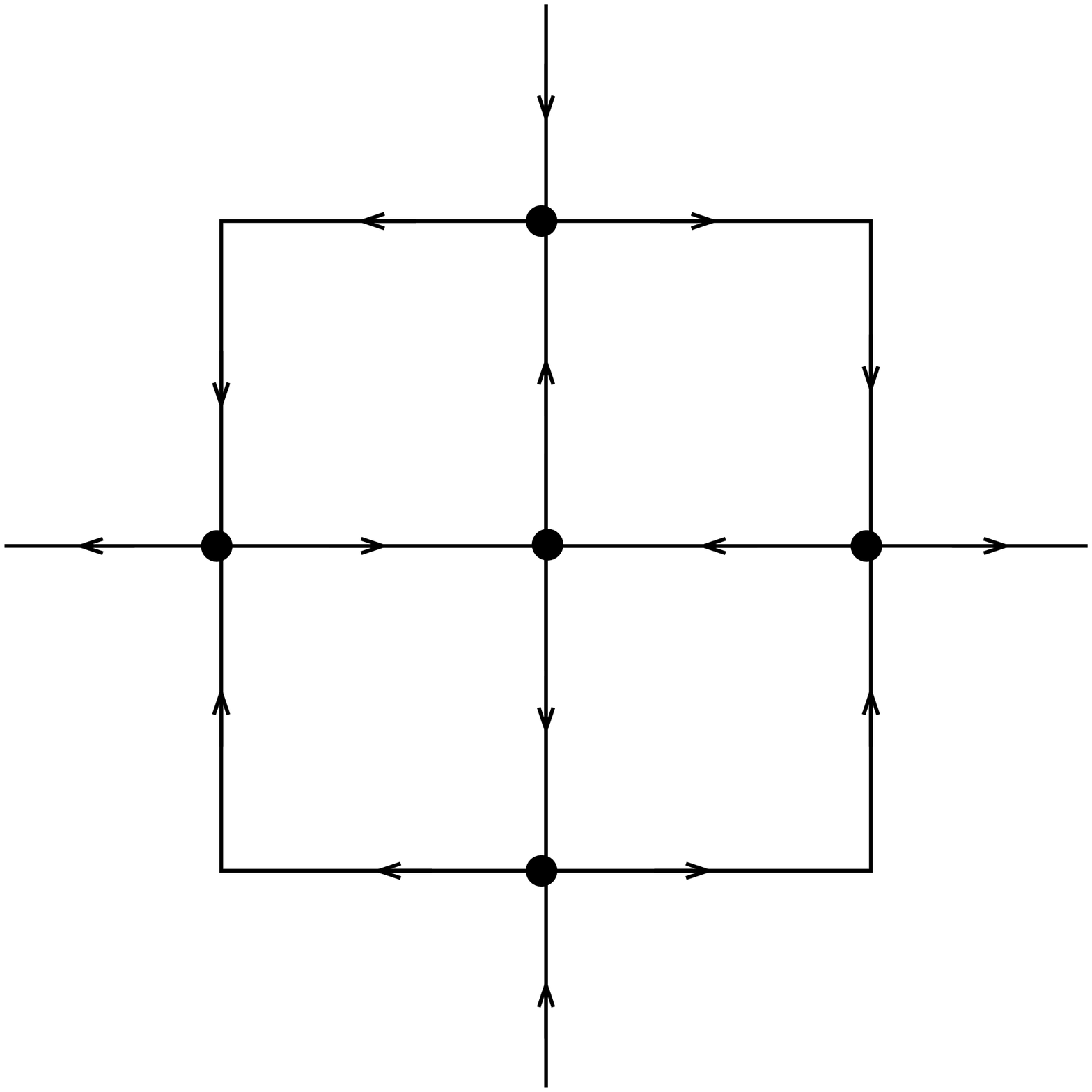}}
\hspace{0.5cm}
\subfigure[]{\includegraphics[width=0.45\linewidth]{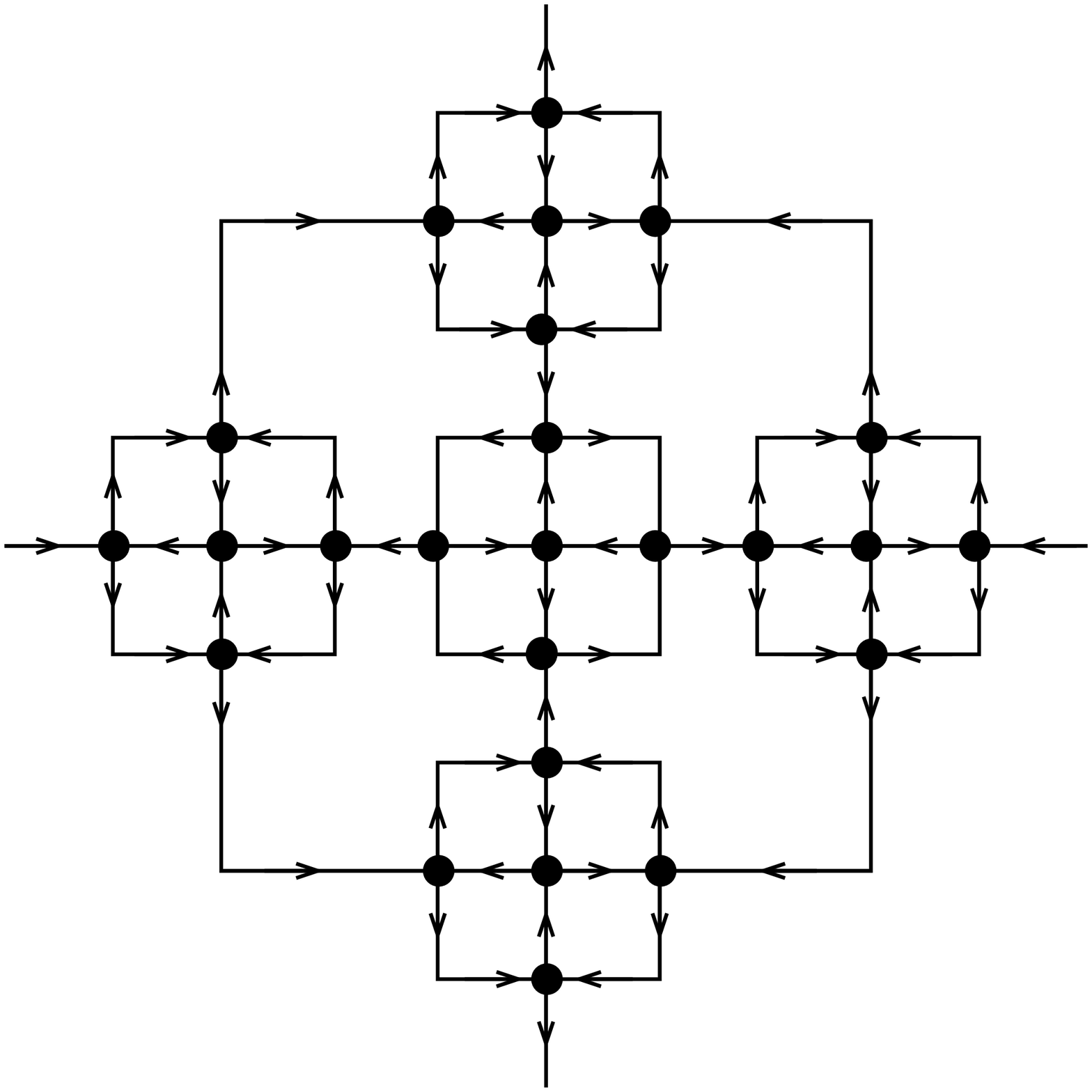}}
\caption{
  Illustration of the renormalization procedure on a hierarchical lattice
  constructed iteratively. (a) Saddle point represented by a dot with two
  incoming and outgoing channels. (b) Second-generation saddle point
  consisting of five original (first-generation) saddle points. A total of
  eight incoming and outgoing channels are here neglected. (c) Second level of
  iteration where five second-generation saddle points are combined to form a
  third-generation saddle point consisting of 25 first-generation saddle
  points. In contrast to the original network (Fig~{\ref{fig:network}}), the
  hierarchical network contains considerably fewer saddle points and links
  [after Ref. \protect\cite{ajs97}].}
\label{fig:rg5hierarc}
\end{center}
\vspace{6mm}
\end{figure}

A different procedure is outlined pictorially in Fig.~\ref{fig:rg5hierarc}.
This is the model of a hierarchical lattice. In the first generation, the
system consists again of the original, statistically independent saddle
points. In a first step, a certain number of these original saddle points
(five in the example of Fig.~\ref{fig:rg5hierarc}) are combined to form a new
scattering center. The corresponding scattering matrix is calculated
approximately as a function of the original ones. This is repeated: the new
scattering centers are combined again into new units
(Fig.~\ref{fig:rg5hierarc}\,c). Their scattering properties are calculated as
functions of the previous ones, and so on and so forth.

The crucial approximation of the procedure is that in each step only two
incoming and outgoing channels are taken into account
(Fig.~\ref{fig:rg5hierarc}\,b) such that the new unit can be considered as a
new, second-generation saddle point. The exact scattering matrix of the units
in each generation would of course contain much more channels, and when
repeating the construction, the number of channels would explode. Neglecting
all of these channels apart from four is in fact the most severe approximation
of the method --- as also made in the previous procedure --- since it cannot
be very well controlled.

However, instead of viewing the hierarchical lattice as an approximation to
the random network model one may also consider it as a model in its own right
for which the critical behavior may be determined exactly. This would yield
only an approximation to the critical behavior of the network model, but if
the critical behavior was universal --- independent of the microscopic details
--- the exponent of this model should be the same as for any other model in
the same universality class. We will come back to this point below in more
detail.

\begin{figure}[hbtp]
\vspace{6mm}
\begin{center}
\includegraphics[width=0.45\linewidth]{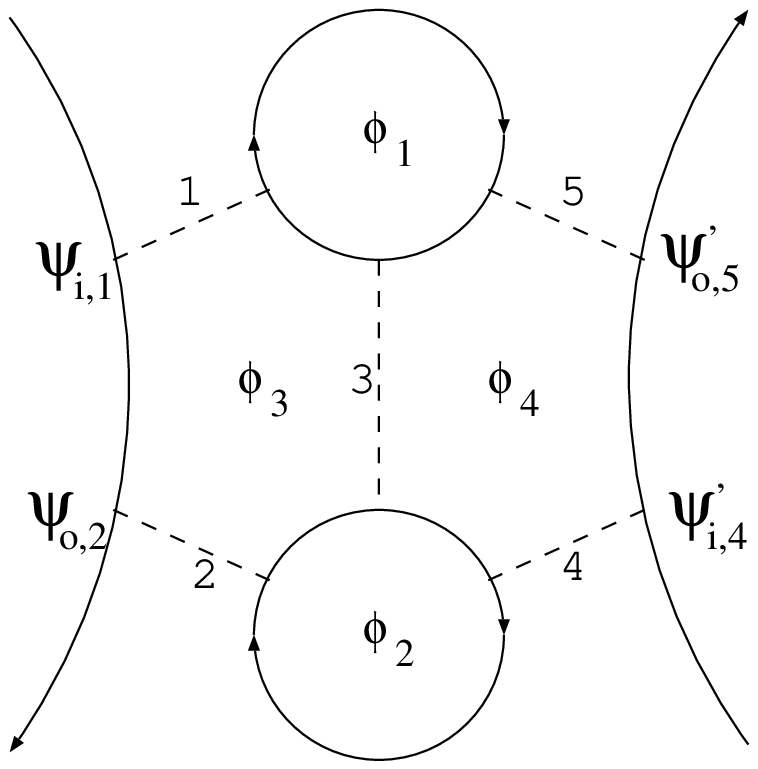}
\vspace{0.5cm}
\caption{
  Illustration of a transformation that can be used in a renormalization group
  transformation. Five original (first-generation) saddle points indicated
  here by dashed lines (right) are renormalized into one single
  second-generation saddle point. This relates $(\psi_{{\rm o},2},\psi_{{\rm
      o},5}')$ to $(\psi_{{\rm i},1},\psi_{{\rm i},4}')$. The quantities
  $\phi_{1}\ldots\phi_{4}$ are random independent Aharonov-Bohm-fluxes that
  penetrate closed loops as indicated.}
\label{fig:rg_saddle}
\end{center}
\vspace{6mm}
\end{figure}

For the derivation of the renormalization group equations of the hierarchical
lattice, we closely follow \cite{gr97}. We consider the five saddle points in
Fig.~\ref{fig:rg_saddle}. The amplitudes must satisfy the five relations
\begin{equation}
\left(\begin{array}{c}\psi_{{\rm o},i}\\ 
\psi_{{\rm o},i}'\end{array}\right)
=\left(\begin{array}{cc}
-r_i & t_i\\
t_i & r_i\end{array}\right)
\left(\begin{array}{c}\psi_{{\rm i},i}\\ 
\psi_{{\rm i},i}'\end{array}\right)
\quad ,\quad (i=1,2,\ldots,5)
\label{eq:rg_saddle}
\end{equation}
where i and o denote the input and output channels, respectively, and $r_i$
and $t_i$, $(=\sqrt{1-r_i^2})$ ($i=1,\ldots,5$) are real quantities. The
first goal is to replace the five saddle points by one.

Some of the amplitudes differ only by a phase factor. For example,
\begin{equation}
\psi_{2,{\rm i}}=\e^{i\theta_{2,1}}\psi_{1,{\rm o}}\,,
\end{equation}
where $\theta_{2,1}$ is the random phase factor gained when traveling from
the saddle point 1 to 2. The phase factors are related to the Aharonov-Bohm
fluxes $\phi_i, \, (i=1,2,3,4)$ obtained when traveling around a closed loop
via
\begin{equation}
\begin{array}{c}
\theta_{5,1}+\theta_{3,5}+\theta_{1,3}=\phi_1\\
\theta_{3,2}+\theta_{4,3}+\theta_{2,4}=\phi_2\\
\theta_{2,1}+\theta_{3,2}+\theta_{1,3}=\phi_3\\
\theta_{3,4}+\theta_{5,3}+\theta_{4,5}=\phi_4.
\end{array}
\end{equation}
Since the phases $\theta$ are assumed to be randomly and uniformly
distributed between $[0,2\pi)$, so are the fluxes $\phi_i$.

>From the relations (\ref{eq:rg_saddle}), one can derive that the total
outgoing amplitudes $(\psi_{{\rm o},2},\psi_{{\rm o},5}')$ are related to the
incoming amplitudes $(\psi_{{\rm i},1},\psi_{{\rm i},4}')$ according to
formally the same relation as Eq.~(\ref{eq:rg_saddle})
\begin{equation}
\left(\begin{array}{c}\psi_{{\rm o},2}\\ \psi_{{\rm o},5}'\end{array}\right)
=\left(\begin{array}{cc}
-\tilde{r} & \tilde{t}\\
\tilde{t} & \tilde{r}\end{array}\right)
\left(\begin{array}{c}\psi_{{\rm i},1}\\ \psi_{{\rm i},4}'
\end{array}\right)\,.
\end{equation}
The new reflection and transmission coefficients, $\tilde{r}$, $\tilde{t}$,
characterize the scattering properties of the renormalized ^^ ^^ super'' saddle
point. The new transmission and reflection amplitudes can be obtained
straightforwardly by solving Eq.~(\ref{eq:rg_saddle}). One obtains after some
tedious algebra
\begin{eqnarray}
\tilde{t}&=&\left|\,
\frac{t_1 t_5 +t_2 t_4 \e^{i(\phi_3+\phi_4)}
-t_2 t_3 t_5 \e^{i\phi_3}
-t_1 t_3 t_4 \e^{i\phi_4}}{D}\right.\nonumber\\
&&\left.\qquad\qquad\qquad+\,\,\frac{t_1 r_2 r_3 r_4 t_5 \e^{i\phi_2}
-r_1 t_2 r_3 t_4 r_5 \e^{i(\phi_3+\phi_4-\phi_1)}}{D}
\,\right|\,.
\label{eq:rg_transmission}
\end{eqnarray}
The corresponding reflection amplitude is obtained from
$\tilde{r}=\sqrt{1-\tilde{t}^{2}}$
\begin{eqnarray}
\tilde{r} &=&\left|\,
\frac{r_4 r_5 -r_1 r_2 \e^{i(\phi_1+\phi_2)}
-r_1 r_3 r_4 \e^{i\phi_1}
+r_2 r_3 r_5 \e^{i\phi_2}}{D}\right.\nonumber\\
&&\left.\qquad\qquad\qquad-\,\,
\frac{t_1 t_2 t_3 r_4 r_5 \e^{i\phi_3}
+r_1 r_2 t_3 t_4 t_5 \e^{i(\phi_1+\phi_2-\phi_4)}}{D}
\,\right|\,,
\end{eqnarray}
where $D$ is the abbreviation for
\begin{eqnarray}
D&=&1-r_1 r_3 r_5 \e^{i\phi_1} -t_1 t_2 t_3 \e^{i\phi_3}
-t_3 t_4 t_5 \e^{i\phi_4}+r_2 r_3 r_4 \e^{i\phi_2}
\nonumber\\
& & \qquad\qquad\qquad\qquad
\qquad+\,t_1 t_2 t_4 t_5 \e^{i(\phi_3+\phi_4)}
-r_1 r_2 r_4 r_5 \e^{i(\phi_1+\phi_2)}\,.
\end{eqnarray}

The transformation (\ref{eq:rg_transmission}) allows one to generate the new
probability distribution of the transmission coefficient $P(\tilde{t})$ from
the distribution $P(t)$. A certain distribution is unchanged by this
transformation, which corresponds to the fixed point distribution. Slight
deviations from the fixed point distribution increase after the
renormalization transformation. This can be used to define the critical
exponent $\nu$.

\begin{figure}[htbp]
  \begin{center}
\includegraphics[width=0.56\linewidth]{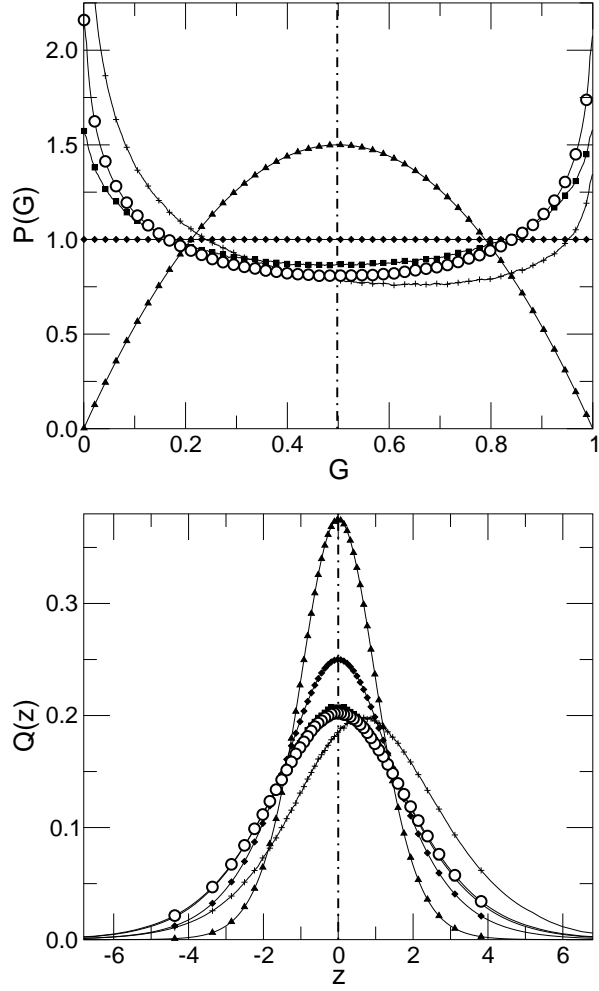}    
    \caption{Top: The fixed point distribution ($\circ$) of the conductance, 
      $P(G)$, at the quantum Hall transition obtained by starting from
      different initial distributions denoted by squares, diamonds, triangles,
      and the distribution after the 16th renormalization step ($+$). Bottom:
      the corresponding distributions of the heights of the saddle points,
      $Q(z)$ [From \cite{c04}]. }
    \label{fig:cainphd1}
  \end{center}
\end{figure}

The renormalization group transformation of the distribution of the
transmission amplitudes has been determined numerically by using this approach
\cite{crsr01,c04}. From $P(t)$ the distribution of the conductance $G=t^{2}$
can be determined,
\begin{equation}
  \label{eq:p(g)}
  P(G)=\frac{1}{2t}P(t)\,.
\end{equation}
This may be transformed to the distribution $Q(z)$ of the heights of the saddle
points $z$ measured relative to the critical energy $\epsilon=0$ (cf.
Eqs.~(\ref{eq:energyparameter}), (\ref{eq:result})). This can be obtained
from Eq.~(\ref{eq:p(g)}) by using the relation 
\begin{equation}
  \label{eq:g(z)}
  G=\frac{1}{e^{z}+1}\,.
\end{equation}
The result is
\begin{equation}
  \label{eq:q(z)}
  Q(z)=P(G)\left|\frac{{\rm d}G}{{\rm d}z}\right|=\frac{1}{4\cosh^{2}{(z/2)}}
\,P\!\left(\frac{1}{e^{z}+1}\right)\,.
\end{equation}
The distributions $P(G)$ and $Q(z)$ are shown in Fig.~\ref{fig:cainphd1}.

\begin{figure}[htbp]
  \begin{center}
    \includegraphics[width=0.6\linewidth]{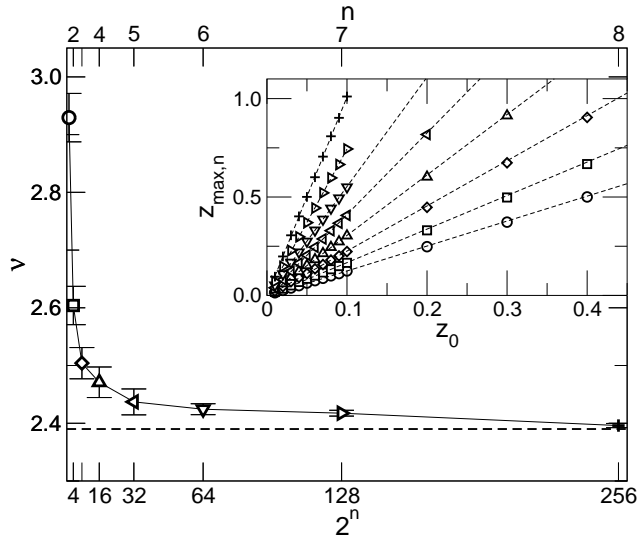}    
    \caption{The critical exponent $\nu$ determined from the scaling 
      behavior of the distribution of the heights of the saddle points $Q(z)$
      with the effective system size $2^{n}$ in the $n$th renormalization
      step. Dashed line: $\nu=2.39$. Inset: maximum $z_{{\rm max},n}$ as a
      function of $z_{0}$. The slopes yield $\lambda^{n}$ and thereby the
      critical exponents $\nu(n)$. }
    \label{fig:cainphd2}
  \end{center}
\end{figure}

The scaling properties of the distribution $Q(z)$ may be used to determine the
critical exponent with very high accuracy. The idea is that an initial shift
of the distribution by an amount $z_{0}$ will be amplified during the
renormalization procedure since the quantum Hall fixed point is unstable.
After $n$ steps, the maximum of the distribution will be shifted by an amount
$z_{{\rm max},n}=\lambda^{n}z_{0}$. After a certain number of steps, $n_{L}$,
the saddle point will be no longer transparent. This limit can be defined by
\begin{equation}
  \label{eq:opaque}
  z_{\rm max,n_{L}}=\lambda^{n_{L}}z_{0}\approx 1\,.
\end{equation}
By defining the localization length $\xi$ with the length $2^{n_{L}}a$ ($a$
lattice constant of the original lattice) one finds that $\xi $ diverges for
$z_{0}\to 0$,
\begin{equation}
  \label{eq:localizationlength}
  \xi\propto az_{0}^{-\nu}
\end{equation}
with the exponent $\nu=\ln{2}/\ln{\lambda}$ (cf. Eq.~(\ref{eq:tileexponent})).
The critical exponent can then be determined from
\begin{equation}
  \label{eq:criticalexp}
  \nu=\frac{\ln{2}^{n}}{\ln{(z_{{\rm max},n})}/z_{0}}\,,
\end{equation}
with $z_{{\rm max},n}/z_{0}=\lambda^{n}$. Figure \ref{fig:cainphd2} shows
results of the numerical calculation \cite{crsr01,c04}.  \vspace{5mm}
The numerical analysis of the data yields the result $\nu=2.39\pm 0.01$, in
excellent agreement with previous direct numerical simulations for the
localization length \cite{crsr01,c04}. As mentioned above, by assuming the
model as an independent realization of the fixed point ensemble, one can
interpret this result as an independent corroboration of the universality
hypothesis for the quantum Hall phase transition.

Though the value of the critical exponent is extremely precise, the form of
the critical distribution of the two terminal conductance is qualitatively
different from the one of the Chalker-Coddington random network. Instead of
$P(G)\rightarrow 0$ in the limit $G\rightarrow 0$ (Fig.\ref{fig:cond_dis}),
the renormalization group approach indicates divergence $P(G)\rightarrow
\infty$. Also near $G=1$, the present approach suggests $P(G)\rightarrow
\infty$, while the actual $P(G)$ is constant. For $G>1$, $P(G)=0$ in the case
of renormalization group, but in the simulation for the original model $P(G)$
shows a tail in this region. This indicates that the renormalization group
approach in fact does {\em not} approximate the Chalker-Coddington network
model although the underlying truncated-network model belongs to the same
universality class.

This renormalization group has recently been extended to the calculation of
the energy level statistics \cite{crr03}. The finite size scaling analysis of
the energy level spacing, Eq.~(\ref{eq:elsscaling}), has been applied to
estimate $\nu$ to be $2.37\pm 0.02$.

\section{Hamiltonians Related to the Network Model}
\label{sec:hamiltonians}

The model defined above is not only suitable for quantitative studies of the
critical behavior. By constructing effective Hamiltonians starting from the
transfer matrix, one can recognize that it is of much more general importance.
In the present chapter we will achieve such a construction. We first introduce
an equivalent tight binding Hamiltonian. In effective mass approximation, this
will be shown to be equivalent to a Dirac Hamiltonian. For the latter, some
analytical results are summarized which put the quantum Hall phenomenon in a
new perspective.

In this and in the subsequent chapters, we will use units such that $\hbar=1$,
$m^{*}=1$, and $e=1$, in order to simplify notation. However, in results for
the conductivity tensors, the correct unit $e^{2}/h$ is reintroduced since it
is of physical importance.  

The above network system has been completely specified in terms of its
transmission or scattering properties at {\em a given energy} of the original
model of an electron moving in a strong magnetic field and a slowly varying
random potential. By using this as a starting point for defining a
Hamiltonian, we fix {\em an entire spectrum of (quasi-)energies} which
contains as a parameter the original energy value. It is not obvious whether
or not there is some arbitrariness in such a procedure \cite{hc96}.

\subsection{A Tight Binding Hamiltonian}
\label{subsubsec:tightbinding}

We first want to establish a connection with a nearest neighbor tight binding
Hamiltonian. We start from the unitary matrices ${\bf S}$ that describe the
scattering at the individual saddle points, Eq.~(\ref{eq:decomposition}). We
first rewrite ${\bf S}$ by redefining phases, $\phi_{1}\to \phi_{1}+\pi$ and
$\phi_{4}\to \phi_{4}+\pi$, again in order to avoid unnecessary notation.
This does not change the physics, as it will be seen below that one of the
relevant system parameters is the total phase accumulated around a unit cell of
the lattice, $\phi=\sum_{j}\phi_{j}$, which is the number of flux quanta in
the unit cell multiplied by $2\pi$.
\begin{eqnarray}
  \label{eq:hochalker}
{\bf S}&=&
\left(
  \begin{array}{cc}
e^{-i\phi_{2}}&0\\
0&-e^{i\phi_{4}}
  \end{array}
\right)
\left(
  \begin{array}{cc}
-r&t\\
t&r
  \end{array}
\right)
\left(
  \begin{array}{cc}
-e^{i\phi_{1}}&0\\
0&e^{-i\phi_{3}}
  \end{array}
\right)\nonumber\\
&&\nonumber\\&&\nonumber\\
&=&\left(
  \begin{array}{cc}
e^{-i\phi_{2}}&0\\
0&-e^{i\phi_{4}}
  \end{array}
\right){\bf S}_{0}\left(
  \begin{array}{cc}
-e^{i\phi_{1}}&0\\
0&e^{-i\phi_{3}}
  \end{array}
\right)
\end{eqnarray}

\begin{figure}[htbp]
\vspace{3mm}
  \begin{center}
    \includegraphics[width=8cm]{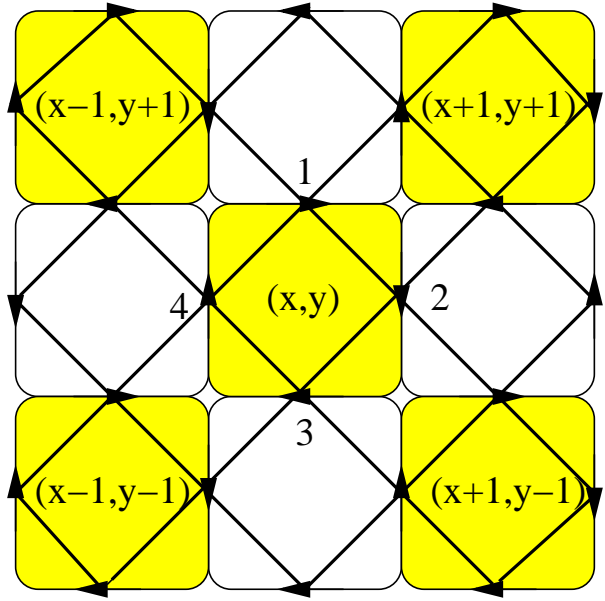}
  \end{center}
  \caption{The network model rearranged as a tight binding
    model on a two dimensional square lattice with four states per site. The
    links between saddle points are associated with the four directed site
    states (arrowheads) in the unit cell. These site states are connected by
    tunneling through the saddle points represented by $\diamondsuit$.}
  \label{fig:tightbinding}
\end{figure}

Then, we rearrange the network in the way shown in
Fig.~\ref{fig:tightbinding}. The loops that are connected by the saddle
points, described by their scattering matrix ${\bf S}_{0}$, are arranged in
the form of a two dimensional square lattice such that the centers of the
loops are associated with the lattice points
\begin{equation}
  \label{eq:squarelattice}
{\vv{R}}_{xy}=x{\vv{e}}_{x}+y{\vv{e}}_{y}
\end{equation}
with $x,y$ integer numbers. This notation might appear unconventional, but is
convenient for emphasizing the connection with the real space. The $x$- and
$y$-directions are assumed to point into the directions of the links that
connect ${\bf S}$ and ${\bf S}'$. The four links between adjacent saddle
points within each of the loops (denoted by arrowheads labeled with 1,2,3,4 in
Fig.~\ref{fig:tightbinding} \cite{hc96}) are now associated with four site
states within a unit cell of the lattice. The site states are characterized by
a lattice vector and a quantum number $\lambda$ in the lattice cell, and they
are assumed to form a complete set,
\begin{equation}
  \label{eq:completeness}
  \sum_{x,y,\lambda=1}^{4}| x,y,\lambda\rangle\langle x,y,\lambda |
={\bf 1}\,.
\end{equation}
These site states are connected via the tunneling and reflection matrix
elements through the saddle points within a given unit cell, and between
nearest neighbor cells.

In order to determine the matrix elements of the effective Hamiltonian, we
adopt the method used in \cite{km95} and \cite{hc96}. We write for the vector
of the amplitudes in the lattice cell at
${\vv{R}}=x{\vv{e}}_{x}+y{\vv{e}}_{y}$ after $M+1$ iteration steps
\begin{equation}
  \label{eq:timeevolution}
  \psi_{{\vt{R}}\lambda}(M+1)=
\sum_{{\vt{R}}'\lambda'=1}^{4}U_{{\vt{R}}\lambda,{\vt{R}}'\lambda'}
\,\psi_{{\vt{R}}'\lambda'}(M)\,.
\end{equation}
The unitary operator ${\bf U}$ can be interpreted as describing the evolution
of the wave function between steps $M$ and $M+1$. The eigenstates of the
$4L$-dimensional matrix ${\bf U}$ ($L$ number of sites)
\begin{equation}
  \label{eq:unitary}
  {\bf U}\,|\,\psi_{\alpha}\rangle=e^{i\omega_{\alpha}(E)}\,
|\psi_{\alpha}\rangle\qquad\qquad(\alpha=1\ldots\ldots4L)
\end{equation}
with eigenvalues 1 ($\omega_{\alpha}(E)=0$) are the stationary states of the
network with an energy parameterized by the energy dependent reflection
parameter of the saddle points, $r:= r(E)=\sqrt{1-t^{2}(E)}$.

Formally, one can interpret the above Eq.~(\ref{eq:timeevolution}) as a
^^ ^^ time dependent'' Schr\"o\-din\-ger equation by writing
\begin{equation}
  \label{eq:schroedinger}
\psi_{{\vt{R}}\lambda}(M+1)-\psi_{{\vt{R}}\lambda}(M)=
\sum_{{\vt{R}'},\lambda'=1}^{4}
({\bf U}-{\bf 1})_{{\vt{R}}\lambda,{\vt{R}'}\lambda'}
\,\psi_{{\vt{R}'}\lambda'}(M)\,.
\end{equation}
This suggests
\begin{equation}
  \label{eq:suggestion}
  {\bf U}={\bf 1}-\frac{1}{i}{\bf H}\,.
\end{equation}
As the Hamiltonian ${\bf H}$ must be self adjoint we define
\begin{equation}
  \label{eq:hamiltonian}
  {\bf H}=\frac{1}{2i}\left({\bf U}-{\bf U}^{\dagger}\right)\,.
\end{equation}
Note that the Hamiltonian is here dimensionless, since it contains a factor
$\Delta t/\hbar$, with $\Delta t=1$ the width of the time step in
Eq.~(\ref{eq:schroedinger}). 

The Hamiltonian has quasi-energy eigenvalues
\begin{equation}
  \label{eq:quasieigenvalues}
 \epsilon_{\alpha}=\sin\omega_{\alpha}\,. 
\end{equation}
Since each lattice cell is connected via four saddle points to its nearest
neighbors, ${\bf U}$ must have a $4\times 4$ block structure. The midpoints of
the links form a bipartite lattice. This means that each of the $4\times 4$
blocks can consist only of two non-zero $2\times 2$ blocks. By arranging the
amplitudes according to the sequence
$(\psi_{{\vt{R}}1},\psi_{{\vt{R}}3},\psi_{{\vt{R}}2},\psi_{{\vt{R}}4})^{\mathrm{T}}$
one gets for the $4\times 4$ block connecting with each other the site states
within a given lattice cell at ${\vv{R}}$, and with the states in the
nearest-neighbor cells,
\begin{equation}
  \label{eq:twoblocks}
  {\bf U}=
\left(
  \begin{array}{cc}
0&{\bf M}\\{\bf N}&0
  \end{array}
\right)\,,
\end{equation}
such that the $2\times 2$ block ${\bf M}$ connects the states
$\psi_{{\vt{R}}1},\psi_{{\vt{R}}3}$ with $\psi_{{\vt{R}}2},\psi_{{\vt{R}}4}$
and the block ${\bf N}$ links the states $\psi_{{\vt{R}}2},\psi_{{\vt{R}}4}$
with $\psi_{{\vt{R}}1},\psi_{{\vt{R}}3}$.

This gives for the $4\times 4$ block of the effective Hamiltonian
\begin{equation}
  \label{eq:twoblocks(h)}
   {\bf H}=
-\frac{1}{2i}\left(
  \begin{array}{cc}
0&{\bf N}^{\dagger}-{\bf M}\\{\bf M}^{\dagger}-{\bf N}&0
  \end{array}
\right)=
\left(
  \begin{array}{cc}
0&{\bf H}_{2}\\{\bf H}_{2}^{\dagger}&0
  \end{array}
\right)
\,.
\end{equation}
The matrices
\begin{equation}
  \label{eq:m}
 {\bf M}=
\left(
  \begin{array}{cc}
te^{i\phi_{1}}\tau_{-}^{x}\tau_{+}^{y}&re^{i\phi_{1}}\\
re^{i\phi_{3}}&-te^{i\phi_{3}}\tau_{+}^{x}\tau_{-}^{y}
  \end{array}
\right)\,,
\end{equation}
\begin{equation}
  \label{eq:n}
 {\bf N}=
\left(
  \begin{array}{cc}
re^{i\phi_{2}}&te^{i\phi_{2}}\tau_{+}^{x}\tau_{+}^{y}\\
te^{i\phi_{4}}\tau_{-}^{x}\tau_{-}^{y}&-re^{i\phi_{4}}
  \end{array}
\right)\,.
\end{equation}
contain the translation operators $\tau_{\pm}$ that connect nearest
neighbor cells,
\begin{equation}
  \label{eq:translations}
  \tau_{\pm}^{x}\psi_{{\vt{R}}\lambda}=\psi_{{\vt{R}}\pm {\bf e}_{x}\lambda}
\qquad\qquad
\tau_{\pm}^{y}\psi_{{\vt{R}}\lambda}=\psi_{{\vt{R}}\pm {\bf e}_{y}\lambda}\,.
\end{equation}
The Hamiltonian is then given by
\begin{equation}
  \label{eq:h2}
   {\bf H}_{2}=
-\frac{1}{2i}\left(
  \begin{array}{ccc}
re^{-i\phi_{2}}-te^{i\phi_{1}}\tau_{-}^{x}\tau_{+}^{y}&\,\,&
te^{-i\phi_{4}}\tau_{+}^{x}\tau_{+}^{y}-re^{i\phi_{1}}\\
te^{-i\phi_{2}}\tau_{-}^{x}\tau_{-}^{y}-re^{i\phi_{3}}&\,\,&
-re^{-i\phi_{4}}+te^{i\phi_{3}}\tau_{+}^{x}\tau_{-}^{y}
  \end{array}
\right)
\end{equation}

This means that within a given cell, say at ${\vv{R}}$, the tunneling matrix
elements are proportional to the reflection amplitudes of the saddle points
$r$
\begin{eqnarray}
  \label{eq:intracell}
  h_{{\vt{R}}1,{\vt{R}}4}&=&\frac{ r}{2i}\,e^{i\phi_{1}}\,,\\
  h_{{\vt{R}}2,{\vt{R}}1}&=&\frac{ r}{2i}\,e^{i\phi_{2}}\,,\\
  h_{{\vt{R}}3,{\vt{R}}2}&=&\frac{ r}{2i}\,e^{i\phi_{3}}\,,\\
  h_{{\vt{R}}4,{\vt{R}}3}&=&-\frac{ r}{2i}\,e^{i\phi_{4}}\,.
\end{eqnarray}
On the other hand, the eight coupling matrix elements to the nearest neighbor
cells are proportional to the transmission amplitude $t$. They are
\begin{eqnarray}
  \label{eq:intercell}
  h_{{\vt{R}}2,[{\vt{R}}+(1,1)]3}&=&\frac{ t}{2i}\,e^{i\phi_{2}}\,,\\
  h_{{\vt{R}}3,[{\vt{R}}+(1,-1)]4}&=&-\frac{ t}{2i}\,e^{i\phi_{3}}\,,\\
  h_{{\vt{R}}4,[{\vt{R}}+(-1,-1)]1}&=&\frac{ t}{2i}\,e^{i\phi_{4}}\,,\\
  h_{{\vt{R}}1,[{\vt{R}}+(-1,1)]2}&=&\frac{ t}{2i}\,e^{i\phi_{1}}\,,
\end{eqnarray}
and the four remaining matrix elements are the conjugates of these.

It is useful to consider as a starting point for the discussion of the
properties of this Hamiltonian the periodic limit. This will also yield the
Dirac Hamiltonian in the limit of small wave number, i.e. in effective mass
approximation. 

\begin{figure}[htbp]
  \begin{center}
\subfigure[]{\includegraphics[width=6cm]{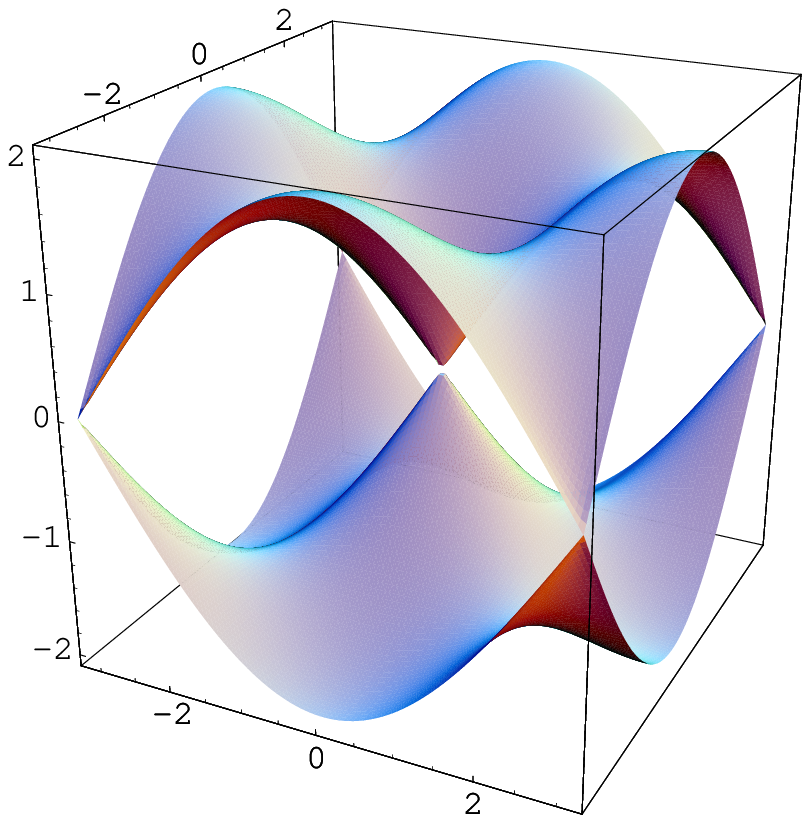}}
\hspace{1cm}
\subfigure[]{\includegraphics[width=6cm]{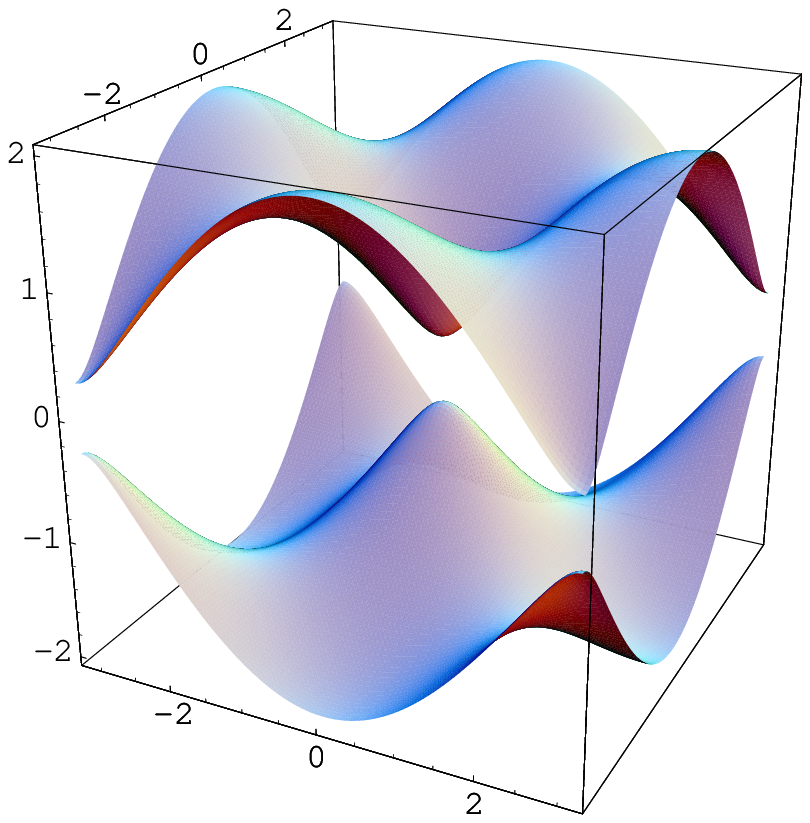}}
  \end{center}
  \caption{The band structure of quasi-energies
    $\epsilon_{j}({\vv{q}}-{\vv{A}})$ of the Hamiltonian representing a
    network with identical saddle points and equal phases in all of the cells
    of the tight binding lattice such that there is half of a flux quantum in
    each unit cell ($\phi =\pi$). In this case, the four bands
    (Eq.~(\ref{eq:bandstructure})) are degenerated with respect to $\pm$. (a)
    The spectrum exactly at the saddle point energy, $r=t=1/\sqrt{2}$, (b) for
    $r=0.5$.}
  \label{fig:bands}
\end{figure}

The periodic case is achieved by assuming the phases in each of the unit cells
and all of the saddle points as identical. The Hamiltonian can be diagonalized
exactly with the Bloch Ansatz
\begin{equation}
  \label{eq:bloch}
  \psi_{\lambda}({\vv{R}})=e^{i{\vt{q}}\cdot{\vt{R}}}u_{\lambda}(\vv{q})\,.
\end{equation}
The resulting band structure is (Fig.~\ref{fig:bands})
\begin{eqnarray}
  \label{eq:bandstructure}
  \epsilon_{j}({\vv{q}})&=&(-1)^{j}\sqrt{2}\,\Bigg[1-2rt\cos{\phi}
\cos{(q_{x}-A_{x})}\cos{(q_{y}-A_{y})}\nonumber\\
&&\nonumber\\
&&\qquad\quad\pm\sin{\phi}
\,\sqrt{1-4r^{2}t^{2}\cos^{2}{(q_{x}-A_{x})}\cos^{2}{(q_{y}-A_{y})}}
\Bigg]^{1/2}
\end{eqnarray}
with $j=1,2$, $A_{x}=(\phi_{1}-\phi_{3})/2$, $A_{y}=(\phi_{4}-\phi_{2})/2$ and
$\phi=\sum_{\lambda=1}^{4}\phi_{\lambda}$ the flux through the cell multiplied
by $e/\hbar$. Exactly at the energy of the saddle points, $r=t=1/\sqrt{2}$,
there is no gap between the bands. For $r\neq t$ a gap
$\Delta=\epsilon_{1}-\epsilon_{2}$ is opened. For
$\phi=0$ it is
\begin{equation}
  \label{eq:gap}
  \Delta= 2\sqrt{2}\sqrt{(1-2rt)}\approx m
\end{equation}
for $r= 1/\sqrt{2}+m/4$ ($m\ll 1$).

The formation of the gap can be discussed in better detail by expanding the
band structure near ${\vv{q}-\vv{A}}\approx 0$. On finds for small $m$
\begin{equation}
  \label{eq:effectivemass}
  \epsilon^{2}({\vv{q}})=m^{2}+({\vv{q}}-{\vv{A}})^{2}\,.
\end{equation}
Thus, the existence of the gap is closely related to the deviation from the
saddle point.

In the following, we discuss briefly the disordered version of the network.

If the reflection amplitude is unity ($r=1$, $t=0$), i.e. if the energy $E$ is
well below the saddle point energy, the unit cells decouple. The system
consists then of independent wave functions localized within the grey unit
cells in Fig.~\ref{fig:tightbinding}.  On the other hand, if $t=1$ ($r=0$),
i.e. for an energy well above the saddle point, the independent cell states
are localized in the complementary lattice of the white unit cells. The
quasi-energy eigenvalues in these decoupled limits are
 \begin{equation}
  \label{eq:quasienergies}
  \epsilon^{\pm}_{j}({\vv{R}},\phi)=
(-1)^{j}\sqrt{2}\left[1\mp \frac{1}{\sqrt{2}}
\left(1-\cos\phi\right)^{1/2}\right]^{1/2}\qquad(j=1,2)
\end{equation}
and $\phi=\phi({\vv{R}})$ is the flux through the loop at ${\vv{R}}$. For
completely independent and random phases, the fluxes in the individual cells,
and thus the quasi-energies $\varepsilon_{\alpha}$ are statistically
independent and random. When varying the energy $E$, i.e. varying $r(E)$ close
to the value of 1, the quasi-energies $\epsilon(E)$ will perform random walks
in the $(E,\epsilon)$-space. The energies of the original problem,
$E_{\beta}(\epsilon_{\alpha}=0)$, will then be statistically independent and
random. They correspond to the localized states associated with the valleys
and hills in the potential landscape.

If we assume that all saddle points have the same parameter $r$ and the phases
are independently and randomly distributed one can see that the localization
lengths of the eigenstates of $H$ are finite and independent of
$\epsilon_{\alpha}$. As $r\to r_{\rm c}=t_{\rm c}=1/\sqrt{2}$, the
localization length diverges uniformly throughout the spectrum \cite{hc96}.
This reflects the singularity of the localization length as a function of the
energy $E$ of the original problem.

We note that with disorder, i.e. assuming the phases to be independently and
randomly distributed, the matrix elements of the Hamiltonian remain
nevertheless strongly correlated. These correlations introduce peculiar
features into the localization behavior of the corresponding eigenstates. In
particular, they are responsible for the localization-delocalization
transitions occurring in the sub-bands induced by the magnetic field. A very
similar observation has been made earlier for the random matrix model where
the matrix elements of the banded matrix are also strongly correlated
\cite{hk89}.

\subsection{The Dirac Hamiltonian}
\label{subsubsec:dirac}

One can also relate the network model with the two dimensional Dirac equation.
This can be done most straightforwardly by considering the effective mass
approximation of the tight binding Hamiltonian. Alternatively, one can
consider the two-step unitary operator \cite{hk89}
\begin{equation}
  \label{eq:twostep}
  {\bf U}^{2}=\left(
    \begin{array}{cc}
{\bf M}{\bf N}&0\\
0&{\bf N}{\bf M}
    \end{array}
\right)=\left(
  \begin{array}{cc}
{\bf V}&0\\
0&{\bf W}
  \end{array}
\right)
\end{equation}
The system of four equations decouples into pairs, and one can deal
with, say, only the upper block ${\bf V}$.

Furthermore, by assuming that the displacement operators act on smooth
functions one can replace
\begin{equation}
  \label{eq:differentials}
  \tau^{x}_{\pm}={1}\pm\partial_{x},\qquad \tau^{y}_{\pm}={1}
\pm\partial_{y}
\end{equation}
For small phases and with the definition $r=r_{{\rm
    c}}+m/4$ ($t\approx t_{\rm c}-m/4$) one can expand
\begin{equation}
  \label{eq:v}
  {\bf V}=e^{-i\tilde{H}}\approx {\bf 1}-i\tilde{{\bf H}}
\end{equation}
with the unit matrix ${\bf 1}$.
This is justified for small $\tilde{{\bf H}}$. From the matrix elements of
${\bf M}$ and ${\bf N}$, Eqs.~(\ref{eq:m}) and (\ref{eq:n}), one finds
\begin{equation}
  \label{eq:v1}
  {\bf V}\approx {\bf 1}+\left(
    \begin{array}{cc}
-\partial_{x}+iA_{x}&\partial_{y}-iA_{y}+m\\
\partial_{y}-iA_{y}-m&\partial_{x}-iA_{x}
    \end{array}
\right)-i\phi{\bf 1}={\bf 1}-i\tilde{{\bf H}}\,.
\end{equation}
With the unitary transformation
\begin{equation}
  \label{eq:transformation}
  {\bf R}=\frac{1}{\sqrt{2}}\left(
    \begin{array}{cc}
i&-1\\
i&1
    \end{array}
\right)
\end{equation}
one can transform
\begin{equation}
  \label{eq:unitarytrnsformation}
  {\bf H}={\bf R}\tilde{{\bf H}}{\bf R}^{-1}
\end{equation}
such that one obtains finally the two dimensional Dirac Hamiltonian
\begin{equation}
  \label{eq:dirac}
  {\bf H}=(p_{x}-A_{x}){\bf \sigma}_{x}+(p_{y}-A_{y}){\bf \sigma}_{y}+
m{\bf \sigma}_{z}+\phi{\bf 1}\,.
\end{equation}
Here, $p_{j}=-i\partial_{j}$ ($j=x,y$) are the components of the momentum
operator and
\begin{equation}
  \label{pauli}
{\bf \sigma}_{x}=\left(
  \begin{array}{cc}
0&1\\
1&0
  \end{array}
\right),\quad
{\bf \sigma}_{y}=\left(
  \begin{array}{cc}
0&-i\\
i&0
  \end{array}
\right),\quad{\bf \sigma}_{z}=\left(
  \begin{array}{cc}
1&0\\
0&-1
  \end{array}
\right)
\end{equation}
are the Pauli matrices. In this Hamiltonian, randomness can be introduced in
different ways. Via randomness in the individual phases one can make the
components of the vector potential random. Randomizing the total Aharonov-Bohm
phases in the loops produces randomness in the scalar potential $\phi$.
Finally, assuming the tunneling parameters of the saddle points to be random
gives fluctuations in the mass parameter $m$.

Introducing disorder in the various parameters of the model leads to breaking
of symmetries, as noted in \cite{lfsg94} and indicated in
Tab.~\ref{tab:diracsymmetries}. The first term in the Hamiltonian, ${\bf
  H}_{0}=\vv{\sigma}\cdot{\vv{p}}$ [$\vv{\sigma}=(\sigma_{x},\sigma_{y})$,
${\vv{p}}=-i\nabla=-i(\partial_{x},\partial_{y})$], is invariant under an
effective {\em time reversal} operation
\begin{equation}
  \label{eq:timereversal}
{\bf H}_{0}=\sigma_{z}{\bf H}_{0}^{*}\sigma_{z}\,.  
\end{equation}
This means that there is a Kramers degeneracy, $\psi$ and $\sigma_{z}\psi^{*}$
are degenerate eigenstates as is easily seen. Similarly, it is easy to see
that $\psi$ and $\sigma_{z}\psi$ are eigenstates to energies $E$ and
$-E$, respectively. This is equivalent to {\em particle-hole} symmetry
\begin{equation}
  \label{eq:particlehole}
-{\bf H}_{0}=\sigma_{z}{\bf H}_{0}\sigma_{z}\,.
\end{equation}
Finally, ${\bf H}_{0}$ is {\em parity} invariant under $x$ and $y$-reflections
$P_{j}jP_{j}=-j,\, (j=x,y)$, for instance
\begin{equation}
  \label{eq:parity}
  {\bf H}_{0}=P_{y}\sigma_{x}{\bf H}_{0}\sigma_{x}P_{y}\,.
\end{equation}
The Dirac mass term ${\bf H}_{m}=m\sigma_{z}$ possesses none of these
symmetries, as is easily seen. The vector potential term ${\bf
  H}_{\vt{A}}=-{\vv{A}}\cdot\vv{\sigma}$ preserves only particle-hole
symmetry, and the random scalar potential ${\bf H}_{\phi}=\phi {\bf 1}$ only
time reversal symmetry. These symmetries are very important for understanding
the quantum phases of the system as will be seen below in more detail.
\begin{table}[htbp]
\vspace{9mm}
\begin{center}
\begin{tabular}{lcccc}
\hline\hline
symmetry&${\bf \sigma}\cdot{\vv{p}}$&$m\sigma_{z}$&$\phi$
&${\bf \sigma}\cdot{\vv{A}}$\\\hline
parity&yes&no&no&no\\
time reversal&yes&no&yes&no\\
particle-hole&yes&no&no&yes\\
\hline
\end{tabular}
\vskip10mm
\caption[]{Symmetry breaking by disorder in the various parameters of
  the Dirac model (${\bf p}$ momentum operator; $m$ Dirac mass parameter;
  $\vv{\sigma}$ vector of the Pauli matrices $\sigma_{x}$, $\sigma_{y}$;
  $\phi$ scalar potential; ${\vv{A}}$ vector potential).}
\label{tab:diracsymmetries}
\end{center}
\vspace{10mm}
\end{table}

\subsection{Some Results for the Two Dimensional Dirac Model}
\label{sec:lsfg94}

In this section we discuss briefly some instructive analytical results
obtained for the Dirac Hamiltonian.

The correspondences between the quantum Hall problem, certain tight binding
models and the two dimensional Dirac model have been noticed by several authors
\cite{lfsg94,ff85,l94,z95}. Fisher and Fradkin \cite{ff85} have obtained the
Dirac model by starting from a two dimensional tight binding model with
diagonal on-site disorder in a perpendicular magnetic field with half a
magnetic flux quantum in the unit cell. They constructed a field theory for
the diffusive modes which was shown to be in the same universality class as
the orthogonal $O(2n,2n)/O(2n)\times O(2n)(n\to 0)$ non-linear $\sigma$-model.
This implies that all states are localized as in the absence of a magnetic
field. It suggests that if delocalization occurs with magnetic field, it must
be a direct consequence of breaking time-reversal symmetry instead of some
other properties of the field. Generalizations to the several-channel
scattering problem have also been discussed \cite{f86}.

Ludwig and collaborators \cite{lfsg94} have used a tight binding model on a
square lattice with nearest and next-nearest neighbor coupling, half a flux
quantum per unit cell and a staggered potential energy $\mu (-1)^{x+y}$ as a
starting point. At low energies, the model was shown to be equivalent to a
Dirac model with two Dirac fields. They found that without disorder this model
exhibits an integer quantum Hall phase transition as a function of some
control parameter which is essentially the mass $m$ of the lighter Dirac
field. This is similar to the transitions of the Hall conductance between
integer multiples of $e^{2}/h$ obtained in the original clean Landau model as
a function of the {\em Fermi energy}, and does not mean that a plateau exists
when the electron density is varied. The transition of the Dirac model has
been shown to belong to the two dimensional Ising universality class. The
associated exponents and the critical transport properties were determined.

\begin{figure}[htbp]
 \begin{center}
   \includegraphics[width=7cm]{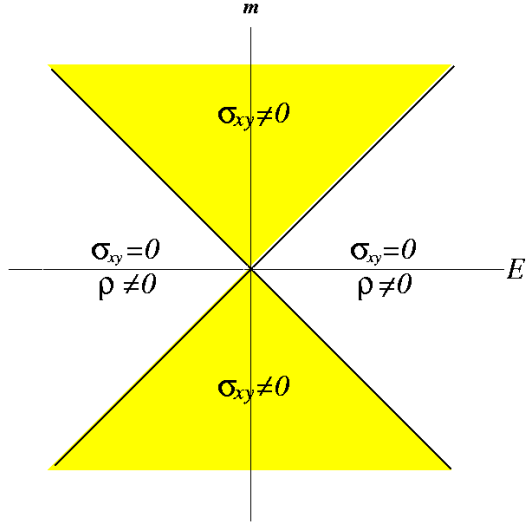}
    \caption{The phase diagram of the ordered Dirac model
      \protect\cite{lfsg94} showing regions of non-zero density of states
      $\rho$ ($|E|>|m|$) and non-zero Hall conductance
      $\sigma_{xy}=\pm\,e^{2}/2h$ ($|E|<|m|$).}
    \label{fig:dirac}
  \end{center}
\end{figure}

The density of states,
\begin{equation}
  \label{eq:diracdos}
  \rho(E)=\frac{|E|}{2\pi}\Theta(|E|-m)\,,
\end{equation}
vanishes at $E=0$.
It can readily be obtained from Eq.~(\ref{eq:effectivemass}) with ${\vv{A}}=0$
using
\begin{equation}
  \label{eq:dosdef}
  \rho(E)=\frac{1}{(2\pi)^{2}}\,\int{\rm d}^{2}\vv{q}\,
\left[\delta\left(E-\sqrt{q^{2}+m^{2}}\right)
+\delta\left(E+\sqrt{q^{2}+m^{2}}\right)\right]
\,.
\end{equation}
Applying linear response theory to the Dirac system one can determine the Hall
conductivity by calculating the ratio of, say, the current density in the
$x$-direction, $j_{x}$ and the electric field, $E_{y}$, in the $y$-direction
\begin{equation}
  \label{eq:sigmaHall}
\sigma_{xy}= \frac{j_{x}}{E_{y}}=\frac{e^{2}}{h}\,\frac{m}{2\pi^{2}}\,
\int\frac{{\rm d}^{2}\vv{q} {\rm d}\omega}
{[(i\omega-E)^{2}-q^{2}-m^{2}]^{2}}\,.
\end{equation}
It is found that at zero energy, where the above density of
states vanishes, the Hall conductance jumps by $e^{2}/h$ at $m=0$
(Fig.~\ref{fig:dirac})
\begin{equation}
  \label{eq:diracHall}
  \sigma_{xy}(m)=\frac{{\rm sgn}(m)}{2}\,\frac{e^{2}}{h}\,.
\end{equation}
The heavier Dirac field contributes towards the Hall conductance with
$e^{2}/2h$ such that the total Hall conductance jumps from $0$ to $e^{2}/h$.
Simultaneously, the magneto-conductivity $\sigma_{xx}$ is non-zero only at
this critical point,
\begin{equation}
  \label{eq:sigmaxx}
  \sigma_{xx}=\sigma_{0}\frac{e^{2}}{h}\delta_{m,0}\,,
\end{equation}
with the Kronecker-symbol $\delta_{m,0}$ equal 1 for $m=0$ and 0 for $m\neq
0$, and the constant $\sigma_{0}$ is $\pi/8$. The critical point shows
time-reversal, particle-hole and parity invariance. Thus, the clean two
dimensional Dirac model exhibits a quantum Hall transition at $E=m=0$, i.e. a
step in the Hall conductivity.

According to our above derivation of the Hamiltonian, introducing spatial
randomness in the Dirac mass, $m=m(x,y)$, is equivalent to introducing
randomness in the saddle point energies. This breaks all of the above
symmetries (cf. Tab.~\ref{tab:diracsymmetries}). However, this kind of
randomness alone, if it is sufficiently weak, does not introduce a
nonvanishing density of states at zero energy. This means that it cannot place
the system into the quantum Hall universality class, in contrast to what one
might suspect. For sufficiently slowly varying mass, the zero-energy wave
functions are argued to be confined along the contours $m(x,y)=0$ since for
any region with a non-vanishing Dirac mass the energy must be non-zero. When
the mass varies slowly in space, the transition can be interpreted in terms of
the percolation of these states. In the absence of randomness of the phases
(i.e. the vector potential), the corresponding critical exponent of the
correlation length is that of the classical percolation model. The analysis in
Ref.~\cite{lfsg94} for weak randomness using the replica trick and
perturbative renormalization group analysis strongly suggests that the random
Dirac mass term is always marginal for the critical properties of the quantum
Hall transition. The system scales to vanishing disorder, and the transition
is described by the free Dirac theory. The replicated effective action is
formally equivalent to that of the random-bond Ising model (cf. next chapter)
with the randomness in the Dirac mass corresponding to the randomness in the
Ising bonds.

Randomness in the scalar potential is equivalent to assuming random fluxes
piercing the plaquettes of the network. For small random scalar potential one
can use the perturbative results obtained for the random Dirac mass. This can
be suspected from the form of the Hamiltonian
\begin{equation}
  \label{eq:scalarandomdirac}
  H={\vv{\sigma}}\cdot\vv{p}+m{\bf \sigma}_{z}+\phi{\bf 1}\,.
\end{equation}
Specifically, it can be shown that the problem with real disorder in the Dirac
mass is formally equivalent to a problem with a purely imaginary random scalar
field which corresponds to a negative disorder strength, $\langle
\phi^{2}\rangle <0$. Equivalently, a positive scalar disorder strength maps
onto a {\em negative} disorder strength for the Dirac mass. Since randomness
in the Dirac mass was found to be marginal, the opposite is true for the
scalar randomness, and the renormalization flow of the scalar disorder will
now be {\em away} from the fixed point instead of towards it. Thus, the scalar
randomness drives the system to some strong coupling regime. Eventually, one
then expects the generation of a non-vanishing density of states. It is
further argued by considering two-particle properties that the transition very
probably has to be described by a symplectic non-linear sigma model due to the
time reversal invariance of every member of the ensemble \cite{lfsg94}.

The case of only a random vector potential was found to be particularly
interesting since it can be treated to a large extent analytically
\cite{lfsg94,ac79,ntw94,ntw95,b95,ckt96,kmt96}. This limit has several
remarkable properties. The Hamiltonian is
\begin{equation}
  \label{eq:randomvectorpot}
  H_{A}=\vv{\sigma}\cdot\vv{p}+\vv{\sigma}\cdot\vv{A}
\end{equation}
and according to what has been said before, the random vector potential is due
to to the random phases along the links of the network. We assume that
$\vv{A}$ satisfies a Gaussian white noise distribution with zero mean and
variance $\Delta _{A}$. The corresponding random magnetic field,
$\vv{B}=\nabla \times \vv{A}$, is then also Gaussian distributed with the
variance
\begin{equation}
  \label{eq:varianceofB}
  \langle B(k)B(k')\rangle=(2\pi)^{2}k^{2}\Delta_{A}\delta(k+k')
\end{equation}
with the Fourier transform $B(k)$ of the magnetic field [perpendicular to the
$(x,y)$-plane]. As the Dirac mass is assumed to vanish, this Hamiltonian is
expected to implement a critical theory. In fact, the model exhibits a fixed
line corresponding to multi-fractal wave functions at $E=0$ \cite{lfsg94}.

For nonvanishing uniform Dirac mass the squared of the Hamiltonian is
\begin{equation}
  \label{eq:mAphi}
  H^{2}=H_{A}^{2}+m^{2}
\end{equation}
for each realization of the random vector potential. Thus, the energy
eigenvalues satisfy $E^{2}>m^{2}$. The gap in the pure Dirac model with $m\neq
0$ is thus not closed by the randomness in the vector potential, at least as
long as the disorder is small.

For this model, several single particle and two-particle properties can be
determined exactly. As an example, we consider the zero-energy wave function.
It is useful to use the Coulomb gauge, $\nabla\cdot\vv{A}=0$. This implies
that $\vv{A}$ can be written in terms of a scalar field $\Phi(x,y)$
\begin{eqnarray}
  \label{eq:coulombgauge}
  A_{x}&=&\partial_{y}\Phi\\
A_{y}&=&-\partial_{x}\Phi\,.
\end{eqnarray}
By inserting the real Ansatz wave functions
\begin{equation}
  \label{eq:E=0wave functions}
  \Psi_{\pm}\propto({\bf 1}\pm\sigma_{z})
 \left( \begin{array}{c}
e^{\Phi}\\
e^{-\Phi}
  \end{array}
\right)
\end{equation}
into the Schr\"odinger equation of the Hamiltonian
Eq.~(\ref{eq:randomvectorpot}) one easily verifies that they are exact
nodeless eigenfunctions corresponding to $E=0$.  Furthermore, it follows from
the symmetry properties of the Hamiltonian and the assumption of zero fluxes
in the plaquettes that these are {\em all} of the zero-energy eigenfunctions.

These random wave functions are {\em not} localized, and {\em not}
normalizable in the thermodynamic limit. In order to quantify their
statistical nature it is useful to introduce their counterparts normalized to
a square of the size $L^{2}$.
\begin{equation}
  \label{eq:normalization}
  \psi(x,y)=\frac{1}{C}\,e^{\Phi(x,y)}
\end{equation}
with 
\begin{equation}
  \label{eq:norm}
  C^{2}=\int_{L^{2}}e^{2\Phi(x,y)}{\rm d}x{\rm d}y\,,
\end{equation}
where the integration is over the square $L^{2}$ and to consider the moments
of the corresponding density, the average inverse participation numbers
introduced in section \ref{subsec:wf_qcp}
\begin{equation}
  \label{eq:participationratios}
  p_{q}(L)=\langle|\psi|^{2q}\rangle_{L^{2}}
\end{equation}
with $\langle\ldots\rangle_{L^{2}}$ denoting the configurational average in
$L^{2}$. 

For the above normalized zero energy wave function,
Eq.~(\ref{eq:normalization}), with Gaussian distributed $\Phi$, the average
inverse participation numbers can be determined
\cite{lfsg94,cmw96,cetal97,cd01,mdh2002}. One finds for $q_{\rm
  c}=\sqrt{2\pi/\Delta_{A}}> 1$
\begin{eqnarray}
  \label{eq:tauexact1}
  \tau(q)= \left\{
    \begin{array}{ll}
2(q-1)(1 - q/q_{\rm c}^{2}),&\quad |q|\leq q_{\rm c}\,,\\
2q\left(1-{\rm sgn}\,q/q_{\rm c} \right)^{2},&\quad |q|>q_{\rm c}\,,
    \end{array}\right.
\end{eqnarray}
For $q_{\rm c}\leq 1$ one gets
\begin{eqnarray}
  \label{eq:tauexact2}
  \tau(q)= \left\{
    \begin{array}{ll}
-2(1 - q/q_{\rm c})^{2},&\quad |q|\leq q_{\rm c}\,,\\
4(q-|q|)q_{\rm c}^{-1},&\quad |q|>q_{\rm c}\,,
    \end{array}\right.
\end{eqnarray}
The nonlinearity in $q$ indicates
multi-fractal scaling of the {\em extended} wave function. The resulting
$f(\alpha)$-spectrum is parabolic with
\begin{eqnarray}
  \label{eq:alphamax}
  \alpha_{0}=\left\{  \begin{array}{ll}
2(1+q_{\rm c}^{-2}),&\quad q_{\rm c}>1\,,\\
4/q_{\rm c},&\quad q_{\rm c}\leq 1\,,
  \end{array}\right.
\end{eqnarray}
depending on the variance of the random vector potential $\Delta_A$, Eq.
(\ref{eq:varianceofB}) (see Sect. \ref{subsec:wf_qcp}).

The density of states was found to show a power law dependence of the energy
near $|E|\approx 0$ with the exponent varying continuously upon moving along
the fixed line \cite{mrf03},
\begin{equation}
  \label{eq:diracdos1}
  \rho(E)\propto E^{(2-z)/z}\,,
\end{equation}
with 
\begin{eqnarray}
  \label{eq:z}
  z=\left\{
    \begin{array}{ll}
1+2/q_{\rm c}^{2},&\quad q_{\rm c}>1\,,\\
4/q_{\rm c}-1,&\quad q_{\rm c}\leq1\,.\\
    \end{array}\right.
\end{eqnarray}
For small disorder, the density of states vanishes. For $\Delta_{A}=\pi$,
$\rho(E\to 0)=\mathrm{const}$, and it diverges for $\Delta_{A}>\pi$. 

The diagonal conductance was found to be $e^{2}/\pi h$ along the fixed line.

The Dirac model with only randomness in the vector potential belongs to the
AIII symmetry class to be discussed in the chapter \ref{sec:symmetries} (see
also Tab.~\ref{table:sym2}). The critical wave functions at $E=0$ for random
$\pi$ phase gauge field \cite{hwk97,mh97,mh98} as well as for arbitrary random
gauge field \cite{rh01} have been investigated in detail.

If the phases along the links of the network are assumed to be independent and
random, the equivalent Dirac model has randomness both in the vector potential
{\em and} in the fluxes through the plaquettes. Such a system has none of the
above symmetries (cf. Tab.~\ref{tab:diracsymmetries}) and obviously belongs to
the original Chalker-Coddington class with a fixed point that corresponds to
the quantum Hall phase transition.

\begin{figure}[htbp]
 \begin{center}
   \includegraphics[width=6cm]{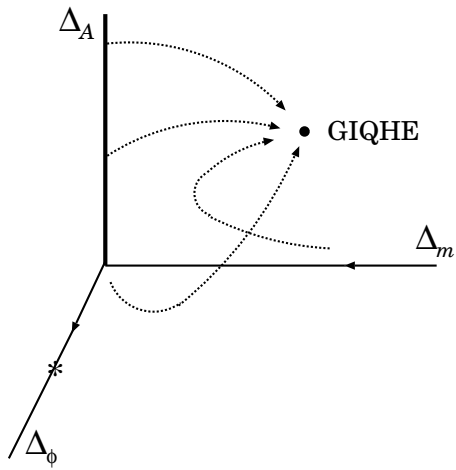}
    \caption{The global phase diagram of the Dirac model including randomness
      in the vector potential, $\Delta_{A}$, the Dirac mass, $\Delta_{m}$, and
      the scalar potential, $\Delta_{\phi}$ (after \protect\cite{lfsg94}). The
      generic fixed point of the Integer Quantum Hall Effect corresponds to
      all of the three different kinds of disorder nonvanishing, and seems to
      be inaccessible analytically.}
    \label{fig:phasediagramme}
  \end{center}
\end{figure}

The scaling picture of the Dirac model with all of the three different kinds
of disorder --- random Dirac mass (parameter $\Delta_{m}$, random saddle point
energies), random scalar potential (parameter $\Delta_{\phi}$, random fluxes
through the plaquettes of the network) and random vector potential (parameter
$\Delta_{A}$, random link phases) --- has been qualitatively sketched by
Ludwig and collaborators on the basis of their analytical results
\cite{lfsg94} (Fig.~\ref{fig:phasediagramme}). The phase diagram consists of a
critical line $(\Delta_{m}=\Delta_{\phi}=0, \Delta_{A}\neq 0)$ which is
unstable with respect to both $\Delta_{m}$ and $\Delta_{\phi}$, a two
dimensional Ising fixed point at $\Delta_{m}=\Delta_{\phi}=\Delta_{A}=0$ which
is only stable for $\Delta_{m}\neq 0$, and a fixed point at some
$\Delta_{\phi}\neq 0$ which was argued to belong to the universality class of
the symplectic non-linear sigma model. According to this scenario, the fixed
point of the genuine integer quantum Hall transition is in a strong coupling
regime with all of the three different kinds of disorder present. Until now,
this has not yet been accessible analytically. It will be the subject of
subsequent chapters to review the approaches to extract nevertheless
analytical information on the GIQHE, and to shed some light on the obstacles
which have still prevented an analytical derivation of its critical parameters.

\section{Relation to the Two-Dimensional Random Bond Ising Model}
\label{sec:ising}

In the last chapter we have seen how mapping to different Hamiltonians can
help to provide new insights into the properties of the quantum Hall phase
transition. In this chapter we will address in addition the opposite question:
given we know the network model and know how to use it for understanding the
quantum Hall critical scenario, can we use this knowledge for understanding
better the nature of phase transitions in two dimensions? As the random two
dimensional Ising model for a system of interacting spins is a fundamental
prototype for phase transitions in two dimensions, for answering this question
it is useful first to establish a relation between the random network and the
Ising model.  Such a mapping has been performed by various authors
\cite{cf97a,rl01,grl01}.  We describe here the elementary route that has been
worked out in the seminal paper by Merz and Chalker \cite{mc02}.

\subsection{The Ising Model}

The Ising model in two dimensions is defined by the Hamiltonian \cite{i25}
\begin{equation}
  \label{eq:ising01}
  H=-\sum_{ij}J_{ij}S_{i}S_{j}\,.
\end{equation}
The exchange integrals $J_{ij}$ connect the sites $i$ and $j$ of a regular
square lattice. The variables $S_{i}$ are the $z$-components of the spin
operators associated with the lattice sites.

We consider the simplest case, where $J_{ij}$ connect nearest neighbors only,
and spin $1/2$. In the ordered limit all of the exchange integrals are the
same. This has been solved exactly by Onsager in classical paper using Lie
algebras \cite{o44}. A comprehensive treatment using the transfer matrix
method is described in \cite{sml64}. The basic reason for the Ising model to
be an exactly solvable many-body problem is that eventually it amounts to
nothing else but the diagonalization of a quadratic form.

\begin{figure}[htbp]
  \begin{center}
   \includegraphics[width=0.6\textwidth]{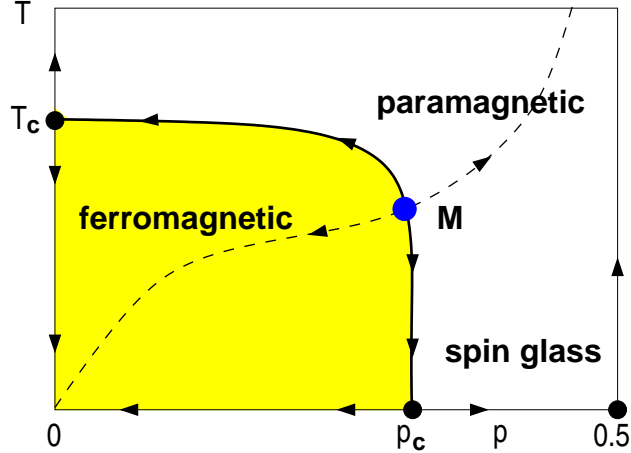}
    \caption{Schematic phase diagram of the Random Bond Ising Model. 
      For small fraction of antiferromagnetic bonds one observes a transition
      from a ferromagnetic to a paramagnetic phase with increasing
      temperature. For larger fraction of antiferromagnetic bonds, the
      ferromagnetic phase is changed to a spin glass. At the multicritical
      point $M$ the three phases coexist (dashed: Nishimori line; arrows:
      renormalization flow).  }
    \label{fig:phasediagramofRBIM}
  \end{center}
\end{figure}

However, in the disordered case, when the exchange parameters are chosen at
random, the model cannot be treated exactly, in spite of being quadratic. We
consider $J_{ij}$ as independent random variables with a distribution
$P(J_{ij})$. Specifically one can assume a two-component distribution,
\begin{equation}
  \label{eq:ising02}
  P(J_{ij})=p\delta(J_{ij}+J)+(1-p)\delta(J_{ij}-J)\,.
\end{equation}
This is the Random Bond Ising Model. For $J>0$, it corresponds to a lattice
containing $p$ antiferromagnetic and $(1-p)$ ferromagnetic bonds. In this
simple case, the model shows a phase transition between an ordered
ferromagnetic phase at low temperatures and a paramagnetic high temperature
phase for small $p$. At larger $p$ the ferromagnetic phase is destroyed in
favor of a spin glass phase in high dimensions, $d\geq 3$, with a
multi-critical point, also called Nishimori point \cite{n81,n86,dh88}, where
the three phases coexist (Fig.~\ref{fig:phasediagramofRBIM})
\cite{by86,ys77,mr82,m84,on87,on87a}. In two dimensions the spin glass phase
becomes unstable for $T\neq 0$. However, the multi-critical point survives.
Along the Nishimori line which crosses the phase boundary at the Nishimori
point, the internal energy can be calculated exactly.

The disordered Ising system has been studied using Monte Carlo simulations and
the transfer matrix method in the spin basis
\cite{m84,on87,on87a,h01,mb80,uo91,ko92,rqs99,hpp01}. Mapping the random Ising
model to fermionic models can have technical advantages: one can avoid the
random-sampling errors of the Monte Carlo technique and it is possible to
avoid the exponential growth of the transfer matrix. Pioneering work in this
direction has been done by starting from the solution of the two dimensional
Ising model using a Pfaffian \cite{b82,bp91,bgp98,sk93,sk94,i95,s00}. In this
approach the statistical properties of the system are written in terms of the
spectral properties of the corresponding matrix. The latter is essentially a
tight binding Hamiltonian on the underlying Ising lattice with random hopping
matrix elements. Thus, a link between the random bond Ising system and the
non-interacting localization problem is established. Alternative approaches
have been formulated by using Dirac fermions which eventually have been
condensed into a random network model with a special symmetry
\cite{lfsg94,cf97a,rl01,grl01,mc02,cetal02}.

The mapping of the Ising model to a random network model is done in three
steps. First, one introduces the conventional description of the partition
function in terms of transfer matrices \cite{sml64}. In the second step, the
transfer matrices are written in terms of fermion operators instead of
spinors. Finally, the network model is introduced by using the equivalence of
second- and first-quantized forms of linear transformations. In the next
sections, we establish the random network version of the Ising model
following \cite{mc02}.

\subsection{Transfer Matrix Formulation of the Ising Model}

\begin{figure}[htbp]
  \begin{center}
   \includegraphics[width=0.9\textwidth]{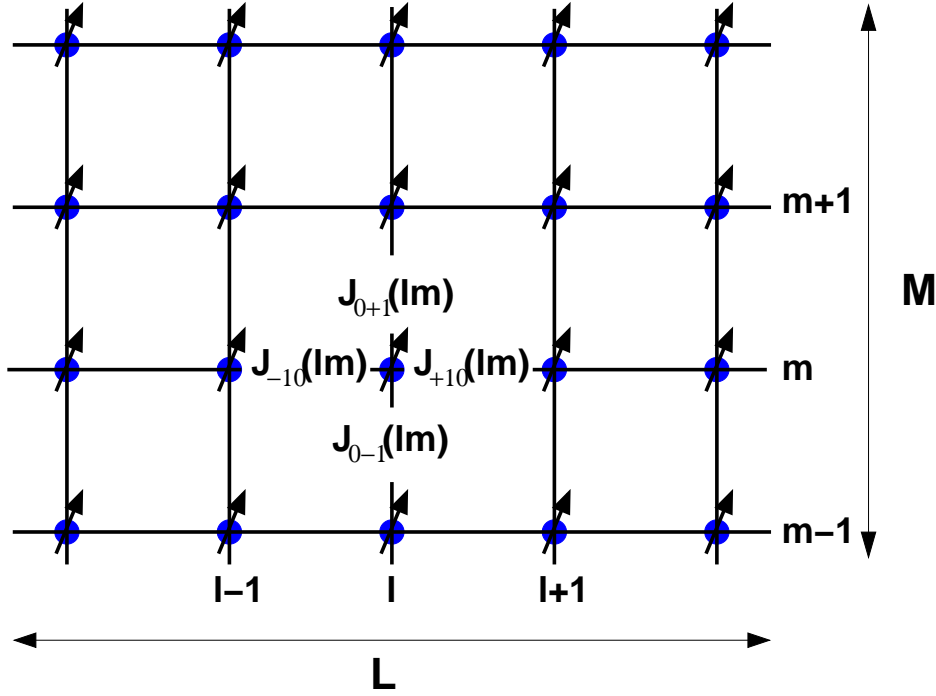}
    \caption{The two dimensional Ising model on a square lattice of 
      length $L$ and width $M$ with exchange matrix elements $J_{\delta}(n,m)$
      between nearest neighbors.}
    \label{fig:isingmodel}
  \end{center}
\end{figure}

We consider the two dimensional Ising model on a square lattice of length $L$
and width $M$ with random nearest-neighbor exchange couplings
$J_{\delta}(l,m)$ (Fig.~\ref{fig:isingmodel}). The pair of integers $(l,m)$
denotes the coordinates of a lattice point in $L$- and $M$-directions,
respectively. The index $\delta=\lambda,\mu$ indicates whether
$J_{\delta}(l,m)$ couples to a nearest neighbor in the directions along $L$ or
$M$, respectively.

In order to introduce the transfer matrix description we start by considering
the one dimensional case $M=1$ \cite{sml64}. The Hamiltonian in suitable
units is
\begin{equation}
  \label{eq:1dising}
H(\sigma_{1}\ldots \sigma_{L})=
-\sum_{l=1}^{L-1}J_{\lambda}(l)\sigma_{l}\sigma_{l+1}\,,
\end{equation}
with independent random $J_{\lambda}(l)$. The partition function is
\begin{equation}
  \label{eq:partition}
  Z={\rm Tr}\,e^{-\beta H(\sigma_{1}\ldots \sigma_{L})}
={\rm Tr}\,\exp{\sum_{l}\kappa_{\lambda}(l)\sigma_{l}\sigma_{l+1}}
={\rm Tr}\,\prod_{l}T_{l}\,,
\end{equation}
with the ^^ ^^ bond strength'' $\kappa_{\lambda}(l)=\beta J_{\lambda}(l)$
($\beta$ inverse temperature) and the transfer matrices
\begin{equation}
  \label{eq:1dtransfer}
  T_{l}=e^{\kappa_{\lambda}(l)\sigma_{l}\sigma_{l+1}}\,.
\end{equation}
Since the spin can take only values $\pm 1$ on the lattice sites the transfer
matrices can be rewritten as $2\times 2$ matrices with the matrix elements
given by $\langle \pm , l+1 \mid T \mid \pm , l \rangle$,
\begin{equation}
  \label{eq:2by2matrix}
  T_{l}=e^{\kappa_{\lambda}(l)}\mbox{\boldmath $\sigma_{0}$}
+e^{-\kappa_{\lambda}(l)}\mbox{\boldmath $\sigma_{1}$}\,.
\end{equation}
With the definition of the Kramers-Wannier dual $\kappa^{*}$ of the bond
strength $\kappa$
\begin{equation}
  \label{eq:kramerswannier}
  \kappa^{*}=-\frac{1}{2}\ln\,\tanh{\kappa}
\end{equation}
the transfer matrices may be written in the diagonal form
\begin{equation}
  \label{eq:pauliform}
 T_{l}=\left[\frac{2}{\sinh 2\kappa^{*}_{\lambda}(l)}\right]^{1/2}
e^{-\kappa^{*}_{\lambda}(l)\mbox{\boldmath $\sigma_{3}$}}\,. 
\end{equation}
Here, \boldmath $\sigma_{j}\,$\unboldmath ($j=0,1,2,3$) are the complete set
of Pauli matrices denoted earlier as ${\bf 1}_{2}$, $\sigma_{x}$, $\sigma_{y}$,
and $\sigma_{z}$ (${\bf 1}_{2}$ two dimensional unit matrix).

In the two dimensional case the Hamiltonian is
\begin{equation}
  \label{eq:2dhamiltonian}
H(\{\sigma_{l,m}\})=
-\sum_{l=1}^{L-1}\sum_{m=1}^{M}J_{\lambda}(l)\sigma_{l,m}\sigma_{l+1,m}
 -\sum_{l=1}^{L-1}\sum_{m=1}^{M}J_{\mu}(l)\sigma_{l,m}\sigma_{l,m+1}\,,
\end{equation}
and we assume periodic boundary conditions in the direction of $M$ such that
$M+1\rightarrow 1$. Also in this case one can write the partition function as
a product of transfer matrices, Eq.~(\ref{eq:partition}), but now these consist
of two factors
\begin{equation}
  \label{eq:2dtransfermatrix}
  T_{l}=V_{l}W_{l}\,.
\end{equation}
The matrix $W_{l}$ is the obvious generalization of Eq.~(\ref{eq:pauliform}),
\begin{equation}
  \label{eq:vau}
  W_{l}=
\prod_{m=1}^{M}\left[\frac{2}{\sinh 2\kappa^{*}_{\lambda}(l,m)}\right]^{1/2}
\exp{\left[-\sum_{m=1}^{M}\kappa_{\lambda}^{*}(l,m)
\mbox{\boldmath $\sigma_{3}^{m}$}\right]}\,,
\end{equation}
but with the $2M\times 2M$ matrix
\begin{equation}
  \label{eq:directproduct}
\mbox{\boldmath $\sigma_{3}^{m}$}=\mbox{\boldmath $\sigma_{0}$}
\otimes \mbox{\boldmath $\sigma_{0}$}\otimes \ldots \otimes
\mbox{\boldmath $\sigma_{3}$}\otimes\ldots\otimes 
\mbox{\boldmath $\sigma_{0}$}\,,
\end{equation}
which is the direct product of $M-1$ unit matrices and \boldmath
$\sigma_{3}\,$\unboldmath at position $m$. The $2M\times 2M$ matrix $V_{l}$
contains the Hamiltonian of the $l$th column of the system
\begin{equation}
  \label{eq:column}
  V_{l}=\exp{\left[ \sum_{m=1}^{M-1}\kappa_{\delta}(l,m)
\mbox{\boldmath $\sigma_{1}^{m}$}\mbox{\boldmath $\sigma_{1}^{m+1}$}
\right]}\,.
\end{equation}

The formulation in terms of transfer matrices has two implications which are
of immense importance for practical purposes. First, one can show that it is
the largest eigenvalue of the total transfer matrix that determines for $L \to
\infty$ the partition function and thus the physical properties of the system
\cite{sml64}. Second, one can write the transfer matrix of a system of length
$L+1$ as a function of the transfer matrix of the system of the length $L$.
We will see below, that these two properties are essential for numerical
evaluation of the phase diagram and critical properties.

\subsection{Transforming to Fermions}

The above transfer matrices involve linear and quadratic forms of the spin
operators. The raising and lowering operators fulfill mixed commutation and
anti-commutation relations
\begin{eqnarray}
  \label{eq:anticomm}
  [\sigma_{\pm}^{m},\sigma_{\pm}^{l}]&=&0\qquad (m\neq l)\,,\nonumber\\
&&\nonumber\\
  \{\sigma_{+}^{m},\sigma_{-}^{l}\}&=&1\,,\\
&&\nonumber\\
  (\sigma_{+}^{m})^{2}=(\sigma_{-}^{m})^{2}&=&0\,.\nonumber
\end{eqnarray}

In one dimension, it is well known how to transform these operators to ones
obeying fermion anti-commutation relations. This is done by the Jordan-Wigner
transformation \cite{jw28,lsm61,lm62,s63}. Annihilation and creation operators
are introduced by
\begin{eqnarray}
  \label{eq:jordanwigner}
  c_{m}&=&\left[\exp{\left(i\pi 
\sum_{j=1}^{m-1}\sigma_{+}^{j}\sigma_{-}^{j}\right)}\right]\sigma_{-}^{m}\,,
\nonumber\\
&&\\
c_{m}^{\dagger}&=&\left[\exp{\left(i\pi 
\sum_{j=1}^{m-1}\sigma_{+}^{j}\sigma_{-}^{j}\right)}\right]\sigma_{+}^{m}
\nonumber\,.
\end{eqnarray}
It is easy to show that these operators obey fermion statistics and
\begin{equation}
  \label{eq:fermions}
  c_{m}^{\dagger}c_{m}=\sigma_{+}^{m}\sigma_{-}^{m}\,.
\end{equation}

The inverse transformation is
\begin{eqnarray}
  \label{eq:jordanwignerinverse}
  \sigma_{-}^{m}&=&\left[\exp{\left(i\pi 
\sum_{j=1}^{m-1}c_{j}^{\dagger}c_{j}\right)}\right]c_{m}\,,
\nonumber\\
&&\\
\sigma_{+}^{m}&=&\left[\exp{\left(i\pi 
\sum_{j=1}^{m-1}c_{j}^{\dagger}c_{j}\right)}\right]c_{m}^{\dagger}\nonumber\,.
\end{eqnarray}
This gives
\begin{equation}
  \label{eq:definitions}
  \sigma_{1}^{m}=\left[\exp{\left(i\pi
  \sum_{j=1}^{m-1}c_{j}^{\dagger}c_{j}\right)}\right]
(c_{m}^{\dagger}+c_{m}), \qquad
\sigma_{3}^{m}=2c_{m}^{\dagger}c_{m}-1\,.
\end{equation}

The fermion representation of the above transfer matrices is 
\begin{eqnarray}
  \label{eq:fermiontransfer}
  W_{l}&=&\prod_{m=1}^{M}
\left[\frac{2}{\sinh 2\kappa^{*}_{\lambda}(l,m)}\right]^{1/2}
\exp{\left[-2\sum_{m=1}^{M}\kappa_{\lambda}^{*}(l,m)
\left(c_{m}^{\dagger}c_{m}-\frac{1}{2}\right)\right]}\,,
\nonumber\\
&&\nonumber\\
&&\nonumber\\
V_{l}&=&\exp{\Bigg[\sum_{m=1}^{M-1}\kappa_{\delta}(l,m)
(c_{m}^{\dagger}-c_{m})(c_{m+1}^{\dagger}+c_{m+1})\Bigg.}\nonumber\\
&&\qquad\qquad\qquad\qquad\qquad  
\Bigg. -\kappa_{\mu}(l,M) e^{i\pi N_{c}}
(c_{M}^{\dagger}-c_{M})(c_{1}^{\dagger}+c_{1})\Bigg]\,,
\end{eqnarray}
with the number operator $N_{c}=\sum_{m=1}^{M}c_{m}^{\dagger}c_{m}$. The last
term in (\ref{eq:fermiontransfer}) represents the periodic boundary condition
in the direction of $M$. 

The matrix $V_{l}$ is biquadratic in the fermion operators. It does not
conserve the number of fermions since it contains terms that create and
annihilate fermions in pairs. Thus, diagonalization of the transfer matrix in
principle should be possible via a unitary Bogoliubov-de-Gennes transformation.

It is illustrative for the disordered case to consider first the ordered limit
in which $\kappa_{\lambda}^{*}(l,m)=\kappa^{*}_{\lambda}$ and
$\kappa_{\mu}(l,m)=\kappa_{\mu}$. First one notes that evenness and oddness of
$N_{c}$ commute with $W$ and $V$ \cite{sml64} since the latter contains only
quadratic forms of the fermion operators,
\begin{equation}
  \label{eq:evenodd2}
  [(-1)^{N_{c}},W]=[(-1)^{N_{c}},V]=0\,.
\end{equation}
Therefore one can classify the eigenstates of $WV$ according to whether they
contain even and odd numbers of fermions. One can then write
\begin{equation}
  \label{eq:vauplusminus}
  V^{\pm}=\exp{\left[\kappa_{\mu}\sum_{m=1}^{M}
(c_{m}^{\dagger}-c_{m})(c_{m+1}^{\dagger}+c_{m+1})
\right]}\,,
\end{equation}
with $c_{M+1}=\mp c_{1}$ and $c_{M+1}^{\dagger}=\mp c_{1}^{\dagger}$ where
$\pm$ denote the operators acting on the even (with anti-cyclic boundary
condition) and odd states (with cyclic boundary condition), respectively.
The task is then to find the eigenvalues and eigenvectors of the transfer
operators
\begin{eqnarray}
  \label{eq:transferoperator}
  T^{\pm}=V^{\pm}W&=&
\left[\frac{2}{\sinh 2\kappa^{*}_{\lambda}}\right]^{M/2}
\exp{\left[\frac{1}{2}\kappa_{\mu}\sum_{m=1}^{M}
(c_{m}^{\dagger}-c_{m})(c_{m+1}^{\dagger}+c_{m+1})
\right]}\nonumber\\
&&\nonumber\\
&&\nonumber\\
&&\,\,\qquad\qquad\quad
\times\exp{\left[-2\kappa_{\lambda}^{*}\sum_{m=1}^{M}
\left(c_{m}^{\dagger}c_{m}-\frac{1}{2}\right)\right]}\nonumber\\
&&\nonumber\\
&&\nonumber\\
&&\qquad\qquad\quad\times\exp{\left[\frac{1}{2}\kappa_{\mu}\sum_{m=1}^{M}
(c_{m}^{\dagger}-c_{m})(c_{m+1}^{\dagger}+c_{m+1})
\right]}\,.
\end{eqnarray}

The diagonalization can be straightforwardly performed by introducing plane
wave operators via the canonical transformation
\begin{equation}
  \label{eq:canonical}
  c_{m}=\frac{1}{\sqrt{M}}\sum_{q}e^{iqm}\eta_{q}\,,
\end{equation}
where $q\leq \pi$ is given by the odd and even multiples of $\pi/M$ for
anti-cyclic and cyclic boundary conditions, respectively ($M$ is assumed to be
even). The resulting quadratic forms in the transfer operator which has four
eigenstates is very reminiscent of the pair Hamiltonian in the BCS theory of
superconductivity. It can be diagonalized by transforming to new variables
\begin{equation}
  \label{eq:bdgtransformation}
  \xi_{\pm q}=\cos{\phi_{q}}\eta_{\pm q}\pm 
\sin{\phi_{q}}\eta^{\dagger}_{\mp q}\,,
\end{equation}
with $\tan{2\phi_{q}}=2C_{q}/(B_{q}-A_{q})$ containing
\begin{eqnarray}
  \label{eq:constants}
A_{q}&=&e^{-2\kappa_{\lambda}^{*}}
(\cosh{\kappa_{\mu}}+\sinh{\kappa_{\mu}}\cos{q})^{2}
+e^{2\kappa_{\lambda}^{*}}(\sinh{\kappa_{\mu}}\sin{q})^{2}\,,\nonumber\\
&&\nonumber\\
B_{q}&=&e^{-2\kappa_{\lambda}^{*}}
(\sinh{\kappa_{\mu}}\sin{q})^{2}
+e^{2\kappa_{\lambda}^{*}}
(\cosh{\kappa_{\mu}}-\sinh{\kappa_{\mu}}\cos{q})^{2}\,,\nonumber\\
&&\nonumber\\
C_{q}&=&(2\sinh{\kappa_{\mu}}\sin{q})
(\cosh{2\kappa^{*}_{\lambda}}\cosh{\kappa_{\mu}}
-\sinh{2\kappa^{*}_{\lambda}}\sinh{\kappa_{\mu}}\cos{q})\,.
\nonumber
\end{eqnarray}
With these operators one obtains the diagonal form of the transfer operators
\begin{equation}
  \label{eq:diagonalform}
  T^{\pm}=\left[\frac{2}{\sinh 2\kappa^{*}_{\lambda}}\right]^{M/2}
\exp{\left[-\sum_{q}\epsilon_{q}\left(\xi_{q}^{\dagger}\xi_{q}
-\frac{1}{2}\right)
\right]}\,.
\end{equation}
The energy dispersion is given implicitly by the solution of
\begin{equation}
  \label{eq:dispersion2}
\cosh{\epsilon_{q}}=\cosh{2\kappa_{\mu}}\cosh{2\kappa_{\lambda}^{*}}
-\sinh{2\kappa_{\mu}}\sinh{2\kappa_{\lambda}^{*}}\cos{q}\,.
\end{equation}
The four eigenstates in terms of the $\xi$-operators are the vacuum
$\psi_{0}$, $\psi_{\pm q}=\xi_{\pm q}^{\dagger}\psi_{0}$ and the pair state
$\psi_{-qq}=\xi_{-q}^{\dagger}\xi_{q}^{\dagger}\psi_{0}$. 

The partition function is given by the $L$-th power of the largest eigenvalue
of $T^{\pm}$.  The critical temperature $T_{c}$ is defined by
$\kappa_{\lambda}^{*}=\kappa_{\mu}$. We do not want to discuss here how to use
the results for obtaining the critical properties. The important lesson to
keep in mind at this stage is the importance of the parity of the fermion
number for the structure of the eigenvalue problem.

\subsection{Mapping to a Localization Problem for Non-Interacting Particles}

In the case of a disordered Ising system, the unitary transformation for
achieving the diagonalization of the transfer matrix depends on the disorder,
and the different transfer steps are not independent. Therefore, a different
approach must be used. This consists of mapping the Ising problem to a
localization problem for non-interacting particles.

Starting point is the rewriting of the transfer matrix in terms of Dirac
fermions fulfilling particle conservation \cite{cf97a,grl01}. In order to
achieve this, an identical copy of the Ising model is introduced. The
corresponding fermions are denoted as $d^{\dagger}_{m}$, $d_{m}$. New Dirac
fermions are then defined by (suppressing the indices for the sake of
simplicity)
\begin{equation}
  \label{eq:transform}
  c=\frac{1}{2}(f + f^{\dagger} + g - g^{\dagger})\qquad\qquad
  d=\frac{i}{2}(f^{\dagger} -f  - g - g^{\dagger})
\end{equation}
and the inverse of these,
\begin{equation}
  \label{eq:inverstransform}
  f=\frac{1}{2}[c+c^{\dagger}+i(d+d^{\dagger})]\qquad\quad
  g=\frac{1}{2}[c-c^{\dagger}+i(d-d^{\dagger})]\,.
\end{equation}

The Hamiltonian of the doubled system is the sum of the two Hamiltonians,
$H=(H^{c}+H^{d})/2$. This implies that one deals eventually with products of
the corresponding transfer matrices. In terms of the new fermion operators
they are given by
\begin{eqnarray}
  \label{eq:productransfer}
  W_{l}^{c}W_{l}^{d}&=&\exp{\left[-2\sum_{m=1}^{M}\kappa^{*}_{\lambda}(l,m)
(g^{\dagger}_{m}f_{m}+f^{\dagger}_{m}g_{m})\right]}\nonumber\\
&&\nonumber\\
&&\nonumber\\
V^{c}_{l}V^{d}_{l}&=&\exp{\left[ 2\sum_{m=1}^{M-1}\kappa_{\mu}(l,m)
(g^{\dagger}_{m}f_{m+1}+f^{\dagger}_{m+1}g_{m})+b\right]}
\end{eqnarray}
and they are particle conserving. The term 
\begin{eqnarray}
  \label{eq:boundaryterm}
  b&=&-\kappa_{\mu}(l,M)\left[(e^{i\pi N_{c}}+e^{i\pi N_{d}})
(g^{\dagger}_{M}f_{1}+f^{\dagger}_{1}g_{M})\right.\nonumber\\
&&\nonumber\\
&&\qquad\qquad\qquad\qquad +\left.
(e^{i\pi N_{c}}-e^{i\pi N_{d}})
(g^{\dagger}_{M}f^{\dagger}_{1}+f_{1}g_{M})
\right]
\end{eqnarray}
is due to the boundary conditions in the direction of $M$. This contains the
two boundary operators
\begin{equation}
  \label{eq:boudaryoperator}
  B^{\pm}=e^{i\pi N_{c}}\pm e^{i\pi N_{d}}\,.
\end{equation}

In order to perform the mapping to the network model it is required that the
transfer operator conserves the number of fermions. This implies that the
Hilbert space for first quantization of the transfer matrix must be
constructed from the Hilbert space of the two copies of the
Ising model in such a way that the term involving $B^{-}$ in Eq.~(\ref{eq:boundaryterm})
vanishes \cite{mc02}.

In particular, one may consider the parity operators $R$ defined by
\begin{equation}
  \label{eq:parity2}
  R=\prod_{m=1}^{M}\sigma_{m}^{3}
\end{equation}
which change the sign of a complete column of the spins,
\begin{equation}
  \label{eq:signchange}
R\sigma_{\mu}^{1}R=-\sigma_{\mu}^{1} \,,
\qquad\qquad R^{2}=\mbox{\boldmath$1$}\,.   
\end{equation}
These obviously commute with the transfer operators and have eigenvalues $\pm
1$.  The transfer matrix of the two copies of the Ising system written in the
basis corresponding to the product space of $R_{c}+R_{d}$ can then be cast
into a block-diagonal form where the four block matrices describe the transfer
of amplitudes in the subspaces corresponding to eigenvalues of the parity
operators operating in the Hilbert spaces of the two copies of the system,
$(1_{c},1_{d})$, $(1_{c},-1_{d})$, $(-1_{c},1_{d})$, and $(-1_{c},-1_{d})$.
Since the eigenvalues of $\exp{(i\pi N)}$ are $\pm 1$ depending on whether $M$
is even or odd, the boundary operators can be written as $B^{\pm}=R_{c}\pm
R_{d}$. Then, in the subspaces corresponding to the same eigenvalues,
$B^{-}=0$. The number of fermions is conserved.

\subsection{The Equivalent Network Model}

In order to construct the first quantized form of the transfer matrix, we
start by comparing first and second quantized forms of an operator in a
Hilbert space, say of the dimension $N=2M$ with a basis $\{|j\rangle \}$.
Assume that the operator can be written in the second quantized form as
$\exp{[\alpha^{\dagger}_{j}\langle j|G|k\rangle\alpha_{k}]}$ with the creation
and annihilation operators $\alpha^{\dagger}_{j}$, and $\alpha_{k}$,
respectively.  The first quantized form of this operator is the $(2M\times
2M)$-matrix $\langle j|\exp{G}|k\rangle$.

\begin{figure}[htbp]
\vspace{2mm}
  \begin{center}
  \includegraphics[width=0.6\textwidth]{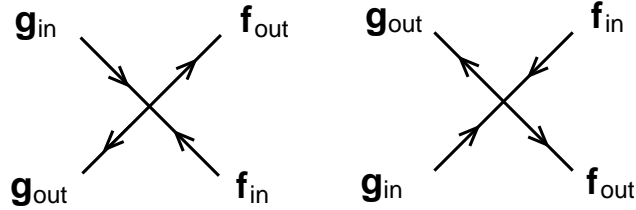}
    \caption{Graphical representation of the scattering nodes of 
      the first quantized form of the transfer matrix of the Ising model.
      Left: $\lambda$-bond; right: $\mu$-bond.}
    \label{fig:nodes}
  \end{center}
\vspace{5mm}
\end{figure}

\begin{figure}[htbp]
\vspace{2mm}
  \begin{center}
  \includegraphics[width=0.9\textwidth]{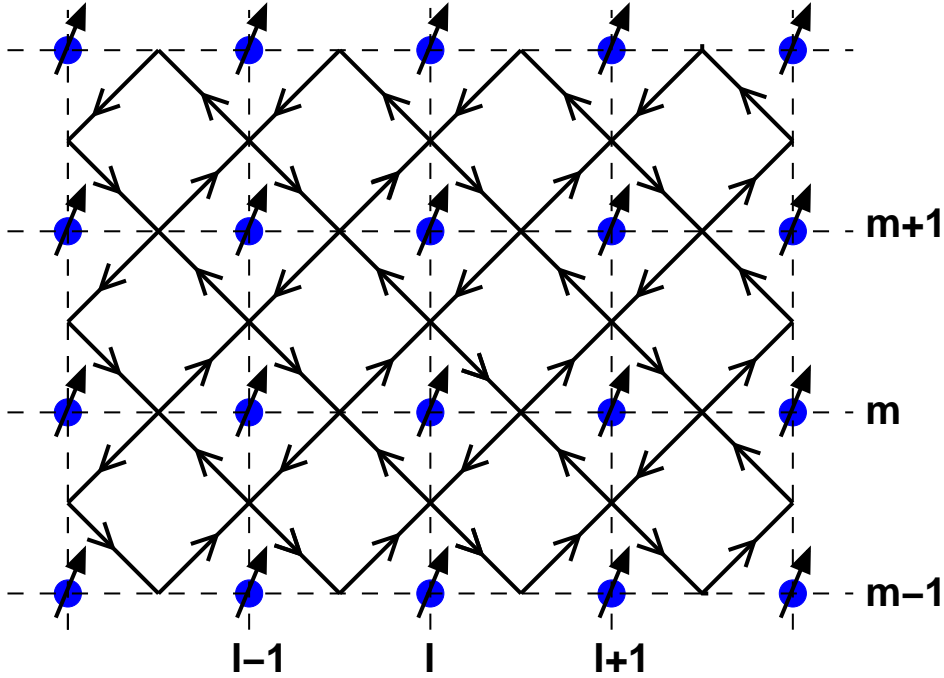}
    \caption{The network model equivalent to the random bond Ising model. 
      Arrows indicate the propagation of flux. The transfer matrix relates
      flux amplitudes from the left to the right. The Ising lattice of spins
      (arrows) with the exchange matrix elements (dashed) are also shown.}
    \label{fig:isingnetwork}
  \end{center}
\vspace{5mm}
\end{figure}

By identifying $\{\alpha_{1},\ldots,\alpha_{2M}\}=
\{f_{1},\ldots,f_{M},g_{1},\ldots,g_{M}\}$ one finds that the first
quantization equivalent of the second quantization transfer operator
corresponding to the $\lambda$-bonds,
$\exp{[-2\kappa^{*}_{\lambda}(l,m)]
  (g^{\dagger}_{m}f_{m}+f^{\dagger}_{m}g_{m})}$, is given by
(Fig.~\ref{fig:nodes})
\begin{equation}
  \label{eq:firstlambda}
W_{lm}=\exp{[-2\kappa^{*}_{\lambda}(l,m)\sigma_{1}]}=
\cosh{[2\kappa^{*}_{\lambda}(l,m)]}-
\sigma_{1}\sinh{[2\kappa^{*}_{\lambda}(L,M)]}\,. 
\end{equation}
By analogy, the first quantized form of the $\mu$-bonds in the transfer
matrix is
\begin{equation}
  \label{eq:firstmu}
V_{lm}=\exp{[2\kappa_{\mu}(l,m)\sigma_{1}]}=
\cosh{[2\kappa_{\mu}(l,m)]}+
\sigma_{1}\sinh{[2\kappa_{\mu}(l,m)]}\,.  
\end{equation}
In terms of right- and left-moving amplitudes $\tilde{f}_{j}$ and
$\tilde{g}_{j}$, respectively, the transfer operators of the lattice sites can
be written as
\begin{eqnarray}
  \label{eq:nodescattering}
\left(
\begin{array}{c}
\tilde{f}_{\rm out}\\
\tilde{f}_{\rm in}
\end{array}
\right)=
\left(
\begin{array}{rr}
\cosh{2\kappa^{*}_{\lambda}}&-\sinh{2\kappa^{*}_{\lambda}}\\
-\sinh{2\kappa^{*}_{\lambda}}&\cosh{2\kappa^{*}_{\lambda}}
\end{array}
\right)
\left(
\begin{array}{c}
\tilde{g}_{\rm in}\\
\tilde{g}_{\rm out}
\end{array}
\right)
\end{eqnarray}
and
\begin{eqnarray}
  \label{eq:nodescattering2}
\left(
\begin{array}{c}
\tilde{f}_{\rm in}\\
\tilde{f}_{\rm out}
\end{array}
\right)=
\left(
\begin{array}{rr}
\cosh{2\kappa_{\mu}}&\sinh{2\kappa_{\mu}}\\
\sinh{2\kappa_{\mu}}&\cosh{2\kappa_{\mu}}
\end{array}
\right)
\left(
\begin{array}{c}
\tilde{g}_{\rm out}\\
\tilde{g}_{\rm in}
\end{array}
\right)\,.
\end{eqnarray}
The relations
\begin{equation}
  \label{eq:conservationlaws}
  \sigma_{3}{\bf W}^{\dagger}\sigma_{3}={\bf W}^{-1}\,,\qquad\qquad
\sigma_{3}{\bf V}^{\dagger}\sigma_{3}={\bf V}^{-1}
\end{equation}
ensure flux conservation (Fig.~\ref{fig:isingnetwork}).

The network model constructed in this way is completely analogous to the
U(1) network of saddle point scatterings used in the previous chapter for
describing the critical localization behavior of the quantum Hall system,
with plaquettes of definite senses of circulation of the fluxes which can be
clockwise and counterclockwise. Disorder can be introduced in the former
network by random quenched phases and random saddle point energies. In the
present case, randomness enters only via the randomness of the exchange
integrals represented in the Ising model by the bonds. The latter have been
mapped to the nodes of the network represented by the parameters $\kappa$ and
$\kappa^{*}$. An antiferromagnetic $\mu$-bond leads to $\kappa <0$. An
antiferromagnetic $\lambda$-bond, however, leads to a complex $\kappa^{*}$,
where  $~^*$ denotes 
 the Kramers-Wannier dual defined by the function, 
  Eq.~(\ref{eq:kramerswannier}). It 
implies the relation
\begin{equation}
(-|\kappa|)^{*}=|\kappa|^{*}+i\frac{\pi}{2}\,.  
\end{equation}
This gives a minus sign in the matrices $W$ representing the scattering at the
nodes. Starting from an Ising system with ferro- and antiferromagnetic bonds
distributed randomly, one arrives at an equivalent network model with nodes
$W$ that have random signs.  In the node matrices $V$, the off-diagonal matrix
elements $\sinh{(2\kappa)}$ acquire random signs. Thus, scattering at the
nodes only causes real phase factors $\pm 1$. Taking into account the topology
of the network (Fig.~\ref{fig:isingnetwork}), this means that a single
antiferromagnetic node introduces phases $\pi$ in both of the anticlockwise
plaquettes that are linked at this node.

This specific property of the disorder determines the symmetry class of the
model. In the classification scheme introduced recently by Altland and
Zirnbauer \cite{z96,az97} this network model belongs to the class D (see
below, Tab.~\ref{table:sym3} in chapter \ref{sec:symmetries}).  Hamiltonian
matrices of this class are purely imaginary, $H^{*}=-H$. The corresponding
time evolution operator $\exp{(-iHt)}$ is represented by a matrix with real
elements and can introduce only scattering phase factors $\pm 1$, in
accordance with the above network model of the random bond Ising model.

The total transfer matrix of the doubled Ising system is then obtained by
combining the node matrices $W$ and $V$. It is a $2M\times 2M$ matrix which
can be arranged in such a way that propagation in one direction is described
by the first $M$ rows and propagation in the other direction is contained in
the $M$ last rows. In the former case of the U(1) network the transfer matrix
was unitary. In the present case it is real and orthogonal. For a random
network, it has the form $\exp{\gamma L}$ with a real diagonal matrix
$\gamma$. The fluctuations are $\propto \sqrt{L}$. It can be calculated
numerically recursively for very long strips \cite{km93,mk81,mk83}, $L\to
\infty$, and converges according to the theorem by Oseledec \cite{oseledec68}
to a diagonal matrix with diagonal elements
$\gamma_{1},\ldots,\gamma_{M},-\gamma_{1},\ldots,-\gamma_{M}$ with the
Lyapunov exponents $\gamma_{j}$.

The above random-bond Ising network model has been analyzed exhaustively
numerically in \cite{mc02} for systems containing as many as $M\times
L=256\times 5\cdot 10^{5}$ lattice sites including scaling analysis along the
phase boundary where systems of size up to $M\times L=32\times 2\cdot 10^{8}$
lattice sites were used. From the scaling behavior of the data a critical
index of $\nu=1.50\pm 0.03$ has been extracted along the Nishimori line at the
multicritical point. On the other hand, along the phase boundary, the scaling
analysis gives a critical exponent $\nu_{T}=4.0\pm 0.5$.
 
Also other network models, but belonging to the same symmetry class D, have
been studied \cite{cf97a,rl01,cetal02}. Most strikingly, these different
choices seem to yield different localization properties. This will be reviewed
in the following chapter.

\section{Symmetry Classes} 

\label{sec:symmetries}

Having established above a network model belonging to a certain symmetry class
corresponding to the quantum Hall transition, one may ask the question if one
can generalize the model to cover {\em all} existing symmetry classes.

Conventionally, random electron systems are classified into three universality
classes \cite{mehta} according to their symmetry with respect to reversal of
time and rotation of spin. Systems belonging to the orthogonal class have both
time reversal and spin rotation invariance. Models belonging to the unitary
class are not invariant under time reversal. The symplectic class contains
models that have time reversal symmetry but the spin rotation symmetry is
broken. This is realized in random electron systems with spin-orbit
interaction. The notation originates from the fact that a distribution of
random matrices ${\bf H}$ of the orthogonal, unitary, and symplectic classes
is invariant under transformations, ${\bf H} \rightarrow {\bf U}^{-1} {\bf H}
{\bf U},\, {\bf O}^{\rm T} {\bf H} {\bf O},\, {\bf W}^{\rm R} {\bf H} {\bf W}$
with unitary, orthogonal, and symplectic (commuting with the antisymmetric
Pauli matrix $\sigma_y$) matrices, ${\bf U},\,{\bf O},\,{\bf W}$,
respectively.  Here, ${\bf O}^{\rm T}$ is the transpose of ${\bf O}$, and $
{\bf W}^{\rm R}$, the time reverse of ${\bf W}$ \cite{mehta}, as defined by $
{\bf W}^R = {\bf K} {\bf W}^T {\bf K}^{-1}$, with the unitary matrix ${\bf K}$
which can be symmetric or antisymmetric, $ {\bf K} {\bf K}^* = \pm 1$,
corresponding to integer spin and half-integer spin respectively.  Any random
matrix in each of the classes can be diagonalized by a unitary, orthogonal,
and symplectic matrix, respectively.

The symmetry properties of a random Hamiltonian have strong impact on the
distribution and statistical correlations between its eigenvalues and the
distribution of moments of local eigenfunction amplitudes.  The level
repulsion of extended states increases as the symmetry is changed from the
orthogonal to unitary to the symplectic class. At the same time the tendency
to localization decreases, accompanied by an according decrease of local wave
function fluctuations \cite{mehta,mirlin}. As reviewed in
chapter~\ref{sec:chalkercoddington}, these properties have been studied in
detail for the Chalker-Coddington network model. By construction, the model
belongs to the unitary symmetry class, since the links are directed, and
correspond to scattering paths which have a well defined chirality
(handedness). This breaks the time reversal symmetry. The question arises,
whether or not the integer quantum Hall transition as modeled by the
Chalker-Coddington network model is uniquely characterized by being identified
as belonging to the unitary symmetry class. This would lead one to the
conclusion that all quantum Hall type transitions had the same critical
exponents.

Since there can be a strong overlap between spin split Landau bands
\cite{pr96}, the question is of great experimental importance
\cite{kucharetal} whether or not the delocalization transition is sensitive to
the spin rotation symmetry. When the time reversal symmetry is broken, the
conventional unitary class makes no distinction whether or not the spin
symmetry is broken. However, for a modified Chalker-Coddington model with two
spin channels, fixing the scattering phases and introducing random SU(2)
mixing between the spin channels --- which corresponds to spin flip scattering
by spin-orbit interaction --- the critical exponent has been found to be $\nu
\approx 1.1$ \cite{kha97,kha98}. This is close to the one of classical
percolation, $\nu =\nu_{\rm p}= 4/3$, see section~\ref{sec:quantumcorrection}. On the
other hand, one recovers the critical exponent $\nu = 2.4(2)$ close to that of
the quantum Hall phase transition \cite{lc94,wlw94,kha97}
(Tab.~\ref{table:sym1}) by choosing in addition the scattering phases at
random.  In the latter case, the unitary matrices mixing the two spin channels
are randomly chosen from the group U(2) = U(1)$\times$ SU(2) where U($n$)
denotes the group of complex unitary $(n \times n)$-matrices ${\bf A}$, and S
stands for the special condition that its determinant is one, ${\rm det} A
=1$.

Recently, it has been realized that there can be two additional discrete
symmetries in Hamiltonians of disordered electrons. These symmetries have been
found to give rise to four distinct transitions of the quantum Hall type with
different critical exponents and different behavior of the quasiparticle
density of states. In the next section, we introduce these symmetry classes
before we review the present knowledge about the properties of the
corresponding random models. In particular, we will concentrate on the nature
of the quantum Hall type transitions.

\subsection{The Additional Discrete Symmetries}

In addition to the invariance under time reversal and spin rotation, there can
be at least two more discrete symmetry operations in condensed matter systems.
One is realized in two-sublattice models \cite{ow1979,g1990,g1993,hsw93}.
It corresponds to an interchange of the two sublattices together with a sign
change of the Hamiltonian. A physical realization is a tight binding lattice
Hamiltonian with randomness only in the hopping amplitude. Disordered systems
with such a symmetry belong to the {\em chiral class}.

The other one is the electron-hole symmetry \cite{o1990} arising from Andreev
scattering in normal metal wires attached to a superconductor \cite{az97} or
in superconductors with a gapless quasiparticle density of states
\cite{o1990}. Disordered systems with such a particle-hole symmetry are in the
Bogoliubov-de-Gennes-Oppermann (BdGO) class.

Both of the discrete symmetries share the property that the eigenenergies of
the corresponding Hamiltonians come in pairs $(-E_n,\,E_n)$, where $E=0$ is
the center of the band of eigenvalues for the chiral models, and the Fermi
energy for the BdGO-models. Random Hamiltonians with either of these two
discrete symmetries show very different behavior in their density of states
and transport properties. Their properties are presently the subject of
intensive ongoing research. In the tables \ref{table:sym1}, \ref{table:sym3},
and \ref{table:sym2}, we have attempted to summarize what is presently known
about the properties of the non-directed and directed network models for all
of the existing symmetry classes.

There are strong arguments, based on a classification of Lie Algebras by
Cartan \cite{helgason1962,helgason2000}, that these symmetries make the
classification of possible random systems complete. In a perturbative
renormalization group study in $2+\epsilon$ dimensions, the corrections to the
conductance in the bosonic and the fermionic replica formulations have been
obtained to 3-loop order for all of the symmetric spaces \cite{hikami1981}.

There is an ongoing debate whether or not this symmetry classification is
sufficient to characterize the quantum critical properties of random systems
\cite{rl01,bfm2003,mdh2002,bc03}. As we will review below, there are some
results which do not seem to fit into this scheme, since the critical
properties sensitively do depend on the spatial correlation of the disorder.
As discussed above, the conventional quantum Hall transition is insensitive to
the correlation length of the disorder potential, unless the disorder has an
extremely long-range correlation length \cite{smk03}.

In the following, we first discuss the properties of quasi-one dimensional
disordered wires, and how their transport properties depend on the symmetry
properties. Then, we consider the critical properties in two dimensions. We
present an overview of the network models studied for the various symmetry
classes.

\begin{table}[htpb]
\begin{center}
\begin{tabular}{ccccc}
\hline\hline
\multicolumn{5}{c}{ ~~~~~~~~~~~~~ Wigner-Dyson Class~~~~~~~~~~~~~~~ } \\
\hline\hline
property&orthogonal&unitary&unitary&symplectic
  \\ \hline
TRS &   yes &  no    & no   &  yes   \\
 SRS  & yes  &   yes &   no & no         \\
 Cartan class &  AI  &  A & A & AII  \\
 $\beta$ & 1 & 2 & 2 & 4 \\ 
 $m_l$ & 1 & 1 & 1 & 1 \\ 
$s$ & 1 & 1 & 2 & 1  \\ 
 NDNM & insulator &  insulator (a)  & insulator   & MIT \cite{hikami} \\
 DOS ($E \rightarrow 0$)   & constant  & constant & constant & constant\\
phase factor  &     &  U(1)       &   U(2) (b)  &  \\ 
in DNM        &     &             &             &  \\    
PT in DNM   & -- &  IQHT  &  IQHT (c)  & --    \\
 $\nu$  & --  & $2.5(5)$ \cite{cc88} & $2.4(2)$ \cite{lc94,wlw94}  
            & 2.8(1) [Sec. \ref{sec:net2ds}]       \\
        &     &         &          \cite{h95} (c)&  \\
 $\alpha_0$  & --  & $2.261(3)$ \cite{emm01}&   & 2.174(3) \cite{mjh98}     \\
$\Lambda_{\rm c}$  & --  &  $1/\pi (\alpha_0-2)$ 
&   $2/ \pi(\alpha_0-2)$ & 1.83(1)[Sec. \ref{sec:net2ds}]
          \\  \hline 
 equivalent models & &  RSVDM  &  RSVDM &  \\ \hline \hline 
\end{tabular}
\vspace{5mm}
\caption{
  Classification of network models (NM) according to their symmetry: the
  Wigner-Dyson class with no discrete symmetry.  Abbreviations: TRS time
  reversal symmetry, SRS, spin rotation symmetry, $\beta$ level repulsion
  exponent, $m_{l}$ multiplicity of long root on symmetric space
  \cite{helgason2000}, $s$ spin factor, IQHT Integer Quantum Hall Transition,
  MIT metal-insulator transition, DOS density of states, (N)DNM (non-)directed
  network model, PT phase transition, $\nu$ critical exponent, $\alpha_{0}$
  position of maximum of $f(\alpha)$, $\Lambda_{\rm c}$ MacKinnon-Kramer
  parameter, RSVDM Dirac model with both, random scalar and vector potentials;
  (a) electrons in a random magnetic field belong to this unitary class. The
  U(2) network model proposed by Chalker and Lee \cite{lc94} describes
  electrons in random magnetic field with large correlation length, and does
  not show a delocalization transition; (b) spin degenerate levels with spin
  orbit interaction \cite{lc94}; (c) two IQHT's at two distinct energies
  \cite{lc94}.\label{table:sym1}}
\end{center} 
\vspace{5mm}
\end{table}

\begin{table}[htpb]
\begin{center}
\begin{tabular}{ccccc}
\hline \hline
\multicolumn{5}{c}{~~~~~~~~~~~~~~~Bogoliubov-de-Gennes-Oppermann Class~~~~~~~~~~~ } \\
\hline \hline
property&orthogonal&unitary&unitary&symplectic\\ \hline
TRS &    yes &
 no & no & yes        \\
 SRS  &  yes  &   yes &   no & no   \\
 Cartan Class &   CI & C & D & DIII \\ 
 $\beta$ & 2 & 4 & 1 & 2 \cite{grv2003}\\ 
 $m_l$ & 2 & 3 & 0 & 0 \cite{grv2003}\\ 
$s$ & 1 & 1 & 4 & 2 \\ 
 transport &SI  &SI & TM& TM (a) \\ 
 DOS ($E \rightarrow 0$)  & $|E|$ \cite{az97,a2002} &
  $|E|^2 $ \cite{az97}&$|E|$ (b)&$|E|$\cite{az97} \\ 
& & & $ \sqrt{\ln (1/|E|)}$&  \\
& & &(c) \cite{sf2000}&\\
phase factor    & & SU(2) & $Z_2$ &            \\  
PT in DNM   &  -- &SQHT  & TQHT  & --   \\
 DOS at CP &   & $|E|^{1/7}$ \cite{glr1999,bcc2002} & &  \\
 $\nu$ &  --  & 4/3 \cite{glr1999} &  (d) &     ?    \\ 
 $\alpha_0$   &-- & $2.137(3)$ \cite{emm03} &  (d) &  ?  \\ 
$\Lambda_{\rm c}$  & --  &   $1/\pi (\alpha_0-2)$ &  (d) &   ?  \\ \hline
 equivalent models   &  & CP &RBIM & 
 \\ \hline  \hline
\end{tabular}
\vspace{5mm}
\caption{
  Classification of network models according to their symmetry: the
  Bogoliubov-de-Gennes-Oppermann class with particle-hole discrete symmetry.
  SI spin insulator, TM thermal metal. (a) critical metal for uncorrelated
  disorder of arbitrary strength, MIT for correlated disorder, (b) DOS in the
  localized phase, (c) DOS in the metallic phase, (d) line of critical points,
  on the Nishimori line $\nu =\nu_{\rm p}= 4/3$ \cite{cf97a}, SQHT spin quantum Hall
  transition, TQHT thermal quantum Hall transition, CP classical percolation
  model, RBIM random bond Ising model \cite{cf97a}. \label{table:sym3} }
\end{center}
\vspace{5mm} 
\end{table}

\begin{table}[htpb]
\begin{center}
\begin{tabular}{cccc}
 \hline\hline
\multicolumn{4}{c}{ ~~~~~~~~ Chiral Class } \\
 \hline\hline
property&orthogonal&unitary&symplectic\\ 
\hline
 TRS    & yes & no  & yes    \\
 SRS & yes  &  yes(no) & no  \\
 Cartan class &   BDI & AIII  & CII  \\
$\beta$   & 1 & 2  & 4  \\
 $m_l$ & 1 & 1 & 1 \\
 $s$ & 1 & 1(2) & 1 \\
 transport($E=0$) & metal  & metal  & ?  \\ 
 DOS ($E \rightarrow 0$)   &$\rho(E)$ & 
$\rho(E)$& ?     \\
phase factor &   2 $\times$ $Z_2$ & 2 $\times$ U(1) & 2 $\times$
  SU(2)  \\ 
  PT in DNM   &  -- &  Yes (a)  & --   \\
 $\nu$  &  -- & &          
 \\
 $\alpha_0$   & --  & &         
 \\
$\Lambda_{\rm c}$   & -- & &           \\ \hline
  equivalent models  & RXY \cite{gll2000} &  RVDM &   \\ \hline  \hline 
\end{tabular} 
\vspace{5mm}
\caption{Classification of network models according to their 
  symmetry: The chiral class with bipartite discrete symmetry (two
  sublattices). Abbreviations: $\rho(E)=\exp ( - c|\ln E|^{1/x})/|E|$
  \cite{g1993} ($x=3/2$ for BDI \cite{mdh2002,mrf03} and possibly also for
  AIII, CII \cite{mrf03}), RXY random XY model, RVDM Dirac model with random
  vector potentials.  (a) The phase diagram is shown in
  Fig.~\ref{fig:diagramAIII}.
\label{table:sym2}}
\end{center}
\vspace{5mm}
\end{table}

For a quasi-one dimensional wire, where the localization length $\xi$ is
larger than the wire width $L_y$, a beautiful formula, valid for all of the
ten symmetry classes has been derived recently by solving the Fokker-Planck
equation \cite{grv2003,bfgm2000}
\begin{equation} 
\label{quasionedimension}
\xi =  l\left[ s \frac{\beta}{m_l}( N-1) +  1 + \frac{1}{m_l} \right].  
\end{equation}
Here, $m_{l}$ is the multiplicity of the root on the symmetric space (see
below), $\beta$ the level repulsion coefficient, $N$ the number of channels
proportional to the width, and $s$ the spin factor. This formula displays
explicitly the importance of the symmetry properties of the disordered
Hamiltonian for the localization.  We have seen already that the transfer
matrix ${\bf T}$ of the Chalker-Coddington model fulfills the commutation
relation Eq. (\ref{eq:condition}), which is a result of current conservation.
Thereby, transfer matrices form a group.  Since they are parameterized by
continuous variables, such groups are called Lie groups. For the
Chalker-Coddington transfer matrix this group is G=U($N,N$), the group of
noncompact $ 2 N$ by $2 N$ matrices, which fulfill Eq. (\ref{eq:condition}).
For the other ensembles additional conditions are imposed, restricting the
corresponding transfer matrices to be in different Lie groups.  Now, all the
powerful tools of representation theory of Lie groups \cite{chen89} can be
used to analyze these transfer matrices systematically.  There is a further
simplification, that a transport property like the conductance
Eq.~(\ref{eq:landauer}) does depend only on products ${\bf T} {\bf T}^+$ and
${\bf T}^+ {\bf T}$ \cite{p84,so05}. As a result, there is a subgroup K of
matrices ${\bf k}$ by which ${\bf T}$ can be multiplied from right $ {\bf T}
\rightarrow {\bf k} {\bf T} $ or left $ {\bf T} \rightarrow {\bf T} {\bf k} $,
without changing the conductance through the system.
For the Chalker- Coddington model this subgroup is found to be K = U($N$)
$\times$ U($N$), the group of $2N$ times $2N$ matrices which consist of two
blocks of unitary $N$ by $N$ matrices.  Thus, this subgroup is not relevant
for a physical property like the conductance, and can be divided from the
group G.  The resulting group G/K is for the Chalker-Coddington model G/K =
U($N,N$)/(U($N$) $\times$ U($N$)).  This is a so called simple Lie group and
is also called a symmetric space. There is a full classification of all
symmetric spaces by Cartan. The way to characterize a symmetric space is to
consider the corresponding Lie algebra, that is the properties of matrices
${\bf A}$, defined by ${\bf T} = \exp ({\bf A})$. Now, one defines
eigenvectors $X$ of $A$, by the commutation relation $[ {\bf A}, {\bf X}] = a
{\bf X}$. The degeneracy of the eigenvalue $a=0$ defines the rank $r$ of the
Lie algebra.  The set of eigenvectors ${\bf H}_i, i=1,...,r$ with eigenvalue
$a=0$ forms an Abelian subalgebra, the so called Cartan Algebra, since $[{\bf
  H}_i, {\bf H}_j]=0$ for all $ i = 1,...,r$.  Since every matrix ${\bf A}$
commutes with itself, $[ {\bf A}, {\bf A}] =0$, it can be written as a linear
superposition $ {\bf A} = \sum_{i=1}^r \lambda_i {\bf H}_i$.  There are $N-r$
eigenvectors ${\bf E}_i$, $i=r+1,...,N$ of ${\bf H}_i$, $i=1,...,r$ with
nonzero eigenvalues $\alpha_1,...,\alpha_r$.  The vector of nonzero
eigenvalues ${\bf \alpha } = (\alpha_1,...,\alpha_r)$ is called the root of
the algebra.  These roots can be characterized by their length, and their
degeneracy, the so called multiplicity.  If there is only one eigenvector to a
root $\alpha$ it is called a simple root.  Now, it can be shown that the
parameters entering in the above formula for the quasi-one dimensional
localization length, Eq.  (\ref{quasionedimension}), $\beta$ and $m_l$ are the
multiplicities of roots on the respective symmetric space G/K
\cite{helgason1962,helgason2000}.
 
We have listed their values in tables \ref{table:sym1},\ref{table:sym3}, and
\ref{table:sym2}. The factor $s$ accounts for the fact that spin flip
scattering from magnetic impurities mixes the two spin channels, and thereby
can double the effective number of channels. Thus, $s =1$, without spin
scattering, and $s=2$ with spin scattering.  The integer $N$ is the number of
transverse channels in the quantum wire, and $l$ the elastic mean free path.
This formula, Eq. (\ref{quasionedimension}) coincides for the ordinary
symmetry classes with the well-known result derived independently by Efetov
and Larkin, and by Dorokhov, namely that the localization length is
proportional to the symmetry parameter $\beta$
\cite{beenakker97,larkin,dorokhov1,dorokhov2,dorokhov3,pichard,kettemann2000}.
For the ordinary and the chiral classes $\beta$ takes the well known
conventional values $\beta = 1,2,4$, $m_l=1$ for the orthogonal, unitary and
symplectic class, respectively.

For the Bogoliubov-de-Gennes-Oppermann class, however, $\beta$ and $s$ have
quite unexpected values. Moreover, in the classes D and DIII, the
multiplicity $m_l$ vanishes. This implies a diverging localization length. For
weak Gaussian disorder the conductance in classes D and DIII has the broad
distribution typical for a critical state \cite{grv2003}. Accordingly, in a
gapless superconductor with broken spin symmetry the quasiparticles are
delocalized for weak disorder. Since both, the charge and the spin, are not
good quantum numbers in such a system, the metallic character of the
quasiparticles can only be measured due to their finite contribution to
transport of heat which can be measured via the thermal conductance
\cite{sf2000}.

In order to understand this better, it is enlightening to generalize briefly
the concept of the quantization of the electrical --- charge --- conductance
to the thermal and the spin conductance.

The defining relations of charge, spin and heat conductivities are
\begin{eqnarray}
  \label{eq:definingequations}
  \vv{j^{\rm c}}&=&\left(\frac{\sigma^{{\rm c}}}{e}\right)\,\nabla(eV)\\
&&\nonumber\\
\vv{j^{\rm s}}&=&\left(\frac{\sigma^{\rm s}}{\mu_{\rm B}\hbar/2}\right)\,
\nabla\left(\frac{\mu_{\rm B}\hbar}{2}B\right)\label{eq:sigmas}\\
&&\nonumber\\
\vv{j^{q}}&=&\left(\frac{\kappa}{k_{\rm B}}\right)\,\nabla(k_{\rm B}T)
\end{eqnarray}
with electron charge $-e$, voltage $V$, Bohr magneton $\mu_{\rm B}$, Boltzmann
constant $k_{\rm B}$, and temperature $T$. The charge, spin and heat
conductivities are denoted as $\sigma ^{\rm c}$, $\sigma ^{\rm s }$ and
$\kappa$, respectively. One can easily verify that these are consistent with
the usual linear response definition apart from rewriting the current
densities as the responses to gradients of energies instead of fields.
We also recall that thermal and electrical currents are related via the
Wiedemann-Franz law
\begin{equation}
  \label{eq:wiedemannfranz}
  \frac{\kappa}{\sigma^{\rm c}T}=\frac{\pi^{2}}{3}\frac{k_{\rm B}^{2}}{e^2}\,.
\end{equation}

The quantization of the charge conductance in terms of $\sigma_{0}=e^{2}/h$
implies that the conductance per charge is given by the ratio of the
elementary charge and the Planck constant, $\sigma^{\rm c}_{0}/e=e/h$.
Analogously, the quantum of the spin conductance will then be given by the
ratio between elementary spin and the Planck constant, $\sigma^{\rm
  s}_{0}/(\mu_{\rm B}\hbar/2)=(\hbar/2)/h$. This implies
\begin{equation}
  \label{eq:quantumofspinconductance}
  \sigma_{0}^{\rm s}=\mu_{\rm B}\frac{(\hbar/2)^{2}}{h}\,.
\end{equation}

The quantum of the heat conductance should be given by the ratio between the
heat per free particle $Q_{0}$ and the Planck constant. The heat per charge
carrier can be deduced from the thermopower $S$
\begin{equation}
  \label{eq:thermopower}
  eST=\frac{\pi^{2}}{3}\,k_{\rm B}T\,
\left(\frac{k_{\rm B}T}{2E_{\rm F}}\right)
\end{equation}
where the last factor accounts for the fact that only electrons in a
temperature window $k_{\rm B}T$ near the surface of the Fermi sea
contribute.
 Note that  the factor $\pi^2/3$ is  the first 
 Sommerfeld expansion parameter in the expansion
 of the energy integral  around the Fermi surface,
 which does not depend on dimensionality. 
 From this one concludes
\begin{equation}
  \label{eq:heatperelectron}
  Q_{0}=\frac{\pi^{2}}{3}k_{\rm B}T\,,
\end{equation}
and from 
\begin{equation}
\frac{\kappa_0}{k_{\rm B}}=\frac{Q_0}{h}
\end{equation}
follows  that
\begin{equation}
  \label{eq:heatconductancequantum}
  \kappa_0=\frac{\pi^{2}}{3}\,\frac{k_{\rm B}^{2}T}{h}\,.
\end{equation}
In the following, we apply this to the case of a quantum Hall system. We use
units such that $\mu_{\rm B}=1$.

When a temperature difference smaller than the Landau gap is applied to two
opposite edges of a quantum Hall bar, there is no thermal transport between
the edges, if the Fermi energy is pinned to localized bulk states. Still, the
total heat current perpendicular to the temperature gradient is non-vanishing,
since the quasiparticles on the hot edge of the sample carry more heat in one
direction than the quasiparticles on the cooler edge in the opposite
direction. This results in a finite thermal Hall conductance $\kappa_{xy}$
defined by
\begin{equation}
I^Q_x = \kappa_{xy}\Delta_{\rm H} T,
\end{equation}
where $\Delta_{\rm H} T$ is the temperature difference between the edges and
$I^Q_x$ the total heat current. One finds that one edge channel contributes
towards the thermal Hall conductance with the quantized amount given in
Eq.~(\ref{eq:heatconductancequantum}) \cite{kf97},
\begin{equation}
\kappa_{xy} = \frac{\pi^2}{3} \frac{k_{\rm B}^{2}T}{h}\,.
\end{equation}
As the {\em charge} Hall conductance, the {\em thermal} Hall conductance is
non-zero only for broken time reversal symmetry and broken parity.
Non-vanishing values can thus only be obtained in {\em directed} network
models.

There has been some debate \cite{cetal02,sf2000,bsz2000,rg2000} if this
critical metal phase of classes D and DIII persists due to the particular
symmetry even at strong disorder as argued in \cite{bfgm2000}, or if there
could be a regime of localized phases characterized by a quantized thermal
Hall conductance parameter, in units of the thermal conductance quantum
$\kappa_{0}$,
\begin{equation}
  \label{eq:thermalHallconductance}
 \sigma^{\rm T}_{xy} = \frac{3h\kappa_{xy}}{\pi^2 k_{\rm B}^2 T}\,,
\end{equation}
with quantum Hall type transitions between phases with $\sigma^{\rm T}_{xy} = 0$ and
$ \sigma^{\rm T}_{xy} = 1$ \cite{kf97}. One could refer to this transition,
accordingly, as the {\rm thermal quantum Hall transition}.

\begin{figure}[btp]
\begin{center}
\leavevmode
\epsfxsize = 0.7\textwidth \epsffile{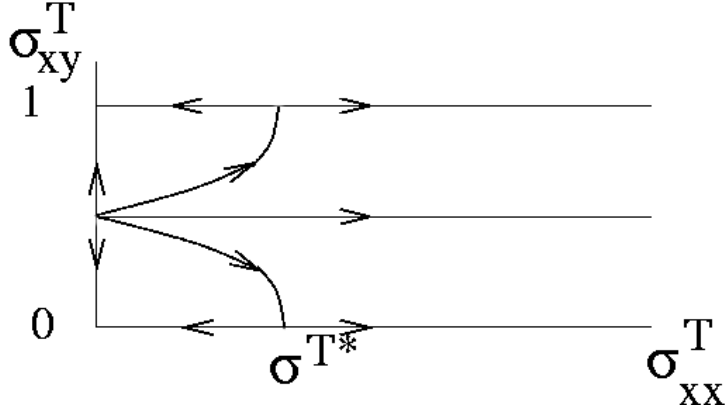}
\caption{
  Scheme of the quantum Hall flow diagram of class D systems in the plane of
  the local thermal conductance and thermal Hall conductance parameters,
  $\sigma^{\rm T}_{xx}$ and $\sigma^{\rm T}_{xy}$, respectively, for correlated disorder
  where an insulating and a quantum Hall phase can exist
  \cite{cetal02,sf2000,rg2000}.  For weak disorder (large $\sigma^{\rm T}_{xx}$),
  the critical metal phase is stable for all values of the Hall conductance.
  For uncorrelated disorder, this is the only possible phase for any value of
  $\sigma^{\rm T}_{xx}$. Correspondingly, $\sigma^{T *} = 0$ for uncorrelated
  disorder.  For strong correlated disorder there is a flow towards an
  insulating phase for $\sigma^{\rm T}_{xy}<1/2$, and a flow to a phase with
  quantized Hall thermal conductance for $\sigma^{\rm T}_{xy}>1/2$,
  respectively.\label{thermalphase} }
\end{center}
\vspace{5mm}
\end{figure}

Recently, several investigations have concluded that the phase diagram of
models belonging to class D critically depends on the spatial correlations of
the disorder potential and that the metal phase persists to strong disorder
only for uncorrelated, white noise disorder
\cite{rl01,bfm2003,grv2003,mdh2001}. If there is at strong disorder a
transition to a localized phase, and the time reversal symmetry is broken, as
it is the case in class D, there can be a finite thermal Hall conductance
$\kappa_{xy}$.  Figure~\ref{thermalphase} shows schematically the suggested
two-parameter flow diagram for correlated disorder
\cite{cetal02,sf2000,rg2000} as a function of the thermal conductance
parameter
\begin{equation}
  \label{eq:thermalconductance}
\sigma^{\rm T}_{xx} = \frac{3 h \kappa_{xx}}{\pi^2 k_{\rm B}^2 T}\,,  
\end{equation}
and the thermal Hall conductance parameter $\sigma^{\rm T}_{xy}$. Such a thermal
metal insulator transition and the corresponding thermal quantum Hall
transition are expected to have very different critical properties from the
integer quantum Hall transition (Tab.~\ref{table:sym3}).

When the time reversal symmetry is broken but global SU(2) spin rotation
symmetry is not, then the quasiparticles in a superconductor can be studied
both by their contribution to thermal and to spin transport.  In two
dimensions this class, denoted as C in Tab.~\ref{table:sym3}, can be studied
in d-wave superconductors.  Here, the order parameter vanishes at four nodal
points on the Fermi surface.  This allows to study the peculiar properties of
the quasiparticles which occur close to the Fermi energy due to the
particle-hole symmetry in class C.  As seen from
Eq.~(\ref{quasionedimension}), the quasiparticles in a quasi-one dimensional
wire belonging to class C are localized.  From results of perturbative
renormalization group studies, one can conclude that they stay localized in
two dimensions \cite{sfbn1998}.

However, in close analogy to the quantum Hall transition, a
localization-delocalization transition is possible in which the quantized Hall
conductance for the {\em spin}, $\sigma^s_{xy}$, is quantized.

The spin transport is caused by a gradient in the perpendicular magnetic field
which results in a gradient in the Zeeman energy. In linear response the spin
current is \cite{smf1999} (Eq.~(\ref{eq:sigmas})),
\begin{equation}
{\vv{j}^s} = \sigma^s {\bf \nabla} B,
\end{equation}
where $\sigma^{\rm s}$ is now a $2\times 2$-matrix with components
$\sigma^{\rm s}_{xx}$ and $\sigma^{\rm s}_{xy}$. When the Fermi energy is
pinned to localized bulk states, the edge states carry a finite spin current
perpendicular to the gradient of the magnetic field,
\begin{equation}
j^s_x = \sigma^{s}_{xy} \nabla_y B. 
\end{equation} 
When the quasiparticles carry the spin $1/2$, the spin Hall conductance,
$\sigma^s_{xy}$ turns out to be quantized in two dimensions in multiples of
the quantum of the spin conductance $\sigma_{0}$
(Eq.~(\ref{eq:quantumofspinconductance})),
\begin{equation}
\sigma^s_{xy} = m \mu_{\rm B}\frac{(\hbar/2)^2}{h},
\end{equation}
where $m$ is found to take only values $0,\pm 2$ \cite{rg2000}, very similar
to the quantization of the charge Hall conductance in units of $e^2/h$. The
spin angular momentum $\hbar/2$ is substituted for the electron charge $e$.
Changing the Fermi energy one finds spin quantum Hall plateau transitions
where the spin Hall conductance changes by two units. This transition is in a
new universality class, and called the {\em spin quantum Hall transition}
(SQHT).  When a strong Zeeman term is introduced which reduces the SU(2)
symmetry of the spin to U(1), this transition splits into two, each of them
being in the usual universality class of the integer quantum Hall transition
\cite{glr1999,bcc2002,sfbn1998,smf1999,khac99}. The spin quantum Hall
transition can also be probed by the thermal conductance, which in a metal
phase is related to the spin conductance by the analogue of the above
Wiedemann-Franz law
\begin{equation}
  \label{eq:swiedemannfranz}
\frac{\kappa}{\sigma^{\rm s}T} = \frac{\pi^{2}}{3} \,
\frac{k_{\rm B}^{2}}{(\hbar/2)^2}\,.  
\end{equation}

Recently, a model belonging to class C has been mapped on the classical
percolation model \cite{glr1999,g1999}. This made it possible to determine the
exact critical exponents of the spin quantum Hall transition, like the
exponent of the divergent localization length, $\nu =\nu_{\rm p}= 4/3$. The mapping is
reviewed below, when the Class C network model is introduced in more detail.
We note that disordered d-wave superconductors also have been argued to be in
the chiral class AIII, yielding delocalized quasiparticles close to the
Fermi energy \cite{a2002}. This situation arises for a slowly varying disorder
which has negligibly small Fourier components for scattering between the four
nodal points where the gap in the quasiparticle density of states is
vanishing. For each node, the quasiparticles independently can be modeled by
Dirac fermions.  In a magnetic field, which breaks the time reversal
invariance, this leads to the sub-lattice class AIII \cite{a2002}.

Sub-lattice models exhibit some similarity to a quantum Hall transition, since
they have one extended state in the middle of a band of localized states.
Their Hamiltonian is given by a tight binding model on a two dimensional
bipartite lattice. There are no on-site potentials, and the hopping matrix
elements are random. Thus, they have perfect particle-hole symmetry. In
contrast to the models for the integer Quantum Hall Effect, two-sub-lattice
models have a divergent density of states in the center of the band where the
extended states appear. For the unitary (AIII) and the orthogonal (BDI) chiral
classes, the density of states diverges,
\begin{equation}
\rho(E) \propto \frac{1}{|E|}\, e^{-c|\ln E|^{1/x}},
\end{equation}
with constant $c$. By using field theory, Gade obtained $x=2$ for both
classes, AIII and BDI \cite{g1990,g1993}. By considering random Gaussian
surfaces, and numerically, the value $x=3/2$ was obtained for class BDI
\cite{mdh2002}. This result for BDI was also derived with the supersymmetry
method \cite{mrf03} and was argued to be a direct consequence of the freezing
transition of the dynamical exponent $z$ \cite{mdh2002,mrf03,hd02,f03,rmf03},
which is related to the fact that the multi-fractal spectrum is bounded
\cite{rmf03}.  While it has been suggested that this freezing transition is
not sensitive to time reversal and spin symmetry and therefore the value
$x=3/2$ could be valid for the chiral classes, AIII and CII, as well
\cite{mrf03}, a derivation is pending.

The model of Dirac fermions in a random vector potential
(Sect.~\ref{subsubsec:dirac}) is another member of the symmetry class AIII.
Although a divergence of the density of states has been found here as well,
the corresponding exponent is varying continuously with the disorder strength
\cite{lfsg94}. When moving away from the band center, the sub-lattice models
acquire the properties of the ordinary models, and the states are localized
for the orthogonal and unitary models. Sub-lattice models with directed bonds
have recently been established \cite{bc03}. Their critical properties still
have to be studied in detail. 
So far, the symplectic sub-lattice class has been
studied only briefly \cite{aso03}. 

According to the above description, the delocalization transition of the
Chalker-Coddington network model without particle-hole symmetry should be
insensitive to the spin rotation symmetry, since the conventional unitary
class makes no distinction whether or not that symmetry is broken.  This
argument does not take into account that the lifting of the spin degeneracy
can result in a change of the statistically relevant density of states and
thereby also a change in the localization behavior. This has been first noted
for localization in quasi-one dimensional disordered wires without time
reversal symmetry, where the localization length is doubled when the spin
degeneracy is broken, Eq.~(\ref{quasionedimension})
\cite{larkin,kettemann2000}.  Recently, it has been argued within a
semi-classical percolation model \cite{am2002}, that spin-orbit scattering can
change the universality class from the one of the conventional quantum Hall
transition to the one of classical percolation.  In short, the argument is
that the Hamiltonian with spin-orbit scattering and random scalar potential
reduces --- in an adiabatic approximation --- to the Hamiltonian of electrons
with opposite spins moving in the effective potential which is the sum of the
scalar potential and the positive/negative locally varying random Zeeman
fields for the electrons with spin up and spin down, respectively. Thus, the
electrons with opposite spins do see different randomness. Close to saddle
points in the effective potential, the electrons have the choice of either to
tunnel while keeping their spin or to flip their spin using the non-adiabatic
part of the spin-orbit interaction, and then continue to propagate in the
potential landscape corresponding to the opposite spin. For a potential with a
sufficiently short correlation length, the potential landscapes for the
electrons with spin up and spin down should be sufficiently different that
this argument applies.  This argument suggests that the critical exponent of
the localization length in that situation is close to the one of classical
percolation, $\nu =\nu_{\rm p}= 4/3$ \cite{am2002}.  The obvious similarity of this model
to models of the Bogoliubov-de-Gennes-Oppermann class, when identifying the
spin flip scattering with the anomalous pairing amplitude $\Delta$, coupling
electrons with holes may be used as a heuristic explanation for the change of
the universality class to the one of the classical percolation transition, in
class~C models. Indeed, the network model of class C with SU(2)-scattering
phases at the links has been first studied as a model of spin degenerate
Landau levels with negligible random potential and strong spin-orbit
scattering \cite{kha97}.  Another realization would be Landau levels with
strong magnetic impurities \cite{hsw93}. This was found to have a single
quantum critical point with $\nu \approx 1.1 $ \cite{kha97}.

In order to study quantitatively the localization properties of the
metal-insulator transitions and the quantum Hall transitions it is necessary
to formulate specific network models for the various symmetry classes. In the
next section, we introduce non-directed network models for the three ordinary
symmetry classes, of which only the symplectic one is expected to have a metal
insulator transition. Next, we review unitary network models with two spin
channels that describe the transition of quantum Hall systems with mixing
between spin split Landau levels. Then, we review the network models for
transitions of the quantum Hall type belonging to the
Bogoliubov-de-Gennes-Oppermann class. Finally, we provide an overview on
the network models for sub-lattice systems.

\subsection{Non-directed Network Models}
\label{sec:net2ds}

In order to construct network models belonging to the orthogonal and
symplectic classes, one should assume links that are not directed. Following
early work on the scattering matrix formulation of the scaling theory of
Anderson localization \cite{ataf80}, an example of such a model has been
established and studied by using the real space renormalization group method
by Shapiro in 1982 \cite{shapiro82}. Recently, such network models have been
explored numerically for all of the three conventional symmetry classes
\cite{mjh98,fjm99}. The results of the scaling theory of localization have
been reproduced: a localized phase exists for the unitary and orthogonal
classes, and a metal-insulator transition occurs in the symplectic class. The
corresponding ${\bf S}$ matrices are unitary, symmetric, and symplectic
\cite{beenakker97}.

The fact that the links are not directed, but nevertheless the time reversal
symmetry is broken for non-directed unitary network models, makes these models
equivalent to two dimensional random fermions in a weak magnetic field that
does not affect the classical motion of the electrons. Accordingly, extended
states are not expected to exist. This has been confirmed by extensive
numerical studies \cite{fjm99}.

\begin{figure}[hbtp]
\vspace{5mm}
\begin{center}\leavevmode
\includegraphics[width=0.8\linewidth]{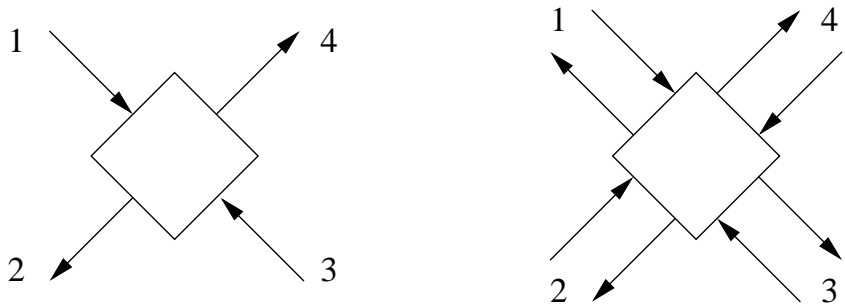}
\caption{
  Comparison of the nodes of a directed (left) and a non-directed network
  model (right).\label{fig:orthouninode}}
\end{center}
\vspace{5mm}
\end{figure}

In non-directed networks, the scattering matrices representing the nodes
relate four incoming to four outgoing currents (Fig.~\ref{fig:orthouninode}),
\begin{equation}
 \left(\begin{array}{c}
\psi_1^\mathrm{o}\\
\psi_2^\mathrm{o}\\
\psi_3^\mathrm{o}\\
\psi_4^\mathrm{o}
\end{array}
\right)
={\bf S}_{\mathrm{orth}}
 \left(\begin{array}{c}
\psi_1^\mathrm{i}\\
\psi_2^\mathrm{i}\\
\psi_3^\mathrm{i}\\
\psi_4^\mathrm{i}
\end{array}
\right).
\end{equation}

With time reversal invariance, the ${\bf S}$ matrix is symmetric
\cite{beenakker97}. By assuming isotropy of the scattering, it can be cast
into the form \cite{mjh98}
\begin{equation}
{\bf S}_{\mathrm{orth}}=\left(
\begin{array}{cccc}
r \,&d\, &t\, &d  \\
d \, &r \, &d \, &t \\
t \, &d \, &r \, &d  \\
d \, &t \, &d \, &r 
\end{array}
\right),
\end{equation}
where $t$, $r$ and $d$ denote transmission, reflection and deflection
amplitudes, respectively.  It is assumed here that the deflection to the left
and to the right are equal, as it should be for a network without chirality.
The scattering parameters are complex and satisfy the conditions
\begin{eqnarray}
 |r|^2+|t|^2+2|d|^2&=&1\,, \nonumber\\
&&\nonumber\\
d^* r+d r^* + d^* t+d t^* &=&0\,, \\
&& \nonumber\\
2|d|^2+r^* t + r t^*&=&0\,.  \nonumber
\label{eq:symplecticrtd}
\end{eqnarray}
>From the 1st and the 3rd condition one finds the relations
\begin{eqnarray}
|r|^2+|t|^2\le 1\, , \quad 1 \le |r|+ |t|\,.
\end{eqnarray}
The second inequality is violated when the time reversal symmetry is broken.

Due to random phases attached to links, we can always set $d$ to be a real
positive number.  Then, once $|r|$ and $|t|$ are given, $r, t$ and $d$ are
uniquely determined by Eq.~(\ref{eq:symplecticrtd}).

Though the transfer matrix, at first sight, seems to be complex, the
calculation can be performed with only real (or quaternion-real) numbers after
a proper unitary transformation \cite{as04}.

In order to construct a model belonging to the symplectic class, we first
establish a scatterer that rotates the direction of the spin \cite{mjh98}. The
${\bf S}$ matrix relates the currents with spin up ($\uparrow$) and spin down
($\downarrow$) according to
\begin{equation}
 \left(\begin{array}{c}
\psi_{1,\uparrow}^\mathrm{o}\\
\psi_{1,\downarrow}^\mathrm{o}\\
\psi_{2,\uparrow}^\mathrm{o}\\
\psi_{2,\downarrow}^\mathrm{o}
\end{array}
\right)
={\bf S}_{\mathrm{symp}}
 \left(\begin{array}{c}
\psi_{1,\uparrow}^\mathrm{i}\\
\psi_{1,\downarrow}^\mathrm{i}\\
\psi_{2,\uparrow}^\mathrm{i}\\
\psi_{2,\downarrow}^\mathrm{i}
\end{array}
\right)\,.
\end{equation}
with 
\begin{equation}
 {\bf S}_{\mathrm{symp}}=\left(
\begin{array}{cc}
 0\,& q \\
\overline{q}\, & 0
\end{array}\right).
\end{equation}

This choice of symplectic scattering matrices mixes only different spin
channels moving in the same direction, but does not have backscattering matrix
elements. Backscattering occurs thus in this model at the orthogonal nodes
only. Here, $q$ is a quaternion-real number,
\begin{equation}
 q=\sum_{k=0}^3 q_k\tau_k \, ,\quad \tau_k=i\sigma_k \,\,
(k=1,2,3)\,\quad \tau_0={\bf 1}\,.
\end{equation}
The coefficients $q_k$ are real numbers that satisfy
\begin{equation}
 \sum_{k=0}^3 q_k^2=1
\end{equation}
and the $2\times 2$-matrices $\sigma_k$ are the Pauli spin matrices. The
quaternion conjugate is denoted by $\overline{q}$.
\begin{equation}
 \overline{q}=q_0 {\bf 1}-\sum_{i=1}^3 q_i\tau_i .
\end{equation}
The strength of the spin orbit scattering is characterized by the distribution
of $q_i$.  When $q_0=1,\, q_1=q_2=q_3=0$ there is no spin orbit scattering.
The spin rotation is most random when
\begin{equation}
\begin{array}{c}
 q_0+i q_3=e^{i \alpha}\cos\beta\\
 q_1+i q_2=e^{i \gamma}\sin\beta\\
 \end{array}
\end{equation}
with $\alpha$ and $\gamma$ distributed uniformly between $[0,2\pi)$, and
$\beta$ distributed according to the probability density,
\begin{equation}
 P(\beta)=\left\{
\begin{array}{cc}
 \sin(2\beta) & \quad 0\le\beta\le\pi/2 \\
0 & \quad \mathrm{otherwise}
\end{array}
\right. \, .
\end{equation}
\begin{figure}[hbtp]
\vspace{5mm}
\begin{center}\leavevmode
\includegraphics[width=0.5\linewidth]{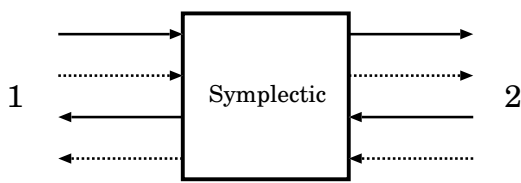}
\caption{
  Schematic view of a scatterer that provides spin rotation. Solid arrows
  correspond to spin-up currents. Broken arrows indicate spin-down
  currents.\label{fig:symplecticnode}}
\end{center}
\vspace{5mm}
\end{figure}

\begin{figure}[hbtp]
\vspace{5mm}
\begin{center}\leavevmode
\includegraphics[width=0.5\linewidth]{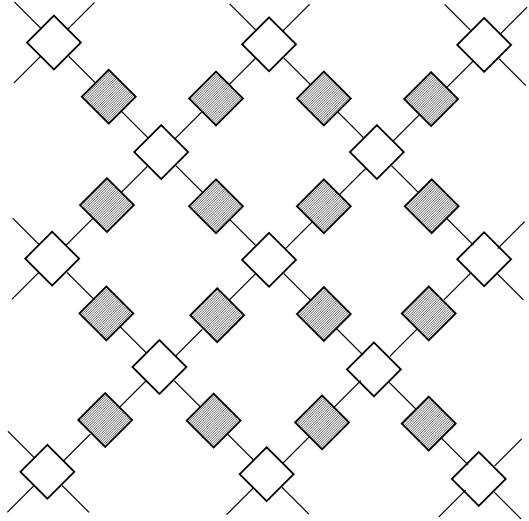}
\caption{
  Network of nodes realizing the symplectic class. White boxes describe
  scattering with orthogonal symmetry. Shaded boxes rotate spin, thus breaking
  spin rotation invariance.\label{fig:symplecticnetwork}}
\end{center}
\vspace{5mm}
\end{figure}

As before, one attaches random phases between the nodes described by the
${\bf S}$ matrix
\begin{eqnarray}
  \label{eq:smatrix}
  {\bf S}_{\rm link}=\left(
    \begin{array}{cc}
0&e^{i\phi}\\
e^{i\phi}&0
    \end{array}
\right)\,.
\end{eqnarray}
The resulting network is shown in Fig.~\ref{fig:symplecticnetwork}. This model
is similar in spirit but different in the details as the tight binding
Hamiltonian on a square lattice including spin-orbit scattering proposed in
\cite{ez87,ando89,aso02}. A similar model also has been proposed in
\cite{minakuchi98}.

\begin{figure}[hbtp]
\begin{center}\leavevmode
\includegraphics[width=0.7\linewidth,angle=270]{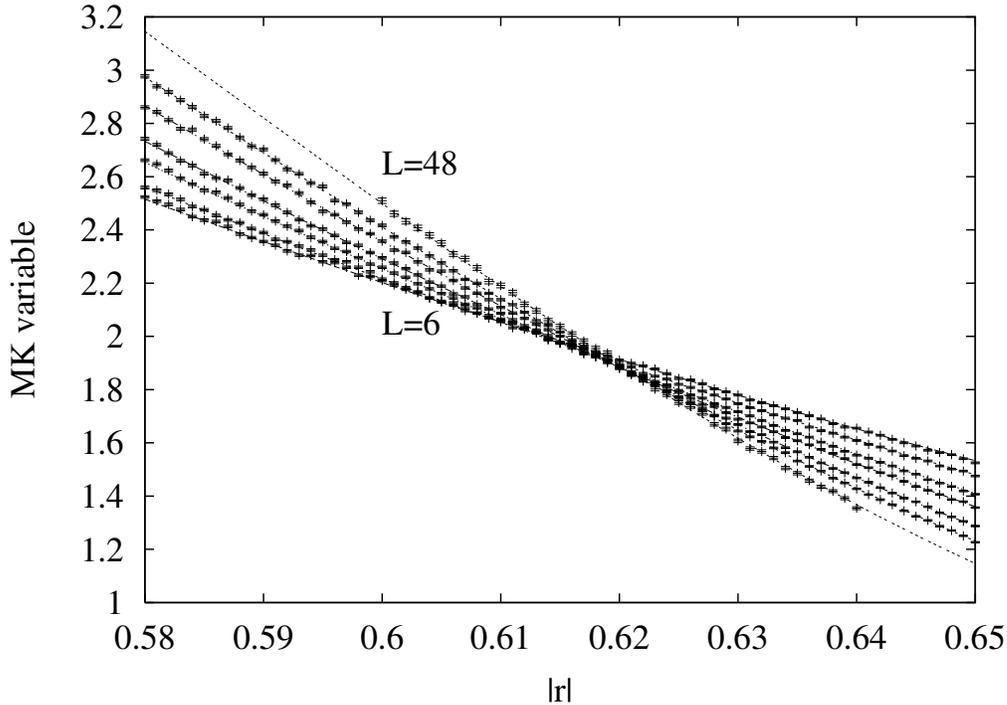}
\caption{MacKinnon-Kramer variable $\Lambda$ as a function 
  of $|r|$ for $|t|=0.6$; $|d|$ is determined according to
  Eq.~(\ref{eq:symplecticrtd}); $L$=6,8,12,16,24,32 and 48; precision of the
  data is 0.2$\%$ except for $L=48$ where it is 0.4$\%$.  Curves intersect
  approximately at $|r|=0.62$, which indicates the presence of an Anderson
  metal-insulator transition in two dimensions.\label{fig:net2dsloc} }
\end{center}
\vspace{5mm}
\end{figure}

Figure~\ref{fig:net2dsloc} shows some results for the localization length
$\lambda(L)$ in a two dimensional strip of the width $L$. The renormalized
localization length $\Lambda=\lambda (L)/L$ as a function of $|r|$ is plotted
for different $L$.  When $|r|$ is smaller than 0.61, $\Lambda$ increases with
increasing $L$.  It is decreasing function of $L$ for $|r|>0.63$. This
behavior indicates a delocalization-localization transition \cite{km93}. For
the symplectic symmetry there occurs a metal-insulator transition in two
dimensions as has been conjectured before \cite{hln80}. By analyzing the data
using the finite size scaling method the critical exponent can be extracted
(cf. Tab.~\ref{table:sym1}). The estimates with 95$\%$ confidence interval,
$\nu=2.81\pm 0.05, \Lambda_c=1.836\pm 0.027$, agree within the uncertainty
with the result of numerical studies done for the tight binding Hamiltonian,
$\nu=2.746\pm 0.009, \Lambda_c=1.843\pm 0.001$ \cite{aso02,aso04}.

To construct orthogonal and a unitary nondirected network models, we assume
in ${\bf S}_{\rm symp}$ $q=1$. In the unitary case, ${\bf S}_{\rm link}$ is
no longer symmetric,
\begin{eqnarray}
  \label{eq:unitaryS}
 {\bf S}_{\rm link}^{\rm unitary}=\left(
    \begin{array}{cc}
0&e^{i\phi}\\
e^{i\phi'}&0
    \end{array}
\right)\,. 
\end{eqnarray}
In both cases, we have spin degeneracy, and the dimension of the transfer
matrix is halved.

\subsection{Network Models of the Bogoliubov-de-Gennes-Oppermann Class}
\label{sec:bdgoc}
{\em Class C}

As mentioned above, systems belonging to the Bogoliubov-de-Gennes-Opper\-mann
class have a particle-hole symmetry, and the spectrum of eigenvalues consist
of pairs, $(-E_n,\,E_n)$.  The zero of energy, $E=0$, is the chemical
potential in the superconductor.  Class C describes quasiparticles in a
spin-singlet superconductor in which time-reversal symmetry is broken, but
spin rotation symmetry is conserved.  This corresponds to systems with
negligibly small Zeeman splitting \cite{az97}. Far away from the Fermi energy,
the transport properties show a crossover to the conventional unitary class A.
Therefore, class C behavior can only be studied in gapless superconductors or
in metals closely attached to superconductors, where quasiparticles close to
the Fermi energy exist. As summarized in table \ref{table:sym3}, the quantum
states of quasiparticles in nondirected models of the class C are localized. This phase is named spin insulator,
 because the insulating behavior reveals itself only as  a vanishing spin or thermal conductance, since the charge is not conserved in a superconductor. When parity
symmetry is broken, as in directed network models, there can  be a quantum Hall type transition to a
phase with a critical  state and an integer-quantized spin Hall conductance $\sigma^s_{xy}$. This is
called the spin quantum Hall transition (SQHT).

Thus, we are now  looking for a generalization of the Chalker-Coddington model
which satisfies the particular symmetries of class C
\cite{glr1999,bcc2002,khac99}. In the simplest case, one has two-component
wave functions, which propagate on directed links through the lattice, and
which scatter at nodes between adjacent links. All the nodes in the lattice
have two incoming and two outgoing links. It turns out \cite{glr1999}, that
the SQHT can be determined from the properties of the perimeters, or hulls, of
classical percolation clusters on a two dimensional lattice constructed of
such nodes and links. This classical problem is much simpler than the original
quantum problem, and can be solved exactly, for example by mapping it on an
integrable super-spin chain \cite{glr1999} which will be introduced below in
the chapter \ref{sec:susy}.
 
In the following, we review the models for class C, as introduced by Beamond
et al. \cite{bcc2002}.  A similar network model with SU(2) scattering phases
at the links has been studied before in the context of the Quantum Hall Effect
of spin degenerate Landau levels \cite{kha97}. We consider a set of nodes $n$
connected by links $l$.  Each node is assumed to be of the degree four, such
that at each node two directed links enter and two leave. A two-component wave
function is assumed to propagate along each link. This propagation may be
described by a unitary evolution operator ${\bf U}$, which describes the
evolution of the wave function one unit forward in ^^ ^^ time'', as the
particle moves from a given link to a neighboring one. The evolution operator
plays a similar role in defining the network model as the time evolution
operator corresponding to a Hamiltonian, for example, of a tight binding
model. This procedure has been discussed in detail before for the U(1)
network model in \cite{km95,hc96} (Sect.~\ref{sec:hamiltonians}).

The model can be constructed by associating each link $l$ with a unitary
$2\times 2$ matrix ${\bf U}_{l}$. This matrix specifies the phase accumulated
when traversing the link. Each node $n$ is assumed to be represented by a
scattering matrix (Sect.~\ref{subsec:saddletransmission})
\begin{equation} 
{\bf S}_n=
{\bf 1}_{2} \otimes \left(\begin{array}{cc}\cos\theta_n & \sin\theta_n \\
                      -\sin\theta_n & \cos\theta_n \end{array}\right)\,,
\end{equation}
where ${\bf 1}_{2}$ is the $2\times 2$ unit matrix. The ${\bf S}$ matrix
describes scattering at the node from the incoming links to the outgoing ones.
If the network has $M$ links (and therefore $M/2$ nodes), then ${\bf U}$ is an
$ N\times N$ matrix, with $ N=2 M$. It consists of $M/2$ blocks, each
associated with a particular node and of size $4 \times 4$. The block at the
node $n$ has the form
\begin{equation}
\left(\begin{array}{cc}{\bf U}_3^{1/2}&0\\0&{\bf U}_4^{1/2}\end{array}\right)
{\bf S}_n \left(\begin{array}{cc}{\bf U}_1^{1/2}&0\\0&{\bf U}_2^{1/2} 
\end{array}\right)
\label{node0}
\end{equation}
where $(1,2)$ and $(3,4)$ label the links which are incoming and
outgoing at the given node.

So far, we only have assumed the links to be directed, so that time-reversal
symmetry is broken. To identify a network model to belong to class C, one can
start \cite{khac99} from the defining property of a Hamiltonian ${\bf H}$
with this symmetry \cite{az97}
\begin{equation}
\label{class-C}
 {\bf H}^* = - {\bf \sigma}_y  {\bf H}{\bf \sigma}_y\,,
\end{equation}
with the Pauli matrix ${\bf \sigma}_y$ acting on the spin variables. The
operator $H^*$ is the complex conjugate of $ {\bf H}$. With this Hamiltonian,
and using that $ {\bf U}$ can be written as
\begin{equation}
  \label{eq:phaseevolution}
 {\bf U} = e^{-i{\bf H}}\,,   
\end{equation}
Eq.~(\ref{class-C}) implies
\begin{equation}
\label{class-C'}
 {\bf U}={\bf \sigma}_y\,  {\bf U}^*{\bf \sigma}_y\,.
\end{equation}
>From this, an equivalent restriction follows for the phases of the links,
\begin{equation}
  \label{eq:phasesoflinks}
{{\bf U}_l}=\sigma_y\, {\bf U}_{l}^*\sigma_y  
\end{equation}
which are therefore unitary Sp(2)-matrices equivalent to SU(2)-matrices, where
$Sp(2 N)$ is defined to be the symplectic group of $2 N$ by $2 N$ matrices
${\bf A}$ which fulfill the condition ${\bf A}^{\mathrm T} {\bf \sigma}_y {\bf
  A} = {\bf \sigma}_y$ \cite{bcc2002}. The wave function consists of two
components in each link. The space of states on each link may be viewed as
consisting of a two-dimensional subspace, within which ${\bf \sigma}_y$
operates.

One possibility to introduce randomness into the model consists of assuming
the link phases to be random variables drawn independently from a distribution
which is uniform on the invariant (Haar-)measure of Sp(2). The ensemble
average of a physical quantity in the network model, denoted by
$\langle\ldots\rangle$, is the mean with respect to this measure.

As reviewed in Sect.~\ref{sec:4.1}, the conductance of an open network of a
finite length can be expressed by the multi-channel Landauer formula in terms
of the transmission probability.  Specifically, the spin conductance measured
between two contacts in units of $(\hbar/2)^2/h$ is given by
Eq.~(\ref{eq:landauer}),
\begin{equation}
\label{conductance}
g = {\rm Tr}\, {\bf t} {\bf t}^{\dagger}\,.
\end{equation}
Here, ${\bf t}$ is the rectangular $N\times N$ transmission matrix containing
the matrix elements for $z=1$ of the propagation operator
\begin{equation}
  \label{eq:C-propagator}
   {\bf G}(z)=\frac{1}{1-z {\bf U}}
\end{equation}
between the incoming and outgoing link states $l_{\rm i}$ and $l_{\rm o}$.

One can now show that the disorder average ${\rm Tr}\,\langle {\bf G}(z)
\rangle$ taken over the Sp(2)-phase factors of the Green function
\begin{equation}
  \label{eq:C-greenfunction}
  G(z;l,l')=\langle l|(1-z {\bf U})^{-1}|l'\rangle
\end{equation}
can be expressed in terms of classical paths \cite{glr1999,bcc2002}.

\begin{figure}[htbp]
\begin{center}
\includegraphics[width=0.7\textwidth]{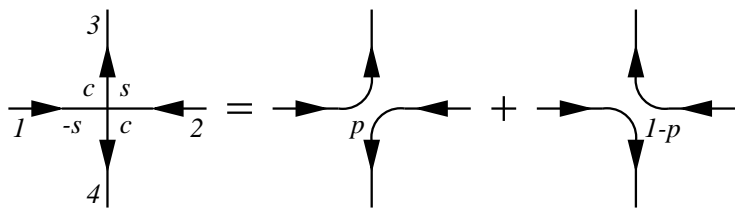}
\caption[decomp]{Decomposition of the scattering at a given node: 
  ${\bf S}$ matrix elements $\cos\theta_n$ and $\pm\sin\theta_n$ are
  associated with the transitions $(1,2)\to(3,4)$ as indicated on the left.
  The two decompositions are weighted with factors $p_n=\cos^2\theta_n$ and
  $1-p_n=\sin^2\theta_n$ [Figure taken from \cite{bcc2002}].\label{decomp}}
\end{center}
\end{figure}

To illustrate this, we define for the same network a {\em classical}
scattering problem as follows. The scattering at each node may be decomposed
into disconnected processes in two different ways, $(1\to 3,2\to 4)$ and
$(1\to 4,2\to 3)$ (Fig.~\ref{decomp}).
\begin{figure}[htbp]
\begin{center}
\includegraphics[width=0.6\textwidth]{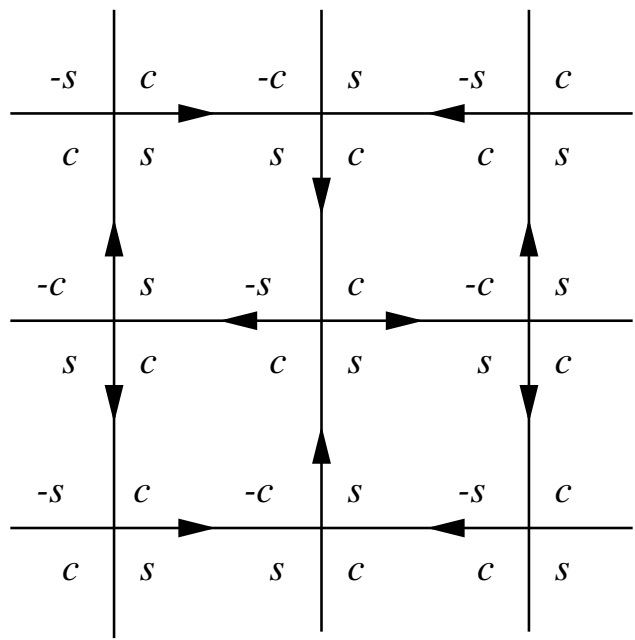}
\caption[Llattice]{The network model for the spin Quantum Hall Effect. 
  Scattering amplitudes $\pm\cos\theta_n$ and $\pm\sin\theta_n$ are indicated
  by $\pm c$ and $\pm s$ [Figure taken from \cite{bcc2002}].\label{Llattice}}
\end{center}
\end{figure}

A two dimensional model exhibiting the spin Quantum Hall Effect is obtained by
taking the network to be the so-called L-lattice shown in Fig.~\ref{Llattice}.
The two possible classical decompositions of a node may be associated with the
presence or absence of a bond, with probabilities $p$ and $1-p$ between
neighboring sites on an associated square lattice \cite{glr1999}. The latter
is rotated by $45^{\circ}$ relative to the L-lattice, and has a lattice
spacing increased by a factor $\sqrt 2$. In this model, closed loops of the
classical problem form interior or exterior hulls of bond percolation clusters
on the larger lattice. It is known that such loops are finite with
characteristic size $\xi(p)<\infty$ except at the critical point, $p=p_c$,
which for bond percolation on the square lattice occurs at $p_c=1/2$. When
approaching the critical point, $\xi$ diverges as $\xi \sim |p-p_c|^{-\nu}$
with $\nu=\nu_{\rm p}=4/3$ (Chap.~\ref{sec:percolationmodel}), while at the critical point
the distribution of the hull lengths is $P(L)\sim L^{-y}$ at large $L$, with
$y=8/7$.

The quantum localization length in this class C model diverges with the same
value of the exponent, $\nu=\nu_{\rm p}=4/3$, as the plateau transition is approached. The
density of states
\begin{equation}
  \label{eq:C-dos}
 \rho(E)=-\frac{1}{\pi}{\mathcal I}m\,{\rm Tr}\langle {\bf G}(E)\rangle 
\end{equation}
 varies for small
$E$ as 
\begin{equation}
  \label{eq:smallE}
\rho(E) \sim E^2  
\end{equation}
in the localized phase, and as
 \begin{equation}
   \label{eq:doscriticalpoint}
\rho(E) \sim |E|^{1/7}   
 \end{equation}
 near the critical point (Tab.~\ref{table:sym3}).

{\em Class D} 

As described in the introduction to this chapter (Tab.~\ref{table:sym3} and
Fig.~\ref{thermalphase}), models in class D have particularly rich phase
diagrams in two dimensions. The symmetry is realized in superconductors with
broken time reversal invariance, and either broken spin rotation invariance,
as in d-wave superconductors with spin-orbit scattering, or spin-less or
spin-polarized fermions, as in certain p-wave states. Since the spin rotation
symmetry is broken, and the charge is not conserved, the quasiparticles can
only be measured by their contribution to thermal transport.
Equation~(\ref{quasionedimension}) shows that quasiparticles are delocalized
in quasi-one dimensional quantum wires belonging to class D. In the two
dimensional case, the flow under the renormalization group is towards larger
thermal conductance \cite{sf2000,bsz2000,rg2000,bcsz1999}. Thus, there is a
phase in which there is a diverging density of extended eigenstates at zero
excitation energy (Tab.~\ref{table:sym3} \cite{sf2000}). A superconductor
described by this model is in a thermal-metal phase. A phase with localized
quasiparticles is a natural possibility, although it has only been found so
far for models with correlated disorder \cite{cetal02} which gives rise to a
phase with a quantized Hall conductance (Fig.~\ref{thermalphase}).
 
In analogy with the network models belonging to class C introduced in the
previous section, one can formulate a directed network model with the
symmetries of class D \cite{cetal02}. Disorder appears in the network model in
the form of random scattering phases, and the symmetries of the class D
restrict scattering phases to the values $0$ and $\pi$, multiplying the wave
function at each node at random with phase factors $\pm 1$. Remarkably, within
this framework, different particular forms of disorder result in quite
different physical behaviors. Depending on the level of correlations between
the random phases three cases have been found so far
\cite{cetal02,bfm2003,jlsc2001}.

The first of these was introduced by Cho and Fisher \cite{cf97a} with the
intention of modeling the two dimensional random bond Ising model (RBIM)
(Chap.~\ref{sec:ising}) which possesses a fermion representation with the
symmetries of class D.  Subsequently, it has been noted
\cite{rl01,cthesis1997} that a precise mapping of the Ising model leads to a
second version of the model which accordingly is denoted as random bond Ising
model. In both of the models, scattering phases $\pi$ appear in correlated
pairs. A third model \cite{cetal02,bsz2000} denoted by O(1), can be
established if one assumes the scattering phases to be independent random
variables. Each model has two parameters: the disorder concentration, $p$
($0\leq p \leq 1$), and a tunneling amplitude $\sin \alpha$ ($0 \leq \alpha
\leq \pi/2$) which controls the value of the thermal Hall conductance at short
distances.

\begin{figure}[htbp]
\begin{center}
  \epsfysize=3.0in \epsfxsize=4.0in \centerline{\epsffile{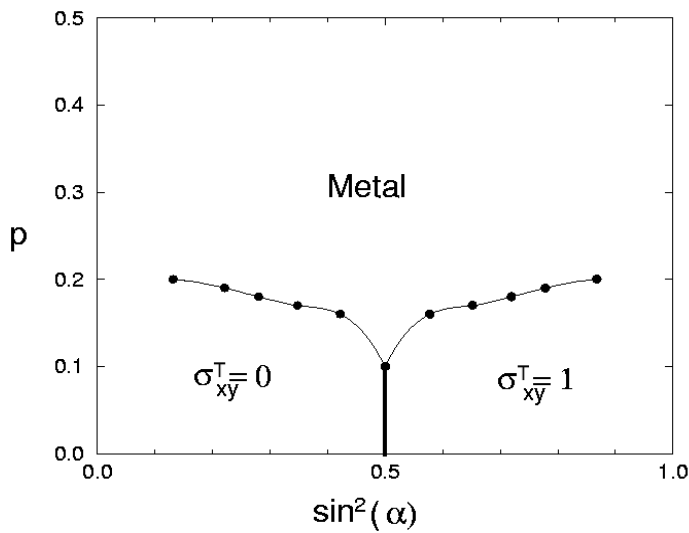}}
  \vspace{0.0in} \caption[phasediag]{The phase diagram in the
    ($p,\alpha$)-plane of the model of Cho and Fisher \cite{cf97a} obtained by
    numerical calculations \cite{cetal02}. Here, $\sigma_{xy}^{\rm T}$ is the
    thermal Hall conductance parameter defined in Eq.
    (\ref{eq:thermalHallconductance}), $p$ the disorder parameter, and
    $\sin^{2}(\alpha)$ is the tunneling probability (Figure taken from
    \cite{cetal02}).
\label{phasediag}}
\end{center}
\end{figure}

A phase diagram for the Cho-Fisher model in the $(p,\alpha)$-plane is shown in
Fig.~\ref{phasediag}. It contains three regions: a metallic phase, a localized
phase, and a thermal quantum Hall phase. While the random bond Ising model has
been shown to have no metallic phase \cite{rl01} the O(1) model has been
argued to have no insulating and thus no quantum Hall phase \cite{bfgm2000}.
Only the Cho-Fisher model has all three phases. It features three potentially
different critical points: a quantum Hall-type transition, an
insulator-to-metal transition, and a multi-critical point at which all three
phases meet. This phase diagram has the form proposed generically for class D
\cite{sf2000}.  The phase diagram is very complex.  The corresponding critical
parameters remain to be determined quantitatively. The network model reviewed
here \cite{cetal02} will certainly be central in achieving a better
understanding of this phase diagram in future investigations. An attempt of
understanding the implications of the all-orders beta functions has been
made in \cite{lc01}.

\subsection{Chiral Network Models}
\label{sec:chiralnetwork}

In the following, we shortly summarize the network models of the chiral
symmetry classes introduced recently \cite{bc03}. The network models presented
here are connected in two distinct ways to the models without chiral symmetry
introduced in the previous sections. Class AIII is connected with the U(1)
network model, while CII and BDI belong to the SU(2) and O(1) network models,
respectively. They describe the plateau transitions in dirty superconductors,
as briefly explained above. In each case, the model with chiral symmetry is
constructed by coupling two copies of the partner model. After a
transformation, this coupling is equivalent to introducing absorption and
coherent amplification in the original models. This equivalence parallels the
established link \cite{msa1998} between the chiral symmetry classes and
non-Hermitian random operators.

We formulate the network models for each of the chiral symmetry classes by
treating in some detail the chiral unitary class (AIII), and outlining the
equivalent steps for the chiral orthogonal (BDI) and chiral symplectic (CII)
classes \cite{bc03}. The strategy is to construct the two dimensional internal
space associated with chiral symmetry using two related copies of a network
model without that symmetry.

The symmetry may be discussed in terms of a Hamiltonian ${\bf H}$, a
scattering matrix ${\bf S}$, or a transfer matrix ${\bf T}$. We consider
systems which conserve probability density. The Hamiltonian is Hermitian and
the scattering matrix is unitary. The equivalent condition for the transfer
matrix involves the operator of the current $J$, and is imposed by current
conservation as in Eq.(\ref{eq:condition}),
\begin{equation}
  \label{eq:hermitian}
 {\bf H}^{\dagger}={\bf H}
\,,\qquad\qquad  
{\bf S}^{\dagger}={\bf S}^{-1}
\,,\qquad\qquad  
{\bf T}^{-1}={\bf J}{\bf T}^{\dagger}{\bf J}  \,.
\end{equation}

Chiral symmetry is implemented on a two dimensional internal space in which
the Pauli operator $\sigma_x$ operates. For the Hamiltonian this is equivalent
to
\begin{equation}
  \label{eq:requirement}
 \sigma_x {\bf H} \sigma_x=-{\bf H} \,.
\end{equation}
By assuming
 the scattering matrix to
have the symmetry of $e^{-i {\bf H}}$, one has
\begin{equation}
  \label{eq:chiralscattering}
  \sigma_x {\bf S} \sigma_x ={\bf S}^{-1}\,.
\end{equation}
Systems in class AIII have no other relevant discrete symmetry.  Those in
classes BDI and CII are also invariant under time-reversal, in the presence
and absence of spin rotation symmetry, respectively.

{\it Class AIII}

\begin{figure}[h]
\begin{center}
\includegraphics[width=0.4\textwidth]{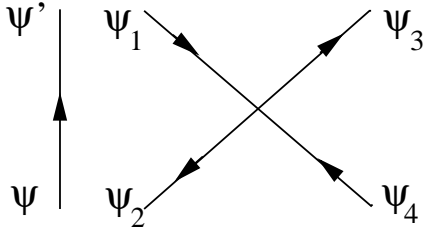}
\caption{The elements of the network model:
  a link (left) and a node (right) [Figure taken from
  \cite{bc03}].\label{buildingunits}}
\end{center}
\end{figure}

\begin{figure}[htbp]
\begin{center}
\includegraphics[width=8cm]{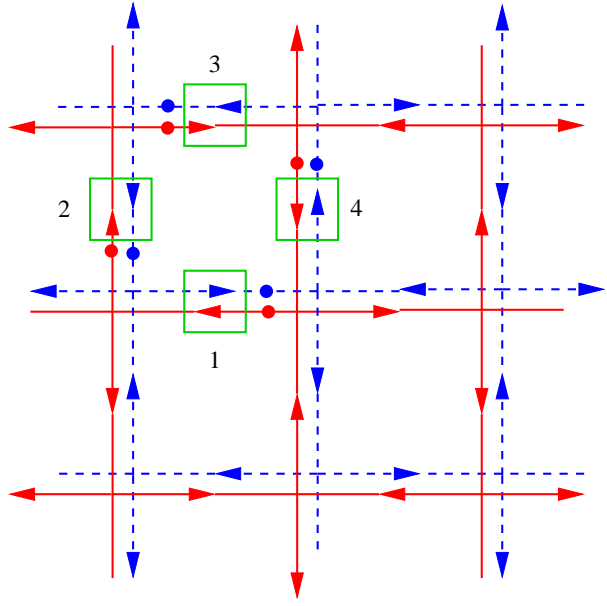}
\caption{The structure of the AIII network model.
  Full and dashed lines indicate the two U(1) models from which the system is
  constructed, with nodes located at the vertices of the two lattices.
  Scattering that couples the sub-systems is represented schematically by
  boxes (upper left plaquette), as parameterized by ${\bf T}_{\rm link}$,
  Eq.~(\ref{link_chiral}).  In one step of time evolution, flux propagates
  between successive points marked with filled circles, in the directions
  indicated by the arrows [Figure taken from
  \cite{bc03}].\label{fig:networkAIII} }
\end{center}
\end{figure}

We recall first the essential features of the U(1) network model for the
integer quantum Hall plateau transition (Chap.~\ref{sec:randomnetwork}), and
put them into the present context. The forms of the transfer matrix and of the
evolution operator follow from the properties of the elementary building
units, sketched in Fig.~\ref{buildingunits}. A wave function in this model
takes complex values $\psi_l$ on the links $l$ of the lattice illustrated by the
full lines of Fig.~\ref{fig:networkAIII}. A particle acquires a phase $\phi_l$
on traversing link $l$, so that in a stationary state amplitudes at either end
are related by
\begin{equation}
\label{link}
\psi' = e^{i\phi}\,\psi\,.
\end{equation}
In a similar way, stationary state amplitudes on the four links which meet at
a node are related by a $2\times 2$ transfer matrix
\begin{eqnarray}
\label{node}
\left( \begin{array}{c} \psi_1 \\  
\psi_2 \end{array} \right) =
\left(
\begin{array}{cc}
\cosh a \,& \sinh a \\
\sinh a \,& \cosh a
\end{array}
\right)
 \left( \begin{array}{c} \psi_3 \\ \psi_4 
\end{array} \right)\,,
\end{eqnarray}
where $a$ is real and all phase factors are associated with links.  This
equation may be rewritten in terms of a scattering matrix as
\begin{eqnarray}
\label{node2}
\left( \begin{array}{c} \psi_3 \\ \psi_2 \end{array} \right) =
\left(
\begin{array}{cc}
\cos \alpha \, & -\sin \alpha \\
\sin \alpha \, & \cos \alpha
\end{array}
\right)
 \left( \begin{array}{c} \psi_1 \\ 
\psi_4 \end{array} \right)\,,
\end{eqnarray}
with $\sin \alpha = \tanh a$. The transfer matrix that results from assembling
these units is described in detail in \cite{cc88}, and the time evolution
operator in \cite{km95,hc96} (cf. Sect.~\ref{sec:randomnetwork}).

Introducing a two dimensional internal space associated with the chiral
symmetry results in a doubling of the number of wave function components.
Thus, starting from two copies of the U(1) model the link amplitudes become
two-component complex numbers ${\psi}_l$. Instead of Eq.~(\ref{link}), the
scattering properties of a link now are characterized by a $2\times 2$
transfer matrix ${\bf T}_{\rm link}$, with
\begin{equation}
\label{chiral_link}
{\vv{\psi}}' = {\bf T}_{\rm link} {\vv{\psi}}.
\end{equation}
Requiring 
\begin{equation}
  \label{eq:classAIII}
 \sigma_x {\bf T}_{\rm link} \sigma_x = {\bf T}_{\rm link}, 
\end{equation}
the transfer matrix must have the form
\begin{eqnarray}
\label{link_chiral}
{\bf T}_{\rm link} = e^{i\phi}
\left(
\begin{array}{cc}
\cosh b & \sinh b \\
\sinh b & \cosh b
\end{array}
\right),
\end{eqnarray}
where $\phi$ is a real phase and $b$ is a real hyperbolic angle. It remains to
discuss scattering at nodes of the doubled system. We replace Eq.~(\ref{node})
by
\begin{eqnarray}
\label{node_chiral}
\left( \begin{array}{c} { \psi}_1 \\ { \psi}_2 \end{array} \right)
= {\bf 1}_{2} \otimes
\left(
\begin{array}{cc}
\cosh a \,& \sinh a \\
\sinh a \,& \cosh a
\end{array}
\right)
\left( \begin{array}{c} { \psi}_3 \\ 
{ \psi}_4 \end{array} \right)\,,
\end{eqnarray}
where ${\bf 1}_{2}$ is here the unit matrix in the two-component space
introduced on links. This choice is the most general one which is compatible
with chiral symmetry, since scattering within the two-component space may be
included in the link transfer matrices ${\bf T}_{\rm link}$.

By combining these elements into a two dimensional system, we arrive at the
model schematically shown in Fig.~\ref{fig:networkAIII}. The transfer matrix
for the system as a whole acts in the $[1,1]$ (or $[1,{\overline{1}}]$)
direction and may be written as a product of factors relating amplitudes on
successive slices of the system. Alternate factors in the product represent
links and nodes, and consist respectively of repeated versions of the $2\times
2$ and $4 \times 4$ blocks appearing in Eqs.~(\ref{link_chiral}) and
(\ref{node_chiral}). Disorder is introduced by assuming the phase $\phi$ in
Eq.~(\ref{link_chiral}) to be independently and uniformly distributed on the
links. The parameter $a$, characterizing scattering at nodes, is assumed to be
non-random (Chap.~\ref{sec:randomnetwork}).

Two possibilities have been considered for setting the value of the coupling
between the chiral subspaces. They are parameterized in terms of either the
hyperbolic angle in Eq.~(\ref{link_chiral}) or the compact angle $\beta$
related to $b$ by $\sin \beta =-\tanh b $. Either $\beta$ is assumed to be
uniformly distributed, or $b$ is assumed to be normally distributed with a
variance $g$. A disorder parameter $\gamma $ can be defined by
\begin{equation}
  \label{eq:disorderparameter}
  \sin \gamma=\tanh\sqrt{g}\,.
\end{equation}
It is guaranteed that the system is statistically invariant under $\pi/2$
rotations of the lattice. This constrains the node parameter $a$, exactly as
in the U(1) model: nodes lie on two distinct sub-lattices, and the node
parameter $a$ on one sub-lattice is related to the parameter $a'$ on the other
sub-lattice by the duality relation (cf. Eq.~(\ref{eq:duality}))
\begin{equation}
  \label{eq:dualityrelation}
 \sinh a \sinh a'=1\,. 
\end{equation}

As before (Sect.~\ref{subsec:wf_qcp}), the system can alternatively be
described using a time evolution operator instead of a transfer matrix. In
order to specify this unitary operator, which has the symmetry of a scattering
matrix, it is convenient first to consider the limit $b=0$. Then, the two
copies of the network model are uncoupled. One denotes the time evolution
operator for one copy by ${\bf U}$. Since from Eq.~(\ref{link_chiral}) link
phases are the same in both copies but propagation directions are opposite,
the evolution operator for the other copy is ${\bf U}^{\dagger}$. The
dimension of ${\bf U}$ is equal to the number of links in the system. It is
useful to define a diagonal matrix of the same dimension, with angles
$\beta_l$ for each link $l$ as diagonal entries: we use $\beta$ to denote this
matrix. The time-evolution operator for the system with chiral symmetry can
then be written
\begin{eqnarray}
\label{TEO}
{\bf U}_{\rm T}=
\left(
\begin{array}{cc}
{\bf U} \cos \beta \,& -{\bf U} \sin \beta \, {\bf U}^{\dagger} \\
\sin \beta  \,& \cos \beta \, {\bf U}^{\dagger}
\end{array}
\right).
\end{eqnarray}
One can straightforwardly verify that $\sigma_x {\bf U}_{\rm T} \sigma_x =
{\bf U}_{\rm T}^{-1}$ and that ${\bf U}_{\rm T}^{\dagger}={\bf U}_{\rm
  T}^{-1}$.

\begin{figure}[htbp]
\begin{center}
\includegraphics[width=8cm]{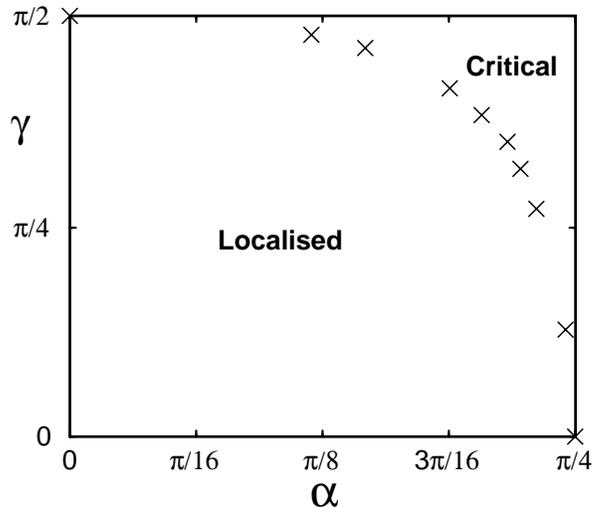} 
\caption{\label{fig:diagramAIII} 
  Phase diagram of the AIII network model in the $(\alpha,\gamma)$-plane
  [Figure taken from \cite{bc03}].}
\end{center}  
\end{figure}

For this model, two regimes of behaviors have been identified \cite{bc03}.
For small $\gamma$, there is a localized phase. For $\gamma$ close to $\pi/2$,
one finds a critical phase. By identifying the value of $\gamma$ which divides
the two regimes and studying its dependence on $\alpha$, one arrives at the
phase diagram shown in Fig.~\ref{fig:diagramAIII}.
The line $\alpha=\pi/4$ corresponds to the critical-metal states studied by
Gade and Wegner \cite{g1993}. On the line without hopping between the
sub-lattices, $\gamma$ has a single critical point at $\alpha=\pi/4$ which is
in the universality class of the Integer Quantum Hall Transition.

{\it Classes BDI and CII}

A model in class BDI can be obtained from one in class AIII simply by imposing
time reversal invariance as an additional symmetry. This condition is
conventionally written in the form ${\bf H}^*={\bf H}$, but for a discussion
based on scattering matrices it is more convenient to transform
\begin{equation}
  \label{eq:transformationtoQ}
{\bf H} \to {\bf QHQ}^{-1} \quad {\rm with} \quad 
{\bf Q}=e^{i\pi \sigma_x/4}\,.
\end{equation}
This transformation leaves the chiral symmetry relation $\sigma_x {\bf H}
\sigma_x=-{\bf H}$ intact. In the transformed basis one has
\begin{equation}
  \label{eq:symmetryequations}
  {\bf H}^*=-{\bf H}\,\qquad {\bf S}={\bf S}^* \,,\qquad 
{\bf T}^*={\bf T}\,. 
\end{equation}
To ensure a real time evolution operator, we restrict the link phases $\phi$
to the values $0$ and $\pi$. Choosing these values randomly, the BDI model
consists of two coupled copies of the class D models reviewed in the previous
section.  Alternatively, one could assume $\phi=0$ on all links, and
introduce disorder only via the chiral couplings $\beta$.

For class CII, Kramers degeneracy is a defining feature of the class and
requires the introduction of an additional two dimensional space arising
from the spin. The time reversal operation includes reversal of spin
direction. Defining ${\mathcal C} = i \sigma_y$, where $\sigma_y$ is a Pauli
matrix acting in the additional space, it is conventionally written in the
form
\begin{equation}
  \label{eq:kramersdegeneracy}
{\mathcal  C}{\bf H}^*{\mathcal C}^{-1}={\bf H}.
\end{equation}
As for the class BDI, it is again convenient to transform according to
Eq.~(\ref{eq:transformationtoQ}). This gives
\begin{equation}
\label{eq:bditransform}
 {\mathcal C}{\bf H}^*{\mathcal C}^{-1}=-{\bf H}\,, \quad 
{\mathcal C}{\bf S}^*{\mathcal C}^{-1}={\bf S}\,,\quad
{\mathcal C}{\bf T}^*{\mathcal C}^{-1}={\bf T} 
\end{equation}
as equivalent expressions of time-reversal invariance. Applying these ideas to
a network model, four channels propagate along a single link which,
generalizing Eq.~(\ref{link_chiral}), is transferred according to a $4\times
4$ transfer matrix ${\bf T}_{\rm link}$ with the generic form
\begin{eqnarray}
\label{link_chiral_su2}
{\bf T}_{\rm link}=
v\otimes \left(
\begin{array}{cc}
\cosh b  & \sinh b \\
\sinh b  & \cosh b
\end{array}
\right) \, ,
\end{eqnarray}
where $v$ is an SU(2) matrix and $b$ a real hyperbolic angle.  Adopting this
form, the time evolution operator for class CII has the structure given in
Eq.~(\ref{TEO}), but with ${\bf U}$ representing a network model of class C,
as reviewed in the preceding section. The links of a such a model of class C
carry two co-propagating channels, coming from two spin components, and the
evolution operator satisfies
\begin{equation}
  \label{eq:anotherone}
 {\mathcal C}{\bf U}^*{\mathcal C}^{-1}={\bf U}\,. 
\end{equation}

\section{Supersymmetry and Localization} 
\label{sec:susy}

In the present chapter, we comment on recent developments which open new
perspectives for better understanding of the field theoretical models that
have been developed during the past two decades for describing the quantum
Hall phase transition. This development is closely related to a mapping to a
chain of superspins starting from the random network model.

While the numerical and experimental evidences for the universality of the
quantum Hall transition appear to be striking, the quantum Hall critical point
has so far successfully escaped any attempts to be placed into a
classification of two dimensional phase transitions, based on the conformal
invariance at the critical point. Recently, it has been suggested that the
quantum Hall critical point may belong to a new class of critical points being
described by a supersymmetric conformal field theory
\cite{zirnconform,tsvelikconform}.  Much progress has been achieved towards
the analytical derivation and characterization of the quantum Hall transition.
Nevertheless, an analytical calculation of its critical exponents is still
missing.

Soon after the discovery of the Quantum Hall Effect a field theory has been
derived from the microscopic Hamiltonian of the random Landau model with
short-range disorder, the correlation length of the disorder potential,
$\ell_{\rm c}$, being much smaller than the magnetic length $\ell_B$
\cite{Pruisken,ef,weiden,wz88}. It has been shown to have two coupling
parameters $\sigma^0_{xx}$ and $\sigma^0_{xy}$, the longitudinal and Hall
conductance as defined on small length scales of the order of the elastic mean
free path $l$. This field theory is based on the theory of localization of
electrons in weakly disordered systems.
 
Subsequently, it has been shown rigorously that this field theory is indeed
critical at half integer Hall conductance parameters $\sigma^0_{xy}$
\cite{Pruisken,affleck,tsvelik}, and that it has a spectral gap to
fluctuations at other values of $\sigma^0_{xy}$. This indicates the
localization of the electron eigenstates of the random Landau model in the
tails of the Landau bands \cite{marikhin}. Since the longitudinal conductance
at the critical point $\sigma^*$ is known to be smaller than $1$ (in units of
$e^{2}/h$), the critical point is located, quite unfortunately, in the strong
coupling limit of the field theory. Thus, it is outside of the validity of the
available analytical methods which can be used to extract quantitative
information on the critical exponents.

\begin{figure}[htbp]
  \center 
\includegraphics[width=\textwidth]{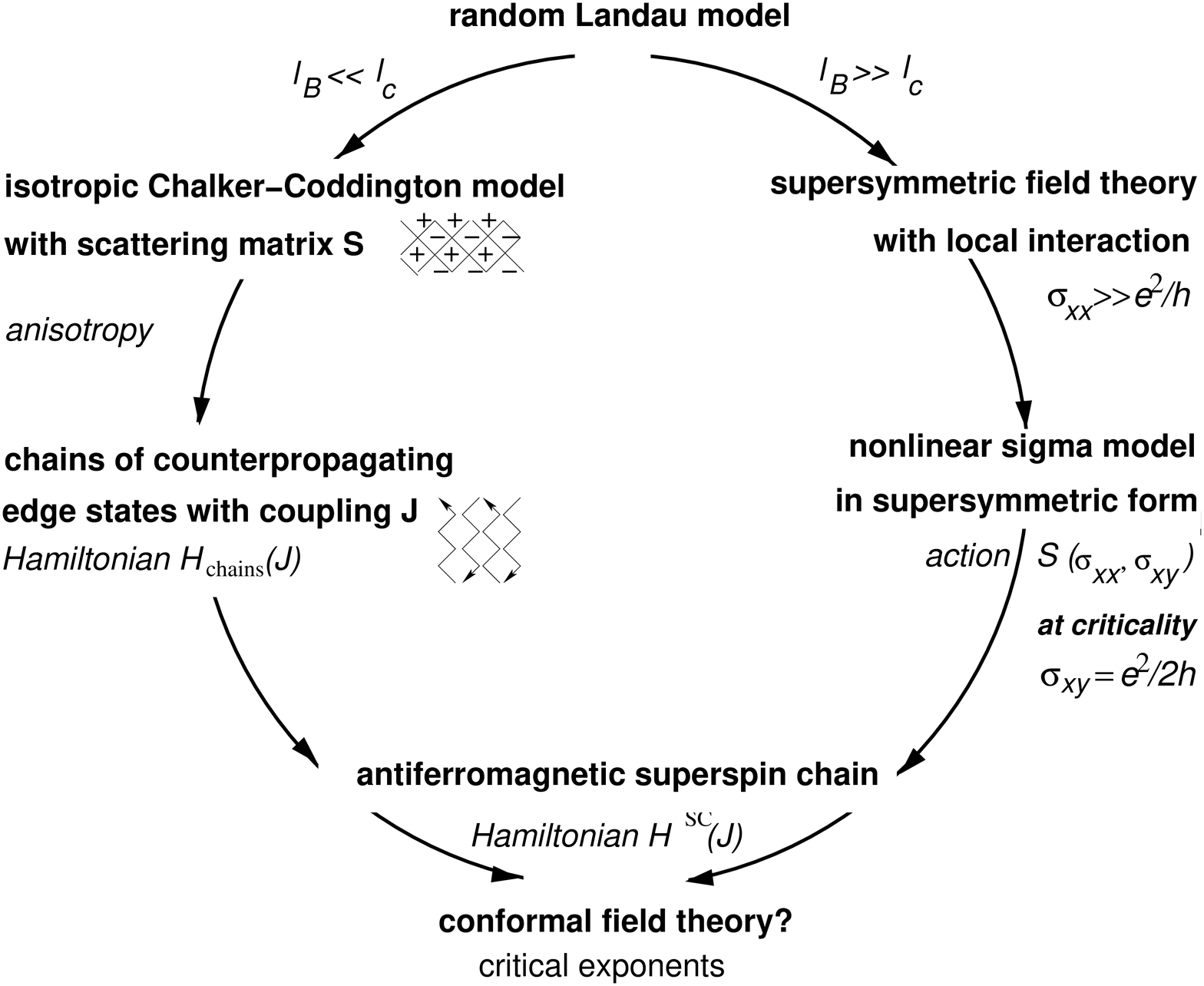} 
\caption{
  Sketch of the development of the analytical proofs for the equivalence of
  the critical theory for the random Landau model and the Chalker-Coddington
  network model.\label{overview}}
\end{figure}

Recently, an anisotropic version of the Chalker-Coddington model has been
mapped directly onto the Hamiltonian of a chain of antiferromagnetically
interacting superspins \cite{kondev,l94,zirn97,tsai}. It had been shown before
that the non-linear sigma model for short-range disorder at the critical
point, $\sigma_{xy} =e^{2}/2h$, and in the strong coupling limit, can also be
mapped onto the Hamiltonian of a chain of antiferromagnetically interacting
superspins \cite{zirnchain}. This provides strong analytical support for the
notion of universality of the quantum Hall transition as sketched
schematically in Fig.~\ref{overview}.
Thereby, the problem of the quantum Hall transition has been transformed to
the task of finding the ground state and the dispersion of the excitations of
the chain of antiferromagnetic superspins \cite{kondev,tsai,balents}.

The model of antiferromagnetically interacting superspins has been shown to be
critical \cite{tsai}. Numerically, the critical exponent $\nu$ has been
obtained from a finite length scaling of superspin chains, by means of the
density matrix renormalization group method, and found to be $\nu = 2.4(4)$
\cite{kondev}.

So far, no analytical information could be directly obtained on the critical
parameters, the localization exponent $\nu$, the MacKinnon-Kramer parameter
$\Lambda_{\rm c}$ of the scaling function, and the multi-critical exponents
$D(q)$.  However, starting from the model of a superspin chain, a class of
supersymmetric conformal field theories has been suggested to be relevant for
the quantum Hall transition, which ultimately should yield the critical
parameters of the quantum Hall transition \cite{zirnconform,tsvelikconform}.
In yet another development, the Bethe-Ansatz method has been applied to the
Hamiltonian of superspin chains. Still, it is not yet clear if that model is
integrable and a closed solution for its ground state and its excitations can
be obtained in this way \cite{gade,efs05}.

In the next section we will provide a derivation of the Hamiltonian of an
antiferromagnetic superspin chain starting from an anisotropic version of the
Chalker-Coddington model, following closely the original derivations by Lee
\cite{l94}, Zirnbauer \cite{zirn97}, and by Kondev and Marston \cite{kondev}.
In the second section, the derivation of Pruisken's non-linear sigma model
in supersymmetric form is summarized \cite{Pruisken,ef,weiden,wz88}. Finally,
the connection between that model and the superspin chain is discussed
following the work of Zirnbauer \cite{zirnconform,zirn97,zirnchain}.

\subsection{From the Network Model to the Antiferromagnetic Superspin Chain}

\begin{figure}
\begin{center}
\includegraphics[width=0.5\textwidth]{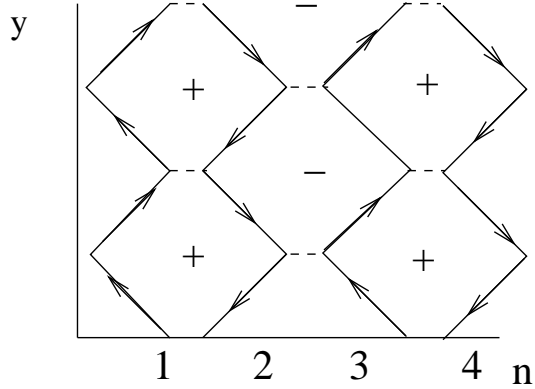} 
\caption{
\label{network_fig}
The network model with counter-propagating chiral fermions with random on-site
potential $w_{n,y}$ describing the propagation along edge states around Hall
droplets with random tunneling amplitudes $t_{n,y}$ between them (dashed).
}
\end{center}
\vspace{5mm}
\end{figure}

We start from a modification of the network model, recently proposed by Lee
\cite{l94}, where the random scattering matrices are replaced by a Hamiltonian
which describes a quasi-one dimensional array of counter-propagating edge
states in the presence of disorder (Fig.~\ref{network_fig}). The
$x$-coordinate is discretized, the integer index $n$ enumerates the discrete
vertices of the network model. The fact that the direction of propagation
alternates between sites $n$ will be shown in the following to result in an
antiferromagnetic interaction between superspins. A one dimensional array of
edge states, which propagate all in the same direction, has accordingly been
mapped to a ferromagnetic super spin chain \cite{balents}. This is an
effective model of the surface states of a layered quantum Hall system, a
so-called chiral metal (Chap.~\ref{sec:higherdimensions}).

In terms of electron creation $\hat{\psi}^\dagger$ and annihilation
$\hat{\psi}$ operators the Hamiltonian may be written as
\begin{eqnarray}
\label{hamilton}
H[\hat{\psi}^\dagger, \hat{\psi}]
&=& \sum_{n,y} \ [\hat{\psi}_{n,y}^\dagger 
((-1)^n i v_{\rm F}\partial_y + w_{n, y}) \hat{\psi}_{n,y}\nonumber\\ 
&&\qquad \qquad
- (t_{n,y} \hat{\psi}_{n,y}^\dagger \hat{\psi}_{n+1,y}
+ t_{n,y}^* \hat{\psi}_{n+1,y}^\dagger \hat{\psi}_{n,y})], 
\end{eqnarray}
where the sum is over the discrete vertices of the network model. This
Hamiltonian reflects the chiral nature and the linear dispersion of the edge
states, with alternating propagation forward and backward in the $y$-direction
with Fermi velocities $\pm v_{\rm F}$. The complex tunneling amplitudes
$t_{n,y}$ between the edge states are assumed to have random phases. These
represent the random Aharonov-Bohm phases of the electrons, accounting for the
fact that in a network of edge states, closed orbits vary randomly in size and
thereby encircle randomly varying magnetic fluxes. There are also random
on-site potentials $w_{n,y}$. These random terms are assumed to be Gaussian
distributed with zero means, and the variances
\begin{eqnarray}
\label{disorder}
\langle w_{n, y}\, w_{n^\prime ,y^\prime} \rangle 
&=& 2 U~ \delta_{n, n^\prime}\, \delta (y -y^\prime) \,,
\nonumber \\ 
&&\nonumber \\ 
\langle t_{n, y}^*\, t_{n^\prime, y^\prime} \rangle &=& 
2 J_{n}~ \delta_{n, n^\prime}\, \delta (y- y^\prime)\,,
\end{eqnarray}
with $J_n = J[1+(-1)^n R]$. The staggered modulation in the parameter $J_n$
allows to trace the differences in tunneling between counter-propagating
electrons in adjacent columns of plaquettes as depicted in
Fig.~\ref{network_fig}.  At $R = -1$, there are disconnected pairs of
counter-propagating edge states, such that electrons circulate only clockwise
around the $(+)$-plaquettes. In the opposite case, $R = 1$, electrons
circulate counterclockwise only around the $(-)$-plaquettes. At $R = 0$, there
is tunneling between all sites, and critical quantum percolation is expected,
corresponding to the transition between the quantum Hall plateaus.

Critical behavior consistent with this scenario was found from the disorder
averaged two-particle Green function describing transport corresponding to the
Hamiltonian Eq.~(\ref{hamilton}) by Wang and Lee \cite{Ziqiang}. From a
Monte-Carlo treatment of the replicated Hamiltonian they obtained a
correlation length exponent $\nu = 2.33(3)$, in good agreement with other
numerical simulations and experiments. It can furthermore be shown that the
Hamiltonian Eq.~(\ref{hamilton}) can be identified in the continuum limit with
the Hamiltonian of two dimensional Dirac fermions with random mass and random
vector fields \cite{l94}, which is in the universality class of the integer
quantum Hall transition, as discussed previously in
the chapter~\ref{sec:hamiltonians}.

It is well known \cite{aalr79,hikami,wegner,w89,elk}, that in weakly
disordered systems, $ k_{\rm F} l \gg 1$, one needs to go beyond lowest order
perturbation theory in the disorder potential in order to describe quantum
localization.  One also has to take full advantage of the symmetries occuring
in the calculation of correlation functions of disordered systems. This can be
traced back to the fact that the disorder averaged electron wave function
amplitude $\langle\psi ({\vv{r}},t)\rangle$ decays on length scales of the
order of $l$, since the random scattering phase shifts associated with the
scattering at the impurities are averaged out. This destroys the information
on quantum coherent multiple scattering, and thus on quantum localization. In
order to describe quantum localization, one needs to consider higher moments
of the wave function amplitudes such as the impurity averaged evolution of the
electron density $n({\vv{r}},t) = \langle |\psi ( {\vv{r}}, t)|^2\rangle$.

The time evolution of the electron amplitudes can be written in terms of the
retarded propagator $G^{\rm R}$\index{$G^{\rm R}$},
\begin{equation} 
\label{qaverage2} 
\psi ({\vv{r}}, t) = \int {\rm d}{\vv{r}'} G^{\rm R}({\vv{r}},t;{\vv{r}'}, t')
 \psi ({\vv{r}'}, t'),
\end{equation}
with $t>t'$. The electron density becomes
\begin{equation} 
\label{qaverage3}
\langle n ({\vv{r}}, t) \rangle =  
\int {\rm d}  {\vv{r}'} \int {\rm d}  {\vv{r}''}
\Gamma ({\vv{r}}, t;  {\vv{r}'},  {\vv{r}''}, t')  \psi ({\vv{r}'}, t')
 \psi^* ({\vv{r}''}, t')\,,
\end{equation}
where the quantum diffusion propagator 
\index{quantum diffusion propagator} 
is
given by
\begin{equation}
 \Gamma ({\vv{r}}, t;  {\vv{r}'},  {\vv{r}''}, t')
\index{\Gamma ({\vv{r}}, t;  {\vv{r}'},  {\vv{r}''}, t')}
 = \langle G^{\rm R}({\vv{r}}, t; {
  \vv{r}'}, t' )  G^{\rm A}({\vv{r}''}, t'; {
  \vv{r}}, t)\rangle .
\end{equation}
After performing a Fourier transformation from time $t-t'$ to energy $E$, a
non-perturbative averaging of products of retarded and advanced propagators,
$\langle G^{\rm R}(E)G^{\rm A}(E')\rangle$ is needed in order to obtain
information on quantum localization.

In a useful analogy to the study of spin systems, the field theoretical
approach contracts the information on localization into a theory of Goldstone
modes $Q$, arising from the global symmetry of rotations between the
functional integral representation of the retarded propagator $G^{\rm R}$
(''spin up'') and the advanced propagator $G^{\rm A}$ (''spin down''). The
field theory can either be formulated by means of the replica trick, where the
$N$ replicas are represented either by $N$ fermionic or $N$ bosonic fields,
yielding a bounded or unbounded symmetric space, respectively, on which the
modes $Q$ are defined \cite{Pruisken}.

Because of the necessity and the difficulty to perform the delicate limit
$N\rightarrow 0$ at the end of the calculation in the replica formulation, 
a more rigorous supersymmetric
field theory has been formulated. This technique represents the product of
Green functions $G^{\rm R}(E)G^{\rm A}(E')$ by functional integrals over two
fermionic and bosonic field components, composing a supersymmetric field
vector $\psi$.  The supersymmetric representation enables one to perform the
averaging over the disorder potential as a simple Gaussian integral
\cite{ef,weiden,wz88}.

Thus, in order to study the localization-delocalization transition in the
network model as described by the random Hamiltonian, Eq.~(\ref{hamilton}), we
consider the average of the quantum diffusion propagator
\begin{equation}
\label{quantumdiff}
K(1,2) = \langle G^{\rm R}_E(1,2)~ G^{\rm A}_{E'}(2,1)\rangle   \,,
\end{equation}
where 
\begin{equation}
\label{raGfun}
G^{\rm R,A}_E(1,2) = \langle 1| \frac{1}{E-H\pm i\eta} |2 \rangle \,,
\end{equation}
are the retarded and advanced Green functions. The parameter $\eta$ is a
positive infinitesimal number. We have introduced here the usual shorthand
notation for the coordinates with $1 := (n_1, y_1)$ and $2 := (n_2,
y_2)$. For non-zero disorder parameter $R$ we expect the quantum diffusion
propagator to decay exponentially
\begin{equation}
\label{shortrange}
K(1,2) \sim e^{-2r_{12}/\xi}\,,
\end{equation}
where $r_{12}$ is the distance between points 1 and 2. This defines the
localization length $\xi \gg l$, reflecting the finite extent of the electron
wave functions near energy $E$. At the critical point $R=0$, the localization
length is expected to diverge like 
\begin{equation}
  \label{eq:divergenceoflocalization}
 \xi \sim R^{-\nu}\,.    
\end{equation}
By introducing a pair of complex scalar fields 
\begin{equation}
\phi(n,y) = \left( 
\begin{array}{c} 
\phi_+(n,y) \\
\phi_-(n,y) 
\end{array} 
\right)\,,
\end{equation}
with $+$ and $-$ denoting the retarded and the advanced sector, respectively,
the advanced and retarded Green functions, Eq.~(\ref{quantumdiff}), can be
rewritten. One obtains for the two-particle propagator
\begin{equation}
\label{gauss}
\langle { K(1,2) } \rangle = \frac{1}{Z}\,\int {\rm D}[\phi] {\rm D}[\phi^*]
\,\phi_+(1) \phi^*_+(2) \phi_-(2) \phi^*_-(1)\,e^{-S[\phi,\phi^*]},
\end{equation}  
with the action
\begin{eqnarray}
S[\phi,\phi^*]&=&\sum_{n,y}\sum_{\alpha =\pm} 
\Big[i \alpha \left\{{\phi}_{\alpha}^{*}(n,y)
\left(-E_{\alpha}+(-1)^n iv_{\rm F}
\partial_y +w_{n,y}\right){\phi}_{\alpha}(n,y)
\right.
\nonumber\\ 
&&\nonumber\\ 
&&\left.\quad
-\left[t_{n,y} {\phi}_{\alpha}^* (n,y) {\phi}_{\alpha} (n+1,y) 
+t_{n,y}^*{\phi}_{\alpha}^* (n+1,y) {\phi}_{\alpha} (n,y) \right]\right\} 
\nonumber \\ 
&&\nonumber \\ 
&&
\qquad\qquad\qquad\qquad +\eta\phi_{\alpha}(n,y) \phi^*_{\alpha} (n,y) 
 \Big] \,.\label{eq:372}
\end{eqnarray}
Here, we have defined $E_{\alpha = +,-} = E,E'$. The choice of the signs
guarantees that the Gaussian integrals are convergent for $\eta > 0$. The
normalization factor
\begin{equation}
\label{norm}
{Z} = \int {\rm D}[\phi] {\rm D}[\phi^*]\,e^{-S[\phi,\phi^*]}\,,
\end{equation} 
is an inverse spectral determinant. Therefore, its inverse can be lifted to
the numerator by introducing another Gaussian integral, this time over {\em
  Grassmann fields} \cite{negele}
\begin{equation}
\label{normGrass}
\frac{1}{Z} = \int {\rm D}[\chi] {\rm D}[\bar{\chi}]\,, 
e^{-S[\chi, \bar{\chi}]} \,,
\end{equation}
with the anti-commuting fields
\begin{equation}
\label{Graspair}
\chi(n,y) = \left( \begin{array}{c} \chi_+(n,y) \\
                                        \chi_-(n,y) 
                       \end{array} \right)\,,
\end{equation}
that are the supersymmetric partners of the scalar bosonic fields. Using
Eqs.~(\ref{normGrass}) and~(\ref{gauss}) the correlation function,
Eq.~(\ref{quantumdiff}), can be written as a combined Gaussian integral over
scalar and Grassmann fields
\begin{eqnarray}
\label{Gaussfinal}
\langle K(1,2) \rangle
&=& 
\int {\rm D}[\phi] {\rm D}[\phi^*] {\rm D}[\chi] 
{\rm D}[\bar{\chi}]\,\phi_+(1) \phi^*_+(2)  \phi_-(2) \phi^*_-(1)
\nonumber\\
&&\nonumber\\
&&\qquad\qquad\qquad\qquad\qquad
\times 
e^{-(S[\phi,\phi^*]+S[\chi,\bar{\chi}])} \,.
\end{eqnarray}
Now the average over the disorder can be performed as a simple Gaussian
integral.
  
Before considering the full Hamiltonian, let us look first at the correlation
function of a single chiral edge state $n$, which is expected to behave
metallic for any disorder strength $U$ since its Hamiltonian contains no
backscattering, $J =0$.
   
Performing the ensemble averaging as the Gaussian integral over the random
potential $w(n,y)$ leads to the functional integral
\begin{equation}
{Z} = \int {\rm D}[\psi,\bar\psi]  
 \exp \int {\rm d}x \left[\bar\psi (\Lambda v_{\rm F}\partial_x - \eta +i
       \omega )
       \psi - U (\bar\psi \Lambda \psi)^2 \right] \,,
\end{equation} 
We have here introduced a four-component superspinor $\psi(x)$ with
components $\psi_{X}$ where $X = R{\rm B}$ (retarded Boson), $X = R{\rm F}$
(retarded fermion), $X = A {\rm B}$ (advanced Boson), and $X = A {\rm F}$
(advanced fermion). The $4\times 4$ supersymmetric $\Lambda$ is defined by
\begin{equation}
  \label{eq:lambda}
\bar\psi\Lambda\psi = \sum_{\sigma={\rm B,F}}
(\bar\psi_{R \sigma}\psi_{R \sigma} -
\bar\psi_{A \sigma}\psi_{A \sigma})\,.    
\end{equation}
Since the energies in the retarded and advanced sectors are different, we have
defined $\omega = E - E' \neq 0$.

\begin{figure}[htbp]
\begin{center}
\includegraphics[width=0.6\textwidth]{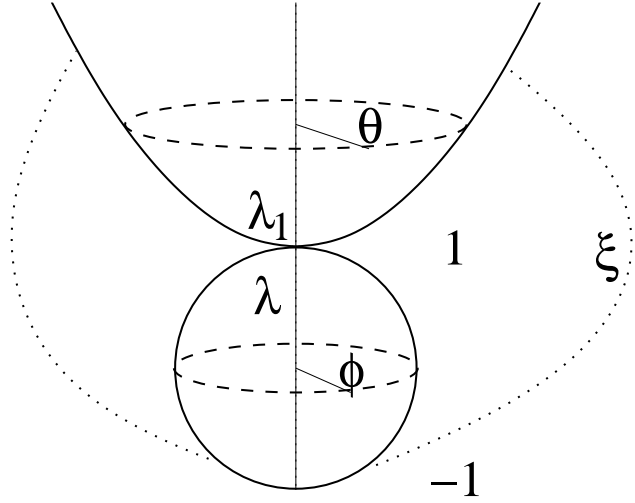} 
\caption{\label{supersymmetricspace}
  Schematic view of the supersymmetric space
  $\mathrm{U}(1,1|2)/(\mathrm{U}(1|1)\times \mathrm{U}(1|1))$
  on which the Hamiltonian of supersymmetric operators $Q$ is defined.  The
  point at which the compact sphere (corresponding to that part of $Q$ which
  arises from decoupling the fermionic $\psi_{\rm F}^4$-term) and the
  hyperbolic (arising from the decoupling of the bosonic $\psi_{\rm
    B}^4$-term) meet, is the classical point which yields the classical
  correlation function. Diagrammatic expansion around that point would miss
  the curvature, and the non-perturbative integral over the whole
  supersymmetric space is needed to describe quantum localization. The compact
  sphere is parameterized by $\Phi \in [0,2\pi)$ and $\lambda \in [-1,1 ]$,
  and the non-compact hyperbolic is parameterized by $\lambda_1 \in [ -1,
  \infty ]$ and $\theta \in [ 0, 2 \pi)$.  The unitary rotations between the
  sphere and the hyperbolic are parameterized by Grassmann variables $\xi$ as
  indicated by dotted lines.}  \vspace{5mm}
\end{center}
\end{figure}

One can decouple the interaction term $(\bar\psi\Lambda\psi)^2$ by introducing
a Hubbard-Stratonovitch field $Q$ coupling to $\bar\psi\psi\Lambda$ and then
integrate out $\psi$ and $\bar\psi$. The resulting effective action for $Q$ is
\begin{equation}
S[Q]= \int {\rm d}x \ {\rm STr} \left[ -\frac{U}{4v^2_{\rm F}} Q^2
 + \ln\left( \partial_x + \frac{i \omega- 
\eta}{v_{\rm F}} \Lambda  + \frac{U}{v^2_{\rm F}} Q\right) \right] .
\end{equation}
The next step is to simplify the $Q$-field functional integral by means of the
saddle-point approximation. As a result, $Q$ gets restricted to the non-linear
space $Q = g\Lambda g^{-1}$. It belongs to the supersymmetric space
$\mathrm{U}(1,1|2)/(\mathrm{U}(1|1)\times \mathrm{U}(1|1))$
which is sketched schematically in
Fig.~\ref{supersymmetricspace}.  Here the notation is reminiscent of the
notation for a 2-dimensional hyperbolic space $\mathrm{U}(1,1)$ and the group of $2$ by
$2$ unitary matrices U(2). Accordingly, the supermatrices $A \in \mathrm{U}(1,1\mid 2)$
have the form 
\begin{eqnarray}
  \label{eq:formofA}
{\bf A}=\left(\begin{array}{cc} {\bf A}^{BB} & {\bf A}^{BF} \\ 
{\bf A}^{FB} & {\bf A}^{FF}
\end{array}\right)  \,,
\end{eqnarray}
where ${\bf A}^{BB} \in \mathrm{U}(1,1)$ and ${\bf A}^{FF} \in \mathrm{U}(2)$ and ${\bf A}^{FB},
{\bf A}^{BF}$ are parameterized by Grassmann variables, and transform between
$\mathrm{U}(1,1)$ and $\mathrm{U}(2)$.
 
This step, being in general only an approximation, here becomes {\it exact} in
the limit $U \to\infty$.  By expanding $\ln \left[ \Lambda g^{-1}(\partial_x +
  A)g +U/v^2_{\rm F} \right]$ to linear order in $g^{-1}(\partial_x + A)g$,
one obtains the action of the Wess-Zumino functional $Z_{\rm WZ}[A]$,
\begin{equation}
{Z}_{\rm WZ}[\omega] = \int {\rm D}Q \ \exp{\left\{
\int {\rm d}x \
{\rm STr} \left[\frac{\Lambda}{2} g^{-1} (v_{\rm F}\partial_x + 
i\omega \Lambda )g\right]\right\}}.
\end{equation}
Higher orders are suppressed by powers of $v^2_{\rm F}/L_x U$, with $L_x$ the
system size.

By interpreting the integral over $x$ as an integral over time, one can
rewrite this in terms of a Hamiltonian
\begin{equation}
{Z}_{\rm WZ}[\omega] = {\rm STr} e^{- L_y H_g(\omega)},
\end{equation} 
where
\begin{equation}
 H_g(\omega) = i \omega   \,{\rm STr} \,\frac{\Lambda}{2} \frac{1}{g}
 \Lambda g .
\end{equation}
This is identical to the super Hamiltonian which one obtains from the unitary
ensemble of random matrices \cite{ef}, typical for a random metallic system
with extended states. Thus, as expected, we recover the result that a single
disordered chiral line supports only metallic states.

Let us next discuss the case of many counter-propagating chiral modes that are
coupled by hopping matrix elements between neighboring modes, with variance
$J$. This corresponds to the anisotropic Chalker-Coddington model. The
Gaussian random hopping matrix elements give rise to an additional term in the
Lagrangian,
\begin{eqnarray}
       \mathcal{L} \mapsto &&\mathcal{L} + 
2 J \sum_n \left( {\bar\psi}_n \Lambda \psi_{n+1} 
       \right) \left( {\bar\psi_{n+1}} \Lambda \psi_n \right) 
       \nonumber \\
       = &&\mathcal{L} + 
2 J \sum_n {\rm STr} \left( \psi_n \bar\psi_n \Lambda \right)
       \left( \psi_{n+1}\bar\psi_{n+1} \Lambda \right)\,.
\end{eqnarray}
This is obtained by averaging Eq.~(\ref{Gaussfinal}) with Eq.~(\ref{eq:372})
over $t$ using Eq.~(\ref{disorder}).  By using the bosonization rule $v_{\rm
  F}\psi{\bar\psi}\Lambda \to Q/2$ for $U \to\infty$, the additional term can
be cast into $(J/2v^2_{\rm F}) \sum_n {\rm STr} (Q_n Q_{n+1})$ . The condition
for the validity of this step is $U \gg J$.  As a result one obtains the
$Q$-field action
\begin{equation}
S[Q] = \int {\rm d}x \sum_n {\rm STr} \left[ 
       (-1)^n {\Lambda\over 2} g_n^{-1}\partial_x g_n^{\vphantom{-1}}
       + {J\over 2v^2_{\rm F}} Q_n Q_{n+1} \right] .
\end{equation}
This is the action \cite{balents,fradkin} of the coherent-state path integral
for a quantum superspin Hamiltonian. The resulting effective supersymmetric
(SUSY) Hamiltonian describes interacting spin-up and spin-down fermions
$c_\sigma$ and bosons $b_\sigma$. The two spin species formally correspond to
the retarded and advanced Green functions introduced above
\cite{kondev,tsai,tm99},
\begin{eqnarray}
H &=& \sum_{j=0}^{L-2}~ J_j~ 
\Big[ \lambda_H \sum_{a=1}^4~  g_a~ S^a_j~ S^a_{j+1}
+ \sum_{a=5}^8 g_a~ S^a_j ~ S^a_{j+1}\nonumber \\
&&\nonumber \\
&&\qquad + \lambda_H  (-1)^j~ \sum_{a=9}^{16}~ g_a~ S^a_j~ S^a_{j+1} \Big]   
 + \eta \sum_{j=0}^{L-1} \Big[ S^1_j + S^2_j + S^5_j + S^6_j \Big]\ . 
\label{H-SUSY}
\end{eqnarray}
The signs $g_a$ are given by
\begin{eqnarray}
g_a = \left\{ \begin{array}{ll} 1 \!\qquad{\rm for}\,\, 
a = 1, 2, 10, 12, 14, 16\\
-1 \quad {\rm for}\,\,
a = 3, \ldots, 9, 11, 13, 15\ .
\end{array}\right.
\label{signs}
\end{eqnarray}
 Equation~(\ref{H-SUSY}) contains 16 spin operators, the components of
a $4 \times 4$ superspin matrix,
\begin{eqnarray}
\begin{array}{l} 
S^1 := b^\dagger_\uparrow b_\uparrow + 1/2 \\ \\
S^2 := b^\dagger_\downarrow b_\downarrow + 1/2 \\ \\
S^3 := b^\dagger_\uparrow b^\dagger_\downarrow \\ \\
S^4 := b_\downarrow b_\uparrow \\ \\
\end{array} \ \ \  
\begin{array}{l}  
S^{5} := c^\dagger_\uparrow c_\uparrow - 1/2 \\ \\
S^{6} := c^\dagger_\downarrow c_\downarrow - 1/2 \\ \\
S^{7} := c^\dagger_\uparrow c^\dagger_\downarrow \\ \\
S^{8} := c_\downarrow c_\uparrow \\ \\
\end{array} \ \ \  
\begin{array}{l}  
S^9 := c^\dagger_\downarrow b_\downarrow \\ \\ 
S^{10} := c^\dagger_\uparrow b_\uparrow \\ \\
S^{11} := b^\dagger_\downarrow c_\downarrow \\ \\
S^{12} := b^\dagger_\uparrow c_\uparrow \\ \\
\end{array} \ \ \  
\begin{array}{l}   
S^{13} := b_\downarrow c_\uparrow \\ \\ 
S^{14} := b_\uparrow c_\downarrow \\ \\
S^{15} := b^\dagger_\downarrow c^\dagger_\uparrow \\ \\
S^{16} := b^\dagger_\uparrow c^\dagger_\downarrow\ . \\ \\
\end{array} 
\label{superspins}
\end{eqnarray}
The boson-valued operators $S^1, \ldots, S^8$ constitute the symmetric sector
of the Hamiltonian. The fermion-valued operators $S^9, \ldots, S^{16}$ form
the anti-symmetric sector. Despite that $H$ is non-Hermitian, it only has
real-valued eigenvalues. For $\lambda_H=0$ one has only fermionic spin
operators. One does arrive at this Fermion model directly from the
Chalker-Coddington network model when calculating the disorder averaged auto
correlation function of spectral determinants \cite{zirnconform,tsvelik}.
Another derivation via the Landauer conductance formula was recently reported
in Ref. \cite{s04}. This yields the antiferromagnetic spin-1/2 Heisenberg
chain, which is critical \cite{tm2000}. The crossover to the integrable
superspin chain of the spin Quantum Hall Effect \cite{glr1999,smf1999}, which
is critical in the universality class of the classical percolation model, can be
studied by freezing out 8 of the 16 spin components \cite{tsaithesis2001}.
This raises the question whether or not the critical parameters vary
continuously, as for the crossover from the antiferromagnetic Heisenberg spin
chain to the XY model \cite{a1990,tm2000}.

The Hamiltonian commutes with four fermion-valued supersymmetry generators,
\begin{equation}
  \label{eq:fermionsusy}
[H,\, Q_{1 \sigma}] = [H,\,Q_{2 \sigma}] = 0  
\end{equation}
with the supersymmetric charges
\begin{eqnarray}
Q_{1 \sigma} &:=& \sum_j \bigg{[} b^\dagger_{j \sigma} c_{j \sigma}
- (-1)^j c^\dagger_{j \sigma} b_{j \sigma} \bigg{]}\,,
\nonumber \\
Q_{2 \sigma} &:=& \sum_j \bigg{[} (-1)^j b^\dagger_{j \sigma} 
c_{j \sigma} + c^\dagger_{j \sigma} b_{j \sigma} \bigg{]}\, .
\label{SUSY-charges}
\end{eqnarray}
One can see that the supersymmetric Hamiltonian must have a unique,
zero-energy ground state. All excited states appear in quartets or larger
multiples of 4, half with odd total fermion content. These cancel out in the
partition function by virtue of the super-trace, yielding the correct value of
1 that we have encountered above when introducing the functional integral over
Grassmann variables to cancel the normalization factor of the bosonic integral
\begin{equation}
Z = {\rm STr} e^{-\beta H} := {\rm Tr} (-1)^{N_c} e^{-\beta H} = 1\,. 
\label{partition}
\end{equation}
Here, $N_c$ is the total number of fermions. The ground state of this
supersymmetric non-Hermitian antiferromagnetic Hamiltonian is very
complicated. The ground state of the {\it Hermitian} supersymmetric
ferromagnet that describes a chiral metal with all edge states propagating in
the same direction is simply the vacuum state \cite{balents}.

The Lieb-Schultz-Mattis theorem proves that for half-odd-integer spin
antiferromagnets on a periodic chain of length $L$ sites ($\vec{S}_L :=
\vec{S}_0$) either the ground state is degenerate or there are gapless spin
excitations in the thermodynamic limit $L \rightarrow \infty$ \cite{lsm61}.
In the supersymmetric problem, Marston and Tsai were able to make even a
stronger statement \cite{tsai,tm99}, because its ground state is unique by
supersymmetry. For $\eta > 0$ they showed that low-energy excitations are
gapless in the thermodynamic limit. This proves that the antiferromagnetic
superspin chain is critical.

In the following section the field theory of the quantum Hall critical point
is approached starting from a model of electrons in uncorrelated disorder in a
strong magnetic field. The resulting supersymmetric field theory, with the
action of a non-linear sigma model was used as a basis of the two parameter
scaling of the Quantum Hall Effect.  It has recently been shown that it can in
a long wave length limit be mapped onto the Hamiltonian of the superspin chain,
Eq.~(\ref{H-SUSY}). This, together with the derivation of the spin chain from
the opposite limit of a model with long-range potential as sketched above,
provides very strong support to the surmise that the Chalker-Coddington model
is a good model for the universal quantum Hall transition \cite{zirn97}.

\subsection{From the Landau Model to Pruisken's Non-linear Sigma Model}

The Hamiltonian of noninteracting electrons in a magnetic field in the
presence of uncorrelated disorder is (Chap.~\ref{sec:introduction})
\begin{equation}
\label{randomlandau}
H= \frac{1}{2m}\left({\vv{p}
}+ e {\vv{A}} \right)^2 + V({\vv{r}})  + V_0({\vv{r}})\,.
\end{equation}
Here, $V({\vv{r}})$ is assumed to be a Gaussian distributed random function
with a distribution
\begin{equation} 
 P([V] ) = \exp \left[-\int\frac{{\rm d} {\vv{r}}}{\Omega} 
\frac{{\rm d} {\vv{r}'}}{\Omega}
 J ( {\vv{r}} - {\vv{r}'}) 
 V ( {\vv{r}} ) V( {\vv{r}'}) \right]\,.
\end{equation}  
where $\Omega$ is the volume of the system. Impurity averaging is thus given
by $\langle...\rangle_V = \int \prod_{{\bf r}} {\rm d} [V] P([V]) ... $ . We
assume
\begin{equation}
J({\vv{r}} - {\vv{r}'}) = {\Omega} \Delta
\frac{\hbar}{\tau}\, \delta ({\vv{r}} - {\vv{r}'})\,,
\end{equation}
for uncorrelated impurities, where $1/\tau$ is the elastic scattering rate and
$\Delta=1/(\rho {\Omega})$ the mean level spacing of the mesoscopic sample with
volume $\Omega$. It is related to the variance of the disorder potential
$V({\vv{r}})$ according to $ W^2 = \Delta \hbar/2 \pi \tau$.  The function
$V_0({\vv{r}})$ is the electrostatic confinement potential defining the width
of the wire $L$.

In order to describe localization, we consider again the correlation function
$K(1,2)$ studied in the previous chapter for the Chalker-Coddington model, but
using this time the Hamiltonian Eq.~(\ref{randomlandau}).  Formulating the
supersymmetric field theory by representing the product of Green functions
$G^{\rm R}(E) G^{\rm A}(E')$ by functional integrals over two fermionic and
bosonic field components, composing a supersymmetric field vector $\psi$, as
in the previous section, the averaging over the disorder potential can again
be performed as a simple Gaussian integral \cite{ef,weiden,wz88}.  As a result
of the averaging one obtains a locally interacting theory of the fields $\psi$
containing an interaction term $\propto \psi^4$, where the {\it interaction
  strength} is proportional to the variance of the disorder potential $W^2$.
This term is next decoupled by introducing another Gaussian integral over
$Q-$matrices.  Clearly, the field $Q$ should not be a scalar, otherwise we
would simply reintroduce the Gaussian integral over the random potential $V$.
Rather, in order to be able to describe the physics of localization, the field
$Q$ should capture the full symmetry of the functional integral representation
of the correlation function. Therefore, the Gaussian integral is chosen to be
over a 4$\times$4 matrix $Q$ which itself is an element of the symmetric space
defined by the matrices ${\bf A}$ that leave the functional integral invariant
under the transformation $\psi \rightarrow {\bf A} \psi$. In the
supersymmetric formulation, this matrix consists of two blocks of 2$\times$2
matrices whose parameter space consists of a compact (bounded) and a
non-compact (unbounded) sector, Eq.~(\ref{eq:formofA})
(Fig.~\ref{supersymmetricspace}). The off-diagonal blocks, so to say the
rotations between the compact and the non-compact sector, are then found to be
parameterized by Grassmann (fermionic) variables. Now, the spatial variations
of $Q$ are governed by the action
\begin{equation} 
\label{exactfree}
S[Q] = 
 \frac{\pi\hbar}{4\Delta \tau}
  \int \frac{{\rm d} {\vv{r}}}{ L^2} {\rm Tr}\, Q ^{2}({\vv{r}}) 
 + \frac{1}{2} \int  {\rm d} {\vv{r}}\,
 \langle{\vv{r}} |{\rm Tr}\, \ln
 G ( \hat{\vv{r}}, \hat{\vv{p}} )| {\vv{r}} \rangle
\end{equation} 
where
\begin{equation}
G^{-1}(\hat{\vv{r}},\hat{\vv{p}} ) =
\frac{(\omega + i \eta) \Lambda_3}{2} - 
\frac{(\hat{\vv{p}}+e \vv{A} )^2}{2 m^{*}} - V_0(\hat{\vv{r}})
+\frac{i\hbar}{2 \tau}
Q (\hat{\vv{r}})\,.
\end{equation}
The 4$\times$4 matrix $\Lambda_3$ is the diagonal Pauli matrix in the
sub-basis of the retarded and advanced propagators, $\eta > 0$ and $\omega =
E-E'$ break the symmetry between the retarded and advanced sector.

It turns out that the physics of diffusion and localization, which arises on
length scales much larger than the elastic mean free path $l$, is governed by
the action of the long wavelength modes of $Q$. Thus, one can simplify and
proceed with the analysis by expanding around a homogeneous solution of the
saddle point equation, $\delta S = 0$.  For $\omega =0$, this is
\begin{equation} 
\label{saddle}
 Q =  \frac{i}{\pi \rho} \langle {\vv{r}}|
\left[E - H_0 - V_0({\vv{r}}) + \frac{i}{2 \tau} \,Q \right]^{-1}
| {\vv{r}} \rangle. 
\end{equation}  
This is solved by $ Q_0 = \Lambda_3 {\bf P} $, which corresponds to the self
consistent Born approximation for the self energy of the impurity averaged
Green function. At $\omega =0$, rotations $ {\bf U}$ which leave the action
invariant yield the complete manifold of saddle point solutions as $ Q =
\bar{\bf U} \Lambda_3 {\bf P U}$, where $ {\bf U} \bar{{\bf U}} = 1$ where
$\bar{{\bf U}}$ is the supersymmetric Hermitian conjugate.

The modes which leave $\Lambda_3$ invariant can be factorized out, leaving the
saddle point solutions in this supersymmetric theory to be elements of the
semi-simple supersymmetric space
$\mathrm{Gl}(2|2)/(\mathrm{Gl}(1|1) \times \mathrm{Gl} (1|1)) $
\cite{weiden,wz88}. Here the notation is chosen in analogy to $\mathrm{Gl} (n)$,
denoting invertible $n$ by $n$ matrices, that is ${\rm det} A \neq 0$ for all
$A \in \mathrm{Gl}(n)$.  Thus, $\mathrm{Gl}(n\mid n)$ denote invertible $2n$ by $2n$
supermatrices 
\begin{eqnarray}
  \label{eq:invertible}
 {\bf A}=\left(
\begin{array}{cc} 
  {\bf A}^{FF} & {\bf A}^{FB} \\ 
  {\bf A}^{BF} & {\bf A}^{BB}
  \end{array}
\right)  \,,
\end{eqnarray}
with 
\begin{equation}
{\rm Sdet} {\bf A} = {\rm det} ({\bf A}^{FF} - 
{\bf A}^{FB} {\bf A}^{BB -1}
{\bf A}^{BF} ) 
{\rm det} {\bf A}^{BB -1} \neq 0\,,
\end{equation}
where ${\rm Sdet}$ is the superdeterminant. See \cite{z96} for a more detailed
definition. In the semi-simple space the subgroup
$\mathrm{Gl}(1|1) \times \mathrm{Gl} (1|1)$
consisting of matrices
\begin{eqnarray}
  \label{eq:h}
{\bf h} = \left(\begin{array}{cc} {\bf h}^{11} & 0 \\ 0 & {\bf h}^{22}
 \end{array}\right)\,,
\end{eqnarray} 
with ${\bf h}^{11}, {\bf h}^{22} \in \mathrm{Gl} (1|1)$, are factorized out.
 
In addition to these gapless modes there are massive longitudinal modes with
$Q^2 \neq 1$ which only change the short distance physics, and not the physics
of localization. They can be integrated out \cite{Pruisken,ef}. Thus, the
partition function reduces to a functional integral over the transverse modes
${\bf U}$.

The action at finite frequency $\omega$ and slow spatial fluctuations of $Q$
around the saddle point solution can be found by an expansion of the action
$S$. Inserting $ Q = \bar{{\bf U}} \Lambda_3 {\bf P} {\bf U}$
into Eq.~(\ref{exactfree}) and
performing the cyclic permutation of ${\bf U}$ under the trace ${\rm Tr}$ allows a
simple expansion to first order in the energy difference $\omega$ and to
second order in the commutator
${\bf U} [ H_0, \bar{{\bf U}} ]$ \cite{Pruisken}. The first
order term in ${\bf U} [ H_0, \bar{{\bf U}} ]$ is proportional to the local current.  It
is found to be finite only at the edge of the wire in a strong magnetic field,
due to the chiral edge currents. It can be rewritten as
\begin{equation}
S_{xy II} = - \frac{1}{8} 
\int {\rm d}x \,{\rm d}y  \,
\frac{\sigma^{0 II}_{xy}({\vv{r}})}{e^2/h}\,{\rm STr} (Q 
\partial_x Q \partial_y Q)  \,,
\end{equation} 
where the pre-factor is the non-dissipative term in the Hall conductivity in
self-consistent Born approximation \cite{ef}
\begin{equation}
\sigma^{0 II}_{xy} ({\vv{r}}) = -\frac{1}{\pi} 
\frac{ \hbar e^2}{m} \langle{\vv{r}}| ( x \pi_y - y \pi_x ) 
 {\rm Im} G^{\rm R}(E) |  {\vv{r}} \rangle, 
\end{equation}
where ${\bf \pi } =(\hbar/i) {\bf \nabla} +e {\vv{A}}$.  This field theory
has now the advantage that one can treat the physics on different length
scales separately: the physics of diffusion and localization is governed by
the action of spatial variations of ${\bf U}$ on length scales larger than the mean
free path $l$.  That is why this field theory is often called diffusive
non-linear sigma model.

The physics on smaller length scales is included in the coupling parameters of
the theory, which is identified in the above derivation as correlation
functions of Green functions in self consistent Born approximation, being
related to the conductivity by the Kubo-Greenwood formula,
\begin{eqnarray}\label{kubo}
\sigma^0_{\alpha \beta } ( \omega,  {\vv{r}} )& =& 
\frac{\hbar e^2}{\pi  m^2}
\langle {\vv{r}} 
| \pi_{\alpha}  G^{\rm R}_{0}(E) {\bf  \pi_{\beta} }   
G^{\rm A}_{0}(E+\omega) | {\vv{r}} \rangle. 
\end{eqnarray} 
The remaining averaged correlators, involve products $ G^{\rm R}_{0}(E) G^{\rm
  R}_{0}(E+\omega)$ and $ G^{\rm A}_{0}(E) G^{\rm A}_{0}(E+\omega)$ and are
therefore by a factor $\hbar/\tau E$ smaller than the conductivity, and can be
disregarded for small disorder. Using the Kubo formula (\ref{kubo}), the
action of $Q$ simplifies to
\begin{eqnarray} \label{free2db}
S &=& \frac{h }{16 e^2}  
\int {\rm d} {\vv{r}} \sum_{i=x,y}  \sigma^0_{ii} (\omega=0,{\vv{r}})
 \,{\rm Tr} \left[ ( { \nabla_i} Q({\vv{r}} ))^2 \right] 
\nonumber \\ 
&&\qquad\qquad\qquad -  \frac{h}{8 e^2}   \int {\rm d} {\vv{r}}  
\sigma^0_{xy} (\omega=0,{\vv{r}}) 
\,{\rm Tr} \left[ Q \partial_x Q \partial_y Q \right]\,,
\end{eqnarray}
where $ \sigma^0_{xy} (\omega=0,{\vv{r}}) = \sigma^I_{xy} (\omega=0,{\vv{r}})
+ \sigma^{II}_{xy} (\omega=0,{\vv{r}})$ and $ \sigma^I_{xy} (\omega=0)$ is the
dissipative part of the Hall conductivity in self consistent Born
approximation Eq.~(\ref{kubo}).

The first term in this action yields localization in two dimensional electron
systems, signaled by the presence of a gap in the field theory. The second
term could not be obtained by any order in perturbation theory. It is of
topological nature.

In two dimensions, and for a homogeneous Hall conductance it can be shown that
this term can take only discrete purely imaginary values,
\begin{eqnarray} 
\label{free2dbtop}
S_{\rm Top} &=&  2 \pi i   \frac{h}{ e^2}    
\sigma^0_{xy} n,
\end{eqnarray}
where the integers $ n $ count how often the field $Q({\vv{r}})$ is winding
around its symmetric space as it varies spatially in two dimensions.  Thus,
disregarding the spatial variation of the coupling functions $\sigma_{ij}
({\vv{r}})$ in Eq.~(\ref{free2dbtop}), and assuming isotropy, one finds in the
two dimensional limit that there are instantons with nonzero topological
charge $q$, which are identical to the skyrmions of the compact $O(3)$
non-linear sigma model, as obtained form the compact part of the supersymmetric
non-linear sigma model \cite{Pruisken,ef}.  Their action is given by
\begin{equation}
F_{q} = 2 \pi|q| \sigma_{xx} + 2 \pi i q \sigma_{x y}, 
\end{equation}
where $\sigma_{xx} = \sigma_{yy}$ and $\sigma_{x y}$ are the spatially
averaged conductivities.

\begin{figure}[htbp]
\begin{center}
    \epsfxsize = 0.9\textwidth \epsffile{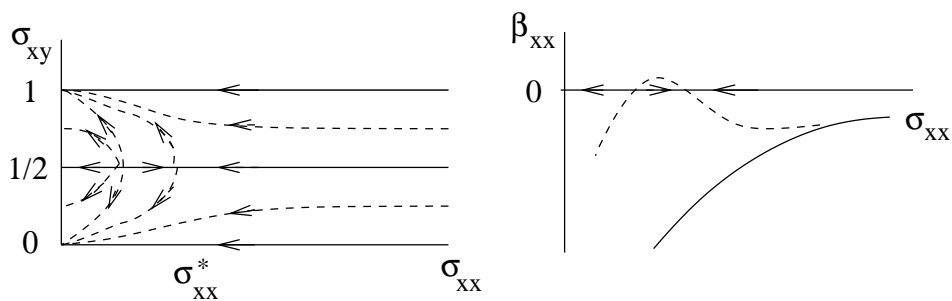}  
\caption[]{ The conjectured  two parameter flow diagram of 
  the integer Quantum Hall Effect (left), and the corresponding beta function,
  $\beta_{xx} = {\rm d} \ln \sigma_{xx} /{\rm d} \ln L$ (right) at
  $\sigma_{yx} =-\sigma_{xy}= 1/2$ (dashed line) and at $\sigma_{xy} = 0$ (full line, units
  of conductivities $e^{2}/h$). \label{fig:flow}}
\end{center}
\end{figure}

Now, we can repeat the derivation of the scaling function, by integrating out
Gaussian fluctuations around these instantons. It is clear, however that the
contribution from instantons with $q \neq 0$ is negligible, as long as
$\sigma_{xx} > e^{2}/h$.  Within the validity of the $1/g$-expansion one does
not find a sizable influence of the topological term on the scaling function
$\Lambda = \xi/L_y$. Still, the tendency is seen that at
$\sigma_{xy}=-e^{2}/2h$ the renormalization of the longitudinal conductance is
slowed down and one may conclude from this observation the two parameter
scaling diagram with a critical state of finite conductance $0 <\sigma_{xx}^*
< e^{2}/h$, which is an attractive fixed point for $\sigma_{xy} = -e^{2}/2h$ as
shown in Fig.~\ref{fig:flow} \cite{Pruisken,khm}. Considering the beta
function of the N-th fermionic replica one notes \cite{Pruisken,affleck} that
only the last one $N=1$, becomes critical and touches $0$ like $\beta \sim
-(\sigma_{xx}-\sigma_{xx}^*)^2$, which is the beta function of the
antiferromagnetic Heisenberg chain \cite{tsvelik}. From that, one can
conjecture that for $N\rightarrow 0$ the maximum of the beta function moves to
positive values, which yields the form drawn in Fig.~\ref{fig:flow} with
another, repulsive fixed point at smaller values of the conductance
$\sigma_{xx}$, below which there is a flow to an insulating phase for all
values of $\sigma_{xy}$. One can also come to that conclusion noting that,
when freezing out the bosonic degrees of freedom from the supersymmetric 
nonlinear sigma model
at $\sigma_{xy} =-1/2$, one arrives again at the beta function of the
antiferromagnetic Heisenberg chain, given above. It remains still to be shown
explicitly that the beta function evolves indeed continuously from that beta
function, as one adds the bosonic degrees of freedom, and to see if it results
in the beta function drawn in Fig.~\ref{fig:flow}.

Subsequently, it has been argued by other means that this field theory is
indeed critical at half integer Hall conductance parameters $\sigma^0_{xy}$
\cite{Pruisken,affleck,tsvelik}, and that it has a spectral gap to
fluctuations at other values of $\sigma^0_{xy}$. This indicates the
localization of the electron eigenstates of the random Landau model in the
tails of the Landau bands \cite{marikhin}. Since the longitudinal conductance
at the critical point $\sigma_{xx}^*$ is known to be smaller than $1$, the
critical point is located in the strong coupling limit of the field theory.
Thus, it is outside of the validity of available analytical methods which can
be used to extract quantitative information on the critical exponents.
Furthermore, it is seen explicitly that in order that the instanton solutions
with nonzero topological charge do exist the system must exceed the
noncritical localization length $\xi_{\rm 2D unit} \sim \exp (\pi^2 g^2)$,
where $g = h\sigma_{xx}/e^2$, when the assumption of uniform coupling
parameters $\sigma_{ij}$ is made. Accordingly, it has been shown that the
Hamiltonian of a chain of antiferromagnetically interacting superspins can be
derived from the non-linear sigma model for short-ranged disorder at the
critical point $\sigma_{xy} =-e^{2}/2h$ on length scales larger than $\xi_{\rm
  2D unit}$ \cite{zirnchain}.  At criticality, $\sigma_{xy}= -e^{2}/2h$, and at
strong coupling, $\sigma_{xx} < e^{2}/h$, Zirnbauer \cite{zirnchain} showed by
discretization, following Shankar and Read \cite{sr90}, that the
supersymmetric nonlinear sigma model can be mapped on the chain of
antiferromagnetic superspin chains in the low energy, long wavelength limit,
meaning that the lattice spacing of the super spin chain is on the order of
the noncritical localization length $\xi_{\rm 2D unit}$.
  
So far, no analytical information has been obtained for the critical
parameters, such as the localization exponent, $\nu$ and the critical value
$\Lambda_{\rm c}$.  However, building on the model of a superspin chain,
supersymmetric conformal field theories have been suggested, which ultimately
are supposed to yield the critical parameters of the quantum Hall transition
\cite{zirnconform,tsvelikconform,kz01}.

Restricting this theory to quasi-one dimension, by assuming a finite width
$L_y$ of the quantum Hall bar, of the order of the unitary noncritical
localization length $\xi_{\rm 2D unit} = l \exp ( \pi^2 \sigma_{xx}^2)$, which
serves as the ultraviolet cutoff of the conformal field theory, one finds that
the critical value of the scaling function $\Lambda_{\rm c}\approx 1.2$, the
ratio of the localization length in a quantum Hall wire and its finite width
$L_y$ when the energy is in the center of the Landau band, see Section
~\ref{sec:localization}, is fixed by the eigenvalues of the Laplace-Beltrami
operator of this supersymmetric conformal field theory
\cite{jmz99,kz01,tsvelikconform}. This is a characteristic invariant of the
theory, arising from the conformal symmetry, just as the quantization of
angular momentum arises from the rotational symmetry of a Hamiltonian.
Furthermore, based on the properties of this constrained class of
supersymmetric conformal field theories, it has been predicted that the
distribution function of local wave function amplitudes is very broadly,
namely log-normally, distributed. This prediction has recently been confirmed
by high accuracy numerical calculations \cite{emm01}.

The quest to derive the critical exponent of the localization length at the
quantum Hall transition from a critical supersymmetric theory has thus
recently gained much progress, by mapping both the Chalker-Coddington model
and the random Landau model on the Hamiltonian of the superspin chain,
Eq.~(\ref{H-SUSY}), which has been shown to be critical at $R=0$
\cite{kondev,tsai,tm99} as defined in Eq.~(\ref{disorder}). Thus, it has been
proven to be a good starting point to continue in the quest for an analytical
derivation of the critical exponent $\nu$ and the multi-critical exponents at
the quantum Hall critical point.

\section{Extension to Higher Dimensions}
\label{sec:higherdimensions}

A natural extension of the two dimensional quantum Hall system is to consider
layers stacked in parallel, where electrons can tunnel from one layer to
another. Such layered systems can be fabricated in experiments and the
magneto-transport properties have been investigated
\cite{segwb86,segbe87,mjl98,kekki04}.

\subsection{Double Layer Network Model}
\label{sec:8.1}

The simplest case is a double layer system. In this case, there appear two
energies where electron states are delocalized \cite{sm96,ook92,gr97b},
instead of a single energy for delocalized states in case of a single layer
system. If the tunneling integral $t'$ between layers is vanishing, the
positions of the delocalized states are both at the Landau band center, while
if the random potential is the same for both layers, delocalized states appear
at $E=\pm t'$. The really interesting situation, however, is none of the above,
i.e., uncorrelated disorder and finite tunneling. The position of the
delocalized states is analyzed in detail in \cite{gr97b}, while the critical
behavior has been discussed numerically in \cite{sm96}.

\begin{figure}[htbp]
\vspace{3mm}
  \begin{center}
    \includegraphics[width=0.6\textwidth]{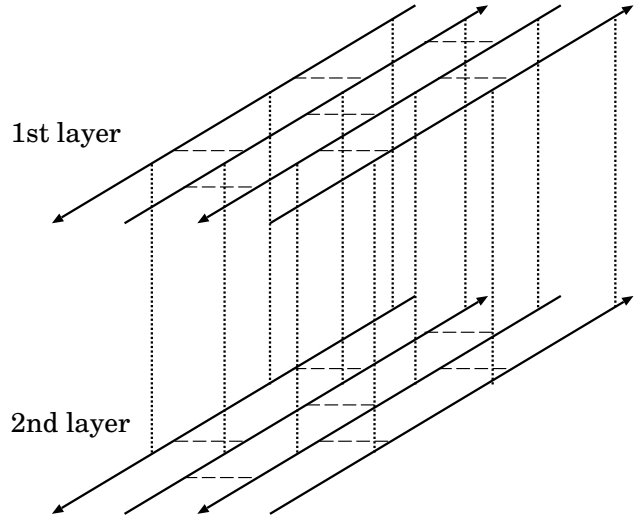}
  \end{center}
  \caption{
    Schematic view of a double layer network model. The dashed lines indicate
    intra-layer saddle points. Dotted lines represent tunneling between the
    layers.  \label{fig:3dnetwork}}
\end{figure}

To incorporate interlayer tunneling in the network model, we introduce
scattering between links in different layers. In Fig.~\ref{fig:3dnetwork}, a
schematic view for double layer system is presented. The dashed lines are the
saddle points that describe intra-layer scattering described by the ${\bf
  S}$ matrix, Eq.~(\ref{eq:decomposition}), and dotted lines represent
tunnelings between the layer. Detailed expressions are presented in Section
\ref{sec:chiralmetal}.  Note that the situation is similar to the class AIII
discussed in Section \ref{sec:chiralnetwork}, but the discrete symmetry
(Eq.~(\ref{eq:classAIII})) is not present in this double layer system.

\begin{figure}[htbp]
\begin{center}
{\includegraphics[width=0.4\linewidth]{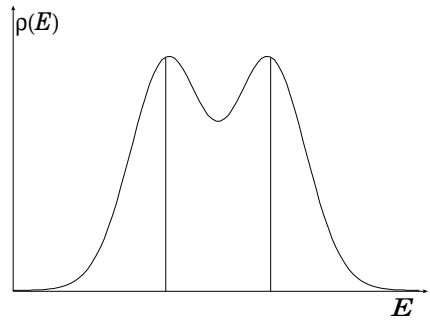}}
\end{center}
\caption{Schematic view of the density of states $\rho(E)$ and
  the positions of delocalized states for a double layer quantum Hall system.
  Delocalized states are indicated by solid lines.}
 \label{fig:2dos}
\end{figure}

The density of states of such a system is sketched in Fig.~\ref{fig:2dos}. The
Landau band splits due to inter-layer tunneling. Whether or not the
localization length exhibits two singularities with the same critical exponent
as a single layer is under debate, but it is likely to be so.

The application of the renormalization group process
(Chap.~\ref{sec:hierarchy}) to two channel network model has been discussed in
detail\cite{mh04,manze05}, where the limitation of the hierarchy model is
clarified.

\subsection{Localization-Delocalization Transition}

With the increase of the number of layers, the positions of energies where
delocalized states appear increase. The number of positions coincides with the
number of layers. In the limit of infinite number of layers, the delocalized
states form a band of energies \cite{oko93}.
\begin{figure}[htbp]
\begin{center}
{\includegraphics[width=0.4\linewidth]{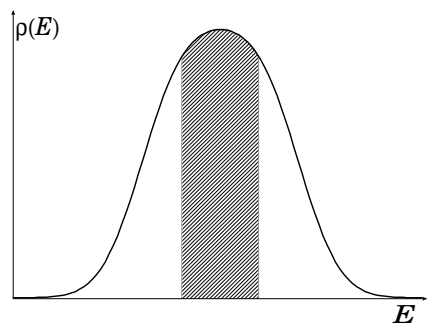}}
\end{center}
\caption{Schematic view of the density of states $\rho(E)$ and
  the positions of delocalized states for a multi-layer system. Delocalized
  states are indicated by the shaded region. \label{fig:layered}}
\end{figure}

As long as the number of layers is finite, the critical behavior of the
localization-delocalization transition is conjectured to be the same as for a
single layer. In the limit of infinite number of layers, the
localization-delocalization transition should become that of the three
dimensional unitary system. For this limit, the localization length exponent
has been estimated numerically in the multi-layer Landau model \cite{oko93}
and the anisotropic tight binding model \cite{wwx99}. In all cases, it has
been found to be consistent with the value of the three dimensional unitary
class \cite{so97,kko98}, $\nu=1.43\pm 0.03$.  This value is not far from 4/3
obtained by mapping the three dimensional layered system onto a spin
Hamiltonian \cite{meir98}.

The numerical estimate of the exponent $\nu$ in the multi-layer network model
via finite size scaling analysis of the MacKinnon-Kramer scaling variable
gives $\nu=1.45\pm 0.2$ \cite{cd95}, which is again consistent with the result
for three dimensional unitary class. The quasi-energy spectral properties of
the multi-layer network model \cite{hk97,metzler98} as well as wave packet
dynamics \cite{hk99} have been studied numerically, which are also consistent
with that for three dimensional unitary class \cite{ok97,bszk96}. These
results suggest that the bulk localization-delocalization properties of three
dimensional layered Chalker-Coddington model (Chalker-Dohmen model) can be
described within the conventional three dimensional unitary class.

\subsection{Chiral Metal}
\label{sec:chiralmetal}

Most of the discussions so far have assumed the periodic boundary condition in
the transverse direction, Eq.~(\ref{eq:pbc}). If we impose the fixed boundary
condition, Eq.~(\ref{eq:fbc}), the electrons travel along the boundary of the
systems and form edge states.

\begin{figure}[htbp]
\vspace{3mm}
  \begin{center}
    \includegraphics[width=0.5\textwidth]{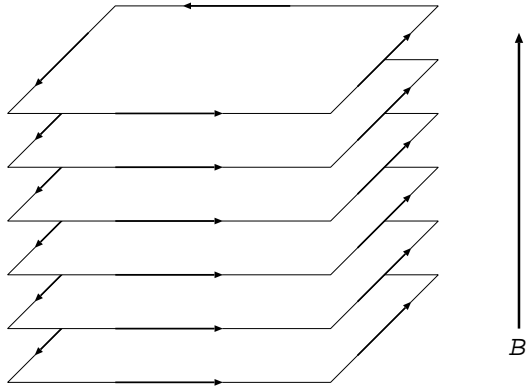}
  \end{center}
  \caption{  \label{fig:chiral}
    The electrons circulate about the two dimensional plane in the case of
    fbc.  They can hop from one layer to another, and form an ^^ ^^ edge
    sheath''.}
\end{figure}

\begin{figure}[htbp]
\vspace{3mm}
  \begin{center}
    \includegraphics[width=0.6\textwidth]{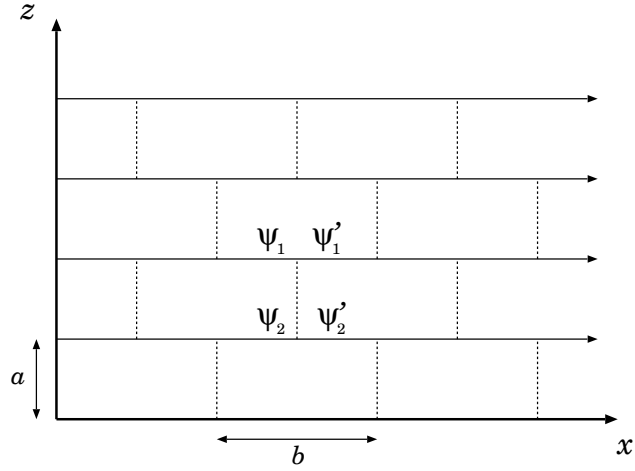}
  \end{center}
  \caption{
    Schematic view of electron motion in the two dimensional sheath.  The
    arrows indicate the edge states circulating around the two dimensional
    plane. It tunnels to that in a different layer through interlayer
    coupling indicated by dotted lines.}
  \label{fig:directed}
\end{figure}

If we consider a multi-layer system where each layer has edge states
circulating around the plane, the edge states are connected due to tunneling.
When the Fermi energy is set between the center of the Landau band, bulk
states are localized and the edge states are decoupled.  In this situation,
they form a sheath \cite{cd95,bf96,k96,cbf97} in which the motion of electrons
is directed (Fig.~\ref{fig:chiral}). The transport properties along the
magnetic field ($z$-direction) is determined by this sheath, since the bulk
states are all localized. We therefore can consider as independent the
electron states in this two dimensional sheath which can be modeled by a two
dimensional directed network model, Fig.~\ref{fig:directed}.

\begin{figure}[htbp]
\vspace{3mm}
  \begin{center}
    \includegraphics[width=0.6\textwidth]{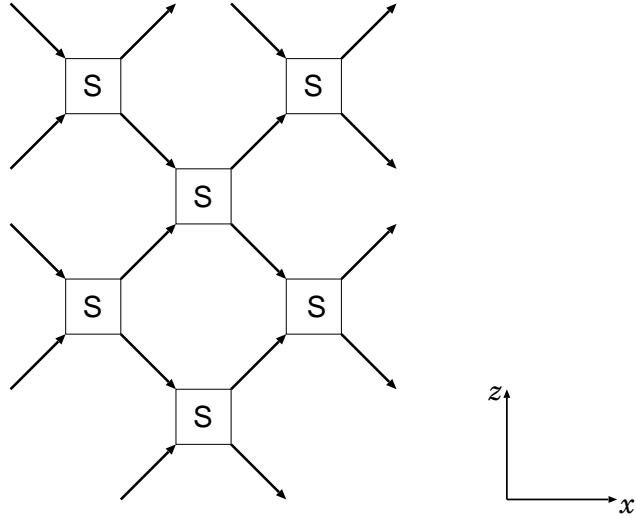}
  \end{center}
  \caption{
    The directed network model. It is similar to Fig.~\ref{fig:network}, but
    different in that the arrows are always towards the right in the
    $x$-direction, and only one type of the ${\bf S}$ matrix appear,
    not ${\bf S}$ and ${\bf S}$' as
    before. \label{fig:network_directed}}
\end{figure}

In this directed network model, at a saddle point, the incoming and outgoing
waves are related via an ${\bf S}$ matrix as
\begin{equation}
 \left(
\begin{array}{c}
\psi_1'\\
\psi_2'
\end{array}
\right)
=
\left(
\begin{array}{cc}
-r' & t' \\
 t' & r'
\end{array}
\right)
\left(
\begin{array}{c}
\psi_1\\
\psi_2
\end{array}
\right)\,.
\end{equation}
In terms of the transfer matrix along $z$-direction, the wave functions are
related via
\begin{equation}
 \left(
\begin{array}{c}
\psi_1\\
\psi_1'
\end{array}
\right)
=
\left(
\begin{array}{cc}
-r'/t' & 1/t' \\
 1/t' & -r'/t'
\end{array}
\right)
\left(
\begin{array}{c}
\psi_2\\
\psi_2'
\end{array}
\right)\,.
\end{equation}
The resulting two dimensional network model is described in
Fig.~\ref{fig:network_directed}.  It is similar to Fig.~\ref{fig:network}, but
in this case the arrows are always rightward and {\bf S}={\bf S}'.

An interesting property of the transport in this edge sheath is that one can
estimate the sheet conductivity along the $z$-axis exactly \cite{cd95,cbf97}.
In the limit of large circumference $C$, the path connecting the bottom edge
to the top cannot circulate along the system. This and the chiral nature of
edge states lead to the fact that the interference between the paths with
different wrapping is absent, i.e., a path does not self-intersect, and we can
estimate the conductivity classically. Let $T=|t'|^2$ be the single layer
transmission probability, and let $T_N$ the transmission probability for $N$
layers ($T_1=T$). Then the following recurrence equation holds
\begin{equation}
 T_{N+1}=T_N \frac{1}{1-R R_N} T\quad,\quad
R=1-T{\ },{\ }R_N=1-T_N
\end{equation}
or
\begin{equation}
 \frac{1}{T_{N+1}}=\frac{1}{T_N}+\frac{1-T}{T}=N\frac{1-T}{T}+\frac{1}{T}.
\end{equation}
This gives
\begin{equation}
 T_N=\frac{1}{N}\frac{T}{1-T(1-1/N)}.
\end{equation}
The conductance along the $z$-axis, $G_z$, is given by
\begin{equation}
 G_z=\frac{e^2}{h}N_c T_N =\frac{e^2}{h} \frac{N_c}{N}\frac{T}{1-T(1-1/N)}
\label{eq:chiralconductance}
\end{equation} 
where $N_c$ is the number of channels. Setting the distance of the saddle
point along the $x$-direction to be $b$ and the layer distance to be $a$, the
expression for the conductivity $\sigma_{zz}$ becomes
\begin{equation}
 \sigma_{zz}=\frac{L}{C}G_z
=\frac{e^2}{h} \frac{a}{b}\frac{T}{1-T(1-1/N)}\,.
\end{equation}
where $L$ is the length of the system along $z$-axis. For a sufficiently large
system, one finds
\begin{equation}
 \sigma_{zz}
=\frac{e^2}{h} \frac{a}{b}\frac{T}{1-T}.
\label{eq:chiralcond} 
\end{equation}
and for perfectly transmitting channels, $T\rightarrow 1$,
$\sigma_{zz}=(e^2/h)L/b$.  Equation~(\ref{eq:chiralconductance}) almost
completely describes the numerical data for $z$-axis conductance \cite{cbf97}.
\begin{figure}[htbp]
\vspace{3mm}
  \begin{center}
    \includegraphics[width=0.8\textwidth]{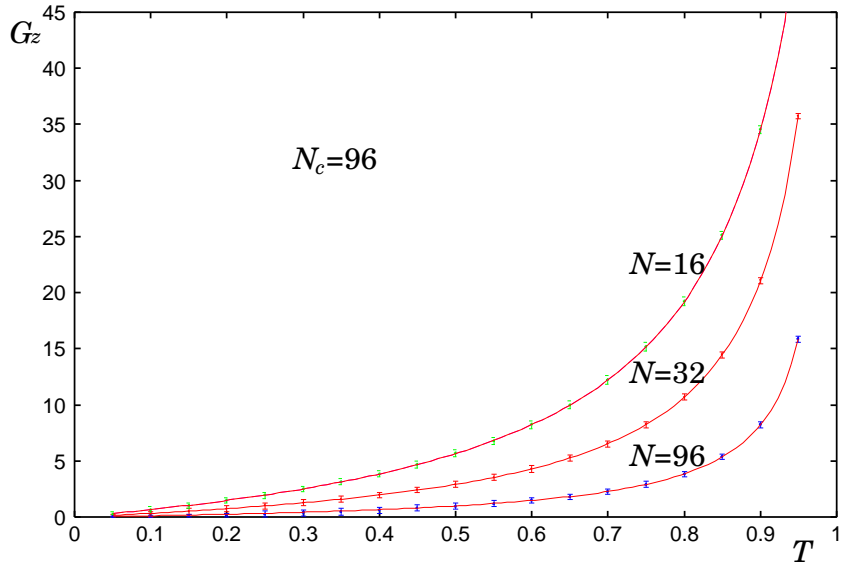}
  \end{center}
  \caption{
    Numerical transfer matrix calculation of conductance $G_z (e^2/h)$ as a
    function of the layer transmission probability $T$.  $N_c=96$ while
    $N=16,32$ and $96$.  Bars around the points indicate the conductance
    fluctuation.  Solid lines correspond to Eq.~(\ref{eq:chiralconductance}).
    \label{fig:chiralcond} }
 \end{figure}
 
 The results indicate than $\sigma_{zz}$ can be much smaller than $e^2/h$, but
 the system is still metallic.  This is in contrast to the conventional two
 dimensional metal-insulator \cite{aks01} or two dimensional
 superconductor-insulator transitions \cite{gm98}, where the transition occurs
 near $e^2/h$. This peculiar property is due to the chiral nature of the
 states in the surface, so that the system is called chiral metal.
 
 The conductivity is related to the localization length along the $z$-axis.
 By parameterizing the transmission eigenvalues, i.e., the eigenvalues of
 $tt^\dagger$, $\tau_n$ via the Lyapunov exponent $\nu_n$ as
 (cf. Eq.~(\ref{eq:taumax}))
\begin{equation}
 \tau_n=\frac{1}{\cosh^2 \left(
{\nu_n N}/{N_c}
\right)}
\end{equation}
we have
\begin{equation}
 \sigma_{zz}=\lim_{L\rightarrow \infty}\lim_{C\rightarrow \infty}
\frac{L}{C}
\frac{e^2}{h}\sum_{n=1}\frac{N_c}{\cosh^2 \left(
{\nu_n N}/{N_c}\right)}
\end{equation}
By noting that $\nu_n\approx n\nu_1$ due to spectral rigidity, the expression
of the conductivity becomes
\begin{eqnarray}
 \sigma_{zz}&=&\lim_{N\rightarrow \infty}\lim_{N_c\rightarrow \infty}
\frac{Na}{N_c b}
\frac{e^2}{h}\sum_{n=1}\frac{N_c}{{\cosh}^2 \left(
{\nu_n N}/{N_c}\right)} \\ \nonumber
&=& \lim_{N\rightarrow \infty}\lim_{N_c\rightarrow \infty}
\frac{Na}{b}
\frac{e^2}{h}
\int_0^1  {\cosh}^{-2}(\nu_1 N x){\ }d x\\ \nonumber
&=&\frac{e^2}{h}\frac{a}{b}\frac{1}{\nu_1}
\end{eqnarray}
Since the smallest Lyapunov exponent is related to the one dimensional
localization length $\xi_z$ via
\begin{equation}
\nu_1\frac{N}{N_c}=\frac{L}{\xi_z}\quad,\quad
\xi_z=\frac{aN_c}{\nu_1},
\end{equation}
in terms of localization length the conductivity is expressed as
\begin{equation}
 \sigma_{zz}=\frac{e^2}{h}\frac{\xi_z}{C}.
\label{eq:chiralcond2}
\end{equation}

>From Eqs.~(\ref{eq:chiralcond}) and (\ref{eq:chiralcond2}) the
MacKinnon-Kramer parameter is obtained \cite{cd95}
\begin{equation}
 \Lambda_c=\frac{\xi_z}{C}
=\frac{a}{b}\frac{T}{1-T}\,.
\end{equation}
It is independent of the size of the system. This means that the wave function
is critical.

With the increase of the system length $L$, the paths of electrons from the
bottom to the top begin to self-intersect. In this case, the transport
properties are conjectured to be metallic.  On the basis of mapping onto a
ferromagnetic super spin chain \cite{bfz97}, this conjecture was
quantitatively discussed in \cite{grs97a,grs97b}, and then numerically
verified \cite{pw98}.

\begin{figure}
  [htbp] \vspace{3mm}
  \begin{center}
    \includegraphics[width=0.6\textwidth]{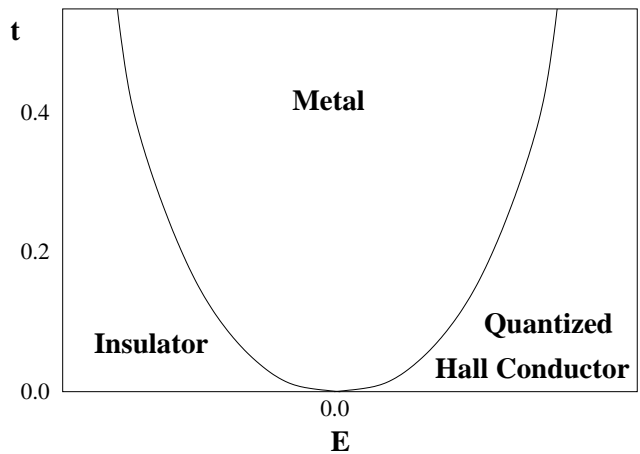}
  \caption{
    Phase Diagram of the layered network model, where $E=0$ denotes the center
    of the lowest Landau band \cite{cd95}.  The width of the metallic region
    for small tunneling amplitude $t$ is expected to increase like $W(T) \sim
    t^{1/\nu_{\rm IQHT}}$.  \label{fig:chiralphase} }
\end{center}
\end{figure}

In Fig. \ref{fig:chiralphase}, the qualitative phase diagram of a layered
quantum Hall system is shown.  The width of the metallic region $W$ is
expected to increase with the tunneling amplitude like $W(t) \sim
t^{1/\nu_{\rm IQHT}}$, where $\nu_{\rm IQHT}$ is the critical exponent of the
two dimensional quantum Hall transition.
This behavior is obtained from the
following argument for the delocalization \cite{cd95}:
When the level spacing in a
localized region $\Delta_c \sim 1/\xi^2$ with localization length $\xi$ is
smaller than the tunneling matrix elements between the localized
wave functions in adjacent layers,
the state becomes delocalized.  This tunneling matrix elements
 is estimated to be $\sim t/\xi$, hence $t\sim 1/\xi$ should hold
 at the mobility edge.
Since in the two dimensional system the
localization length diverges at $E=0$ like $\xi \sim|E|^{-\nu_{\rm IQHT}}$,
one obtains the above dependence of the width $W(t)$.

The chiral metal has been theoretically predicted in 1995 \cite{cd95,bf96}.
Soon, it has been verified experimentally in an organic conductor
\cite{honold97} and in semiconductor quantum well structures
\cite{dtmgg98,zhang99,keki99,druist03,walling04}. Magneto-conductance
\cite{cs99}, conductance fluctuations and the effect of interactions
\cite{bc00,bc01,tcc04} have been discussed in relation to the experiments.

Recently, layered network models have been considered for other symmetry
classes as well. Since, as reviewed in the previous chapter 9, classes C and D
(disordered superconductor with broken time reversal symmetry,
see Tab.~\ref{table:sym3})
exhibit a rich phase diagram already in two dimensions, the corresponding
phase diagram of the layered network can have additional phase boundaries to
the one shown in Fig.~\ref{fig:chiralphase} \cite{kha04,kha04a}. Still, the
width of the metallic region is expected to be for small tunneling amplitudes
$t$,  $W(t) \sim t^{1/\nu}$, where $\nu$ is now the critical exponent of the
corresponding quantum Hall transition at $t=0$.

\section{Conclusion}
\label{sec:conclusions}

We have attempted to describe the development during the past decade of the
random network model originally designed by Chalker and Coddington for the
critical behavior of the quantum Hall phase transition. As the field is
presently in a transient state with new ideas and developments appearing very
rapidly, we cannot hope to have covered all of the different facettes
completely. We can only hope that we have been successful in sketching at
least the most important aspects such that a newcomer to the field can get an
idea about what is going on.

Two distinct and characteristic features of the model have been very important
during the development. The first is of great practical importance.  Similar
to the tight binding Anderson model for localization, the network model is
perfectly suited for numerical studies since the defining scattering operator
is represented by a {\em sparse} matrix. Thus, quantitative numerical studies
of the fundamental quantum critical properties of the model have been the
subject of uncountably many works. These include not only the critical
exponents but also the quantum fluctuations of the multi-fractal wave functions
and the statistics of the conductances.

The second property is perhaps of more fundamental nature. The network model
can be mapped onto a great variety of Hamiltonians ranging from a bipartite
tight binding Hamiltonian --- of which the Dirac model is a limiting case, the
Ising model to an antiferromagnetic chain of superspins. The versatility of
the model invented by Chalker and Coddington for combining results from
different areas seems to be truly unique. Using this, and the already
mentioned remarkable practical flexibility, reliable quantitative information
about quantum phase transitions in very different kinds of disordered systems
can be obtained which include all of the ten presently known universality
classes of disordered quantum systems in two dimensions.

Moreover, the mapping to the antiferromagnetic superspin chain and applying
field theoretical methods has opened novel possibilities of putting the
quantum Hall phase transition in a much wider context. Several new phenomena,
such as the thermal Quantum Hall Effect, the spin Quantum Hall Effect, and the
chiral metal, have been predicted and are waiting for more theoretical and
experimental efforts. Eventually, this also may contribute to explaining the
universality of the quantum Hall phenomena which forms the underlying basis
for the exactness of the quantization of the Hall conductance.

Thus, the model can be considered as paradigmatic. It seems to us that the
development has not yet come to an end. Many of the questions that have been
raised during the development are still waiting for answers
(Tabs.~\ref{table:sym1}~to~\ref{table:sym2}). These are especially the
quantitative investigations of the critical properties at the boundaries of
the various novel quantum phases that are predicted to occur in the models
belonging to the different symmetry classes.

As a major challenge, it remains to be explored how the model can be
generalized to include eventually interactions and correlation effects. There
is evidence that electron-electron interaction is of great importance for
understanding the properties of the two dimensional electron system in the
region of the integer Quantum Hall Effect \cite{sak97,lw96,yacoby}. Also,
generalizations to the regime of the Fractional Quantum Hall Effect should be
desirable \cite{pc98}. Therefore, the generalization of the model towards
including Coulomb interaction will eventually be crucial for getting insight
into the physics behind the quantum critical phenomenon which seems to be of
central importance in modern condensed matter physics.

\section*{Acknowledgments}
We thank Ferdinand Evers, Rochus Klesse and Philipp Cain for providing the
originals of the figures \ref{fig:falphaspectrum} (F. E.),
\ref{fig:multifractalwave}, (R. K.) and \ref{fig:cainphd1} (P. C.),
\ref{fig:cainphd2} (P. C.). We are also grateful to Kai Dittmer for plotting
the figures \ref{fig:bands}.  We want to thank all the colleagues who have
been contributing with useful discussions and enlightening remarks, especially
on those issues for which we do not feel to be the real experts. In
particular, we would like to mention John Chalker, Ferdinand Evers, Bodo
Huckestein, Rochus Klesse, Hidetoshi Nishimori, Shinsei Ryu, Keith Slevin, and
Shan-Wen Tsai. We thank Hajo Leschke for providing mathematical references
concerning the theorem of Oseledec. One of us (T.O.) wishes to thank Kosuke
Yakubo and Tsuneyoshi Nakayama of Hokkaido University, the
Sonderforschungsbereich 508 ^^ ^^ Quantenmaterialien'' and the
Graduiertenkolleg ^^ ^^ Nanostrukturierte Festk\"orper'' of the Universit\"at
Hamburg for kind hospitality during this work. He also wishes to thank
Yoshiyuki Ono and J\`anos Hajdu for guiding him to this interesting field of
research.  S. K. gratefully acknowledges the hospitality of the Max-Planck
Institut f\"ur die Physik Komplexer Systeme in Dresden where several issues
concerning especially the chapters 9 to 11 have been discussed with visitors
especially attending the workshops {\it Quantum Phase Transitions} and {\it
  Quantum Transport and Correlations in Mesoscopic Systems and QHE}.  B. K.
acknowledges gratefully financial support from the Japan Society for the
Promotion of Science, and the hospitality of Sophia University, Tokyo, during
a stay as a JSPS fellow, during which this work has been completed.  Financial
support of the Schwerpunktprogramm ^^ ^^ Quantum Hall Effect'' of the Deutsche
Forschungsgemeinschaft is gratefully acknowledged. The work has been
performed within the TMR and RTN programmes of the European Union (contracts
FMRX-CT96-0044, FMRX CT98-0180, and HPRN-CT2000-00144).

\bibliography{qherevbib}

\end{document}